\documentclass{aa}  

\pdfoutput=1

\usepackage{xcolor}
\usepackage{graphicx}
\usepackage{comment}
\usepackage{txfonts}
\usepackage{multirow} 
\newcommand{\teff}{\mbox{$T_{\mathrm{eff}}$}}

\begin{document}

   \title{A comparison between X-shooter spectra and PHOENIX models across the HR-diagram}


   \author{A. Lan\c{c}on \inst{1}
          \and
          A. Gonneau \inst{2}
          \and
          K. Verro \inst{3}
          \and 
          P. Prugniel \inst{4}
          \and
          A. Arentsen \inst{5,1}
          \and 
          S. C. Trager \inst{3}
          \and 
          R. Peletier \inst{3}
          \and
          Y.-P. Chen \inst{6}
          \and
          P. Coelho \inst{7}
          \and 
          J. Falc\'on-Barroso \inst{8,9}
          \and
          P. Hauschildt \inst{10}
          \and
          T.-O. Husser \inst{11}
          \and 
          R. Jain \inst{1}
          \and 
          M. Lyubenova \inst{12}
          \and
          L. Martins \inst{13}
          \and
          P. S\'anchez Bl\'azquez \inst{14}
          \and
          A. Vazdekis \inst{8,9}
          }
   \institute{Universit\'e de Strasbourg, CNRS, Observatoire astronomique de Strasbourg, UMR 7550, 67000 Strasbourg, France;
              \email{ariane.lancon@astro.unistra.fr}
         \and
         Institute of Astronomy, University of Cambridge, Madingley Road, Cambridge CB3 0HA, United Kingdom
         \and
         Kapteyn Astronomical Institute, University of Groningen, Postbus 800, 9700 AV Groningen, the Netherlands
         \and
         Centre de Recherche Astrophysique de Lyon (CRAL, CNRS, UMR 5574), Universit\'e Lyon 1, Ecole Nationale Sup\'erieure de Lyon, Universit\'e de Lyon
         \and
         Leibniz-Institut f\"ur Astrophysik Potsdam (AIP), An der Sternwarte 16, D-14482 Potsdam, Germany
         \and
          New York University Abu Dhabi, P.O. Box 129188, Abu Dhabi, United Arab Emirates
          \and
          Universidade de S\~ao Paulo, Instituto de Astronomia, Geof\'{\i}sica e Ciencias Atmosf\'ericas, Rua do Mat\'ao 1226, 05508-090 S\~ao Paulo, Brazil
          \and 
          Instituto de Astrof\'isica de Canarias, V\'ia L\'actea s/n, E-38200 La Laguna, Tenerife, Spain
          \and
          Departamento de Astrof\'isica, Universidad de La Laguna, E-38205 La Laguna, Tenerife, Spain
          \and
          Hamburger Sternwarte, University of Hamburg, Gojenbergsweg 112, 21029 Hamburg, Germany
          \and
          Institut f\"ur Astrophysik, Georg-August-Universit\"at G\"ottingen, Friedrich-Hund-Platz 1, 37077 G\"ottingen, Germany
          \and
          European Southern Observatory, Karl-Schwarzschild-Strasse 2, D-85748, Garching, Germany
          \and
          NAT - Universidade Cruzeiro do Sul, Rua Galv\~ao Bueno, 868, S\~ao Paulo, SP, Brazil
          \and 
          Departamento de F\'isica Te\'orica, Universidad Aut\'onoma de Madrid, 28049 Cantoblanco, Spain    
             }

   \date{Received 9 September 2020; accepted 24 November 2020}


\abstract
{}
{The path towards robust near-infrared extensions of stellar population
models involves the confrontation between empirical and synthetic stellar 
spectral libraries across the wavelength ranges of photospheric emission. 
Indeed, the theory of stellar emission enters all population synthesis models, 
even when this is only implicit in the 
association of fundamental stellar parameters with empirical spectral library stars. 
With its near-ultraviolet to near-infrared coverage, the X-shooter Spectral Library (XSL) 
allows us to examine to what extent models succeed in 
reproducing stellar energy distributions (SEDs) and stellar absorption line spectra
simultaneously.}
{ As a first example, this study compares the stellar spectra of XSL with those of the G\"ottingen Spectral Library, which are based
on the PHOENIX synthesis code. 
The comparison is carried out both separately in the UVB, VIS, and NIR arms 
of the X-shooter spectrograph, and jointly across the whole spectrum.
We do not discard the continuum in these comparisons; only reddening is allowed to modify the SEDs of the models.}
{When adopting the stellar parameters published with data release DR2 of XSL, 
we find that the SEDs of the models are consistent with those of the data
at temperatures above 5000\,K. Below 5000\,K,
there are significant discrepancies in the SEDs. When leaving the stellar parameters 
free to adjust, satisfactory representations of the SEDs are obtained down
to about 4000\,K. However, in particular below 5000\,K and in the UVB spectral range, 
strong local residuals associated with intermediate resolution spectral features are 
then seen;  the necessity of a compromise between reproducing the line spectra 
and reproducing the SEDs leads to dispersion between the parameters 
favored by various spectral ranges.
We describe the main trends observed and we point out localized offsets between 
the parameters preferred in this global fit to the SEDs and the parameters in DR2.
These depend in a complex way on position in the Hertzsprung-Russell diagram (HRD).
We estimate the effect of the offsets 
on bolometric corrections as a function of position in the HRD and use this
for a brief discussion of their  impact on the studies of stellar populations.
A review of the literature shows that comparable discrepancies are mentioned
in studies using other theoretical and empirical libraries.}
{}


   \maketitle
%

\section{Introduction}

Stellar population studies based on the integrated spectra or colors 
of galaxies have a long history. Together with the studies
of nearby resolved populations, they have allowed us to reach a 
broad understanding of the stellar contents
of galaxies of different types or environments and of their components. 
While it may not be critical to characterize each galaxy in
detail to study the average evolution of the stellar
mass across cosmic time 
\citep[e.g.,][]{Madau_Dickinson_2014, Moutard_etal16},
age and metallicity estimates
become essential when studying the assembly of galaxy components;
accurate interpretations of colors and spectral features are
necessary to disentangle otherwise degenerate properties such as 
age, metallicity, and extinction, to obtain absolute rather than 
relative values for these quantities, or to constrain the stellar
initial mass function based on integrated light. 
Such detailed investigations require population
synthesis models to rest on robust stellar evolution calculations
and on libraries of stellar spectra with both accurate spectral features
and accurate spectral energy distributions (SEDs).
Indeed, the bolometric corrections associated with the SEDs 
determine the relative contributions of any type
of star at any particular wavelength, and these contributions
are key ingredients of the interpretation of the summed spectral
features of the stellar population under scrutiny.

In this context, it is disturbing that it remains so difficult 
to combine optical and near-infrared (near-IR) studies of stellar populations.
Attempts to match the colors of large samples of galaxies from
the near-ultraviolet (near-UV) to the near-IR have been found to leave significant 
residuals, sometimes leading authors to discard near-IR data
in part of their analysis (Taylor et al. 2011). Star formation histories
derived from optical and near-IR data separately, or from
a given data set with different population synthesis models,
still differ significantly
\citep{Powalka_etal2017, Baldwin_etal2018, 
Dahmer-Hahn_etal18, Dametto_etal19, Riffel_etal19}.
A part of the problem certainly lies in the theoretical modeling
of advanced phases of stellar evolution, but it is also worth 
questioning the stellar spectral libraries. Indeed, if the balance 
between optical and near-infrared
flux, for a given pattern of spectral features, is incorrect for individual 
stars, biases in the population synthesis models are inevitable.

Stellar spectral libraries can either be theoretical or empirical.
In an ideal world, both would be equivalent. But this is a far-off target,
and we are in the middle of a slow converging process that implies 
progress on both sides. Indeed, theoretical libraries cannot be tested
without extensive empirical libraries, and empirical libraries cannot
be used without stellar parameters, which are themselves 
estimated via comparisons with theoretical spectral properties.
For the purpose of checking the overall consistency of 
spectral features and SEDs, continuity across the wavelength
range of photospheric emission, a reasonable spectral resolution
and a good spectrophotometric calibration are keys. 

The above requirements can be fulfilled with instruments such as
the X-shooter spectrograph on the Very Large Telescope of the
European Southern Observatory. The three arms of X-shooter, 
known as UVB, VIS, and NIR, together cover wavelengths
from the near-UV well into the near-IR. The instrument was used to construct
the X-shooter Spectral Library \citep[XSL;][hereafter Paper\,I]{Chen_etal2014, 
Gonneau_XSLDR2}, with the dual purpose of testing synthetic 
spectral libraries and serving as a direct ingredient for stellar population 
modeling. In this article, we provide the results of a first confrontation
of these spectra with an extensive collection of synthetic spectra, namely 
the G\"ottingen Spectral Library \citep[GSL;][]{Husser_etal2013}. Comparisons
with a few other model sets will follow and we encourage model-builders 
to repeat similar studies independently. 
Companion papers will present population synthesis
models based on XSL, and the de-reddened, merged UVB, VIS, and NIR
spectra constructed for these (Verro et al., in preparation).

Our immediate aims are (i) to
investigate to what extent stellar parameters based on optical absorption
line spectra \citep[][hereafter Paper\,II]{Arentsen_PP_19} lead to good matches between
theoretical and empirical energy distributions over the wavelength range
of X-shooter data and (ii) to determine to what extent the models are able
to account for the global energy distribution of the data when the assumption that
the parameters are those of Paper\,II is relaxed. 
Side products of this study are (iii) a
comparison between the parameters obtained from individual spectral
ranges (i.e., the UVB, VIS and NIR arms of X-shooter), (iv) a
validation of the relative flux calibration of the X-shooter data,
and (v) a quantitative estimate of the errors on bolometric
corrections that may result from inconsistencies between SED-based 
and line-based parameter-estimates.

Considering the challenges of the modeling of cool stars, we
expect to find the largest discrepancies between empirical 
and theoretical spectral energy distributions 
at low effective temperatures. For the practical purpose of population 
synthesis, any such discrepancies translate into uncertainties 
in the fundamental parameters associated with the empirical spectra. 
Issues that can be neglected to some extent when  
dealing with star and galaxy spectra in only 
a restricted spectral range, or with only low resolution SEDs, 
become more important to quantify when flux-calibrated spectra are
used across optical to near-infrared wavelengths.

The main features of the GSL collection of synthetic spectra
are recalled in Sect.\,\ref{sec:models} and those of the XSL
data in Sect.\,\ref{sec:data}. We then specify the two methods
used to confront the empirical and theoretical data sets with each
other in Sect.\,\ref{sec:method}, and describe the results as 
a function of the position in the HR diagram in Sect.\,\ref{sec:results}.
For reasons that will become clear later, we focus on the
most significant trends and on the temperature regime between
4000 and 5000\,K. Section\,\ref{sec:discussion} provides 
a comparison with other relatively recent confrontations between
empirical and theoretical stellar spectra, a review of the
limitations of GSL (many of which are common to a number
of synthetic spectral libraries), and a brief discussion of 
the potential impacts of the trends we found, via
bolometric corrections. The appendices provide a
selection of additional figures; XSL-GSL comparison-figures for
all the XSL spectra can be requested from the first author.

\section{The models}
\label{sec:models}
The synthetic spectra used in this article are 
taken from G\"ottingen Spectral Library \citep[GSL,][]{Husser_etal2013} in its
version v2\footnote{\small http://phoenix.astro.physik.uni-goettingen.de/}. 
The comparison with the X-shooter Spectral Library
was one of the motivations for the computation of that grid,
which provides a dense coverage of the HR-diagram at several
compositions and covers the near-UV to near-IR range
of X-shooter spectra at an adequate spectral resolution.
This series of synthetic spectra provides a good representation of
broad-band color-color relations of stars in the Milky Way \citep[figures 2 and 22
of][]{Powalka_etal2016}; it also provide good
matches to optical absorption line spectra at medium resolution
\citep[e.g.,][]{Roth_etal18, Husser_etal2016}.

The underlying model atmospheres
are obtained with version v16 of the PHOENIX code 
\citep{Hauschildt_etal99}, in spherical symmetry.
This geometry is important in the low-gravity regime,
where an atmosphere's extension is not negligible compared
to the stellar radius \citep{Scholz85, Plez_etal92_spherical, 
HeiterEriksson_2006}.
The models and synthetic spectra adopt the equation of
 state and opacities referred to as PHOENIX-ACES, the calculation
of which includes measured and theoretical lines lists by R.Kurucz as 
available in 2009 ($>$88 million metal lines and $>$1 billion molecular 
lines). PHOENIX models with these inputs were found to have a structure similar
to that of MARCS models of the same period 
\citep[see Sect.\,7 of][]{Gustafsson_etal08}.
Like many other large grids in the
literature, the model spectra assume Local Thermal Equilibrium (LTE), 
which is a recognized limitation \citep{Short_Hauschildt_2003,LanzHubeny07}.

The synthetic spectra are available on a grid of parameter space
of which the four axes are the effective temperature (\teff), 
the surface gravity log($g$) ($g$ in cm.s$^{-2}$),
the metallicity [Fe/H] \citep[with respect to the solar abundances of][]{Asplund09},
and the $\alpha$-element enhancement [$\alpha$/Fe]. 
The boundaries and sampling steps are summarized in Table~\ref{tab:modparams}; 
we note that a few [$\alpha$/Fe] ratios higher than listed are available but were
not considered here. 
$\alpha$-enhancements affect the abundances of O, Ne, Mg, Si, S, Ar, Ca and Ti. 
They are introduced at a given [Fe/H], hence 
they affect the overall metallicity $Z$ of the models.

\begin{table}
\caption[]{Parameter coverage of the GSL grid of synthetic spectra.}
\label{tab:modparams}
\begin{tabular}{lccl}
\hline \hline
Variable & Range & Step size & \\ \hline
\teff [K] & 2300 .. 7000 & 100 &  \\
          & 7000 .. 12000 & 200 & \\
          & 12000\,..\,15000 & 500 &  \\
log($g$) & -0.5 .. +5 & 0.5 & no $\leqslant \! 0$ at high \teff \\
$\mathrm{[Fe/H]}$  & -4.0 .. -2.0 & 1.0 &  \\
        & -2.0 .. +1.0 & 0.5 &  \\
$\mathrm{[\alpha/Fe]}$ & -0.2 .. +0.6 & 0.2 & full grid only at\\
             &                  &   & [$\alpha$/Fe]=0 and +0.6\\                 
\hline
\end{tabular} 
\end{table}

Spherical geometry introduces
an extra degree of freedom compared to plane-parallel models. The adopted 
model masses are expressed as a simple function of log($g$) and T$_{\mathrm{eff}}$ 
\citep[figure~1 of][]{Husser_etal2013}; they range from about 0.5 to about 
5\,M$_{\odot}$ along the main sequence, and reach values 
$>10$\,M$_{\odot}$ at high temperatures and low gravities. Then 
surface gravity $g$ determines the radius.
%
The stellar structures account for convection 
using the mixing length theory of turbulent transport, 
with a mixing length parameter in the range [1, 3.5] that depends on 
T$_{\mathrm{eff}}$ and log($g$).

Between 300 and 2500\,nm, the synthetic spectra are computed with a sampling
step $\Delta \lambda$ such that $\lambda/\Delta \lambda \simeq 500\,000$.
That such a high resolution is necessary for a good representation of 
empirical spectra of cool stars even at $R=3000$ has been known for some time
(e.g., figure~1 of Lan\c{c}on et al. 2007).
The micro-turbulent velocities adopted to determine individual line profiles
in the calculations are taken proportional to the average 
turbulent velocities of convection zones. They are typically around
1\,km/s for $T_{\mathrm{eff}}<9000$\,K and reach a maximum of about
3.5\,km/s between 5000 and 6000\,K at the lowest gravities 
\citep[figure~3 of][]{Husser_etal2013}. They drop to essentially zero above 10\,000\,K.
The wavelengths of the synthetic spectra were converted from vacuum to air
for comparison with the empirical spectra described below.

\section{The X-shooter spectra}
\label{sec:data}

\subsection{The sample}
\label{sec:sample}

The X-shooter spectra used in this article are those of Data Release 2 
of XSL (XSL-DR2, Paper\,II). 
Once carbon stars and incomplete spectra are excluded,
the release contains data for 730 observations of 598 stars. 
Giants (on the red giant branch, on the asymptotic giant branch 
or in the red supergiant phase) represent
$\sim$\,55\,\% of these observations, dwarfs $\sim$\,35\,\%, and the remainder of the
spectra belong to subgiants, to horizontal branch or blue loop stars, 
or in a few cases to 
objects in other short-lived transition phases.

\begin{figure}
\begin{center}
\includegraphics[clip=,width=0.45\textwidth]{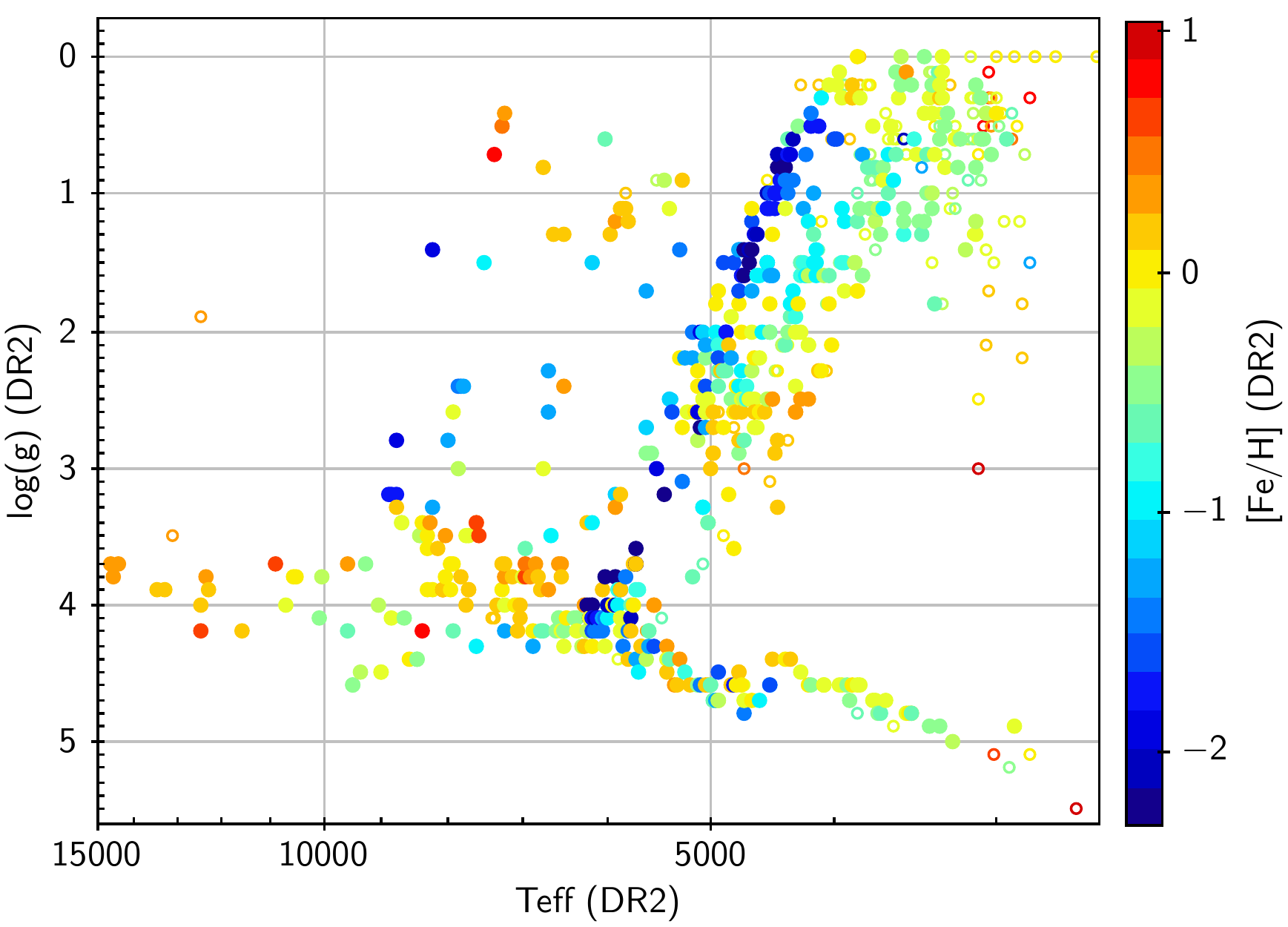}
\end{center}
\caption[]{Parameters of the XSL dataset, as evaluated by 
\citet{Arentsen_PP_19}. Open
symbols correspond to spectra for which at least one of the three X-shooter 
arms was {\em not}\ corrected for slit-losses.
}
\label{fig:HRdiag0_byFeH0}
\end{figure}

%
Fundamental stellar parameters have been estimated for 713 of 
these spectra in Paper\,II. The effective temperatures range from
$\sim 3500$\,K to $\sim 38\,000$\,K and the metallicities [Fe/H] from
$\sim -2.7$ to supersolar (Fig.\,\ref{fig:HRdiag0_byFeH0}). 
The 17 stars for which Arentsen et al. did
not derive parameters are two late M dwarfs and 
some of the coolest Mira-type variables in XSL. 

\citet{Gonneau_XSLDR2} flagged a number of spectra as peculiar. We exclude three
of these from the sample (X0214, X0248, X0424) because their energy distributions
are obviously incompatible with the models used here : X0214 and X0248 correspond
to an Ae/Be star with a near-infrared excess due to circumstellar material, and X0424 
to the combined emission of two stars. We further exclude the 6 hottest spectra of the remaining collection, whose estimated effective temperatures ($>21000$\,K) are 
well beyond those available in the model grid.


%
%
%

In the end, we use a total of 704 spectra with parameter estimates from Paper\,II and we 
extend the sample to 721 spectra (704+17) when
the parameter estimates are not critical.
We have not rejected long-period variable stars (LPVs) from our sample and figures,
but they are clearly special cases for which we cannot expect good 
representations with synthetic spectra based on static atmosphere 
models \citep[see ][for preliminary results of a dedicated study]{LanconIAU18}. 
The results highlighted in this paper focus on regions of stellar parameter
space where these stars have little or no impact.

\subsection{Properties of the spectra}

The properties most relevant to the analysis in this article are
the wavelength coverage of the spectra, the spectral resolution, the 
flux calibration, and the signal-to-noise ratio.
The XSL-DR2 spectra are published in the rest-frame of each star,
with wavelengths in air,
and the exact wavelength coverage depends on the star observed. 
Before correction for radial velocity, 
the spectral ranges sampled by the UVB, VIS, and NIR arms of the X-shooter instrument
are, respectively, 300-556\,nm, 533-1020\,nm, and 994-2480\,nm. In practice, our
analysis systematically excludes rest-frame wavelengths shorter than 400\,nm for
cool stars (T$_{\rm eff}<4200$\,K) and above 2.3\,$\mu$m for all
stars; other masked regions are mentioned in Sect.\,\ref{sec:method}. 

As described in Paper\,I, the resolving power 
$R=\lambda/\Delta \lambda$ is approximately constant
across each arm, with $R\simeq 9800$, $11\,600$, and $8000$, respectively in the UVB, 
VIS,  and NIR arms. Because this study focuses on spectral
energy distributions and major spectral features, we choose to perform 
all comparisons at a constant, reduced resolving power of either 
$R=500$ or $R=3000$. 
This smoothing also has the advantage of reducing any sensitivity of the 
analysis to local residual wavelength calibration errors or
variations in $R$.
In practice, we first project the spectra onto an (oversampled) logarithmic
wavelength scale and then convolve each arm with an adequate gaussian broadening
function. To reduce computation times, we project the empirical and 
synthetic spectra, after smoothing, 
to a regular log-wavelength scale with $\sim$\,3 pixels per resolved element. 

\medskip
 
The XSL-DR2 data are available in two subsets, depending on whether or
not a correction for slit-losses was applied to the fluxes. Among the 721
observations we consider, 597 are corrected for slit-losses
in the UVB arm of X-shooter, 660 in the VIS arm, and 652 in the NIR arm.
We focus on those spectra in our analysis.
In Paper\,I synthetic broad band photometry was used to compare
colors of the XSL spectra with those of the MILES and IRTF spectral
libraries, and with colors in the 2MASS point-source catalog.
For the colors measured, the distributions of the color-differences between
the slit-loss corrected XSL spectra and other data typically 
have standard deviations of $\sim$\,0.03\,mag
(Table~\ref{tab:DR2photom}), with a varying fraction of outliers
(typically 2 to 7 percent 4-$\sigma$ outliers). 
The systematic offsets compared to the MILES and IRTF colors are 
smaller than 1\,\%. The color-offsets with respect to 2MASS, also 
discussed in Paper\,I, are between 3 and 5.6 percent, 
but these values depend sensitively on the selected subsample, on the modeled 
contribution of telluric absorption to the filter transmission functions, 
and also on the adopted Vega spectrum; they are compatible with
2MASS photometric errors and systematics in the near-infrared 
synthetic photometry.

\begin{table}
\caption[]{Standard deviations of the color-differences between
XSL and external datasets \citep[from][]{Gonneau_XSLDR2}.}
\label{tab:DR2photom}
\small
\begin{tabular}{lccl}
\hline \hline
External & color  &  std. dev. & Comments \\ 
reference &  &  $\sigma$ [mag]   & \\ 
\hline
MILES & (420$-$490) & 0.03 & 0.056 incl. 7 outliers \\ 
MILES & (580$-$670) & 0.02 & 0.032 incl. 3 outliers \\
IRTF \rule[0pt]{0pt}{12pt} &  $(J\,-\,H)$ & 0.023 & 0.033 incl. 6 outliers \\
IRTF & $(H\,-\,K_s)$ & 0.019 & 0.029 incl. 4 outliers \\
2MASS \rule[0pt]{0pt}{12pt} & $(J\,-\,H)$ & 0.032 & 0.050 incl. 2MASS errors \\
2MASS & $(H\,-\,K_s)$ & 0.038 & 0.056 incl. 2MASS errors \\
\hline \hline
\end{tabular} \\
\tablefoot{The synthetic colors in the first two lines are measured
in 600 or 800\,nm-wide rectangular filters centered on the wavelengths 
indicated in the color-name, in nanometers. The 2MASS standard 
deviations listed each include six 4-$\sigma$ outliers. The outliers
are not the same for all colors.}
\normalsize
\end{table}

\medskip

The signal-to-noise ratio (S/N) of the spectra are diverse, with typical values
around 90 per original spectral pixel and above 100 after smoothing. 
We note that the computation of the initial noise spectra by the standard
X-shooter pipeline, starting from the original 2-dimensional echelle images, 
does not keep track of correlations introduced by the 
rectification and wavelength calibration. This impacts the noise-propagation 
through subsequent transformations such as our smoothing procedure\footnote{Some 
of the issues with X-shooter noise spectra
are described for the UVB and VIS data in \citet{Schoenebeck_etal14};
the (partial) corrections suggested there are not implemented in the XSL pipeline, whose
structure depends on the architecture of the original X-shooter instrument pipeline 
(distributed by the European Southern Observatory) until after the flatfielding and rectification
of the spectra.}. 
Also, we are aware of a number of cases in which the absolute level of 
the noise spectrum in one or the other X-shooter arm is dubious for
unknown reasons (this is most obvious when the noise is overestimated). 
We must therefore interpret any result that depends on the absolute noise levels 
with care. However, the noise spectra do contain the signature of the blaze 
function of the spectrograph, they show the lower S/N ratio at both ends of
the arm spectra, and they include a representation of the errors induced by the
correction for telluric absorption (which is important in the near-infrared). 
Hence the inverse-variance based on the noise spectra provides useful, though
not statistically optimal, weights for the pixel-by-pixel comparison of 
a given XSL spectrum with several synthetic  spectra.

\section{Data-model comparison methods}
\label{sec:method}

Our data-model comparisons aim at evaluating in which regions
of stellar parameter space the theoretical spectra are
able to reproduce the (low or intermediate resolution)
spectral energy distributions across the
wavelength range of X-shooter spectra. We do this first by
adopting the stellar parameters of XSL-DR2 from Paper\,II
as they are; then we repeat the exercise without the strong assumption
that the parameters are known {\em a priori}. 

An essential point to remember is
that only optical absorption line
spectra were exploited in Paper\,II: while applying the full-spectrum fitting code ULySS
\citep{Koleva_etal09}, the authors discarded any information potentially carried by the SED,
which was absorbed in a high order multiplicative polynomial of which
the properties were not exploited. Discarding continuum information is a common
procedure in stellar parameter estimation work and also in full-spectrum fitting
methods for the analysis of the integrated light of stellar populations. 
Here, on the contrary, the low resolution energy
distributions of the XSL spectra is an essential  
piece of information we examine. Hence our data-model comparisons
do not allow for any polynomial corrections. Only extinction
is allowed to modify energy distributions.

\subsection{Data-model comparison with the parameters of XSL-DR2}
\label{sec:method_forced}

The first data-model comparison is performed for the XSL spectra 
for which the three parameters $T_{\rm eff}$, log($g$), and [Fe/H] were 
estimated in Paper\,II, and we adopt these values.
As explained in detail in that article, these  so-called ``DR2-parameters" are 
tied to the estimated fundamental parameters of stars in previous empirical,
optical spectral libraries 
\citep[namely MILES and initially ELODIE;][]{Prugniel_Soubiran_2001,
Sanchez-Blazquez_etal06}, which are themselves 
the result of a detailed literature compilation followed by extensive work to
homogenize values obtained by different methods and authors
\citep{Cenarro_etal07, Prugniel_etal11, Soubiran_etal16_PASTEL, Sharma_etal16}.

Estimates of the [$\alpha$/Fe] ratios are not systematically
available for the XSL stars and literature values are dispersed. 
After a first exploration using four values of [$\alpha$/Fe],
we chose to base our analysis
on synthetic spectra at [$\alpha$/Fe]=0 and $+$0.4, values
that roughly represent the bulk of observed abundances in the
Milky Way and Magellanic Clouds 
\citep{Gonzalez_etal2011,   
Nidever_etal2020}. 
The conclusions based on these two chemistries are similar, and 
the discussion further on explains why a
more detailed approach would not be justified in this paper.

The discrepancies between the empirical and theoretical energy distributions
are measured at low resolution ($R=500$), using a normalized r.m.s difference to
quantify distance:
\begin{equation}
\qquad D = \sqrt{ \frac{1}{N_{\lambda}}\ 
                 \sum_{\lambda} \ \left( 
                 \frac{F^{GSL}_{\lambda} - F^{XSL}_{\lambda} }{
                         F^{XSL}_{\lambda} } \right)^2 }.
 \label{eq:D}
\end{equation}
In this definition, $N_{\lambda}$ is the number of wavelengths accounted for in the sum,
taking into account that a mask is used to exclude troublesome wavelength ranges
(for instance regions with residuals from telluric absorption, ends of X-shooter arms 
that are affected by large flux calibration errors, and sometimes regions affected by line emission). 
$F^{XSL}$ is the smoothed XSL spectrum and $F^{GSL}$ is the smoothed, optimally reddened and optimally rescaled GSL spectrum under consideration. The extinction law
of \citet{Cardelli89} is used with a standard ratio of total to differential extinction,  $R_V=3.1$. For each GSL model, reddening is optimized by minimizing $D$ with
respect to the extinction parameter $A_V$. We allow only values of $A_V\geqslant -0.1$,
the negative values being tolerated to limit edge effects that could be due to 
flux calibration errors in the XSL data. For each $A_V$, the optimal rescaling
is re-evaluated.

In practice, the procedure is implemented in four variants that
exploit four wavelength ranges. Each of the first three use only one of the X-shooter
arms (UVB, VIS or NIR) to determine the best $A_V$, while the fourth exploits all
available wavelengths jointly (it will be labeled ALL). In the fourth case, we allow the optimization 
procedure to select the rescaling factors for the UVB, VIS, and NIR ranges 
independently of each other. 
If the collection of model spectra contains a perfect match to the data, reddening
included, then all the four variants of the adjustment procedure
should point to the same A$_V$.

For very cool or highly reddened stars, the flux drops to zero at the
blue end of the UVB arm and flux calibration errors may even produce negative fluxes. 
To avoid that $D$ becomes exceedingly sensitive to regions of very low flux,
we weight down the wavelengths ranges where the flux drops below 
15\,\% of the maximum flux of the arm.

The values of $D$ can be interpreted as typical fractional differences between
the best reddened model and the data. 
Flux calibration errors with 1-$\sigma$ amplitudes equal to those listed 
in column 3 of Table \ref{tab:DR2photom}, if they are assumed 
to correspond to simple errors in the 
general slope of the spectra of one X-shooter arm, translate into 
changes in $D$ of about $0.03$ for each arm (with a slight dependence 
on the masked wavelength ranges used when computing $D$).
Poor matches will be characterized by values of $D$ larger
than $\sim 0.06$ as well as by inconsistent A$_V$ between arms.

For the comparisons described in this section, we implemented an
interpolation in the GSL grid before selecting the model nearest the
parameters of a given XSL spectrum. The aim of the interpolation is 
to ensure that the discrepancy $D$\ between neighboring {\em models}\ 
in the theoretical grid remains smaller than the $\sim 0.03$ that can result 
from flux calibration errors in the observations.
Interpolation does not change the main trends discussed in this paper which,
when we choose to mention them, are larger than a model grid step on average; but
it reduces the dispersion due to finite sampling of parameter space and hence
raises the level of significance of the trends.
Our interpolation reduced the step in $T_{\rm eff}$ by 
a factor of 2 (typically from 100\,K to 50\,K), the step in log($g$) by 
a factor of 2 (from 0.5 to 0.25), and the step in [Fe/H] by a factor of 5
(from 0.5 to 0.1, or at low metallicity from 1 to 0.2). We used spline interpolation,
in a local volume around the target parameters typically including 27 original
grid points at a given [$\alpha$/Fe]. We did not interpolate in [$\alpha$/Fe].

\subsection{The search for best-fit models}
\label{sec:method_bestfit}

In the search for a best-fit model, \teff, log($g$), [Fe/H], and [$\alpha$/Fe] are 
free parameters and the DR2-values of Paper\,II are used
only to initiate the optimization algorithm. Where useful, we include
the 17 spectra with no associated parameters that were 
identified in Sect.\,\ref{sec:sample}, and we assign them initial-guess effective 
temperatures between 2300\,K and 3000\,K, a surface
gravity log($g$)=0 or 4.5, and a metallicity [Fe/H]=0 (based on spectral type).

At any given [$\alpha$/Fe], a subset of synthetic spectra 
centered on the DR2-parameters is 
selected. For each of these model spectra separately, a best-match 
extinction parameter $A_V$ is obtained, using the standard extinction law of 
\citet{Cardelli89} with $R_V=3.1$.  $A_V < -0.1$ is prohibited.
The fit-quality is measured at resolution $R=500$
or $R=3000$ with a classical inverse-variance weighted $\chi^2$-difference 
between the reddened
model and the XSL-spectrum, or alternatively with the quantity $D$ of the 
previous section (Eq.\,\ref{eq:D}). If the best-fitting reddened model lies 
on the edge of the subset of models first selected,
the subset is extended in the adequate direction until this is not the case (except 
on edges of the available model grid). 

As in the previous section, the fitting procedure is implemented in four variants that
exploit four wavelength ranges (UVB, VIS, NIR, and ALL). 
If the collection of model spectra contains a perfect match to the data, reddening
included, then all these variants point to the same set of parameters.

When the figure of merit for each arm is the inverse-variance weighted $\chi^2$,
the fourth variant of the fitting procedure, which we refer to as the ``global fit", uses
a combined-$\chi^2$ based on the values obtained
separately in each arm. There are numerous ways to proceed and we have implemented
several, but since we restrict our discussion to major trends any choice 
that ensures a relatively even balance between the three arms 
leads to similar conclusions. The spectra of the three arms each contain 
comparable numbers of resolved spectral elements 
($\sim 10^4$) and we choose to give all three similar importance
by calculating
\begin{equation}
  \chi^2_{\mathrm{global}} = \frac{1}{3}
   \left(\  \frac{\chi^2_{\mathrm{UVB}}}{\min\, (\chi^2_{\mathrm{UVB}}) } \ + \
       \frac{\chi^2_{\mathrm{VIS}}}{\min\, (\chi^2_{\mathrm{VIS}}) } \ + \
       \frac{\chi^2_{\mathrm{NIR}}}{\min\, (\chi^2_{\mathrm{NIR}}) } \
   \right),
 \label{eq:chi2}
\end{equation}
where the minima are taken over the subset of all reddened models to which a given XSL
spectrum was compared. To this latter purpose, the $\chi^2$-values per arm are 
initially saved for each synthetic spectrum in the subset
{\em and} for a broad range of values $A_V$ including the values
found best in each arm; the $A_V$ of the best global fit, like the other
parameters, can differ from each of the by-arm values.
An advantage of this choice of $\chi^2_{\mathrm{global}}$
is its low sensitivity to errors in the scaling of the noise spectrum that may
affect one or the other arm; a caveat is the excessive
weight that might be attributed to a spectral arm with poorer signal-to-noise ratio.

The interpolation scheme of the previous section is also
implemented here. We recall however that our point
is not to re-evaluate stellar parameters {\em per se}, but to determine where 
in the HR-diagram good matches are possible or otherwise what the main 
trends in the discrepancies are. These conclusions are already 
apparent without interpolating. 
The procedure above produces a number of figures for eye inspection, of which
examples are provided in Appendix \ref{app:fits_with_bestfit}.

\section{Results}
\label{sec:results}

\subsection{The SEDs for the parameters of DR2}
\label{sec:results_forced}

\begin{figure*}
\begin{center}
\includegraphics[clip=,width=0.47\textwidth]{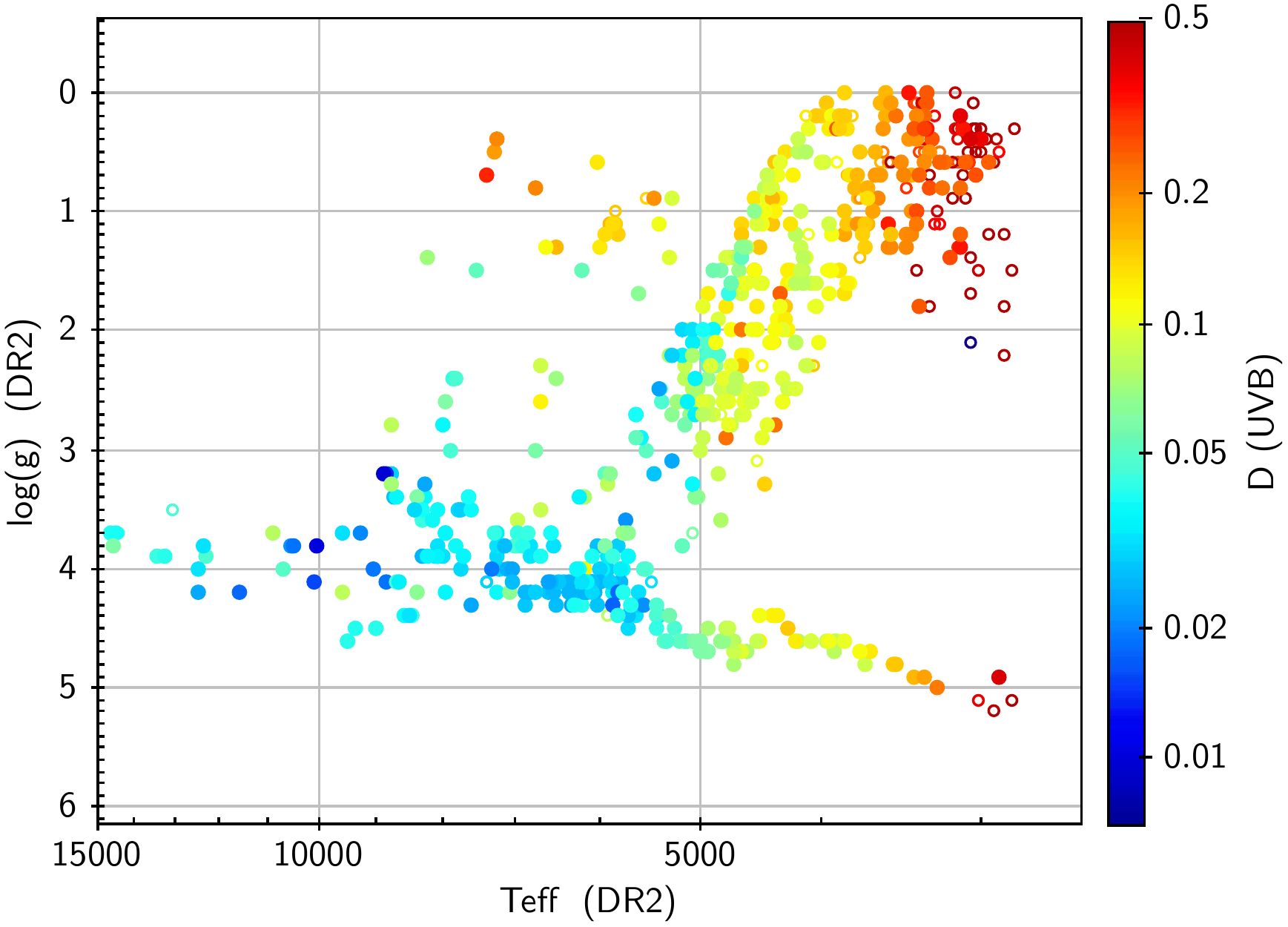}\hfill
\includegraphics[clip=,width=0.47\textwidth]{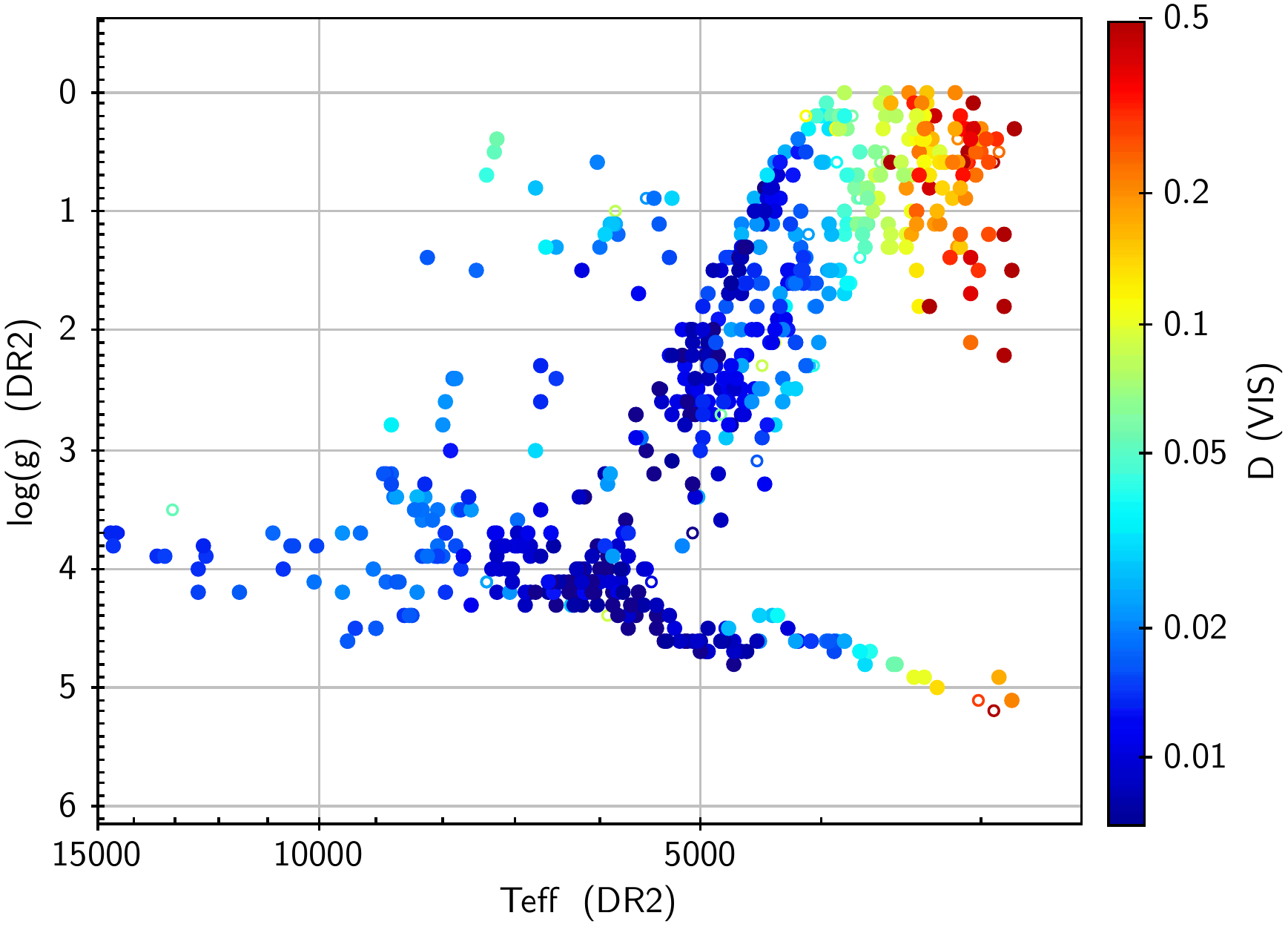}\\
\vspace{-12pt}
\includegraphics[clip=,width=0.47\textwidth]{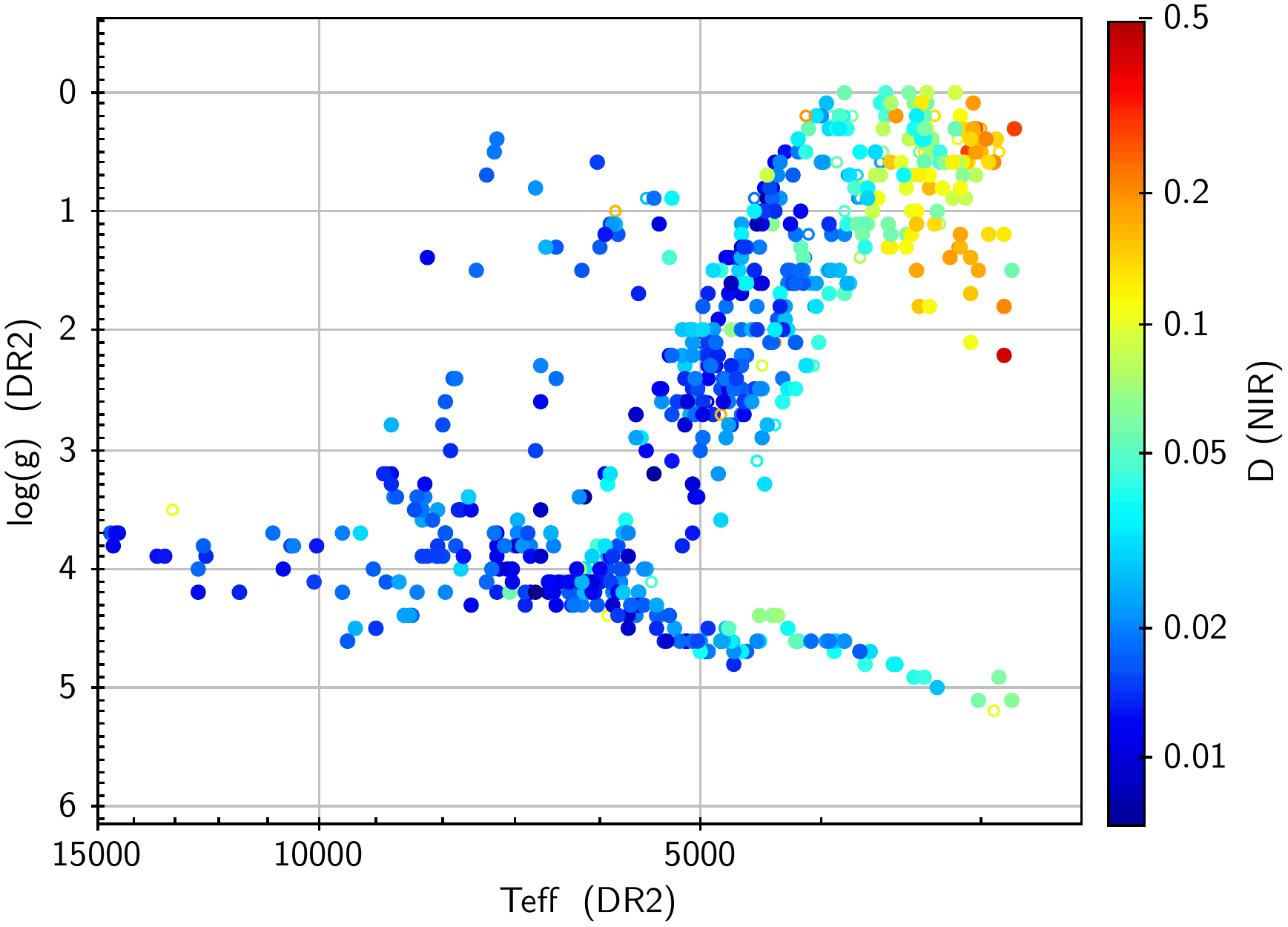}\hfill
\includegraphics[clip=,width=0.47\textwidth]{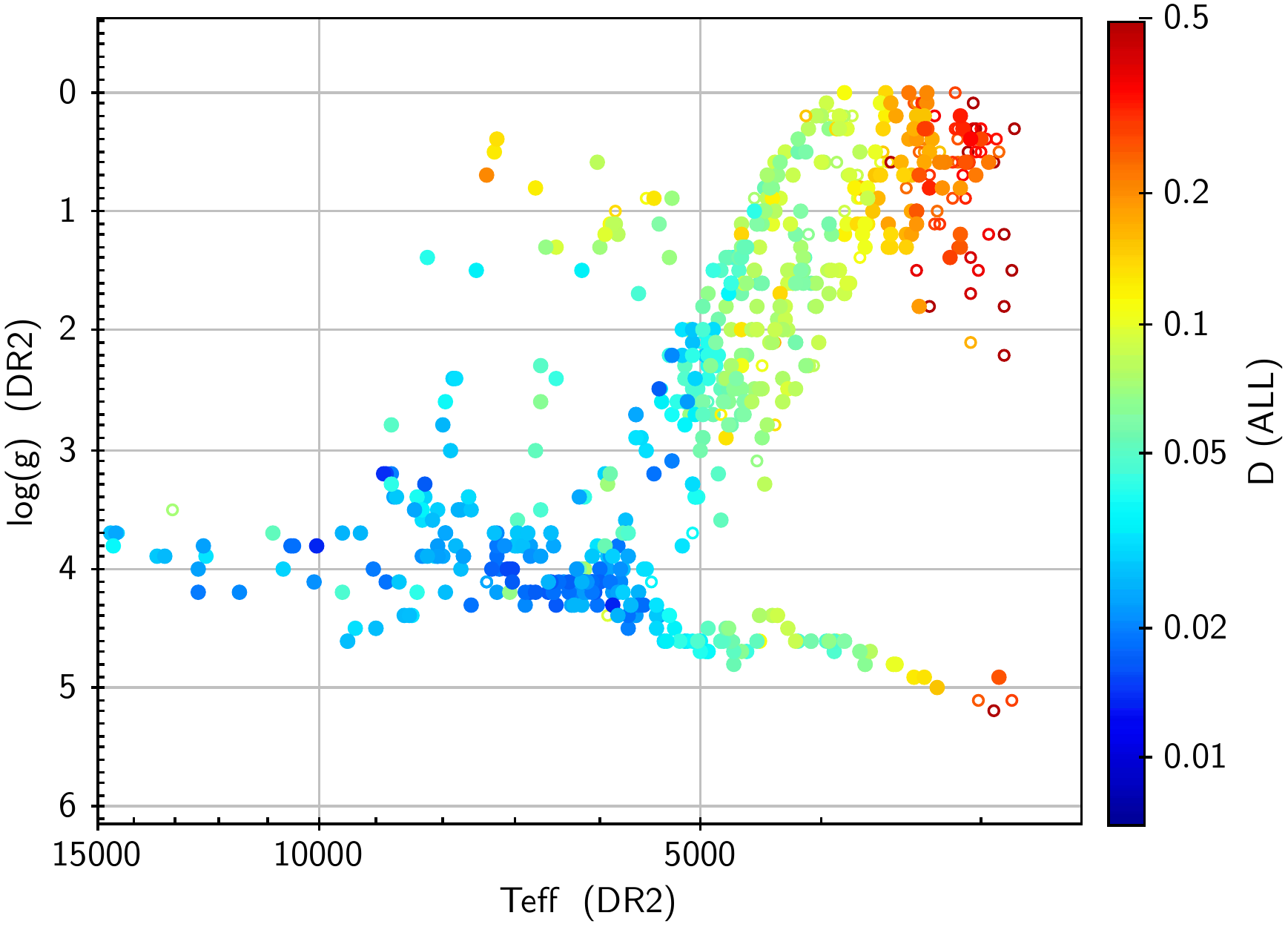}
\end{center}
\caption[]{Comparison between the low-resolution SEDs ($R=500$) 
of XSL spectra and the nearest GSL spectrum, when adopting 
the fundamental parameters of \citet{Arentsen_PP_19}. 
Only extinction is allowed to modify the energy distribution of
the theoretical spectra. The color-coding shows $D$ (Eq.\,\ref{eq:D}),
as measured in the UVB, VIS, and NIR arms and 
across the whole spectrum. Open symbols correspond to spectra for
which no slit-loss corrections were applied. 1-$\sigma$ flux calibration
errors correspond to $D\simeq 0.03$ in each of
the UVB, VIS, and NIR arms.}
\label{fig:forcedfits}
\end{figure*}

The differences between empirical and theoretical energy distributions
as measured by $D$ (Eq.\,\ref{eq:D}) display systematic trends as a 
function of effective temperature, surface gravity, and metallicity, and we 
have therefore chosen to present them in HR-diagrams directly comparable 
to Fig.\,\ref{fig:HRdiag0_byFeH0}. The maps for the four wavelength 
ranges we have considered (UVB, VIS, NIR, and ALL) are shown
in Fig.\,\ref{fig:forcedfits}. For comparison, the values of $D$ 
resulting solely from the finite sampling of parameter space
in the theoretical spectral library can be examined in Appendix
\ref{app:TOHvsTOH}. A selection of direct comparisons of the 
XSL and GSL low resolution spectra, for the parameters of
Paper\,II as adopted in this section, can be found in Sect.\,\ref{app:sec:fits_with_DR2params} of 
the Appendix.


Strong systematic trends dominate the aspect of the four panels of
Fig.\,\ref{fig:forcedfits}, with comparatively
small local scatter. This provides evidence that we are indeed measuring 
differences between empirical and theoretical stellar energy distributions rather
than any dominant random errors in the XSL SEDs that previous examination of 
these data would have missed, or random errors due to the finite sampling
of the model grid. 
\smallskip

In the VIS and NIR arms, the theoretical SEDs are in excellent agreement 
with the empirical SEDs over vast regions of the HR diagram, corresponding
roughly to $T_{\rm eff} > 4000\,$K. This is a remarkable achievement
of the GSL collection. The consistently low values of $D$ over this part
of parameter space ($D\leqslant 0.03$) indicate that the photometric errors recalled in
Table~\ref{tab:DR2photom} are not underestimated.

In the UVB arm, the model SEDs for the upper main sequence (\teff $>5000$\,K) also agree with the data, 
albeit with somewhat larger discrepancies $D$ than in the VIS and NIR. With respect to
the standard deviation $\sigma$ associated with flux calibration errors, the discrepancies 
are distributed between essentially 0 and about 2-$\sigma$ ($D\simeq 0.06$).
However, below $\sim 5000$\,K the models disagree significantly with the UVB SEDs 
($D>0.06$) for essentially all stars. They also disagree for the warm low-gravity stars in XSL. 
As a consequence, the energy distributions across all wavelengths are also in 
disagreement for these regimes. 
\smallskip

Before discussing the general trends further, we briefly examine the most 
extreme outliers in Fig.\,\ref{fig:forcedfits}. 
Among the three data points
near $T_{\rm eff}=7500$\,K and log($g$)\,$=0.5$, the two lower gravity
ones are two observations of the same stars, a high metallicity star with
one of the highest extinction estimates in the sample ($A_V \simeq 3$); 
for the third data point, the literature and our own estimates in Sect.\,\ref{sec:results_best}
suggest the $T_{\rm eff}$ of Paper\,II is too high by more than
1500\,K (the DR2-parameters had been correspondingly flagged). Between 
4000\,K and 5000\,K, the spectra with the largest values of $D$(UVB) or $D$(ALL)
are among those for which the observed fluxes are near zero below 400\,nm,
and the exact definition of the denominator in Eq.\,\ref{eq:D} matters. This
is one of our reasons for focusing on typical situations and general trends instead
of individual objects in this article.
\smallskip

Below 4000\,K, it is notorious that
stellar spectra are difficult to model, for instance because of large surface convection cells, 
of variability, of the formation of dust grains. 
In the following, we choose to
concentrate on temperatures between 4000\,K and 5000\,K, where these
issues are in principle less severe. This regime contains stars with strong 
contributions to the integrated light of stellar populations over a broad range
of optical and near-infrared wavelengths, which is a further reason for our
interest.
\smallskip

The nature of the disagreement for stars between 4000\,K and 5000\,K is 
illustrated with typical cases in Figs.\,\ref{fig:X0705etal_forced_R500} to
\ref{fig:X0314etal_forcedVIS_R500}. There are systematic differences 
in spectral features (in particular in the UVB range), but our focus here is
on energy distributions. Whatever wavelength range is used to estimate
$A_V$, discontinuities appear in the figures at the limits between X-shooter's spectral
arms, showing that the energy distributions are not matched properly.
Moreover, the steps are in the same direction for all the spectra shown (the
examples are selected to be representative, only rare cases deviate
from this behavior). Let us for instance
examine Figs.\,\ref{fig:X0705etal_forcedVIS_R500} and 
\ref{fig:X0314etal_forcedVIS_R500},
where the extinction applied to the synthetic spectra is 
optimized using the VIS arm: the empirical spectra in
the UVB and NIR fall off faster than the theoretical
spectra when moving away from the VIS range. If the
empirical UVB and NIR spectra were rescaled to connect to the
VIS data smoothly, the merged spectrum would lack flux both 
in the UVB and NIR compared to the model. This trend is 
more obvious at low metallicities (Fig.\,\ref{fig:X0705etal_forcedVIS_R500})
than at near-solar metallicity (Fig\,\ref{fig:X0314etal_forcedVIS_R500}).

\begin{figure}
\begin{center}
\includegraphics[clip=,trim=0 39 0 0, width=0.48\textwidth]{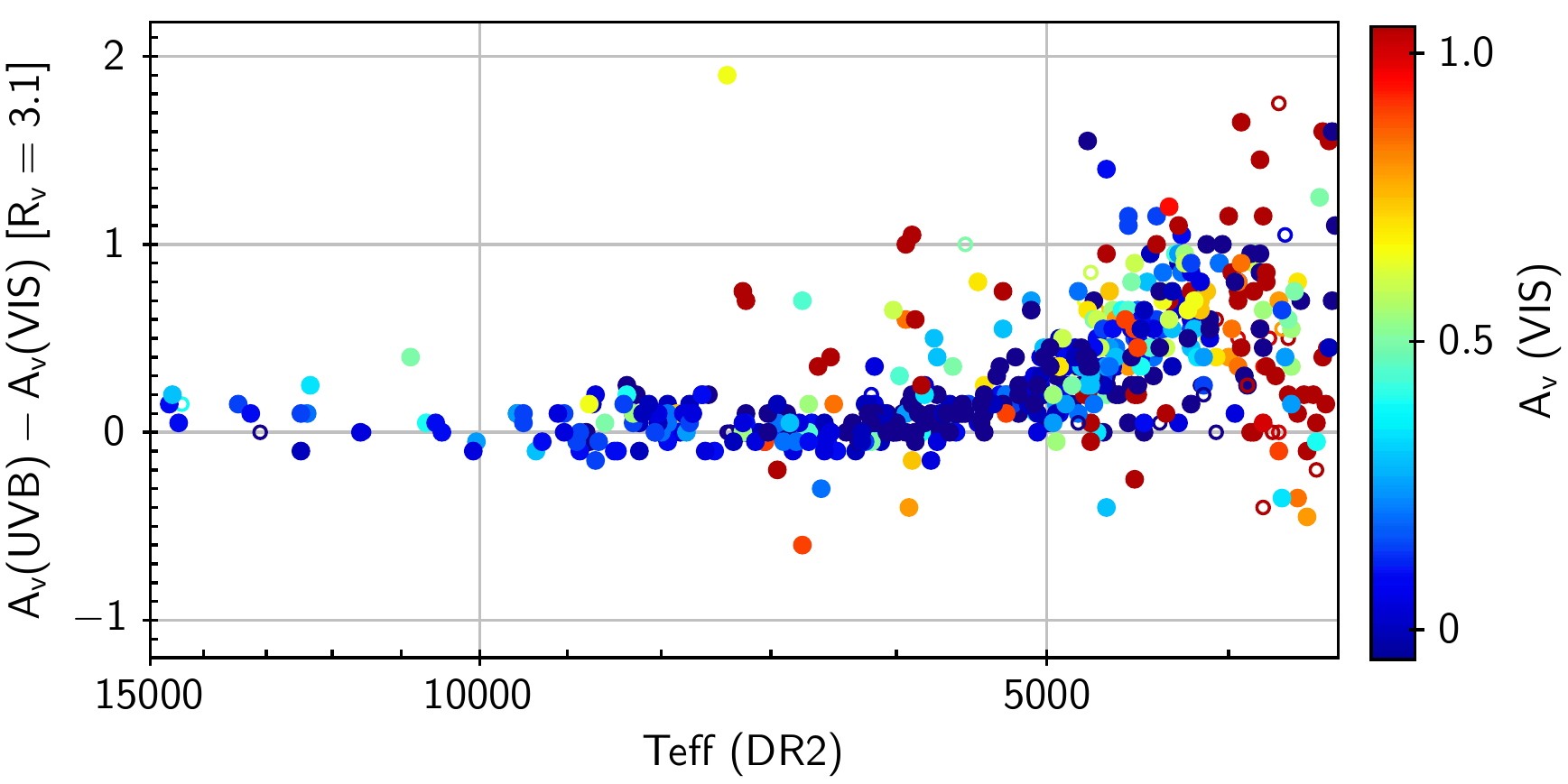}
\includegraphics[clip=,width=0.48\textwidth]{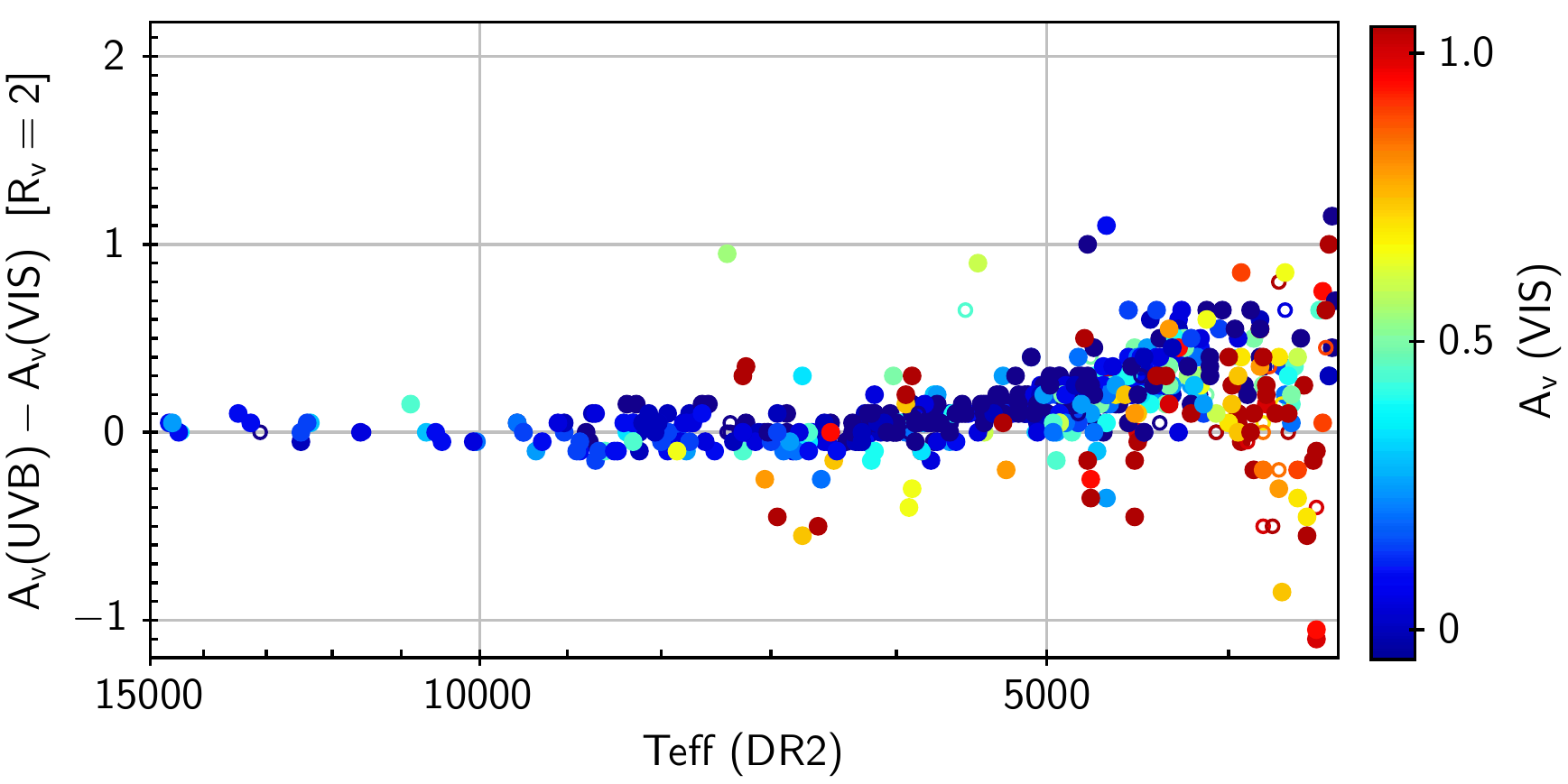}
\end{center}
\caption[]{Difference between the extinction estimates obtained with the UVB
and VIS segments of the XSL data, from the comparison
with GSL energy distributions
when adopting the stellar parameters of \citet{Arentsen_PP_19}. Top: standard
extinction law (Rv=3.1). Bottom: extreme extinction law (Rv=2).
}
\label{fig:deltaAv_vs_teff_forced}
\end{figure}

Another manifestation of the mismatch between empirical and theoretical SEDs
is seen when comparing the extinction estimates obtained
separately from the UVB and the VIS data: Figure\,\ref{fig:deltaAv_vs_teff_forced} 
shows that below about 5000\,K, $A_V$\,(UVB) is in general larger than $A_V$\,(VIS). 
Both also tend to exceed $A_V$\,(NIR), but because
of the wavelength dependence of extinction laws, $A_V$\,(NIR) is  
more sensitive to flux calibration errors and that trend is more dispersed.
In brief: when using the UVB arm for the fit, a large $A_V$ is preferred and the
theoretical spectrum has excess flux in the NIR compared to the observations; 
when using the NIR arm for the fit, a small $A_V$ is preferred and the 
theoretical spectrum has excess flux in the UVB compared to the observations.

Three hypotheses come to one's mind
immediately: (i) either the extinction law is inadequate (it should rise more
steeply towards the ultraviolet than the steepest law we explored), 
(ii) or the models lack opacities with a more severe deficit at 
shorter wavelengths, (iii) or the parameters in Paper\,II are 
systematically offset for some other reason
from those that would provide good fits with the GSL
library.

Proposition (i) is unlikely, considering what is known about extinction
in the Local Universe \citep{Schlafly_etal16}. It would also imply a correlation between 
the difference  $A_V$(UVB)$-A_V$(VIS) and a mean estimate of extinction 
which, if present at all, is weak in our data. 
Propositions (ii) and (iii) on the other hand are reminiscent 
of previous studies with various collections of synthetic and empirical spectra.
To test propositions (ii) and (iii) further, the results of the free search for 
the best-fit GSL match to each XSL observation are needed and we
therefore postpone discussions to Sections \ref{sec:results_best} 
and \ref{sec:discussion}.

\smallskip

We complete this part with a brief comment on the [$\alpha$/Fe] ratio. 
The results shown in Figs.\,\ref{fig:forcedfits} and \ref{fig:deltaAv_vs_teff_forced} 
combine estimates obtained with [$\alpha$/Fe]=0 and
[$\alpha$/Fe]=$+0.4$. Outside the range of parameters covered by the $\alpha$-enhanced models, that is for $T_{\rm eff}>8000$\,K, $T_{\rm eff}<3500$\,K, log($g$)\,$<-0.1$ or [Fe/H]$>+0.1$,
solar $\alpha$-abundances were used. Elsewhere, we decided that the $\alpha$-enhanced
model was favored over the solar one, when that change in abundances reduced
$D$\,(UVB) by at least 0.01, without degrading $D$\,(VIS), $D$\,(NIR), $D$\,(ALL)
or $\left| \, A_V\,\mathrm{(UVB)}- A_V\,\mathrm{(VIS)} \, \right|$ significantly.
The XSL-spectra favoring $\alpha$-enhanced models are those of
metal-poor giants, as expected from statistics in the Milky Way. Eye-inspection
of the superimposed empirical and theoretical spectra with solar and with
$\alpha$-enhanced abundances, be it at $R=500$ or $R=3000$,
then confirms the better match of the  Ca\,II lines or the Mg\,I triplet 
(which are deeper in $\alpha$-enhanced models)
and the CH and CN bands (which are shallower in $\alpha$-enhanced 
models because 
enhanced O captures a larger fraction of the available C). However, the
effect of accounting for $\alpha$-enhancement produces only very small
changes in Fig.\,\ref{fig:forcedfits}, that the untrained eye would not even notice.
The trends discussed above are unchanged. 

\bigskip
In summary, the SEDs of the GSL models agree well
with the SEDs of XSL, for the parameters of Paper\,II,
over wide parts of the HR-diagram. But there is statistically significant
systematic disagreement below $\sim$5000\,K, 
and for luminous warm stars (though with lower significance due to
smaller numbers).

In the following section, we relax the assumption that the
fundamental parameters are known and we search for the best-fitting model
with that extra freedom. If models can be found that
match the SEDs to within the errors (in the data
and due to the discrete grid), we would be facing a classic 
parameter calibration problem: the reference libraries
used in Paper\,II and GSL are different, and 
this may lead to offsets in derived parameters. 
It would remain to be determined which
calibration is more robust. If adequate models cannot be found 
in certain parts of the HR-diagram, then these should be regions
on which to focus future efforts in stellar spectral synthesis.

\subsection{Best-fit model SEDs for each XSL spectrum}
\label{sec:results_best}

\begin{figure}
\begin{center}
\includegraphics[clip=,width=0.45\textwidth]{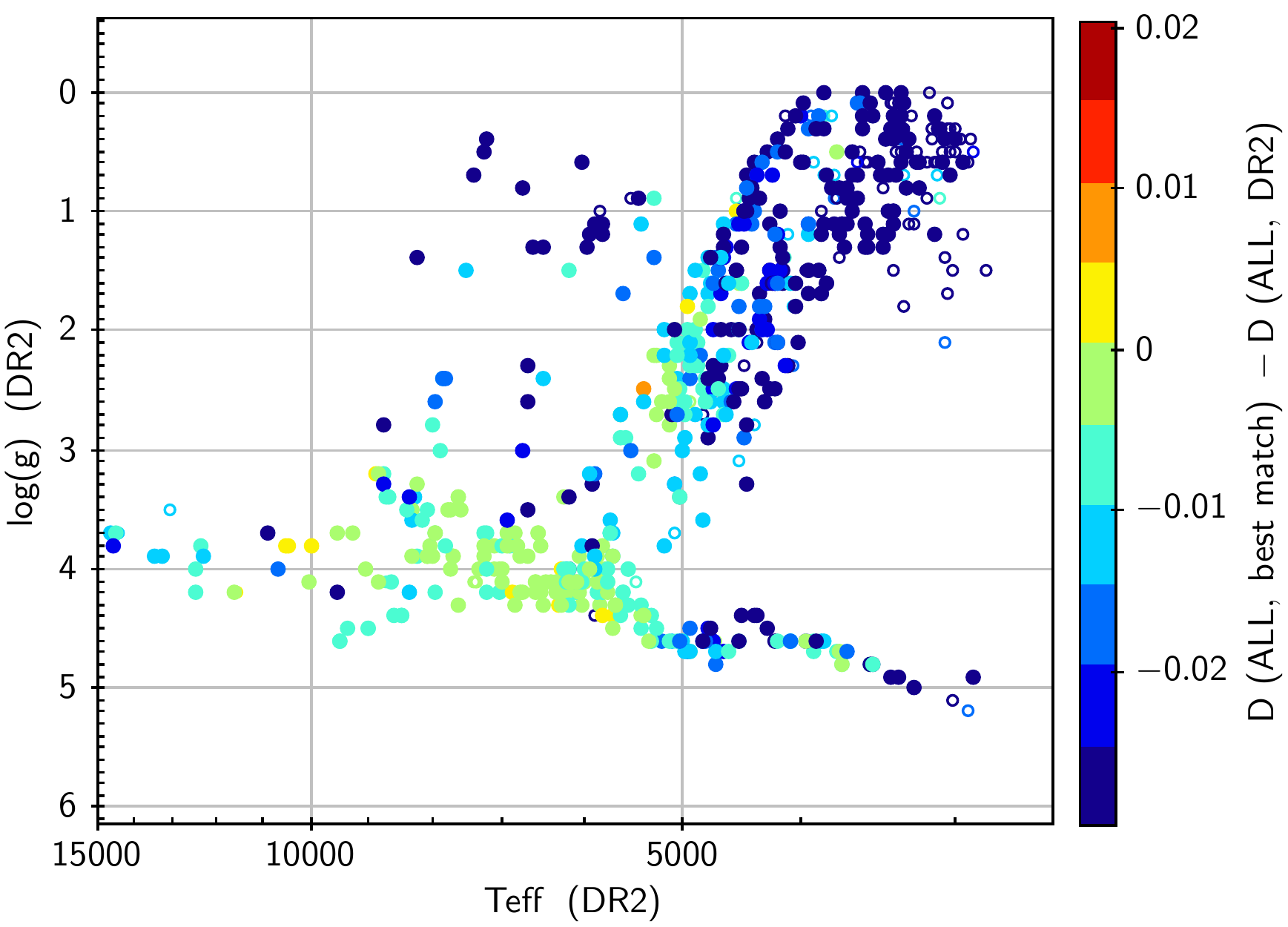}
%
%
%
\includegraphics[clip=,width=0.45\textwidth]{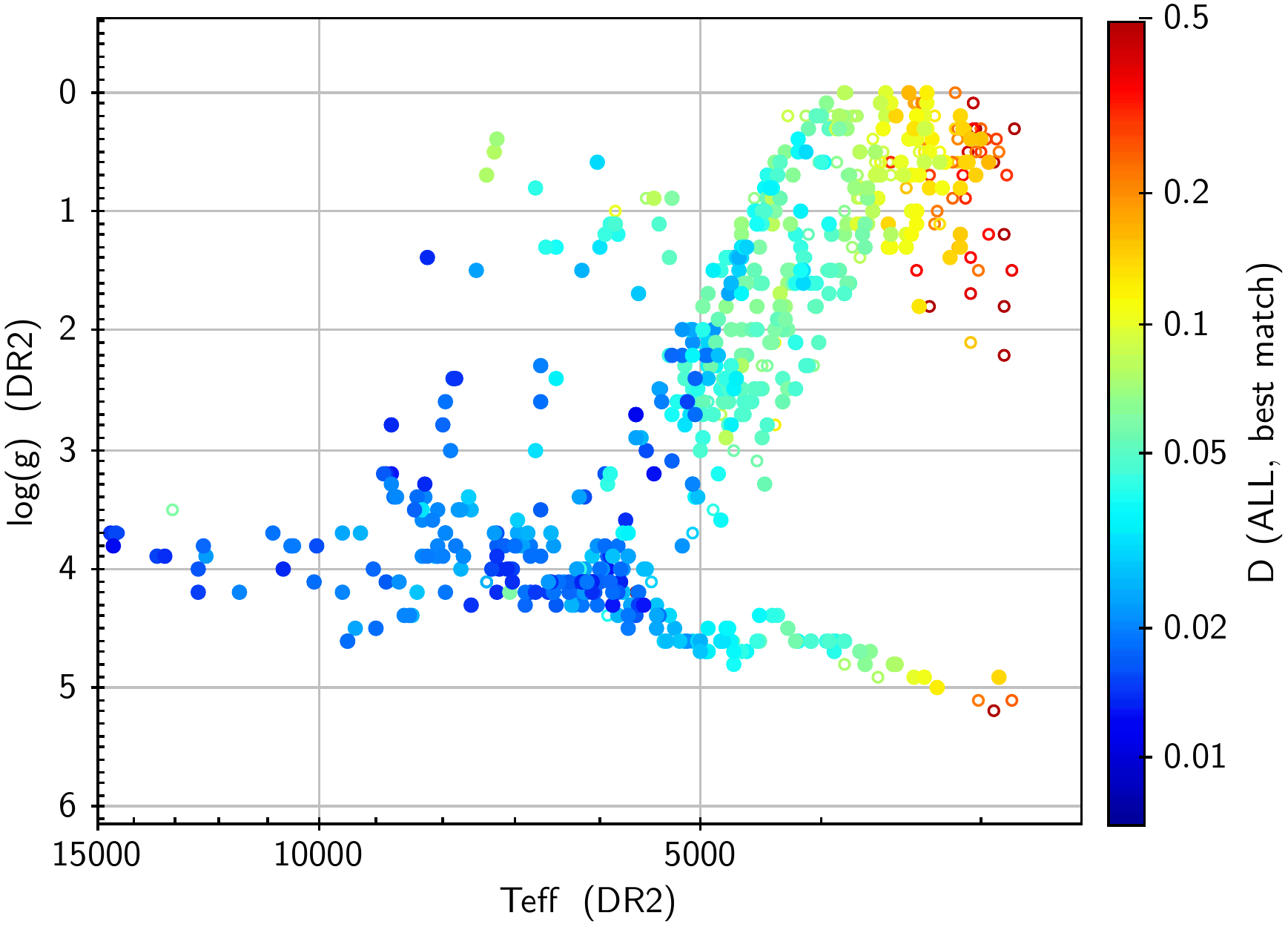}
\end{center}
\caption[]{Top: Change in the discrepancy measure $D$ (Eq.\,\ref{eq:D}) when, 
instead of adopting stellar parameters from \citet{Arentsen_PP_19},
the parameters are freely optimized. 
The data points are located in the diagram according to the DR2-parameters 
of Arentsen et al. The symbol color maps the difference
between the best-fit value of $D$ (Eq.\,\ref{eq:D}) and its value for 
the DR2-parameters. Bottom: discrepancy $D$
for the best-fit stellar parameters (to be compared to the
bottom right panel of Fig.\,\ref{fig:forcedfits}, where the parameters of
Arentsen et al. were assumed). Complete versions of these figures,
with 1 panel per arm, are available in Figs.\,\ref{fig:HRdiag_Dimproved_allArms} and \ref{fig:HRdiag_D_best_allArms} of the appendix. 
}
\label{fig:HRdiag_D_best}
\end{figure}

When the stellar parameters are free, better matches with GSL 
energy distributions can be obtained
in the critical regions of the HR-diagram identified above.
$D$ now takes values between 
0.03 and 0.06 for effective temperatures between 4000 and 5000\,K
(Fig.\,\ref{fig:HRdiag_D_best}). 
Such values, taken individually, would be consistent with
flux calibration errors at the 2\,$\sigma$ level. In fact
a part of the discrepancies are due to local spectral features,
and hence the part of $D$ associated with the general low resolution 
energy distribution is even smaller than these 2\,$\sigma$. 

However the homogeneous behavior of the numerous data points
below 5000\,K in Fig.\,\ref{fig:HRdiag_D_best} 
tells us that the values of $D$ are not the result of random errors.
Systematic discrepancies are still present. Despite the
free exploration of parameter space, we must conclude that
it remains difficult to reproduce both the spectral features 
and the low-resolution SEDs
of the cool stars simultaneously below 5000\,K. Below, in  Sect.\,\ref{sec:results_best_residualDiscrepancies}, we clarify the nature of the systematic
discrepancies seen between the models with the best-fitting SEDs
and the observations, where such discrepancies are present. Subsequently,
in Sect.\,\ref{sec:results_best_vs_DR2}, we quantify how
the parameters of the best-SED models obtained here compare to 
those of Paper\,II.
\medskip

\subsubsection{Residual differences between best-SED models and XSL spectra}
\label{sec:results_best_residualDiscrepancies}

When the purpose is to find the best theoretical match to a given observation
(as opposed to comparing the quality-of-fit for different observations), 
inverse-variance weighted $\chi^2$ values are a more appropriate figure of merit 
than $D$, which gives low-flux regions at short wavelengths too much weight. For the 
sake of validation, we have performed all calculations with the 
two methods. The directions of the trends are unchanged, 
but amplitudes of the differences between
the parameters from optimization and the parameters in
Paper\,II depend on the weighting, 
in the sense that slightly larger differences are
found on average when using $D$, than when using the inverse-variance 
weighted $\chi^2$. In the following, we restrict the figures presented to those
obtained with the weighted $\chi^2$. We focus on temperatures between 4000 and 5000\,K.

\begin{figure*}
\parbox[b]{0.49\textwidth}{
\includegraphics[clip=,width=0.48\textwidth]{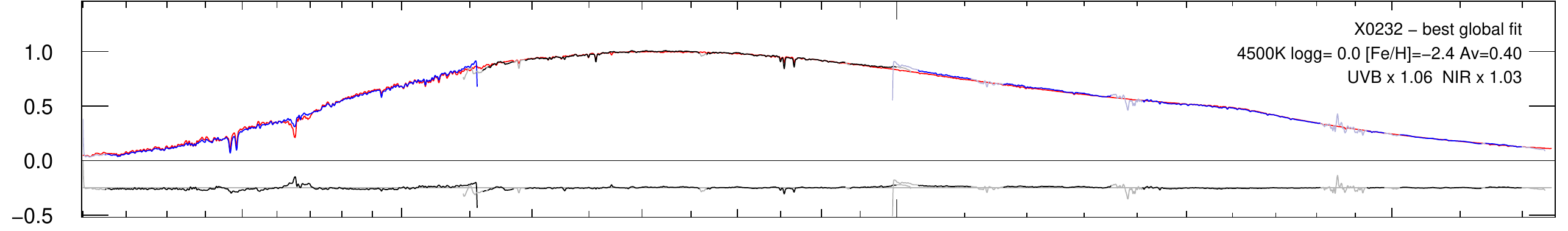} 
\includegraphics[clip=, width=0.48\textwidth]{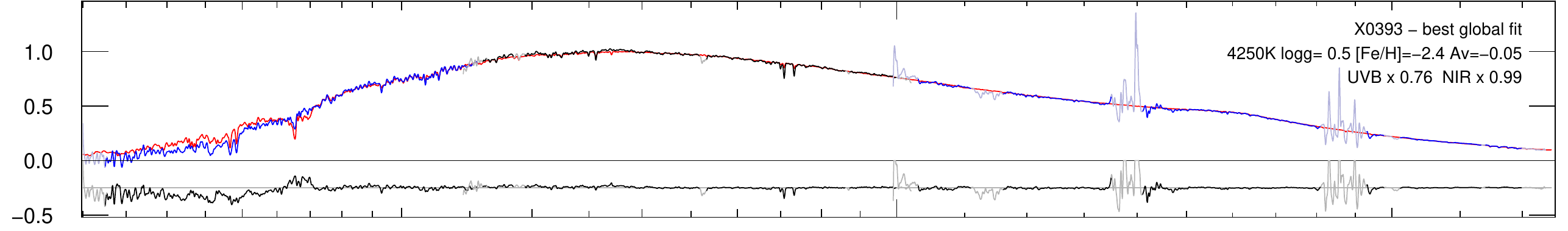} 
\includegraphics[clip=, width=0.48\textwidth]{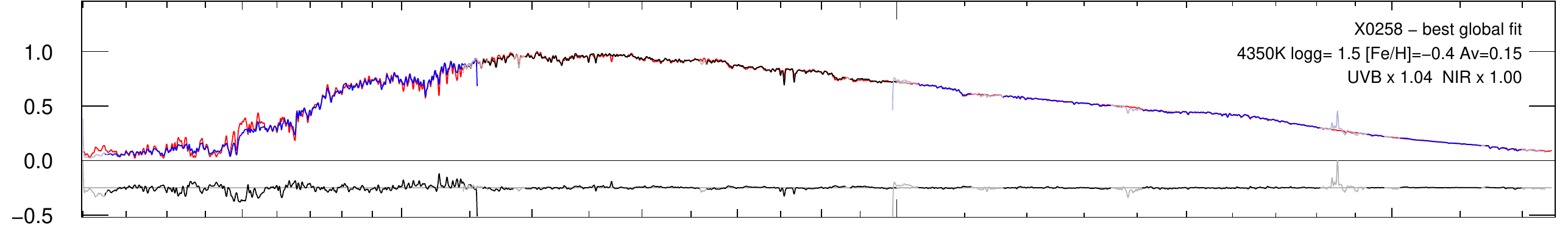} 
\includegraphics[clip=, width=0.48\textwidth]{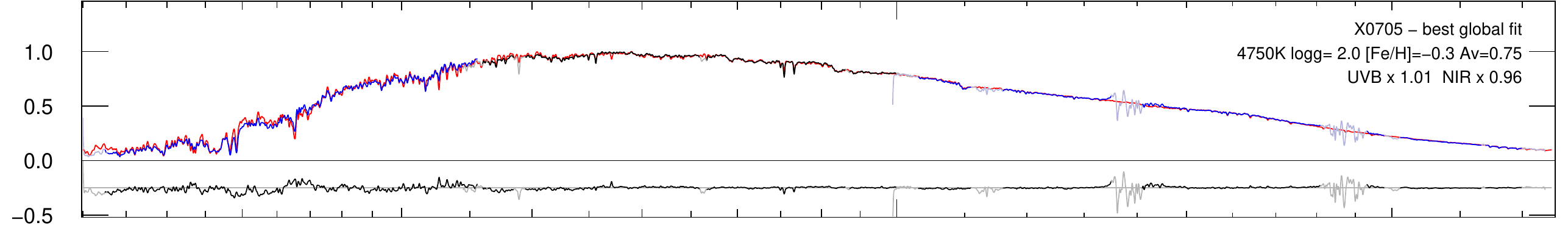} 
\includegraphics[clip=, width=0.48\textwidth]{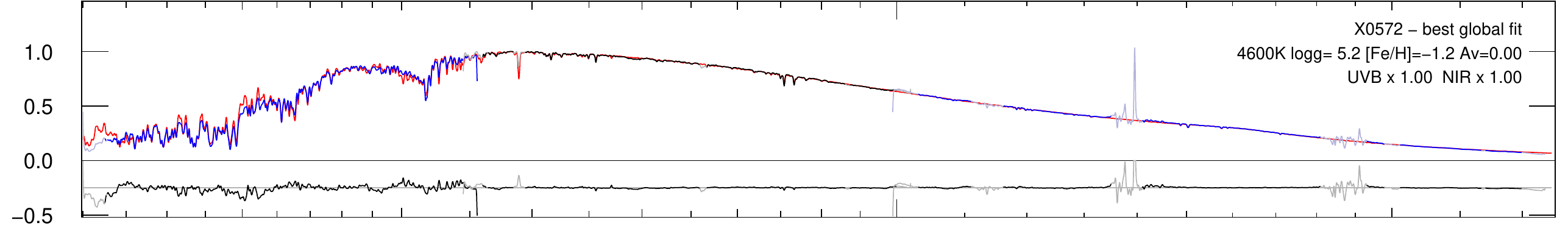} 
\includegraphics[clip=, width=0.48\textwidth]{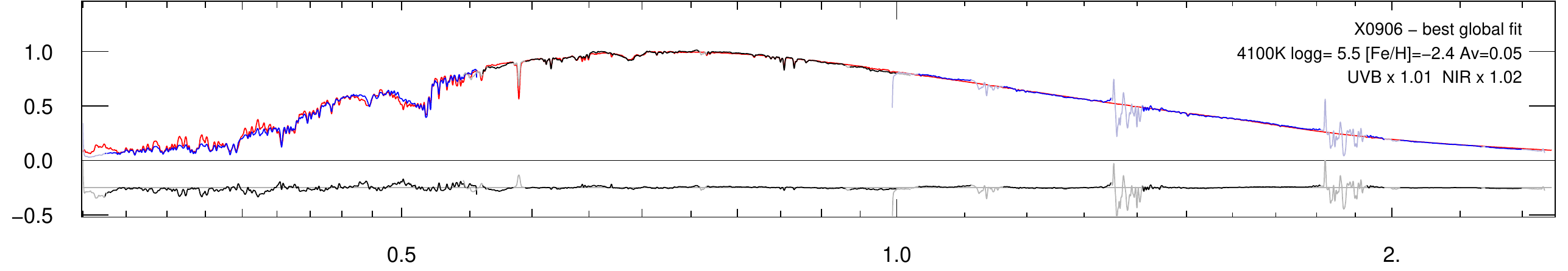} 
\vspace{-3pt}

\includegraphics[clip=, trim=0 0 0 1, width=0.48\textwidth]{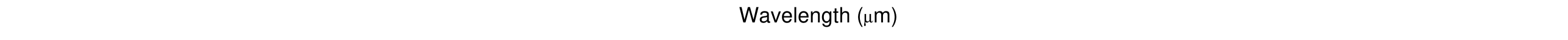}
}
%
\parbox[b]{0.49\textwidth}{
\includegraphics[clip=,width=0.48\textwidth]{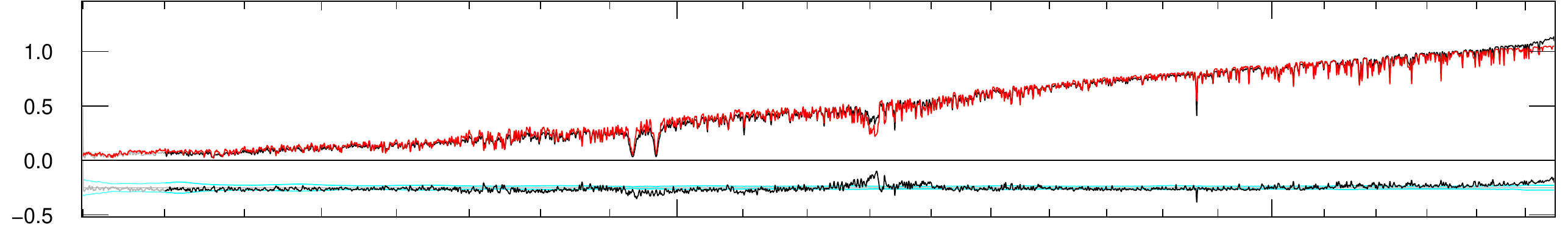}
\includegraphics[clip=, width=0.48\textwidth]{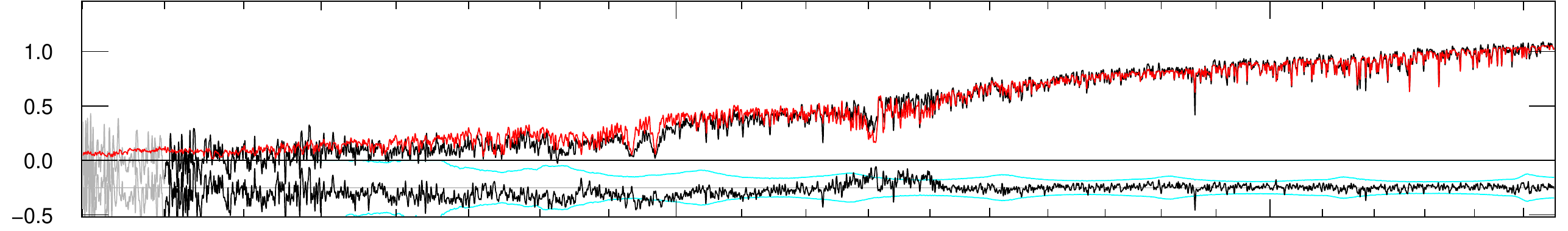}
\includegraphics[clip=, width=0.48\textwidth]{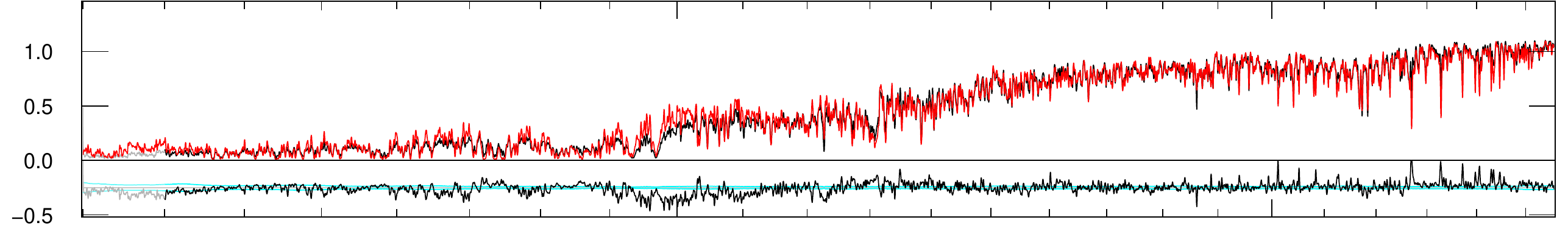}
\includegraphics[clip=, width=0.48\textwidth]{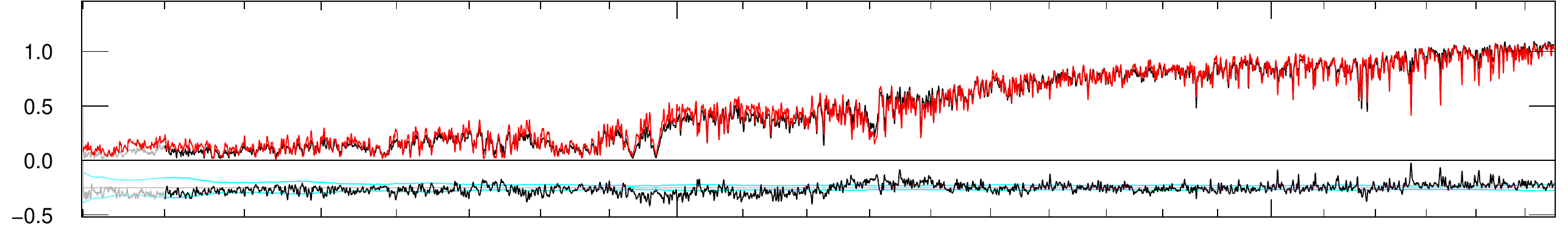}
\includegraphics[clip=, width=0.48\textwidth]{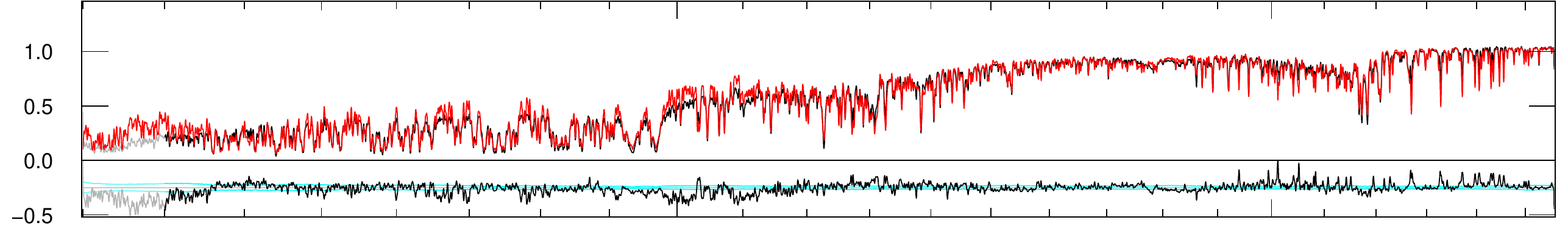}
\includegraphics[clip=, width=0.48\textwidth]{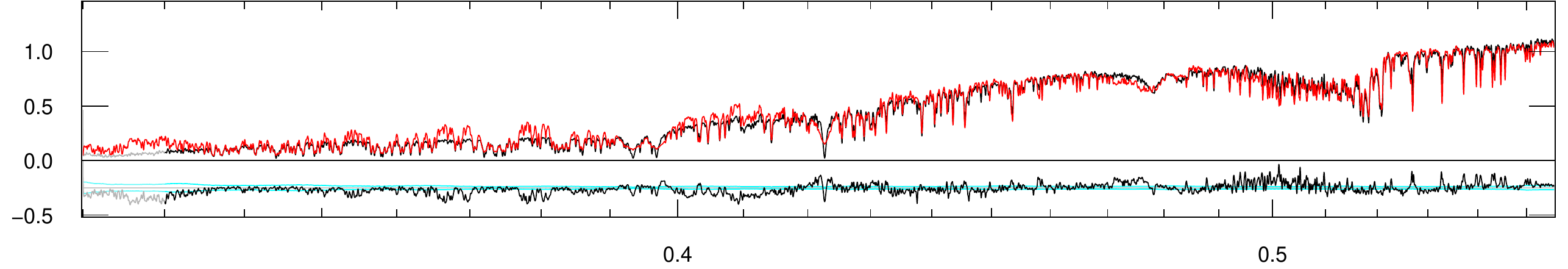}
\vspace{-3pt}

\includegraphics[clip=, trim=0 0 0 1, width=0.48\textwidth]{figs_spectra_refined/waveBanner.pdf}
}

\parbox[b]{0.49\textwidth}{
\includegraphics[clip=, width=0.48\textwidth]{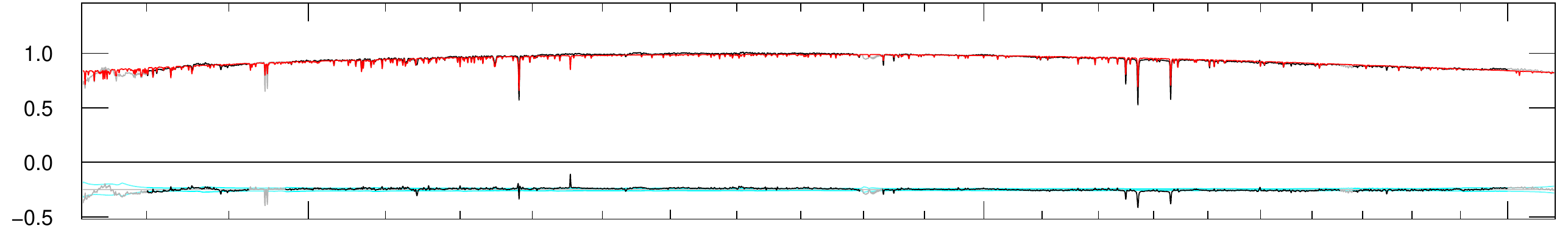}
\includegraphics[clip=, width=0.48\textwidth]{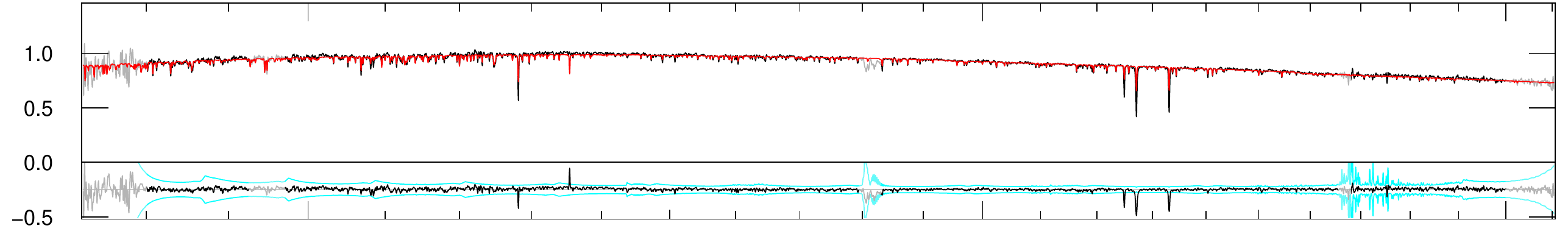}
\includegraphics[clip=, width=0.48\textwidth]{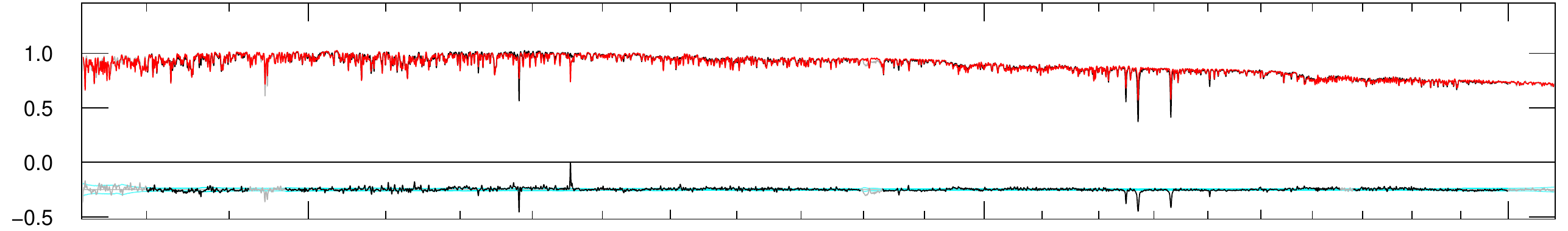}
\includegraphics[clip=, width=0.48\textwidth]{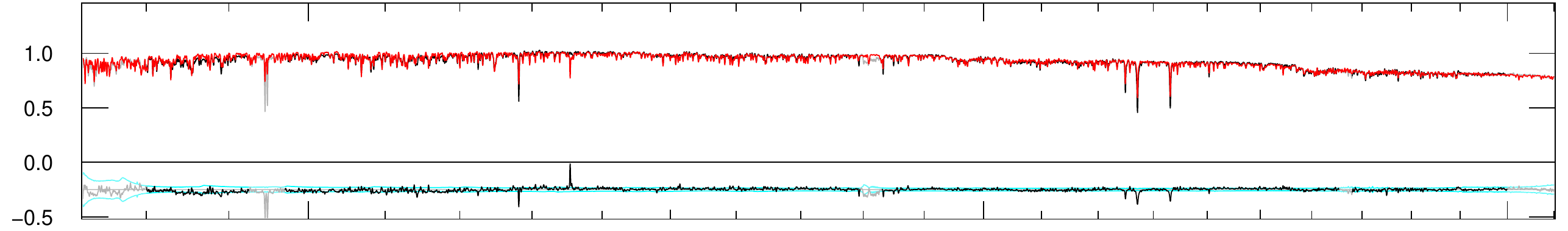}
\includegraphics[clip=, width=0.48\textwidth]{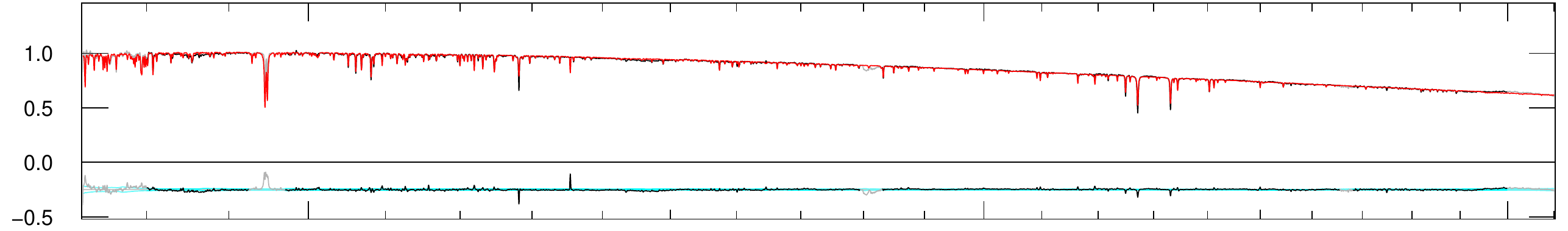}
\includegraphics[clip=, width=0.48\textwidth]{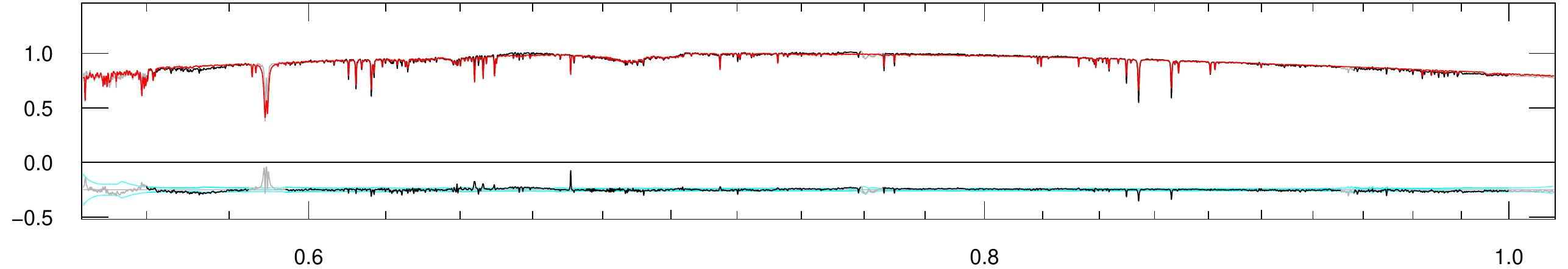}
\vspace{-3pt}

\includegraphics[clip=, trim=0 0 0 1, width=0.48\textwidth]{figs_spectra_refined/waveBanner.pdf}
}
%
\parbox[b]{0.49\textwidth}{
\includegraphics[clip=, width=0.48\textwidth]{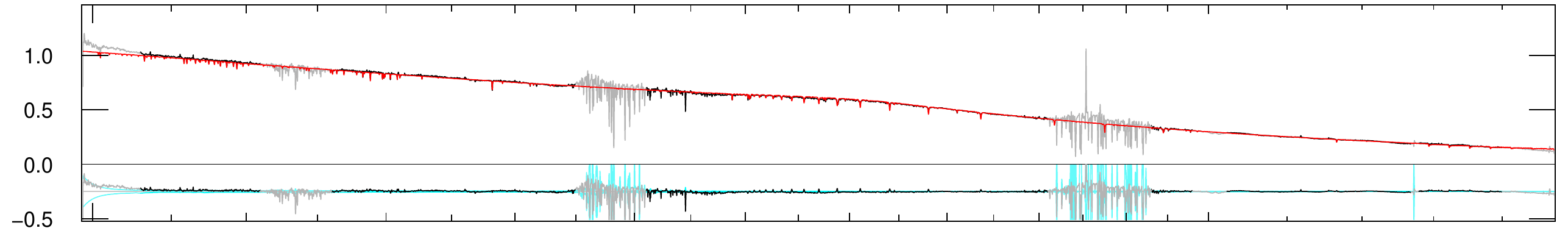}
\includegraphics[clip=, width=0.48\textwidth]{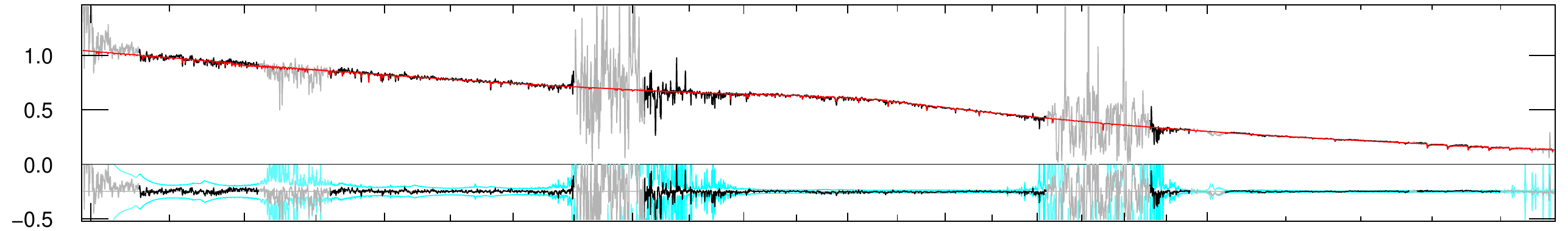}
\includegraphics[clip=, width=0.48\textwidth]{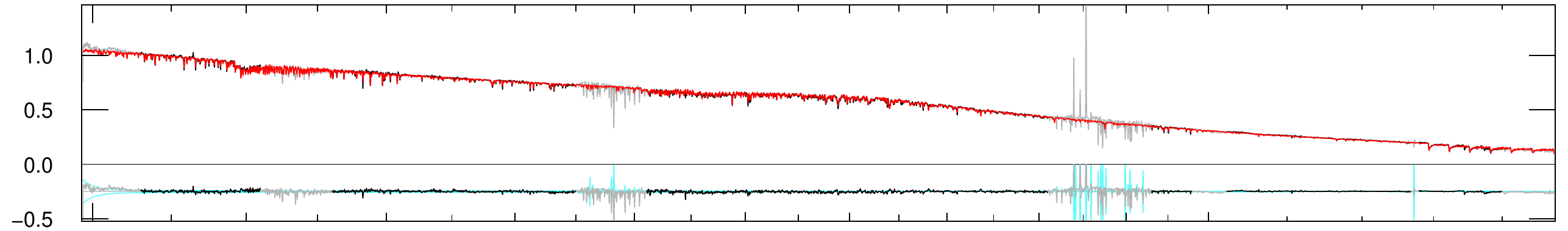}
\includegraphics[clip=,width=0.48\textwidth]{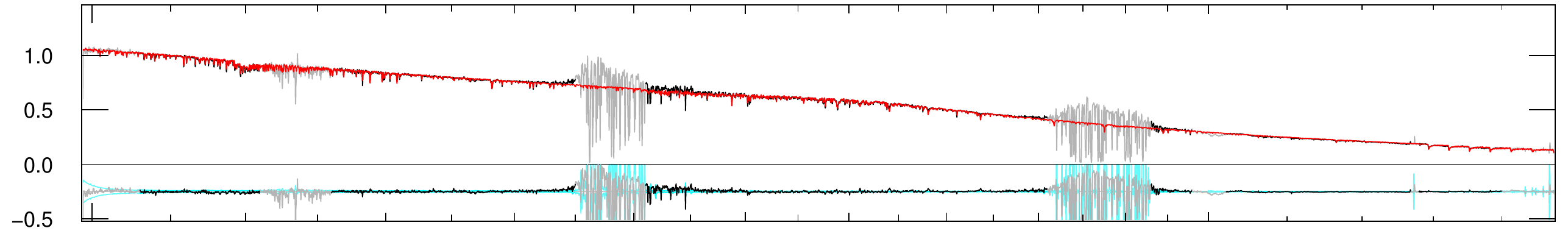}
\includegraphics[clip=, width=0.48\textwidth]{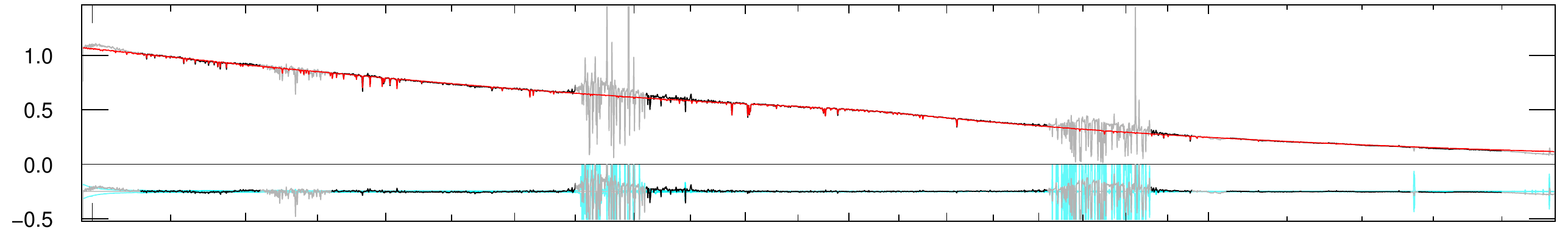}
\includegraphics[clip=, width=0.48\textwidth]{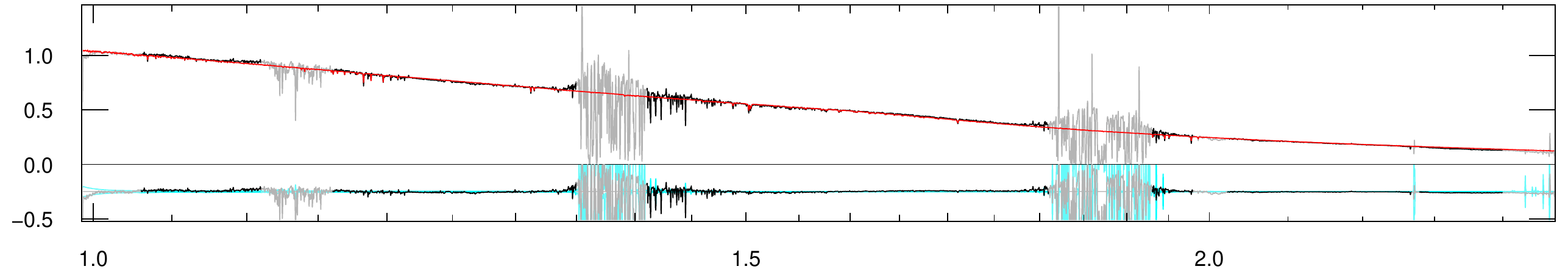}
\vspace{-3pt}

\includegraphics[clip=, trim=0 0 0 1, width=0.48\textwidth]{figs_spectra_refined/waveBanner.pdf}
}
%
\caption[]{
Typical comparisons between XSL and best-match GSL energy
distributions for stars with relatively low metallicities according to
XSL DR2 and with DR2-\teff\ between 4000\,K and 5000\,K.
The comparisons are shown for the parameters that minimize 
the combined inverse-variance weighted $\chi^2$ over all wavelengths (Eq.\,\ref{eq:chi2}).
The upper left panels show all wavelengths at $R=500$, the other panels
are zooms into these same comparisons, at $R=3000$.
The best-match synthetic spectra are in red; the empirical spectra in black, except in the
upper left panel where they are shown in blue, black, and blue for the UVB, VIS, and NIR arms of X-shooter.
Gravity increases from top to bottom: HD\,165195 (X0232), HD\,1638 (X0258),
NGC\,6838\,1037 (X0705), LHS\,1841 (X0572), LHS\,343 (X0906).
Compare with Fig.\,\ref{fig:X0705etal_forced_R500}.
}
\label{fig:X0705etal_refined_R500_withnoise}
\end{figure*}

\begin{figure*}
\parbox[b]{0.49\textwidth}{
\includegraphics[clip=, width=0.48\textwidth]{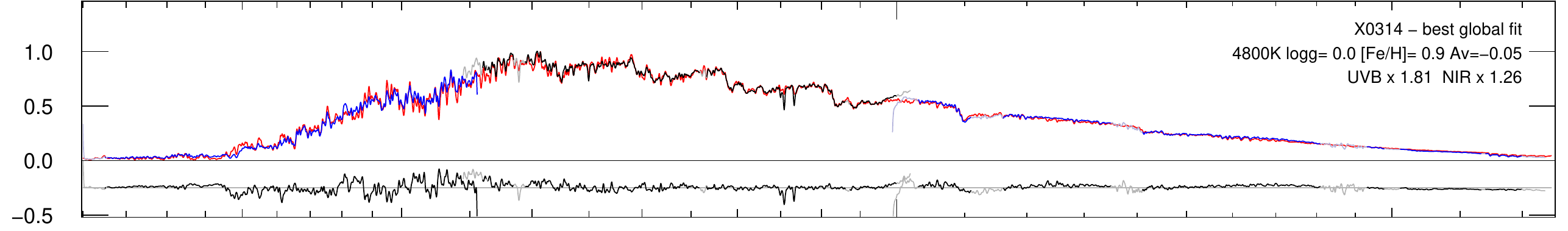} 
\includegraphics[clip=, width=0.48\textwidth]{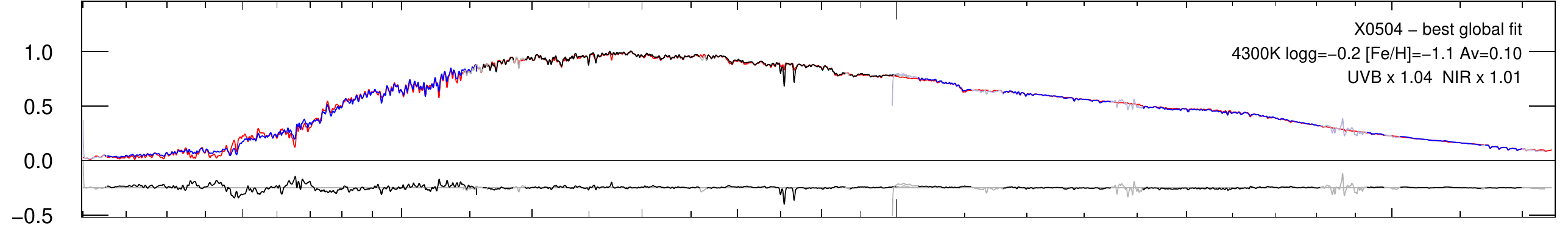} 
\includegraphics[clip=, width=0.48\textwidth]{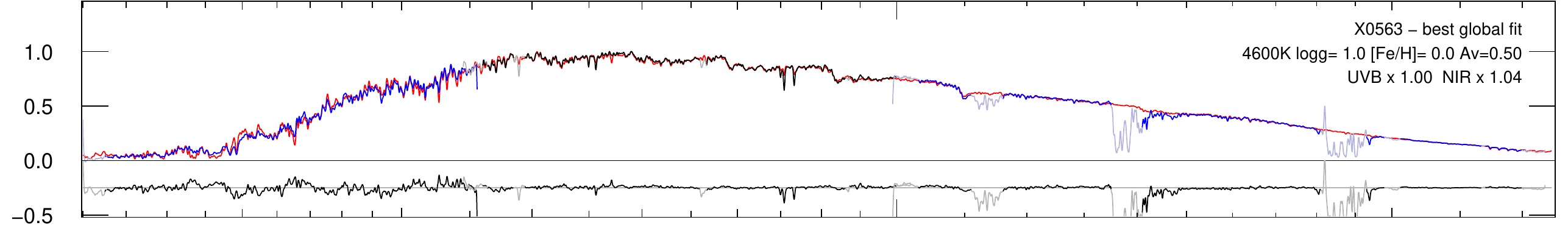} 
\includegraphics[clip=, width=0.48\textwidth]{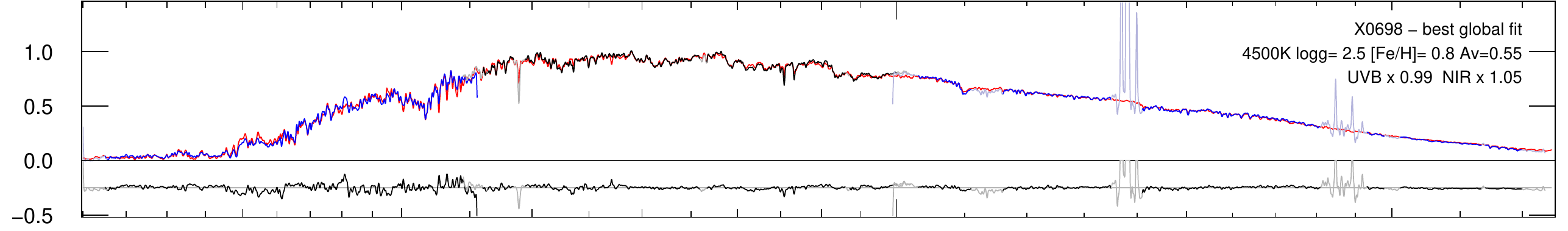} 
\includegraphics[clip=, width=0.48\textwidth]{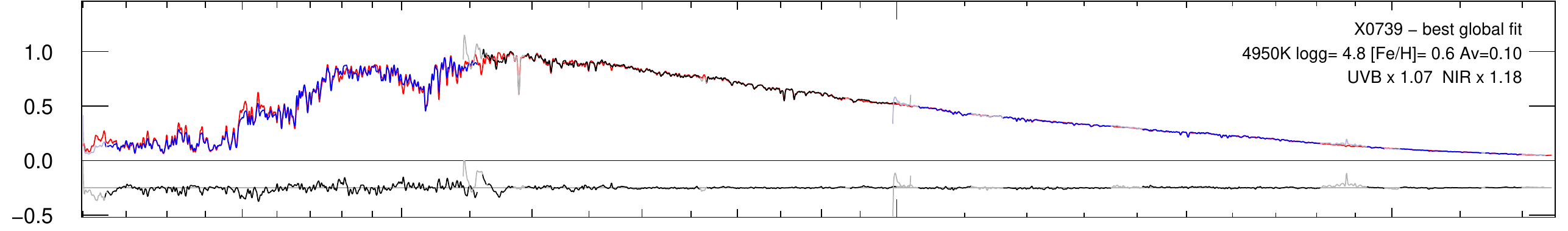} 
\includegraphics[clip=, width=0.48\textwidth]{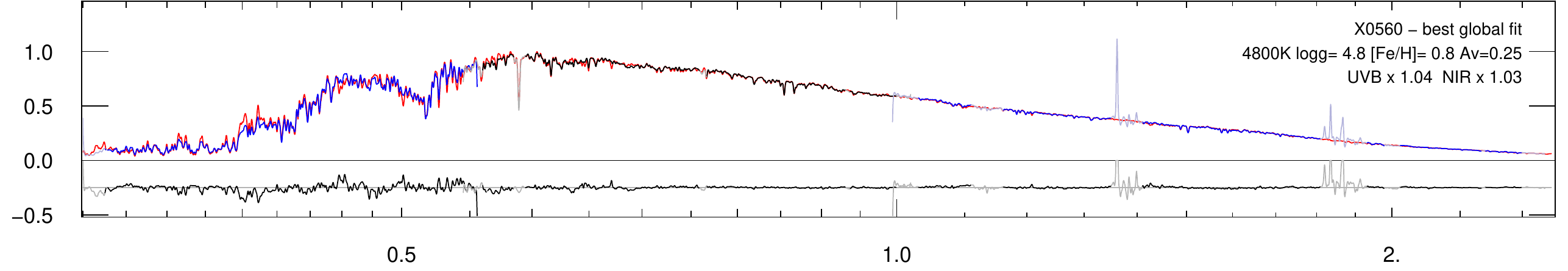} 
\vspace{-3pt}

\includegraphics[clip=, trim=0 0 0 1, width=0.48\textwidth]{figs_spectra_refined/waveBanner.pdf}
}
\parbox[b]{0.49\textwidth}{
\includegraphics[clip=, width=0.48\textwidth]{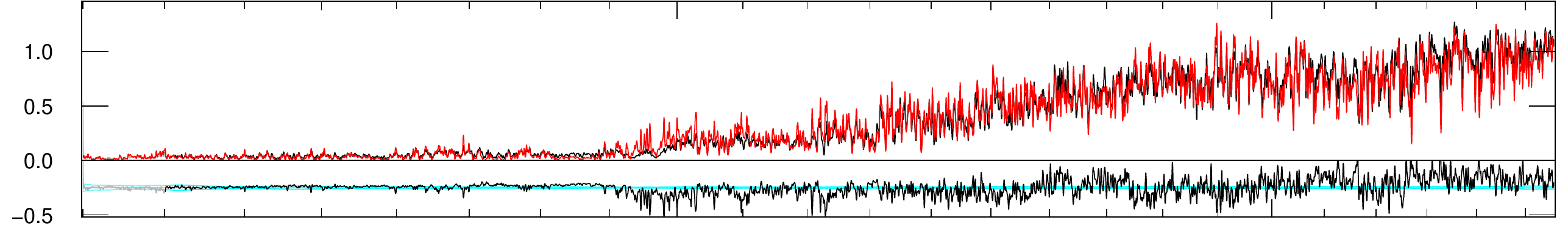}
\includegraphics[clip=, width=0.48\textwidth]{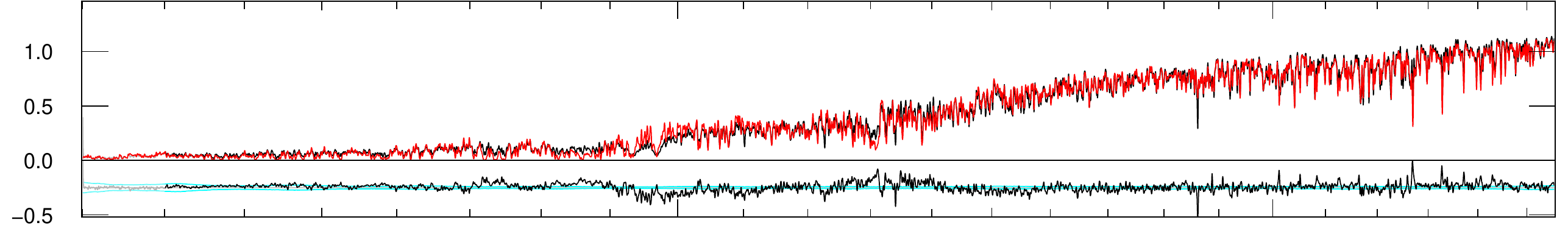}
\includegraphics[clip=, width=0.48\textwidth]{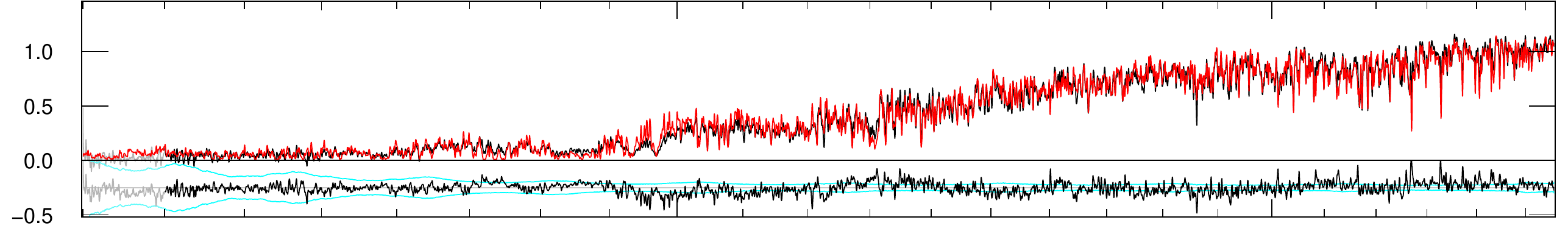}
\includegraphics[clip=, width=0.48\textwidth]{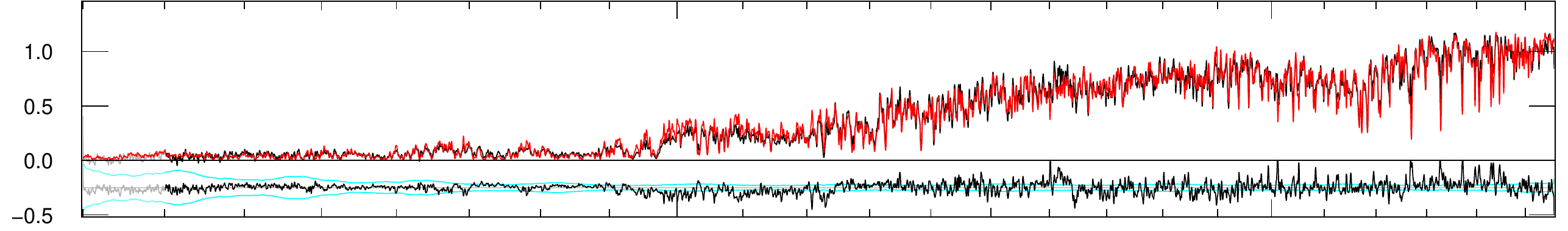}
\includegraphics[clip=, width=0.48\textwidth]{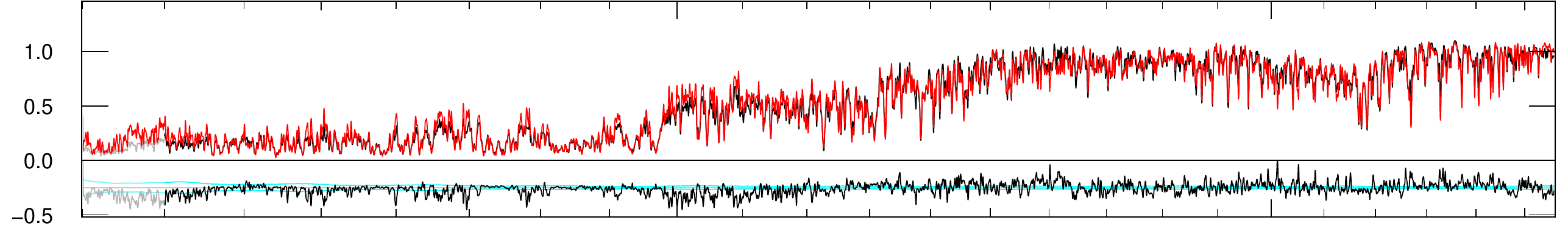}
\includegraphics[clip=, width=0.48\textwidth]{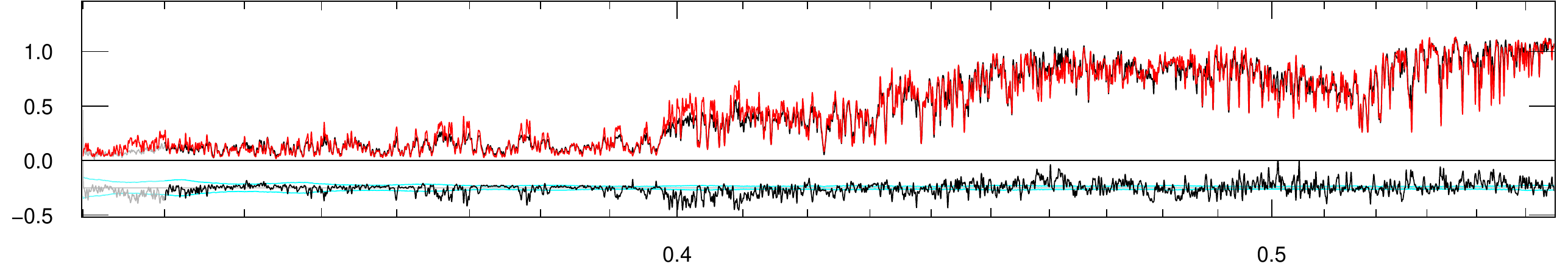}
\vspace{-3pt}

\includegraphics[clip=, trim=0 0 0 1,width=0.48\textwidth]{figs_spectra_refined/waveBanner.pdf}
}
\caption[]{Typical comparisons between XSL and best-match GSL energy
distributions for stars with relatively high metallicities according to
XSL DR2 and with DR2-\teff\ between 4000\,K and 5000\,K.
The comparisons are shown at $R=500$, for the parameters that minimize 
the combined inverse-variance weighted $\chi^2$ over all wavelengths (Eq.\,\ref{eq:chi2}).
Gravity increases from top to bottom: HD\,50877 (X0314),
BBB\,SMC\,104 (X0504), HD\,44391 (X0563), 2MASS\,J18351420-3438060 (X0698),
HD\,218566 (X0739), HD\,21197 (X0560). In the case of X0504, the best-match 
metallicity is lower than the DR2-value ($-0.6$), which had justified its presence is 
this subset of spectra.
Compare with Fig.\,\ref{fig:X0314etal_forced_R500}.}
\label{fig:X0314etal_refined_R500_withnoise}
\end{figure*}

Examples of matched spectral energy distributions, based on fits performed at $R=500$
by minimizing the inverse-variance weighted $\chi^2$ across all available
wavelengths (Eq.\,\ref{eq:chi2}), are shown 
in Figs.\,\ref{fig:X0705etal_refined_R500_withnoise}
and  \ref{fig:X0314etal_refined_R500_withnoise}
respectively for relatively metal-poor and
relatively metal-rich stars, with effective temperatures between 4000 and 5000\,K 
according to \citet{Arentsen_PP_19}.
The energy distributions from the near-UV to 2.4\,$\mu$m are
reproduced well for the new sets of stellar parameters indicated
in the panels (compare with
Figs.\,\ref{fig:X0705etal_forced_R500} and \ref{fig:X0314etal_forced_R500}).
The residuals are roughly flat on average and the three arms in general connect
naturally. In the near-IR, where the main low-resolution features are
the shape of the H-band (set by continuous H$^-$ opacities) and the
CN band near 1.1\,$\mu$m, excellent agreement with the observed shapes
are obtained.

\smallskip

However, in this temperature range 
it is clear that the improvement in the SEDs does not
always come with a good match of all the spectral features.
The local features in the residuals are of varying amplitude
but are generally highly significant (the signal-to-noise ratio
per resolved element is of several hundred at R=500). They
are not random but correspond to features in the spectra. 
There is some natural dispersion in the properties
of the residuals and we cannot provide an exhaustive
description; the points we mention are those for which we
have enough cases to consider they are systematic trends.

A striking first impression is that the UVB residuals would
not satisfy any expert of stellar parameter estimates. They 
are clearly larger than those one can expect to achieve
when fitting spectra over the traditional optical range
(400-700\,nm) when allowing for a polynomial correction
of the continuum. They are also significantly
larger than those we obtain with GSL when fitting only
the UVB range with reddened models, instead of 
ALL wavelengths.
In the VIS and NIR ranges, the features
are intrinsically weaker than in the UVB,
and the physical natural variance is smaller
between 4000 and 5000\,K. The tension between the SED and the
spectral features is still present to some extent, but weaker.

In luminous metal-poor giants (top rows of each panel of 
Fig.\,\ref{fig:X0705etal_refined_R500_withnoise}), 
the best-SED models tend
to display a deeper G-band (CH molecule, 0.55\,$\mu$m) than the
observations. A fit restricted to the UVB arm would be able to eliminate
this issue, but at the cost of degrading the panchromatic SED. 
At warm enough temperatures in the range considered,
the metal-poor models display hydrogen lines, which for 
the best-SED tend to be too weak in the Balmer series but too deep
in the near-IR Brackett series. 

At higher giant-branch gravities (log($g$)$\simeq$2), 
an interesting trend appears
across the UVB spectrum of metal-poor giants (middle rows in the 
upper right panel of Fig.\,\ref{fig:X0705etal_refined_R500_withnoise}): 
the largest residuals take the shape
of positive features at wavelengths longer than 0.49\,$\mu$m,
while it is the opposite below 0.43\,$\mu$m. 
A closer look shows that the largest residual differences at the
red end of the UVB arm correspond to strong metal lines, that 
are too deep in the best-SED models; this suggests
that the constraints from the panchromatic SED pull the fit
towards high metallicities.
At the blue end of the UVB, the largest residuals are not 
associated with strong lines, but are more broadly distributed. 
For main sequence stars (last rows in
the panels of Fig.\,\ref{fig:X0705etal_refined_R500_withnoise}),
the largest residuals below 0.43\,$\mu$m correspond to
regions in between the deepest spectral features.

Moving to higher metallicities (Fig.\,\ref{fig:X0314etal_refined_R500_withnoise}),
the trends just described for gravities log($g$)\,$\gtrsim 2$ remain present. 
The multitude of lines of
molecular and atomic species makes it difficult to emphasize any
other feature specifically. By letting the eye slide over the residuals
presented, it can be seen that their features repeat (within a given
regime of the HR diagram). The discrepancies are mostly systematic,
rather than random. The calcium triplet for instance, around 0.86\,$\mu$m,
is usually too strong in the best-SED model for giants (the 
SED favoring [$\alpha$/Fe]=0 at solar-like metallicities), while it is
well matched in the dwarfs.

\smallskip

\begin{figure}
\includegraphics[clip=,width=0.48\textwidth]{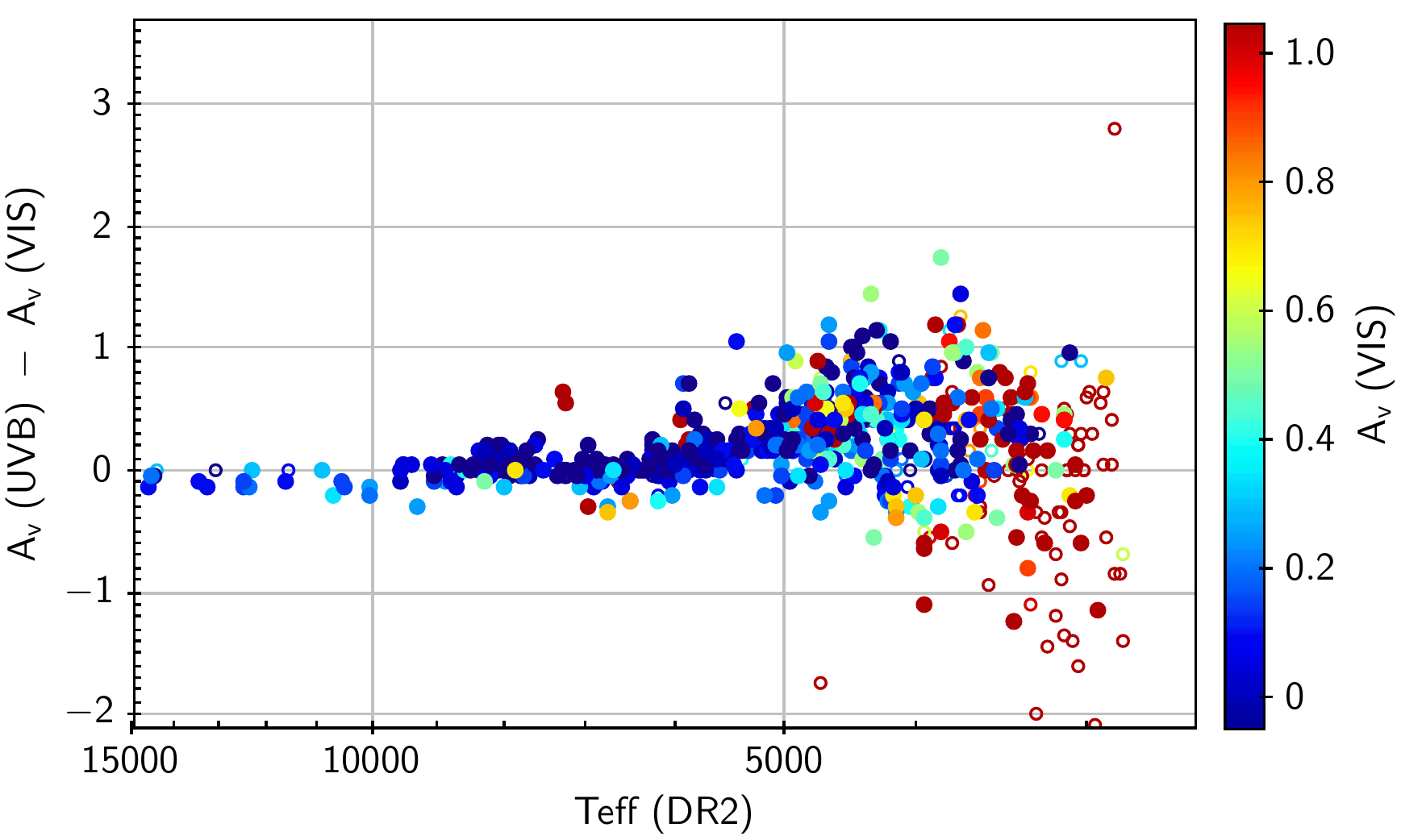}
\caption[]{Same as Fig.\,\ref{fig:deltaAv_vs_teff_forced}, but
for the parameters and extinction that minimize the inverse-variance weighted 
$\chi^2$-difference between models and data 
in the UVB and VIS arms of X-shooter.
\normalsize}
\label{fig:deltaAv_vs_teff_bestfit}
\end{figure}

The tension between the SED and the spectral features
also manifests in the sensitivity of the best-fit parameters
to the wavelength range considered in the comparison. For
instance, $\Delta A_V \equiv A_V$(UVB)$-A_V$(VIS) 
still is positive on average between 4000 and 5000\,K, 
with 80\,\% of the values spread between 
0 and 0.8 (Fig.\,\ref{fig:deltaAv_vs_teff_bestfit}; compare
with Fig.\,\ref{fig:deltaAv_vs_teff_forced}). 
This apparent temperature-dependence of $\Delta A_V$ is due primarily 
to mismatches in the UVB arm, where the
density of strong spectral lines is largest. As described above,
the simultaneous fit to all three arms produces residual
features in the UVB
with a systematic sign-difference between the blue end and the
red end of that arm. A larger extinction
in the UVB would help reduce these residuals, and this
happens in fits that do not use the VIS and NIR constraints.
On average, the extinction estimate 
based on the VIS arm compares well with the extinction evaluated from the
three arms together. 

In the comparison between parameters preferred by the
different arms of the spectrograph, the effects of degeneracies
are seen strongly. For the sample as a whole, 
the difference $\Delta A_V$ 
correlates positively with the corresponding differences 
in \teff\ and in [Fe/H] obtained using the UVB or the VIS arm,
as expected from the notorious effects on \teff\ and $A_V$ on spectral slopes, 
and from the need to compensate a higher \teff\ with higher metallicity in order
to obtain spectral features of similar strength. 
However, this hides a more complex dependence on position in the
HR diagram and on metallicity. A few examples are given
in Appendix \ref{app:diffs_between_arms}. 

\smallskip

We find no significant correlation between $\Delta A_V$ and $A_V$(ALL) or
[$A_V$(UVB)+$A_V$(VIS)]/2 for the sample as a whole, 
which would have been the most evident indication 
of an inadequate extinction law. Nevertheless, we repeated
the fits with an extinction law with a steeper rise in 
the UV, using $R_V=2$ instead of $R_V$=3.1 in the parametric description
of \citet{Cardelli89}. As expected, $R_V=2$ reduced $\Delta A_V$ 
on average for giants between 4000 and 5000\,K\footnote{Not below
4000\,K, where the fits are poor and dispersion washes out any trend.}, 
but without eliminating the positive average.  The other region
of the HR diagram where our sample has high extinctions corresponds 
to luminous warm stars. Here, switching to the steeper $R_V$ tends to increase
the discrepancies $\Delta A_V$ between UVB and VIS estimates.

\smallskip


\subsubsection{Systematic differences between best-SED parameters 
and those based on ULySS and MILES or ELODIE}
\label{sec:results_best_vs_DR2}

As discussed above, the models that reproduce the SEDs of the XSL-spectra
best are not always those with the parameters derived by 
comparison of the absorption line spectra with the empirical 
libraries MILES or ELODIE. In a small number of cases, this can be traced back
to an outlier-type failure of the analysis of the optical spectra, but
this is not our main point here. On the contrary, we focus
on generic trends, that systematically affect stars of a given
part of parameter space and that withstand errors
on individual parameters.

\medskip

\paragraph{Gravity and temperature --} 

\begin{figure*}
\includegraphics[clip=,width=0.45\textwidth]{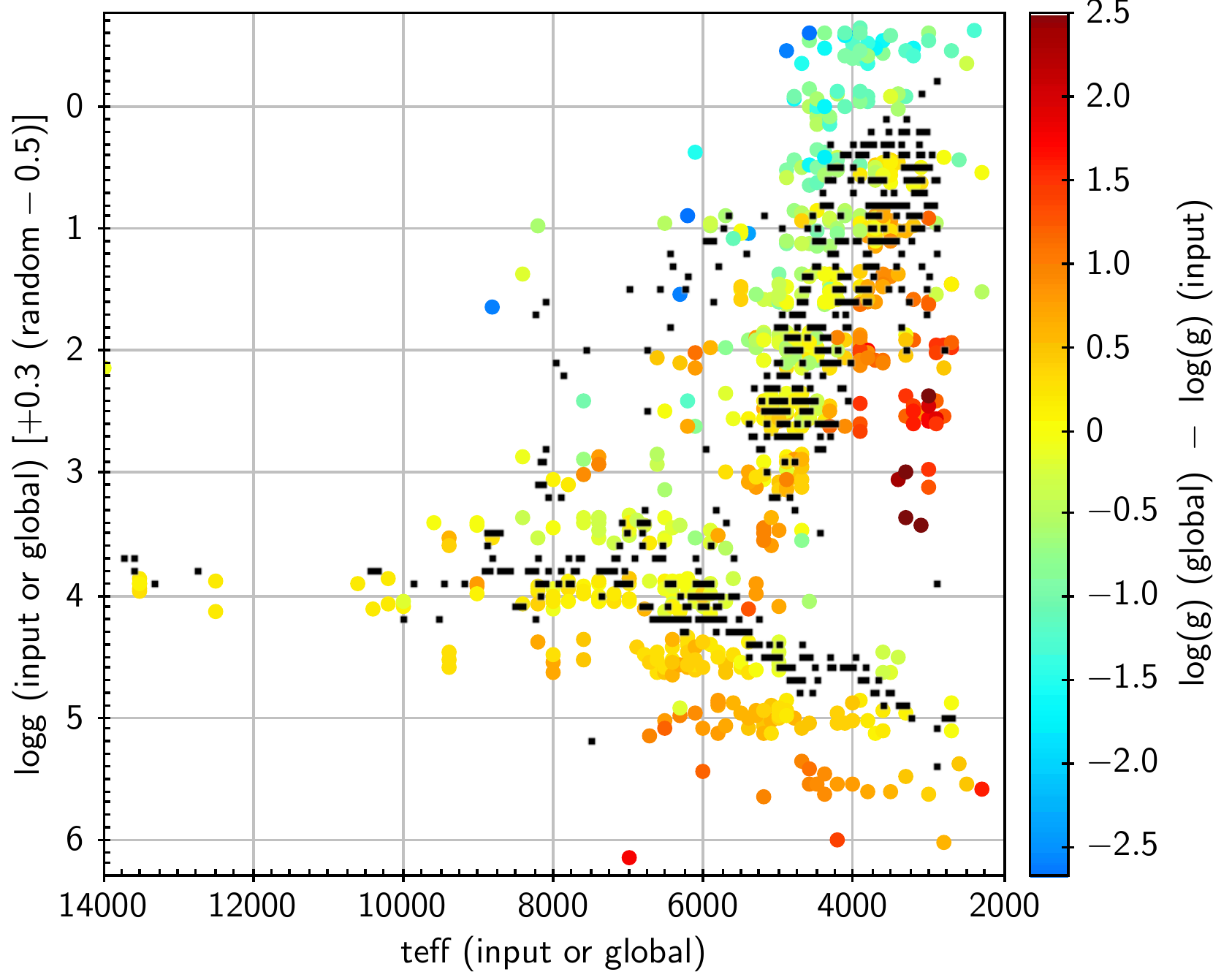} \hspace{0.5cm}
\includegraphics[clip=,width=0.45\textwidth]{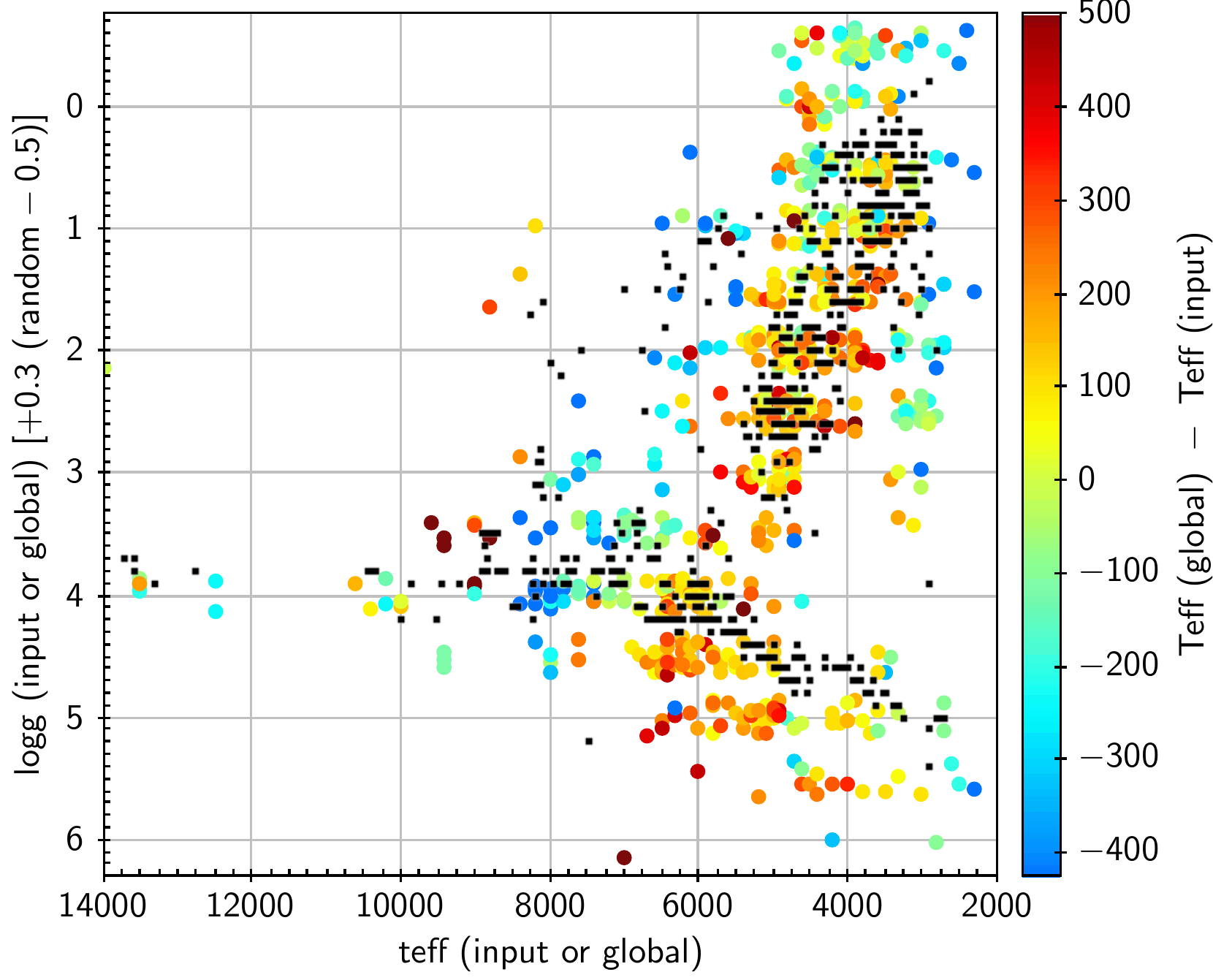}
\caption[]{Differences between input and output gravities and 
temperatures across the HR diagram.
The small black dots locate the XSL data according to our initial parameters, 
derived from the optical absorption line spectra using ULySS+MILES
\citep{Arentsen_PP_19}.
The larger dots locate the same XSL data using the best-fit parameters 
obtained from the global comparison of the empirical spectra
(UVB+VIS+NIR) with GSL models, reddened as necessary, at $R=3000$.
In the {\em left panel} color codes the differences in log($g$) between
the two estimates; in the {\em right panel} it codes the differences 
in \teff.  }
\label{fig:HRDbyDTorDg_double}
\end{figure*}

\begin{table*}
\caption[]{Effects of the fitting method on estimated parameters.}
\label{tab:param_errors}

\begin{tabular}{lcccccc} \hline \hline
  & \multicolumn{2}{c}{\em ~ R=500 vs. R=3000 ~} 
  & \multicolumn{2}{c}{\em ~ R=500, original model grid ~}
  & \multicolumn{2}{c}{\em ~ Weighted $\chi^2$ vs. $D$, at R=500 ~} \\
  &  \multicolumn{2}{c}{~} 
  & \multicolumn{2}{c}{\em vs. interpolated grid}
  & \multicolumn{2}{c}{\em ~ (with slightly different masks)} \\ \hline
Quantity & ~ Mean & Std. dev. ~ & ~ Mean & Std. dev. ~ & ~ Mean & Std. dev. ~ \\ \hline
$\delta$\,\teff\ (K) & 22 & 64 & 0.7 & 123 & 72 & 239 \\
$\delta$\,log($g$) \ (cm.s$^{-2}$) ~  & 0.04 & 0.22 & -0.03 & 0.5 & 0.02 & 0.9 \\
$\delta$\,[Fe/H] \ (dex) & 0.007 & 0.15 & 0.015 & 0.34 & 0.07 & 0.65 \\
$\delta$\,A$_V$ \ (mag) & 0.001 & 0.08 & 0.03 & 0.34 & 0.16 & 0.49 \\
\hline
%
\end{tabular}
\tablefoot{Each subsection of the table lists mean parameter
differences and standard deviations obtained when switching between
the two methods indicated above the column titles.
The listed values are based on fits to
data from ALL wavelengths of the XSL spectra (UVB+VIS+NIR).
%
}
\normalsize
%
%
\end{table*}

\begin{figure*}[h]
\begin{center}
\includegraphics[clip=,trim=0 35 0 0, width=0.32\textwidth]{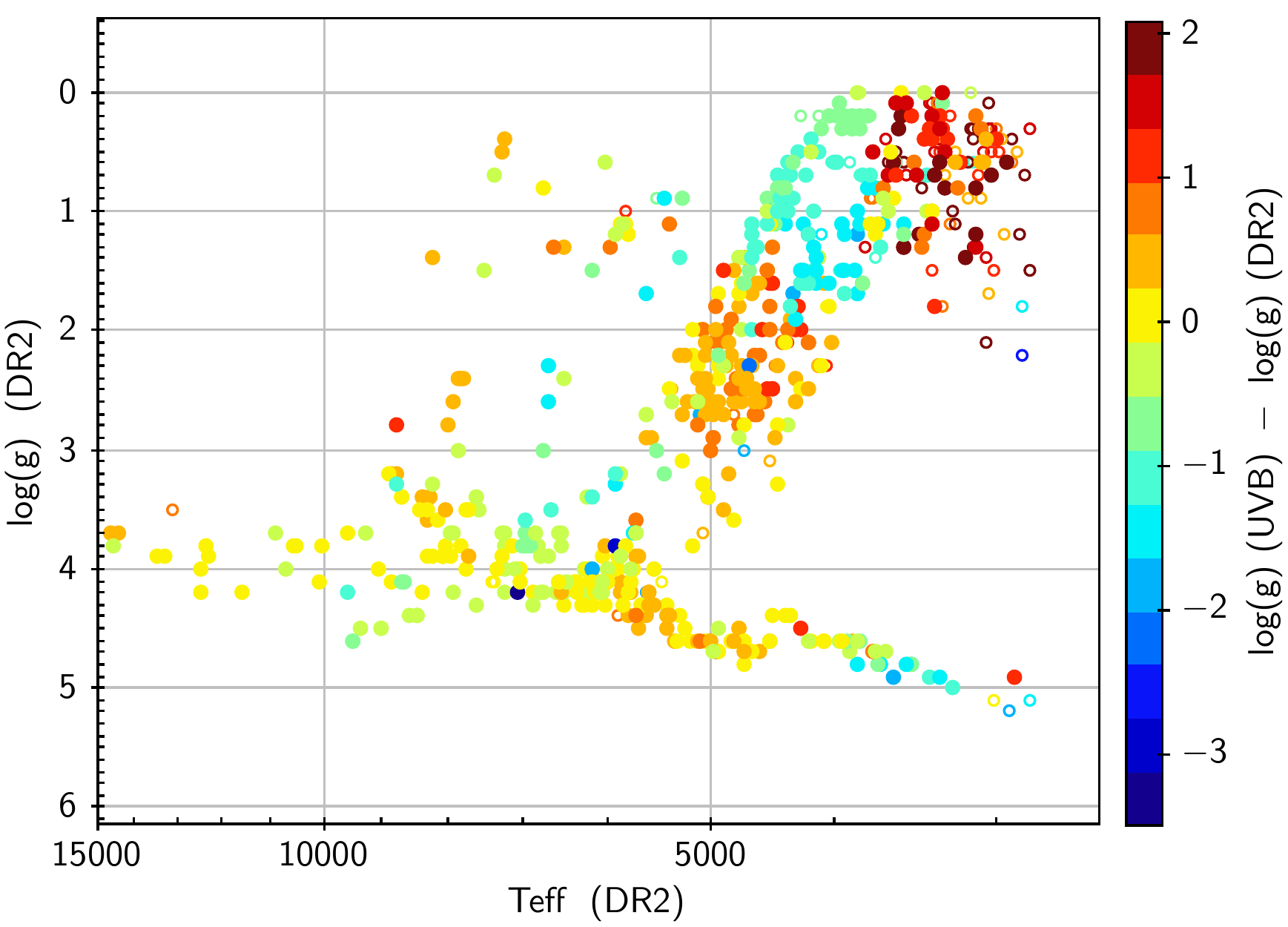}
\includegraphics[clip=,trim=0 35 0 0,width=0.32\textwidth]{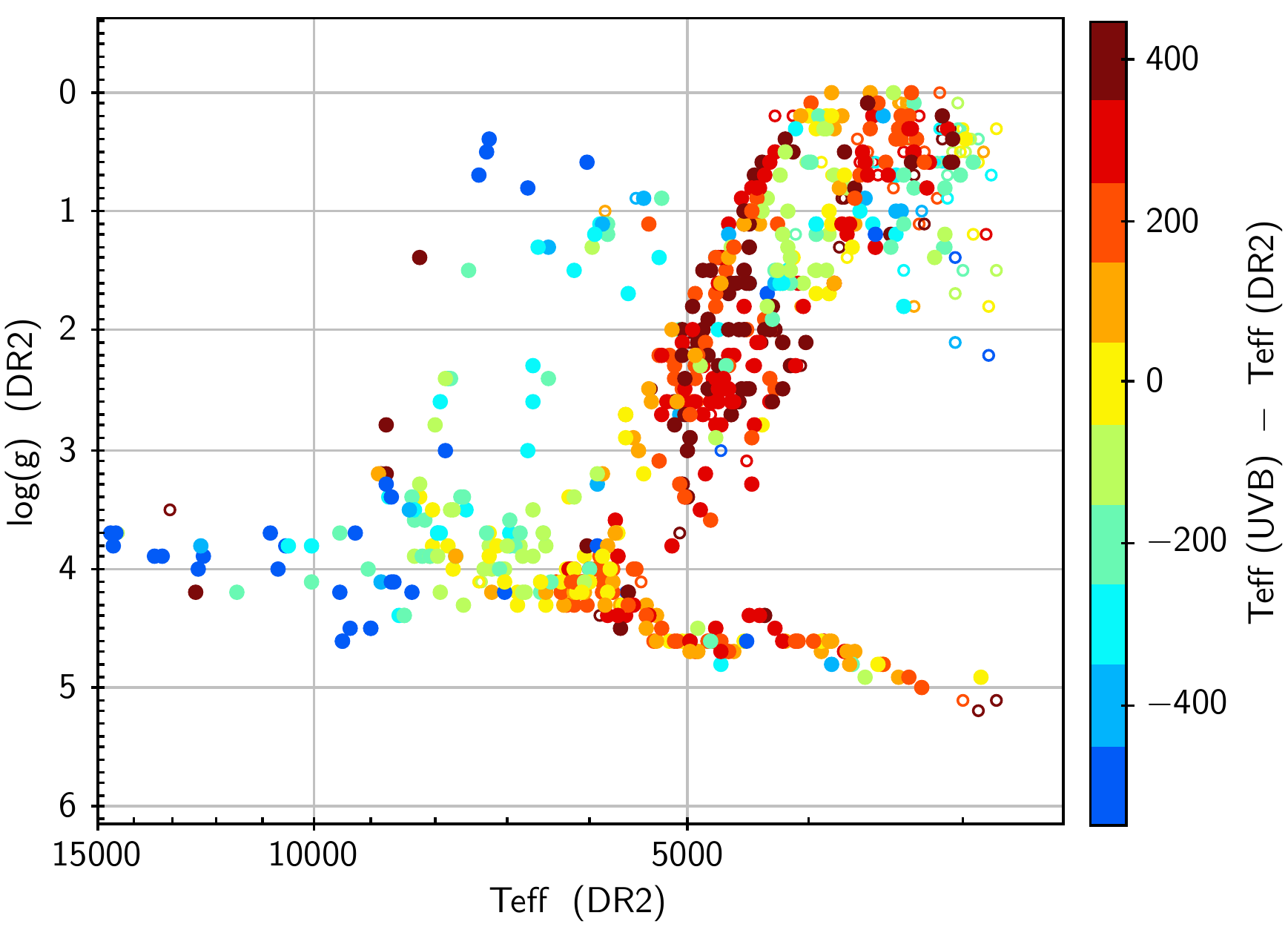}
\includegraphics[clip=,trim=0 35 0 0,width=0.32\textwidth]{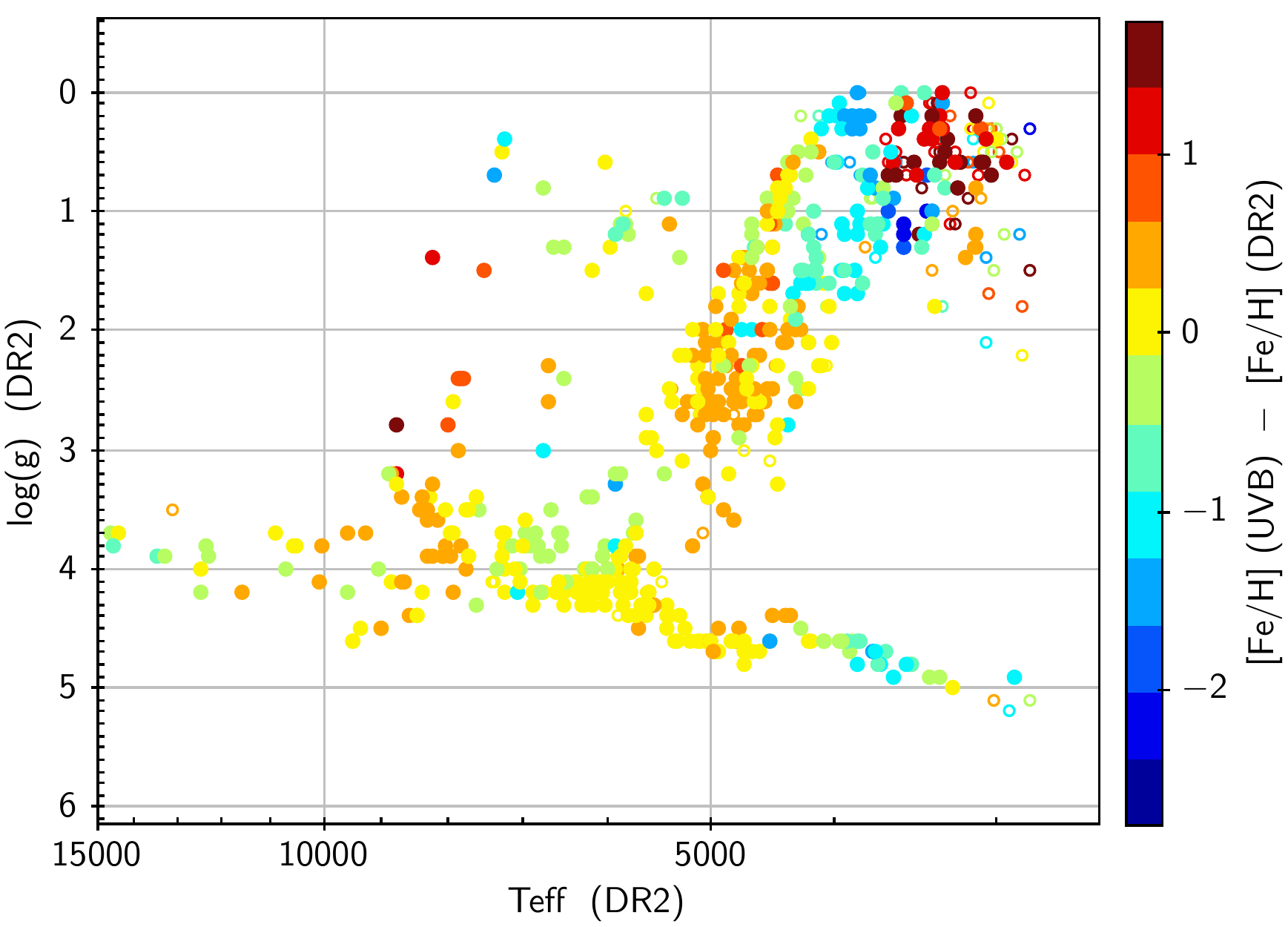} \\
\includegraphics[clip=,trim=0 35 0 0,width=0.32\textwidth]{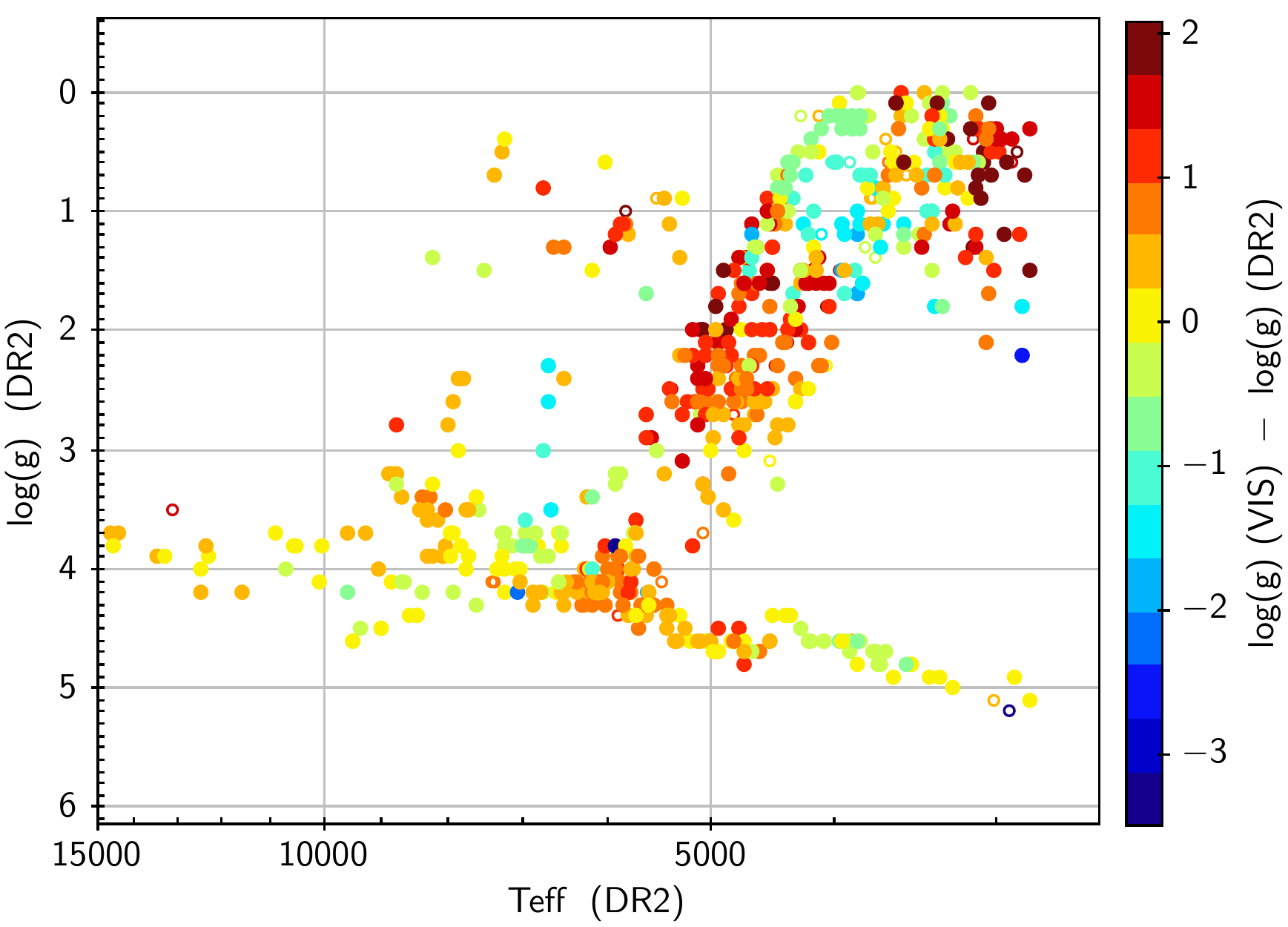}
\includegraphics[clip=,trim=0 35 0 0,width=0.32\textwidth]{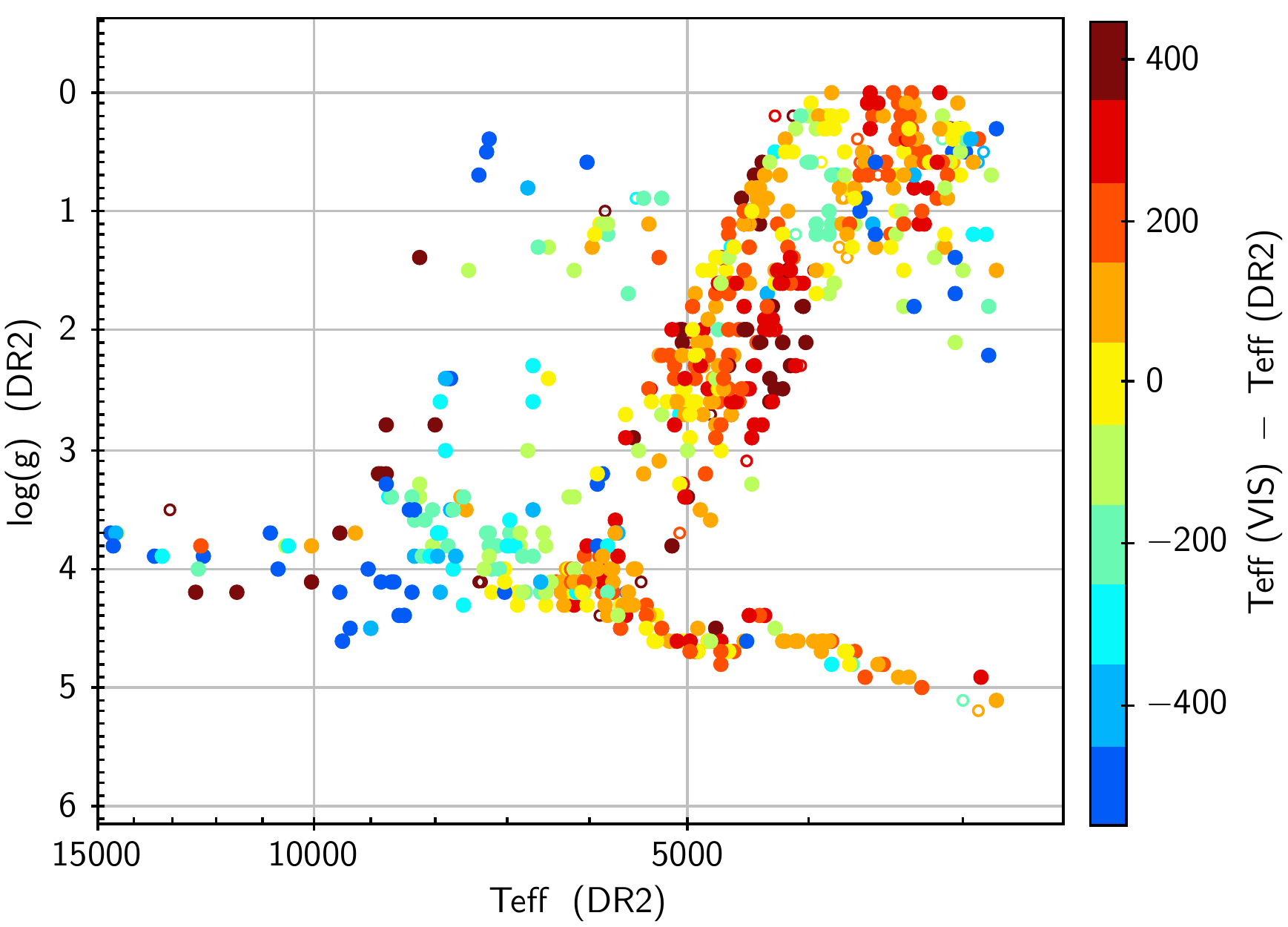}
\includegraphics[clip=,trim=0 35 0 0,width=0.32\textwidth]{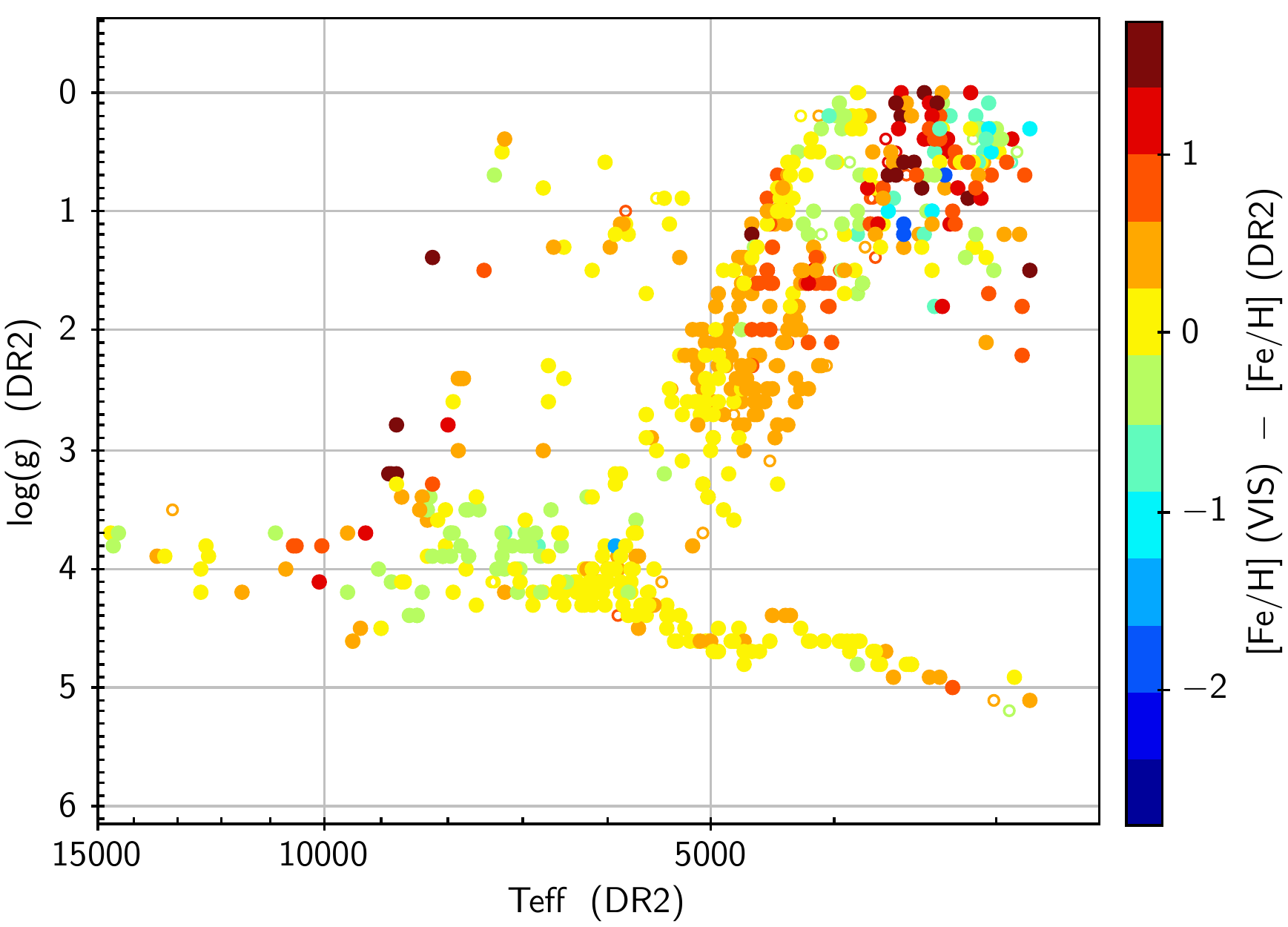} \\
\includegraphics[clip=,trim=0 35 0 0,width=0.32\textwidth]{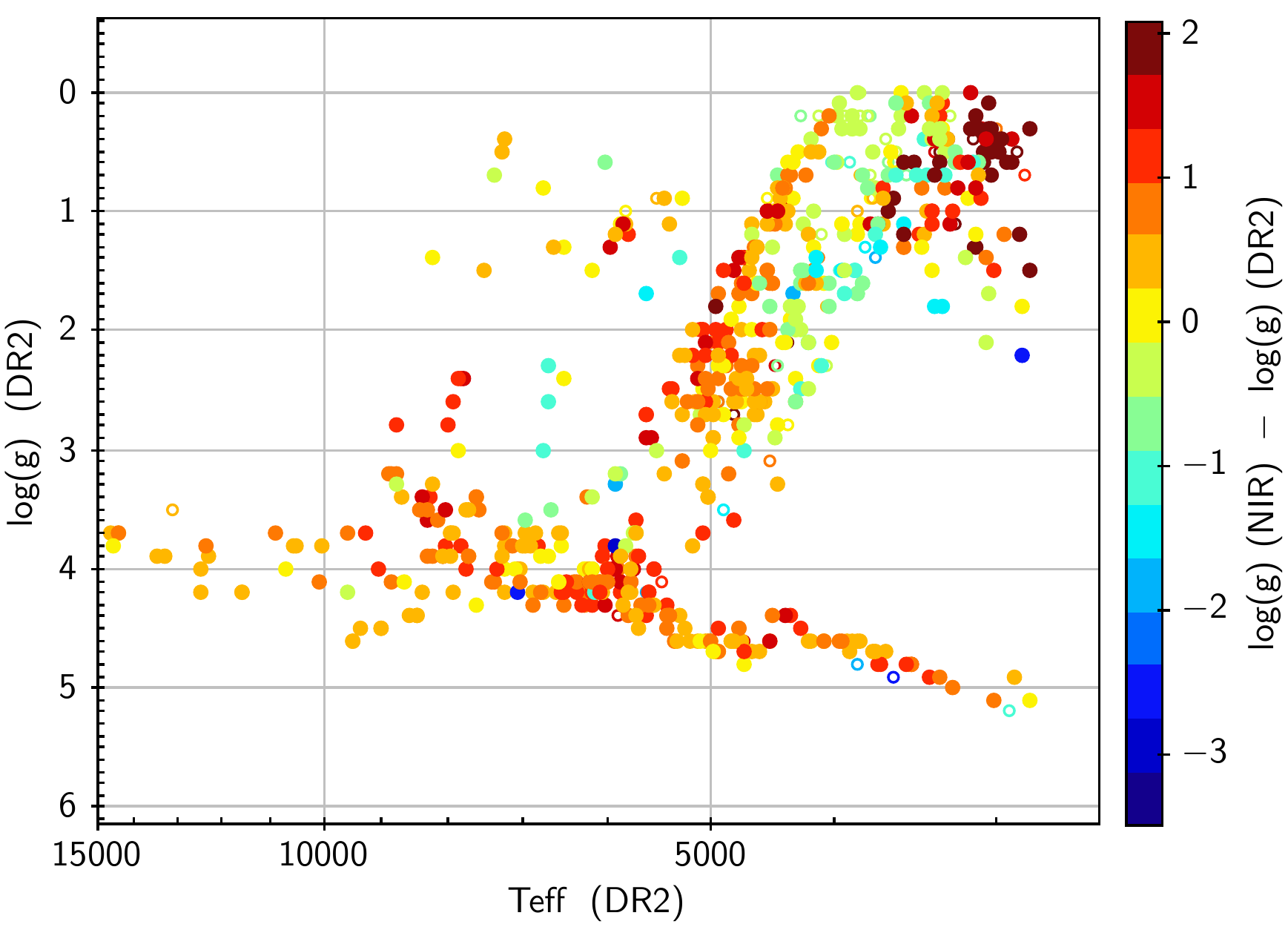}
\includegraphics[clip=,trim=0 35 0 0,width=0.32\textwidth]{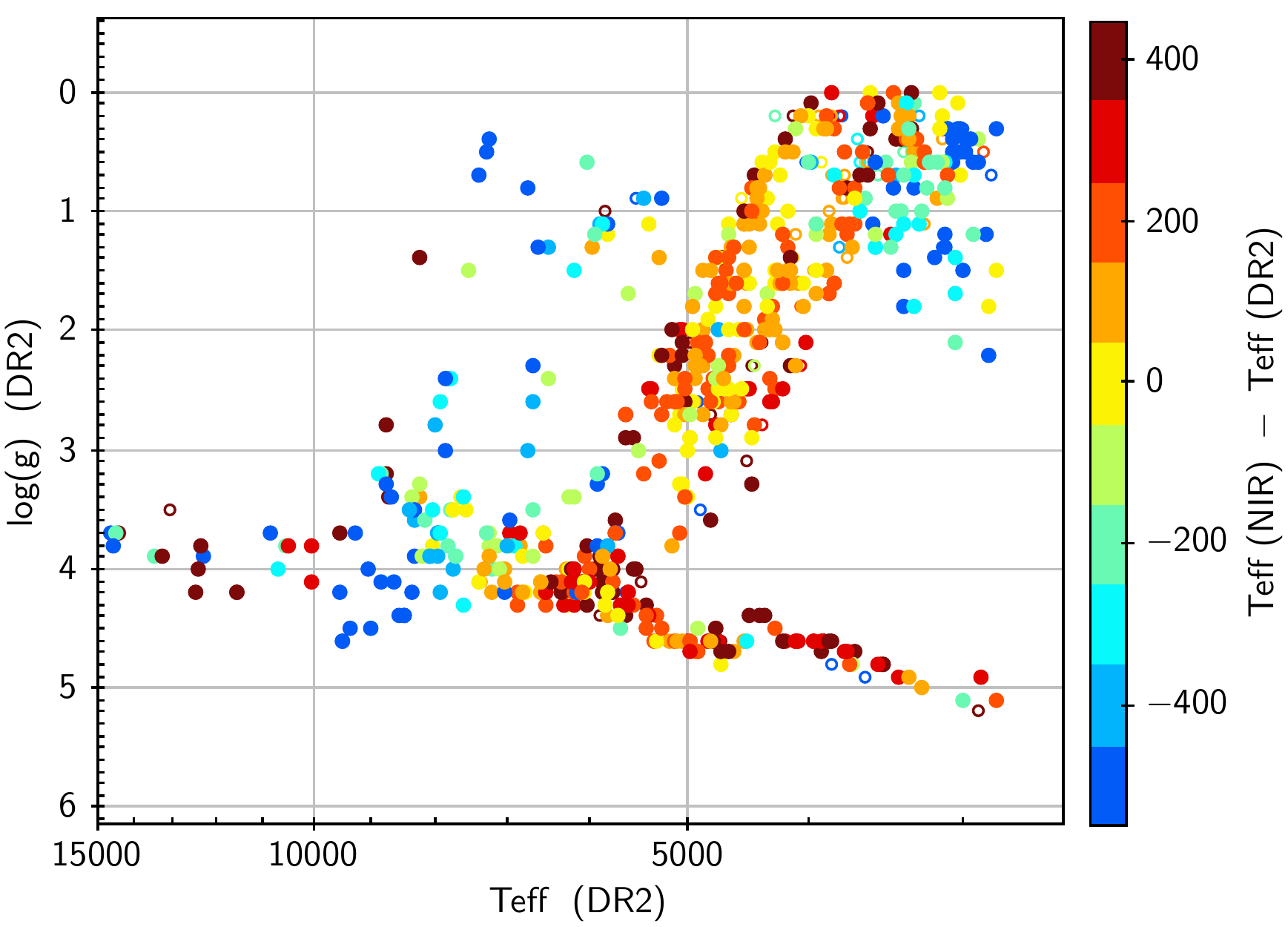}
\includegraphics[clip=,trim=0 35 0 0,width=0.32\textwidth]{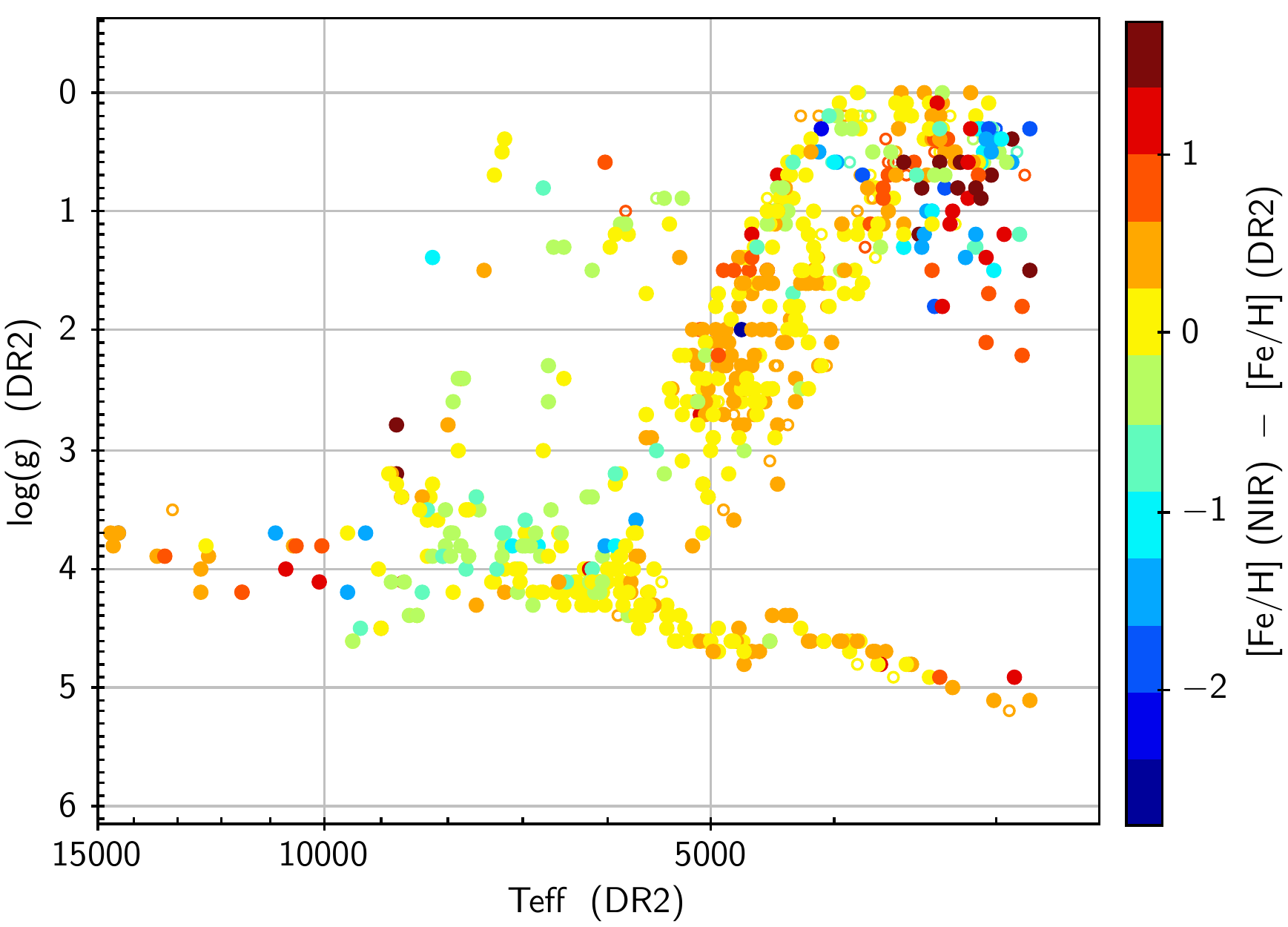} \\
\includegraphics[clip=,width=0.32\textwidth]{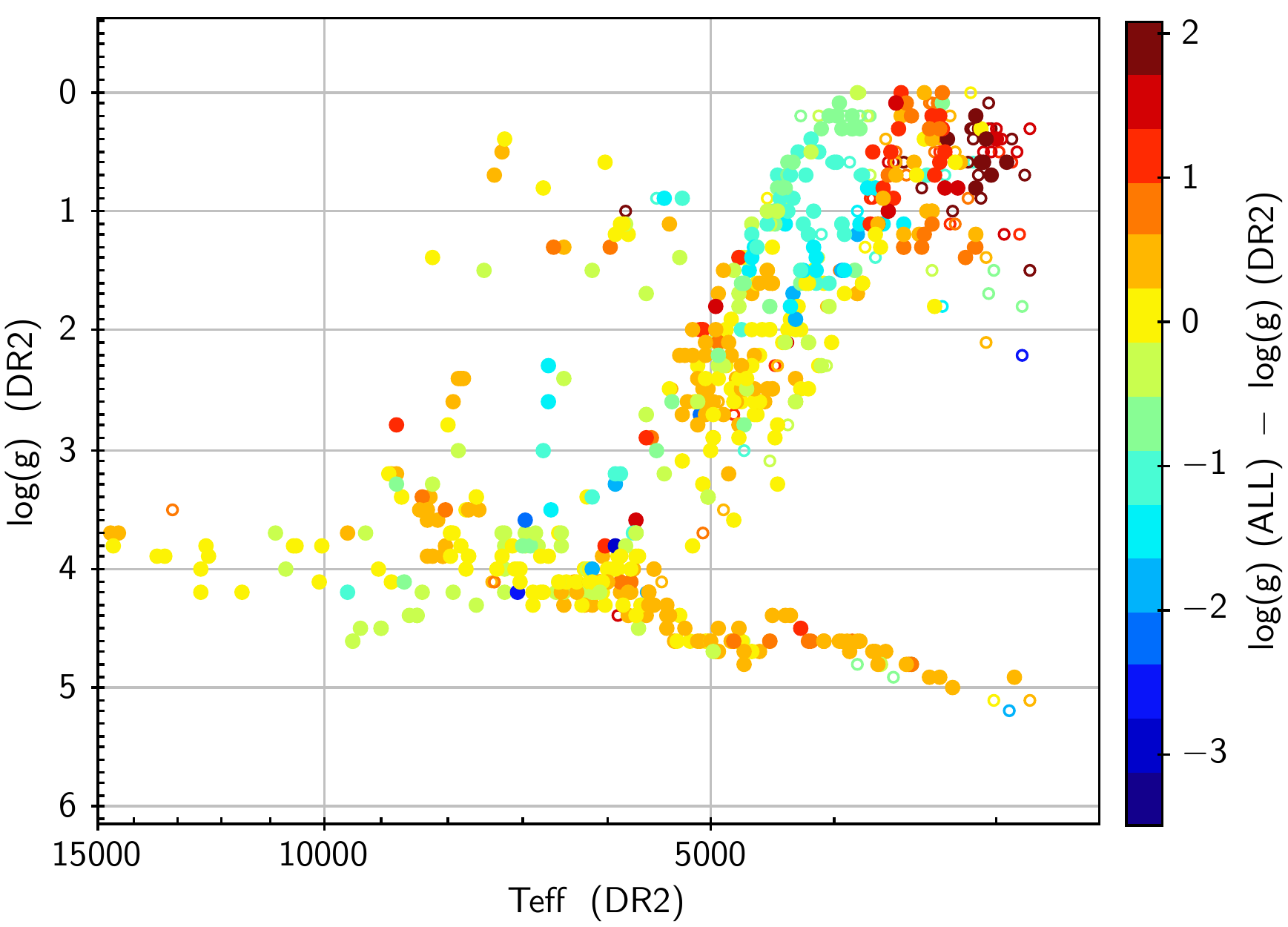}
\includegraphics[clip=,width=0.32\textwidth]{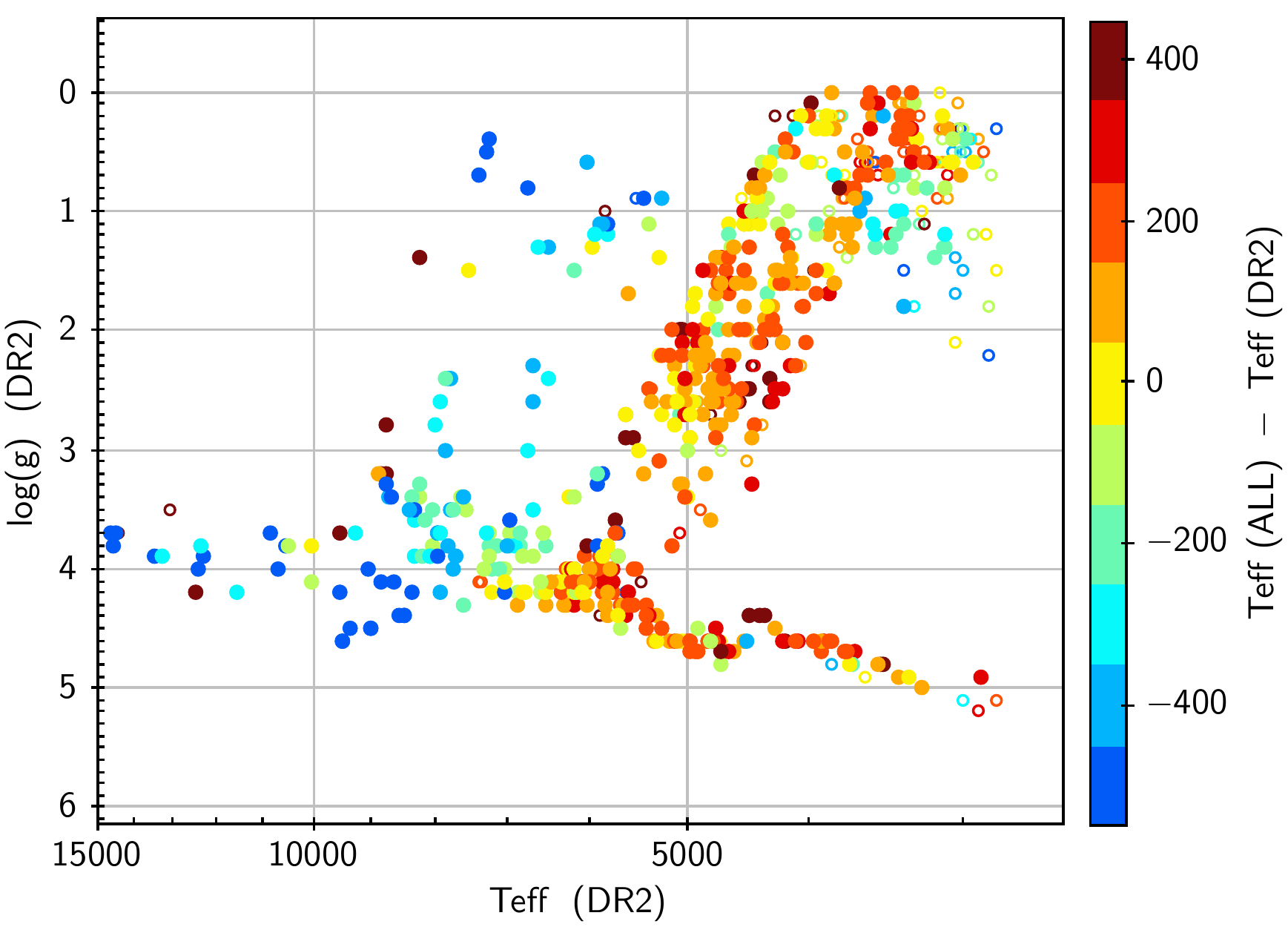}
\includegraphics[clip=,width=0.32\textwidth]{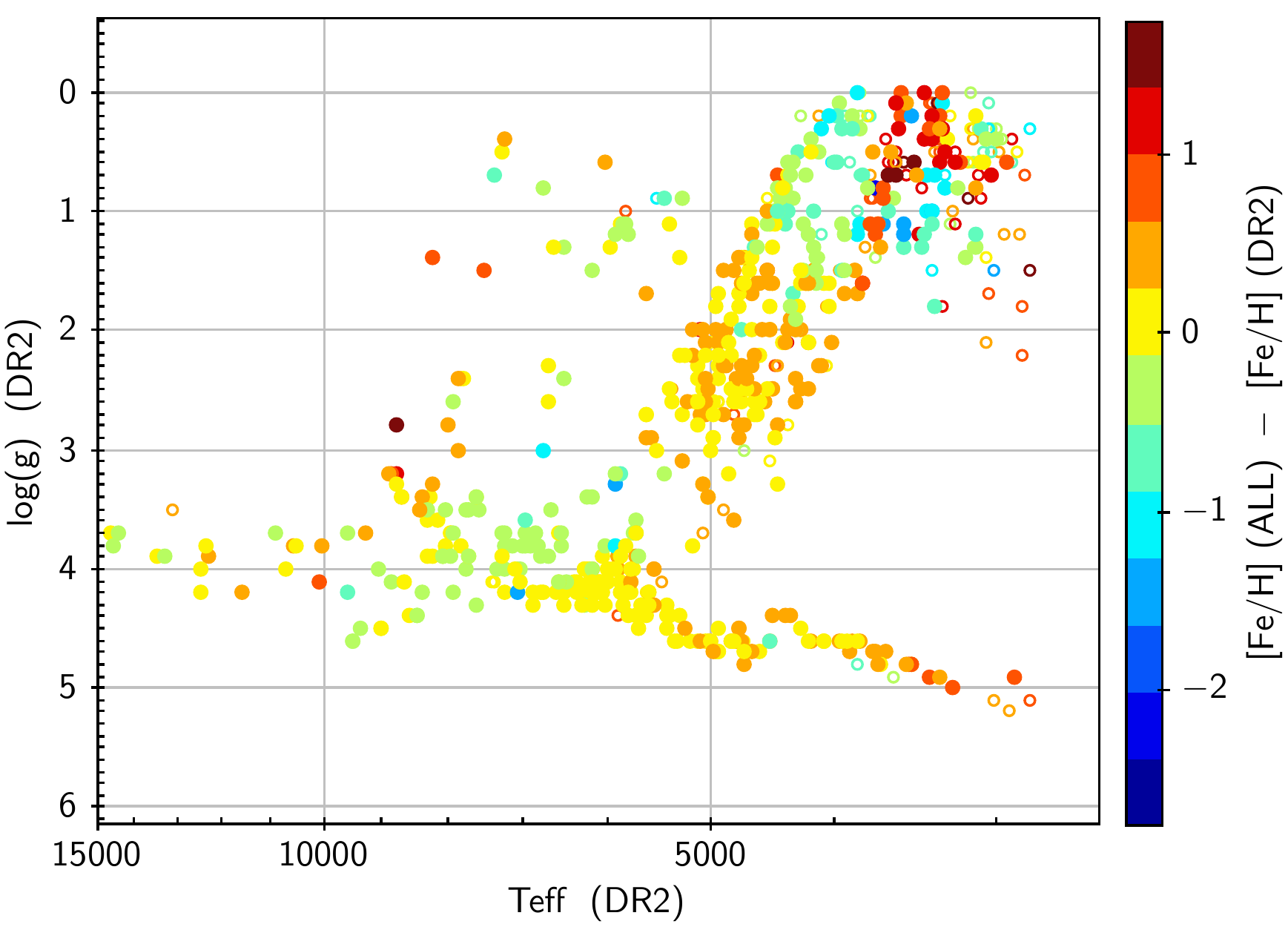}
\end{center}
%
\caption[]{Trends in the differences between the parameters of DR2 and those from 
best-matching GSL SEDs as a function of position in the HR diagram.
The data in the {\em left column} are colored by difference in log($g$),
the {\em middle column} by difference in \teff, and the {\em right column} by 
difference in [Fe/H]. From top to bottom, the spectral ranges used for the
comparison with GSL are the UVB, the VIS, the NIR, and all three arms of X-shooter.
}
\label{fig:HRDbyDTorDg}
\end{figure*}

A synthetic view of the differences between the parameters \teff\ and log($g$)
of Paper\,II and those obtained from the 
comparison with (reddened) theoretical spectra (at R=3000),
is presented in the HR-diagrams of Fig.\,\ref{fig:HRDbyDTorDg_double}. 
Compared to the positions
assigned in Paper\,II in \teff\ and log($g$) (black dots),
the parameters assigned via the joint comparison of the three arms
to GSL-spectra are systematically more 
dispersed (colored points). More dispersion is expected
as a result of flux calibration errors in the XSL data combined with
degeneracies between parameters, and is exacerbated
at low temperatures by the absence of good matches in the model collection
(due to the unique adopted extinction law, to the constant 
abundance ratios in the models, and to numerous other potential
discrepancies between the physics implemented in the models and reality). 
Despite the dispersion, the color-coding
in the figure shows systematic trends that are highly significant. 

The left panel of Fig.\,\ref{fig:HRDbyDTorDg_double}
highlights differences between the surface gravities
derived from the fits to GSL models and those of the initial guesses.
They are responsible for most of the broadening of the 
main sequence and of the giant branch seen in the GSL-based 
HR-diagram. 
For main sequence stars below $\sim 6000$\,K, the gravities preferred
by the comparison to GSL models are larger than the initial guesses
(the largest of these values exceed expectations from 
stellar evolution models). For
the red giants between 4000 and 5000\,K, the GSL-based gravities 
are on average lower and this trend is carried mostly
by the metal-poor giants ([Fe/H]\,$\leqslant \! -0.5$).
On the other hand, between 5000 and 5500\,K,
the gap seen in the initial parameters between
the main sequence and the giant branch is filled, 
when using GSL-based parameters, with objects
from the lower-luminosity giant branch that are assigned
higher gravities. 
The stars for which GSL-based effective temperatures
and gravities are around 3000\,K and log($g$)=2 to 3, are LPVs,
for which inspection of the empirical and theoretical 
spectra shows the models are obviously inadequate
\citep{LanconIAU18}. We do not discuss 
these objects further in this paper.

The main trends seen in the comparison between our initial
effective temperatures and those from the global GSL-fit,
in the right panel of Fig.\,\ref{fig:HRDbyDTorDg_double}, display a pattern as 
a function of position that differs from the one seen for the surface gravities.
For warm stars along the main sequence or in the transition 
region between the main sequence and the giant branch (blue loop,
blue horizontal branch), the \teff\ from GSL-fits is systematically 
lower than obtained from optical line studies, 
while differences in the opposite direction are found 
mostly along the giant branch and at
intermediate temperatures on the main sequence (4000 to 7000\,K). 
A comparison between the two panels
shows that there is no systematic correlation or anticorrelation 
between the effects on log($g$) and T$_{\mathrm{eff}}$
that would be valid across the whole HR diagram. 
Locally, correlations can be found, for instance, at intermediate 
temperatures on the main sequence, higher gravities are 
compensated with higher temperatures.

Because the direction
of the systematic offsets depend on the area of the HR diagram,
we can be confident that they reflect a real difference
between observed stars and the models, rather than errors in the data
(which should not correlate with position in the HR diagram).
The patches over which a coherent trend is observed do not
have shapes that would suggest an effect of the sampling of
parameter space in the model grid (the general 
aspect is identical whether or not interpolation is implemented
within the grid).

\smallskip

We now repeat the previous comparison, but using best-fits 
carried out in individual arms (Fig.\,\ref{fig:HRDbyDTorDg}). 
To ease comparisons between panels and with 
figures in previous sections,
the location of the points in the diagram is now again
based on the parameters of Paper\,II (not on the comparison with GSL);
the bottom panels repeat the data from Fig.\,\ref{fig:HRDbyDTorDg_double},
in this preferred format.

A striking result from this exercise is the similarity
between the trends found, whether the fits are based on
the UVB, the VIS, the NIR or all three spectral ranges.
It shows that certain major systematics are very robust.

\smallskip

While the trends are similar for the different fitting methods
(usage of $D$ or of the weighted $\chi^2$, areas in the UVB
arm excluded or not by a mask, resolution adopted for the fits),
the individual estimated parameters depend on these.
The errors also depend on the area of the HR diagram considered, 
with larger errors where the general quality of the fit is poorer,
but we provide only a global summary in Table\,\ref{tab:param_errors}.
The method-induced changes 
$\delta$\,\teff,  $\delta$\,log($g$),  $\delta$\,[Fe/H], and $\delta$\,A$_V$
are all positively correlated with each other, 
within a given comparison-experiment.
None of the method-changes erases the trend seen in 
A$_V$(UVB)$-$A$_V$(VIS) versus \teff.

\medskip

\paragraph{Metallicity --} 

The differences between the best-fit metallicities found here and the
DR2 values from Paper\,II are displayed in the 
right-hand panels of Fig.\,\ref{fig:HRDbyDTorDg}. 
Over vast areas of the HR diagram, the average local offsets are smaller than
0.5\,dex. In particular, main sequence metallicities between 5000 
and 7000\,K are similar to those in DR2, despite the systematic offsets
in temperature and log($g$) shown in the left-hand panels. 

Along the giant branch, the offsets in [Fe/H] evolve from typically positive
values at high gravity (low luminosity) to typically negative values at low gravity
(high luminosity). The higher metallicities in the low luminosity regime
contribute to the excessive strength of the metal lines between  490 and 
550\,nm (UVB arm) that were highlighted as a systematic feature of the
spectral residuals in Sect.\,\ref{sec:results_best_residualDiscrepancies}.
It is noteworthy that these higher metallicities did not suffice to solve the 
apparent issue of lacking absorption at the blue end of the UVB arm,
in this regime. At higher luminosities, lower metallicities are selected 
with our SED-fits; the metal-lines at the red end of the UVB wavelength 
range tend to match those of the observed spectra better, but on 
average the synthetic lines remain deeper than those observed even there.

In view of the complex patterns in Fig.\,\ref{fig:HRDbyDTorDg}, an 
exhaustive analysis would require separate in-depth studies in each
subarea of the HR diagram, with tunable models. This is an enormous task, 
that we postpone until the global study presented here has been
repeated with other model collections of the literature.

\bigskip

\subsection{Varying the [$\alpha$/Fe] ratio}

In the above, we have not insisted on the [$\alpha$/Fe] ratio.
When searching for the best-fit with the methods of this paper, 
the uncertainties on individual derived parameters
are too large to allow a robust estimate of [$\alpha$/Fe] per star. In particular,
the preferred [$\alpha$/Fe] ratio depends on the wavelength range 
used (left panels of Fig.\,\ref{fig:effect_aFe_x16} of the 
appendix). The UVB range is in principle sensitive to [$\alpha$/Fe],
via CaII, CN, and CH features that are obvious in the residuals.
In practice, the UVB range favors GSL models with a 
super-solar [$\alpha$/Fe] almost everywhere in the HR diagram.
The VIS range is sensitive to [$\alpha$/Fe] in certain regions
of parameter space; for instance the CN bands that are prominent
in giants around 4500\,K react to changes in
oxygen-proportions via the chemical networks involving C,N, and O
\citep[see][for examples of the complex behavior
of the CN bands]{Lancon_etal07_RSG, Aringer_etal2016}. 
The near-infrared calcium triplet ($\sim 880$\,nm)
is also located in the VIS range of X-shooter. 
The VIS-fits point to super-solar
[$\alpha$/Fe] mostly for low-metallicity giants and for intermediate
temperature metal-poor main sequence stars, but also for about
half the stars above [Fe/H]$=-0.5$ (Table~\ref{tab:pref_aFe_bestfit}).
The NIR arm, or the SED as a whole, either show no preference
or weakly favor solar models in some parts of the HR diagram.
Hence, the expected anticorrelation between [$\alpha$/Fe] 
and [Fe/H] characteristic of the Milky Way 
is recovered only when applying a relatively fine-tuned
selection to define the ``preferred" [$\alpha$/Fe]. 

In the above sections, we have adopted solar-scaled
models by default, except for stars for which $\alpha$-enhanced were 
favored in the UVB and not significantly disfavored in the
VIS and NIR. The $\alpha$-enhanced subset with this definition
contains less than 15\,\% of the metal-rich stars ([Fe/H]$>-0.5$ in
Arentsen et al. 2019), and about 75\,\% of the metal-poor stars
([Fe/H]$<-1.5$). We refer to Appendix \ref{app:aFe_bestfit} for details.
An attempt to extract [$\alpha$/Fe] estimates
for the XSL stars from the literature rather than perform
our own evaluation was abandoned because of the heterogeneity
of the results found in publications.
The uncertainty on [$\alpha$/Fe] is one of the reasons why we 
have restricted our discussion to major global trends.

\section{Discussion}
\label{sec:discussion}

\subsection{Comparison with trends exposed in the literature}
\label{sec:comparison_with_lit}

One of the earliest studies comparable to the one we
exposed was published by 
\citet{Bertone_etal04}, who compared ATLAS 
and PHOENIX models of that time with each other
and with more than 300 empirical low-resolution spectra 
that extended from 350 to 1050\,nm. Their method was similar to ours, 
although they chose to compare logarithms of fluxes, 
and to correct for Galactic extinction a priori. They concluded
that  the fit quality was good for spectral type F and earlier 
and degraded drastically
with later spectral types. Today we would soften this
judgement to include at least the early G type stars
in the ensemble of stars that can be represented 
with average spectral residuals of a few percent. 
This was also the conclusion of 
\citet{Coelho_2014} based on ATLAS and MARCS
model atmospheres and ATLAS9/SYNTHE
synthetic fluxes. In the remainder of this section,
we focus on temperatures below this threshold, with
an emphasis on giants between 4000 and 5000\,K.

It is well known that cool stellar spectra are difficult to
model and analyze \citep{Bessell_etal89_static, 
Westera_etal2002, Martins_Coelho_2007, Lebzelter_etal2012, Short_etal2012, 
Davies_etal2013, Coelho_2014, Franchini_etal18_Intrigoss}. 
As we cannot possibly review the vast body of pre-existing 
work exhaustively, we concentrate on comparing the directions 
of the trends found here with those found previously. 
%

\citet{Bertone_etal04} compared the temperatures they
derived from the low resolution spectra to empirical temperature 
calibrations as a function of spectral type (based in
particular on angular diameter measurements and surface 
brightnesses). For K and M giants, 
their fit-\teff\ were warmer than those of the empirical calibrations
by 4 to 8\,\%, for both giant and dwarf stars. This trend seems 
to have withstood the passing of time, though its amplitude
may be decreasing. 
Our Fig.\,\ref{fig:HRDbyDTorDg_double} 
shows that the temperatures obtained with more recent 
PHOENIX models, after a significant update in line opacities
in 2008, still lead to higher {\teff}-values than those of the
literature for similar stars (as traced by the 
parameters of Paper\,II).
We find that this remains true for any of the UVB, VIS, NIR and
ALL fitting-ranges. Our study shows that the discrepancy
cannot be blamed on a detail, but has a generic cause
that will be fundamental to elucidate.

Similar offsets have been found in other recent studies.
For instance,
\citet{Husser_etal2016} and \citet{Jain_etal2020} analyzed low-resolution
optical spectra of hundreds of stars in the globular cluster
NGC 6397 ([Fe/H]\,$\simeq -2$), by comparing them on one hand
with PHOENIX spectra (the GSL collection also used here), on the other with
the empirical collections MILES and ELODIE 
(as was done for XSL in Paper\,II). Compared
to our study, two differences in methodology are 
noteworthy: stellar gravities were derived prior to the 
spectroscopic study, from the comparison of the 
broad band photometry with a theoretical isochrone; 
and a multiplicative polynomial was allowed to absorb 
extinction and flux calibration errors. Despite these
differences, the trends resemble those obtained here with XSL:
the temperatures assigned to giant stars based on GSL are
higher by 150 to 200\,K than those obtained 
with the empirical libraries. The studies in NGC 6397
also found that the [Fe/H] assigned with GSL were higher,
as expected from the classical temperature-metallicity 
degeneracy. In our study, this correlation is recovered when
fitting only the VIS arm of X-shooter, but it is lost in dispersion
when the other wavelength ranges are used. 

A comprehensive re-analysis of giant stars in 16 Milky Way
clusters, with a range of metallicities, was carried out by
\citet{Mucciarelli_Bonifacio_2020}. They compared stellar parameters
from high resolution optical lines on one hand (excitation and ionization 
balance, for spectral lines between 480 and 680 nm), 
with parameters from dereddened broad-band colors on
the other (metallicity-dependent color-\teff\ 
relations based on the infrared flux method).  
They found increasing discrepancies between spectroscopic and 
photometric parameters, with decreasing cluster metallicity. This 
larger tension at low metallicity between the SEDs and the spectral 
features is a trend we have also noted in the comparison 
between XSL and GSL. It is difficult to compare the trends 
between the two studies more quantitatively, because none of the methods
used by \citet{Mucciarelli_Bonifacio_2020} corresponds to those
used here precisely. The strongest of these trends, seen at
low metallicity ([Fe/H]$<-1.7$ in their sample), is a shift to 
lower temperatures and lower gravities
when switching from photometric to spectroscopic parameters.
The signs of these offsets correspond to our 
results with XSL (though with some dependence on the X-shooter arm
and on luminosity in the case of gravity), if we assimilate the parameters 
of Paper\,II with Mucciarelli \& Bonifacio's 
``spectroscopic parameters'' and those 
from the fits to the GSL-SEDs to the "photometric" ones (Figs.\ref{fig:HRDbyDTorDg_double} 
and \,\ref{fig:HRDbyDTorDg}).  This pairing of analysis methods is certainly
not perfect, but it is more reasonable than the alternative pairing would be. 
%

\medskip

The studies above focus on individual stars or on stellar populations that can be resolved
into stars. The two recent articles we now discuss compare empirical and theoretical stellar
spectra in the context of the calculation of the integrated spectra of unresolved stellar populations. 

\citet{Maraston_etal20} exploit the empirical library MaStar
\citep[][$\lambda \lambda$ 362--1035\,nm, $R\simeq 1800$]{Yan_etal19_MaStar}, 
and discuss parameters from two estimation methods.
One is very similar to the method adopted in
Paper\,II and rests on the MILES library
\citep{Chen_etal2020}. 
The other is based on a comparison with
theoretical spectra taken mostly from the BOSZ-ATLAS9 calculations 
of \citet{Meszaros_etal12} and \citet{Bohlin_etal17}.
For simplicity, we refer to estimates from the latter method as theoretical 
parameters, as in the original article. Both methods 
discard information present in the stellar continua, by allowing 
for a multiplicative polynomial in the fits.

A number of trends found by Maraston et al. are similar to the ones 
we find here with XSL and GSL, despite different fitting methods 
and reference models. 
For instance, the theoretical surface gravities are lower than
those based on MILES for luminous red giants,
while they tend to be higher for main sequence stars.
As in our case, the dispersion of the data points in log($g$)-{\teff}\ 
planes is much larger when using theoretical parameters than when
using parameters based on MILES (figures 3, 7 11 of Maraston et al.). 

\citet{Coelho_BC_2020} confront the
dereddened empirical spectra of the MILES library 
($\lambda\lambda$ 354--741\,nm) to those they computed 
for atmosphere models constructed 
with ATLAS or MARCS, assuming spherical symmetry 
for giants below 4000\,K.  They adopt stellar 
parameters from the literature, from references
that mostly match those adopted in Paper\,II. 
Hence their results can be compared to our discussion in 
Sect.\,\ref{sec:results_forced}, albeit with a difference in
the extinction correction.
Their goodness-of-fit results are similar to those we obtain 
here: the fit-quality starts to degrade below 5000\,K. Interestingly, the 
two fit-discrepancy measures they define 
reach their minimum at different effective temperatures.
One is a weighted $\chi^2$ that exploits the empirical noise spectrum,
the other a summed fractional difference $\tilde\Delta$ similar in essence to our $D$.
Like our $D$, their $\tilde\Delta$ rises quickly below 5000\,K
while the $\chi^2$ only picks up large values below about 4000\,K.
This confirms our impression that $D$ or $\tilde\Delta$
might be too sensitive to regions of low flux in the blue part of the spectrum
of cool stars. 

For more detailed comparisons, \citet{Coelho_BC_2020} 
refer to \citet{Coelho_2014}. The spectral residuals shown
there for temperatures below 5000\,K are of amplitudes 
that seem roughly comparable to those found here between XSL
and GSL. \citet{Coelho_2014} experiment in using MILES to
re-evaluate parameters starting from their synthetic optical
spectra. The results however are presented (and were
archived) in a very 
compact way that does not allow us to directly compare trends
with those found here (their Fig.\,12). We intend, instead,
to compare XSL and the models of \citet{Coelho_2014}
directly in the future.

\medskip

\subsection{Limitations of the models}
\label{sec:limitations_of_models}

The spectral properties of cool stellar models depend on numerous
parameters and even extensive grids are inevitably limited. 
The systematic offsets between our best-SED parameters
and those of Paper\,II are likely due to a 
combination of issues in the model ingredients and
systematics in the parameters assigned to MILES and ELODIE
spectra in the literature. The complex discrepancies between 
empirical and best-fit theoretical spectra, in particular below 5000\,K,
demonstrate that the models carry their share of responsibility,
and Sect.\,\ref{sec:comparison_with_lit} shows that this is not an
isolated problem of the GSL collection. Here, we examine
a few of the likely issues that affect GSL in particular, and 
a number of other collections by generalization.

In spherical symmetry, the assumed relation between mass and position in
the HR-diagram sets the extension of the 
atmosphere models, which affects molecular band depths
and colors \citep[figure \,8 of][]{Aringer_etal2016, Hauschildt_etal99, 
Gustafsson_etal08}. This is important only at low gravities.
In GSL, the masses assumed at low temperatures do not exceed
a few solar masses. This leads to relatively large atmospheric 
extensions, to relatively low luminosities, and it is more adequate for giants than
for red supergiants. Models more specifically designed for 
red supergiants, with masses of 15\,M$_{\odot}$ were confronted
with observations across the optical and near-infrared spectrum 
by \citet{Lancon_etal07_RSG} and \citet{Davies_etal2013}. 
The former are based on PHOENIX models
and adopt surface abundances specifically taylored for red supergiants
but with line lists that are obsolete compared to those used here;
the latter are based on MARCS models with solar scaled abundances
and prior assumptions on metallicity and gravity. Apart from the
general conclusion that the features and energy distributions are
difficult to match simultaneously for cool red supergiants, it is difficult
to compare the two studies. In particular, the ``optical" wavelength 
range used by Davies et al. for their comparisons differs from our
VIS or UVB range: the range they select is specifically sensitive to 
TiO absorption and excludes the red end of our VIS range where
CN bands can be dominant (around 4500\,K). The two molecules
form bands at different depths in the atmosphere (TiO further
out), and this makes the comparison sensitive to the (different)
extensions of the atmospheres in the two collections. Future
comparisons should exploit independent measurements of luminosity
and include models with a range of masses in the low-gravity
regime.

The assumption of Local Thermal Equilibrium
(in GSL and in many synthetic spectral
libraries) impacts the stratification of atmospheres
and subsequently the line strengths, line profiles, 
and SEDs. 
According to \citet{Short_Hauschildt_2003}
and \citet{Short_etal2012}, the tendency between
4000 and 5000\,K is for 
non-LTE spherical models to have slightly higher 
gas temperatures in their atmospheres (i.e., at optical
depths smaller than one), but with a complex 
{\teff}-dependence in the very outer layers. This 
makes it difficult to identify "rule-of-thumb" behaviors
for molecular bands.
The studies also suggest that non-LTE models 
produce more short-wavelength flux than LTE models,
which we anticipate would increase the SED-issues
that we have illustrated in this paper via the discrepancies between 
extinction  estimates in the UVB and in the VIS range of the spectra.

Although atmosphere models generally include turbulent 
energy transport via the mixing-length theory, the effects of 
turbulence are by no means treated completely in 
current calculations.  Some of the additional velocity dispersion 
is accounted for in the spectral synthesis via a micorturbulence parameter. 
Its values affect the importance of line blanketing, 
as well as relative strengths of saturated and unsaturated 
lines in spectra \citep{Tsuji_1976, Short_Hauschildt_2010,
Lancon_Hauschildt_2010}. But the surfaces of red giants
and supergiants have convective patterns on larger
scales that translate into distributions of surface temperatures
and effective gravities, and their dynamical behavior
may lead to more extended atmospheres, outer molecular
envelopes, or the formation of dust. Further developments
of 3D-calculations \citep[e.g.,][]{Chiavassa_etal18, 
Hoefner_Freytag_19}
will probably be necessary to assess
the relevance of these processes to the relation between
SEDs and spectral features. In particular it is unclear 
whether significant systematic effects could be produced
above \teff$\simeq$4000\,K.
\smallskip

We now move from physical to chemical considerations. 
In the GSL library, the metal abundance ratios are fixed 
except for [$\alpha$/Fe]. In this article, 
we have explored only two values of this ratio, but 
based on this exploration
we do not expect [$\alpha$/Fe] by itself to resolve the tensions
between data and models that we have discussed.
It seems necessary to also account for dredge-up, or
more generally to use surface abundances that account for stellar evolution.
Dredge-up in particular affects C,N,O abundances
\citep[e.g.,][]{Georgy_etal2013_III}, 
and these provide some of the dominant molecular 
features between 4000 and 5000\,K
\citep{Lancon_etal07_RSG, Aringer_etal2016}.
The clear $^{13}$CO bandheads seen beyond 2.3\,$\mu$m
in many XSL spectra of giants also signal dredge-up
and are not reproduced by the models. At low 
metallicity, carbon-enhancements are common in the Milky Way;
this may contribute to explain the tension we observed between
spectral features and SEDs in this regime, and
it should be accounted for in future synthetic spectral libraries.

Another common particularity of synthetic spectral libraries,
including GSL, is a constant value of the He/H ratio at all metallicities.
This is at odds with the positive $\Delta Y/ \Delta Z$
representative of the global chemical evolution of a galaxy, that
is usually assumed in stellar evolution calculations
\citep{Bressan_etal12, Georgy_etal2013_I}.
%
The helium fraction impacts the stellar structure 
more than it affects the calculated spectrum for a given 
stellar structure, hence the latter is often neglected 
\citep{Girardi_etal2007, 
Dotter_etal2008, Chantereau_etal2016, Chung_etal2017}. Indeed, at 
temperatures above about 4500\,K and in all but the bluest photometric
passbands, moderate He-enhancements ($\Delta Y \simeq +0.1$ at a given [M/H])
change bolometric corrections by less than 1\,\% in plane-parallel
models \citep{Girardi_etal2007}. However changes induced by such 
He-enhancements on the bolometric corrections and colors 
can be larger than 5\,\% below 4000\,K,
and it seems that systematic studies with spherical models remain to be
carried out. 

Finally, we note that even with the correct abundances mismatches
at some level are expected from incomplete opacity data. 
The PHOENIX calculations include lists of empirically validated spectral lines
but also theoretical lists, in which line wavelengths and strengths
are given with poor precision. These so-called ``predicted lines"
are mostly weak ones, but are very numerous. Once the spectra, 
initially calculated with a spectral sampling resolution 
near 500\,000, are smoothed to $R=3000$, they manifest to
first order as
a pseudo-continuous absorption. As shown by \citet{Coelho_2014},
albeit with a different spectral synthesis code than used here
and not with exactly the same opacity input, this effect 
increases towards blue wavelengths and towards low effective
temperatures. Considering the uncertainties associated with 
the predicted lines, it is very likely that shortcomings in current
lists contribute to the tension that, in our study, led to
the apparent need for more reddening in the UVB arm of X-shooter
than in the VIS arm. In Sect.\,\ref{sec:results_best_residualDiscrepancies}
and Fig.\,\ref{fig:X0705etal_refined_R500_withnoise} we noticed
that between 4000 and 5000\,K, and at wavelengths shorter
than about 430\,nm, the best-fit models for dwarfs or low-luminosity
giants seemed to have excess flux  in between the strongest lines; 
this might argue in favor of missing pseudo-continuous opacity
(although considering the degeneracies in the fitting exercise,
other explanations are also possible). The effect is not as clear
at higher metallicities (Fig.\,\ref{fig:X0314etal_refined_R500_withnoise}),
where uncertainties in the very numerous molecular
lines, combined with degeneracies in the model parameter selection,
may mask it.

\subsection{Anticipated effects on the integrated colors of simulated SSPs}
\label{sec:BCs}
In this section, we briefly discuss how errors on fundamental parameters, 
of amplitudes comparable to 
the differences between the parameters of Paper\,II
and those we obtained
from matching the global SEDs with GSL models, may affect the contributions of 
various types of stars 
to the integrated light of a stellar population.  Population synthesis models
based on XSL spectra are in preparation (Verro et al.). Here
we approach the question by comparing the bolometric corrections of the synthetic spectra
that correspond respectively to the DR2-parameters  and to the SED-fit parameters
(fits to all wavelengths). 
This avoids any calculation of bolometric corrections from the XSL-spectra themselves,
which despite their broad spectral range do not cover the full spectrum of 
photospheric emission.
We define the bolometric correction for photometric passband $X$ as 
$BC_X = M_{\mathrm{bol}}-M_X$.

The synthetic spectra distributed in GSL are wavelength-dependent surface fluxes, 
integrated over all directions. 
After integration over the photometric passband of interest, 
the fluxes must be multiplied by the area of the stellar surface
to allow a comparison with the total luminosity of the model. 
For stars with extended atmospheres, the 
relevant outer radius is larger than the effective radius 
 (which relates luminosity to effective temperature
via the Stefan Boltzmann law), but only the latter is present 
in the default headers of the data files. 
The correct outer radii were extracted from intermediate data files of the PHOENIX archives\footnote{Unfortunately a few
of the intermediate files are lost, so the parameter space covered here is slightly more restricted 
than in previous sections.} and bolometric corrections computed. The procedure does not
include extinction, because most population synthesis codes use extinction-free libraries 
of spectra or colors; spectra such as those of XSL would
be de-reddened before implementation in a population synthesis code. 

 \begin{figure*}
\includegraphics[clip=,width=0.45\textwidth]{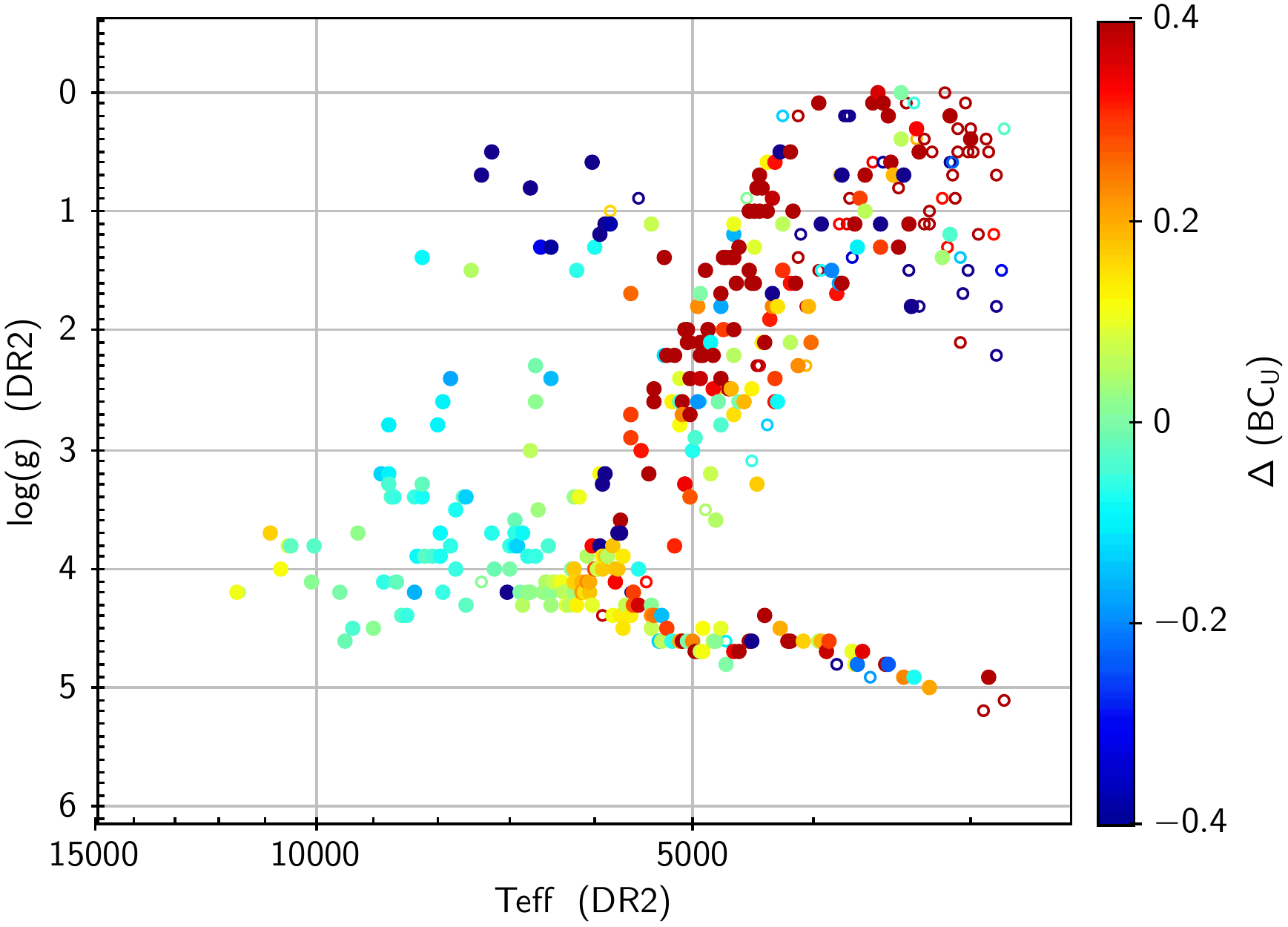} \hfill
\includegraphics[clip=,width=0.45\textwidth]{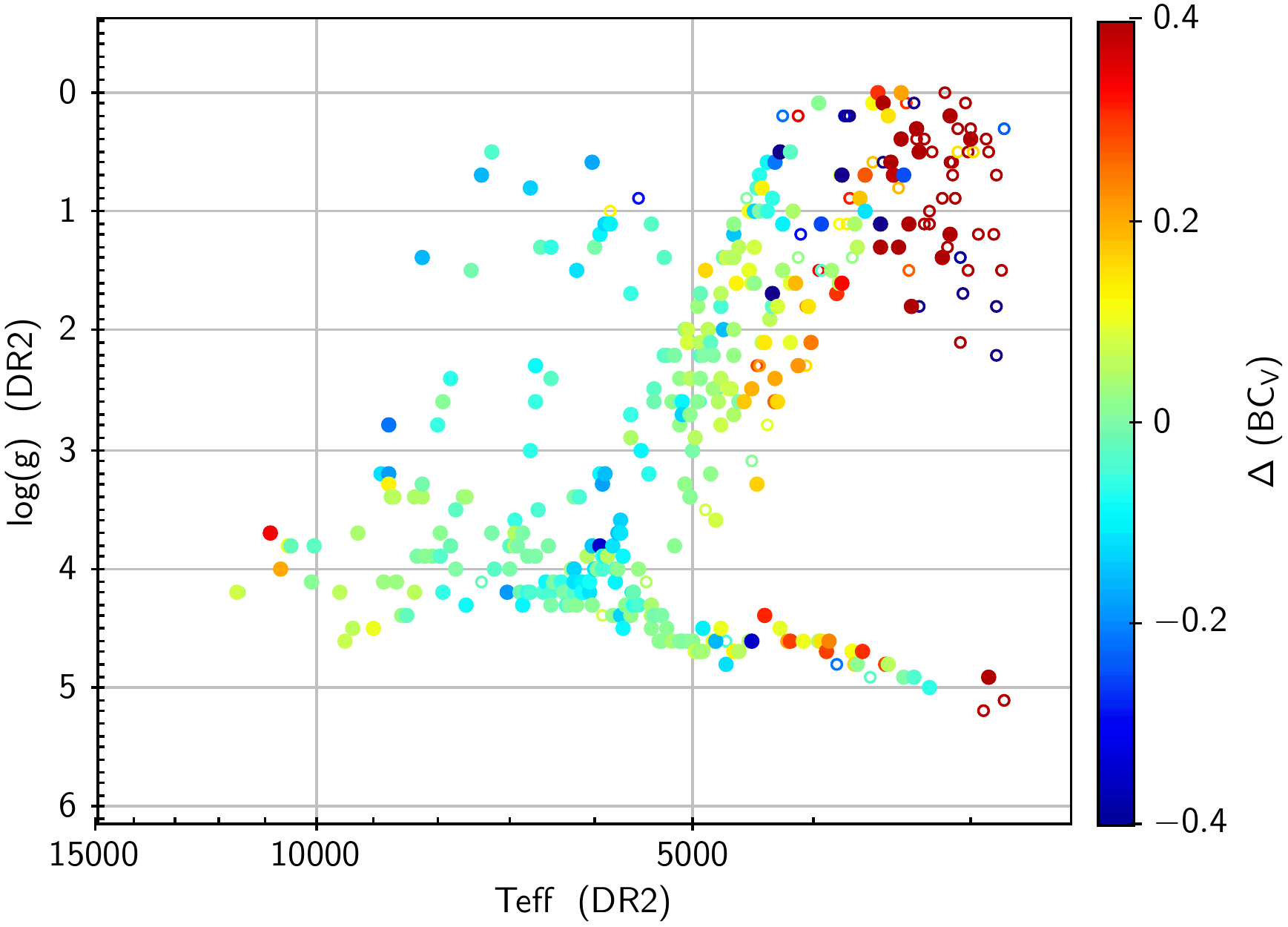}\\
\includegraphics[clip=,width=0.45\textwidth]{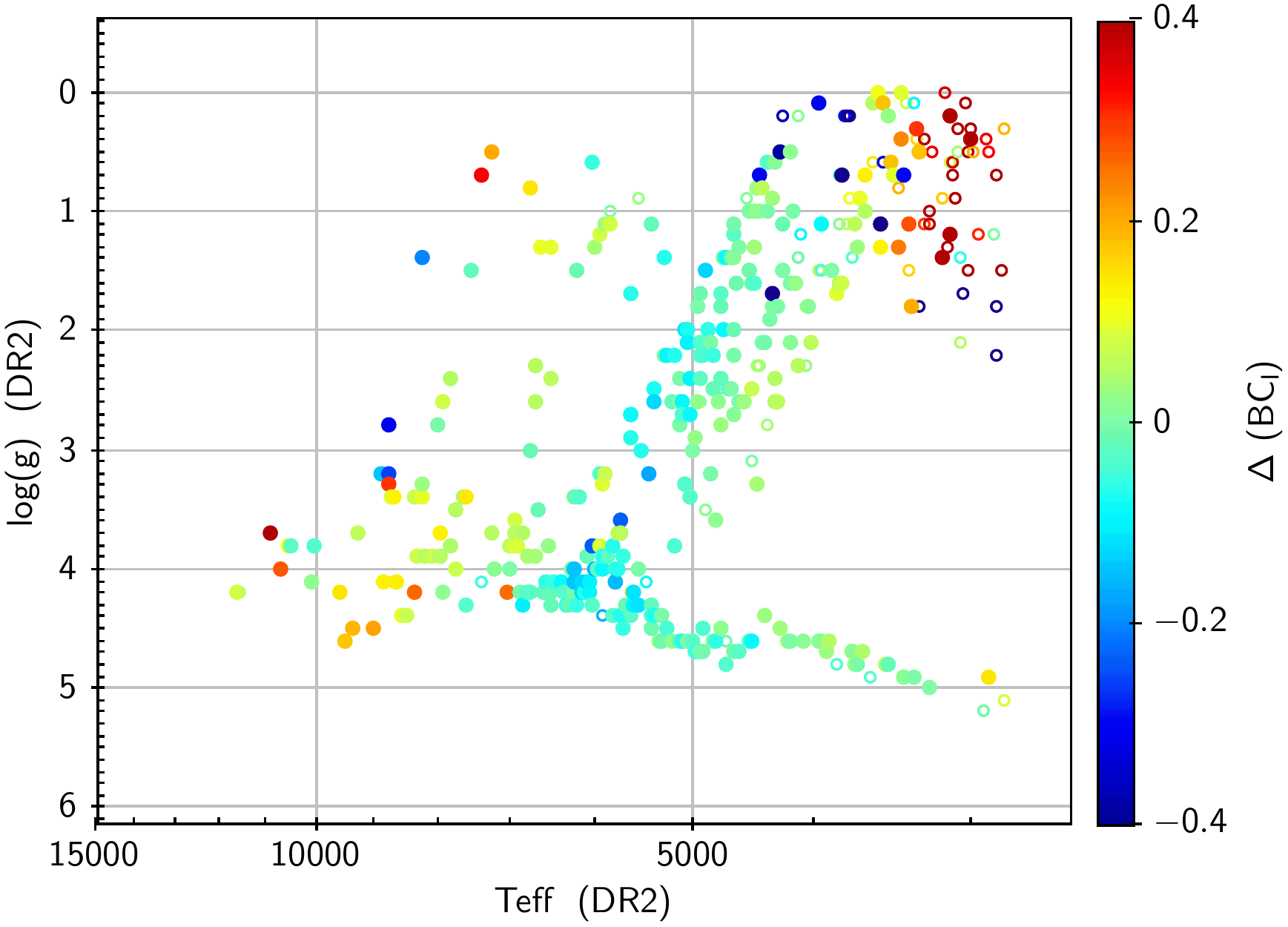} \hfill
\includegraphics[clip=,width=0.45\textwidth]{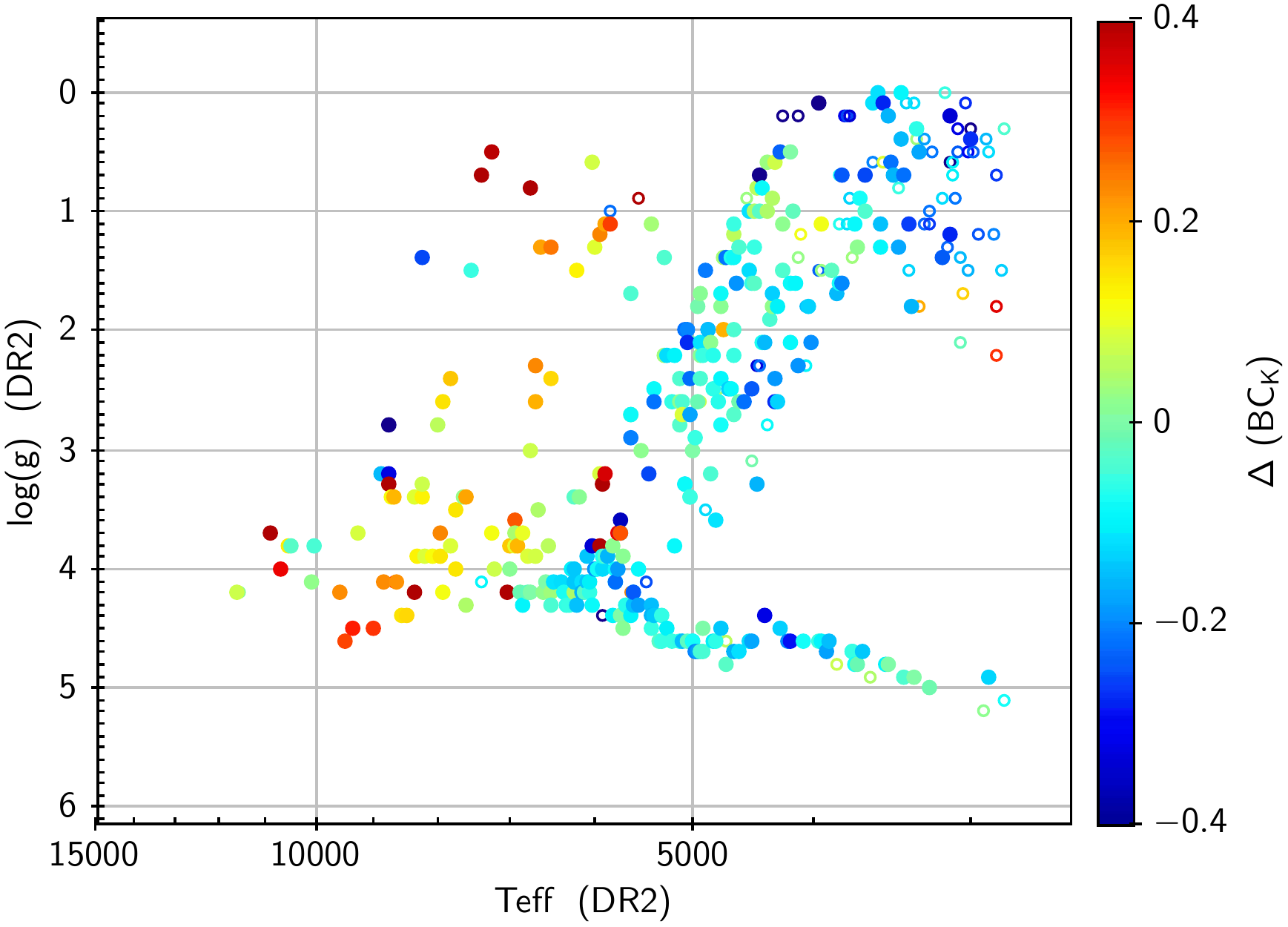}\\
\caption[]{Effects of parameter-differences between \citet[][DR2]{Arentsen_PP_19} and
SED-fits with GSL spectra (ALL wavelengths) on bolometric corrections:
$\Delta \mathrm{BC}_X = \mathrm{BC}_X \mathrm{(ALL)}-\mathrm{BC}_X\mathrm{(DR2)}$. The four panels are for the bolometric corrections to classical U, V, I, and K bands,
respectively.
The layout of the figure is as in Fig.\,\ref{fig:HRdiag0_byFeH0}.}
\label{fig:HRDbyDeltaBC}
\end{figure*}
f
As illustrated in Fig.\,\ref{fig:HRDbyDeltaBC}, the effects of changes in 
fundamental parameters of the amplitude discussed in this paper are of variable importance depending on the passband selected and 
on the region in the HR diagram.
In the $U$ and $K$ bands (top left and bottom right panels), 
a simple dichotomy appears between 
main sequence stars below 6500\,K and giants on one hand,
and warmer stars on the other. This essentially mirrors the 
distribution of \teff(ALL)-\teff(DR2) across the HR diagram 
(Fig.\,\ref{fig:HRDbyDTorDg}). In the analysis of the 
integrated light of a galaxy, the dichotomy
would directly affect the proportions of young and 
old stellar subpopulations.

The bolometric corrections vary less across the HR diagram
in the optical passbands $V$ and $I$. The dichotomy seen in $U$ and $K$
is weak or absent, and replaced with more subtle local effects.
For instance, a gradient in 
$\Delta \mathrm{BC}_X \equiv \mathrm{BC}_X \mathrm{(ALL)}-\mathrm{BC}_X\mathrm{(DR2)}$ is seen across the red giant branch in the $V$ band, and with
lower contrast in the $I$ band (or in the $z$ band; not shown). 
The temperature of the giant branch depends on metallicity, and metallicity
estimates are usually undertaken at optical wavelengths. Hence, 
the gradient in $\Delta \mathrm{BC}_V$ across the giant branch may
translate into uncertainties in proportions of metal poor and metal rich stars.

Among the four panels of Fig.\,\ref{fig:HRDbyDeltaBC}, 
the $I$-band is most relevant to current studies of the stellar
Initial Mass Function (IMF) of galaxies based 
on their integrated light (bottom left panel;
the corresponding diagram for the $z$-band, not shown,
is very similar).
Such studies measure gravity-sensitive features of
cool stars to constrain the  ratio of dwarfs to giants,
and focus on the red to near-infrared wavelengths at which
the contribution of cool dwarfs to the total
emission of a population is maximal 
\citep{vanDokkum_Conroy_2011, LaBarbera_etal17}.
The systematic offsets in parameters that we have discussed in this 
paper would affect such studies if $\Delta \mathrm{BC}_I$ differed
between giants and dwarfs. No strong difference of that type 
is seen, but only complete population synthesis models could 
quantify the small effects that may still be present and that may
be relevant considering the high accuracies required in IMF-studies.

\section{Conclusions}

We have compared the spectra of XSL-DR2 \citep{Gonneau_XSLDR2}
with the theoretical spectra of the G\"ottingen Spectral Library
\citep{Husser_etal2013}, at low and intermediate resolution
(R=500 and 3000), for wavelengths extending from the 
near ultraviolet to the near-infrared ($\sim$350\,nm to 2.45\,$\mu$m), with the 
main aim of determining whether a good match to both the energy
distributions and the spectral features can be obtained simultaneously.

(i) When adopting the stellar parameters derived by \citet{Arentsen_PP_19},  
the low resolution SEDs of XSL spectra and those of (reddened) 
GSL PHOENIX-spectra are comparable over a wide part
of the HR diagram, and the comparison of spectral features at 
R=3\,000 is very good overall at warm effective temperatures. However,
systematic discrepancies between the empirical and theoretical energy 
distributions appear below 5000\,K. Between 4000 and 5000\,K, when extinction
is optimized for a best match in the VIS range of 
X-shooter (0.55--1\,$\mu$m), the empirical spectra in the UVB range 
($<$550\,nm) tend to be redder than
the models and the empirical spectra in the NIR range tend to be bluer 
than the models. The effect is more obvious at low metallicity than 
at near-solar metallicity.
We recall that Arentsen et al. compared the optical part of the XSL 
absorption line spectra, at R\,$\simeq$\,2000, with existing empirical libraries 
(ELODIE/MILES), for which published parameters are the result of a 
compilation of varied detailed studies in the literature. 
Those libraries are a frequent reference 
in stellar population studies \citep{LeBorgne_etal2004,
FalconB_etal2011,Coelho_BC_2020, Chen_etal2020}.
\smallskip

(ii) When the GSL library is searched for the best-fitting reddened model for each XSL data-set, 
good matches can be found down to about 4000\,K. By ``good matches", we mean 
that the residuals (XSL$-$GSL) at R=500 
are  on average flat across the whole spectral range of XSL. However, spectral features, in particular in the UVB arm, are not always reproduced well when the SED is. This is in particular true between 4000 and 5000\,K, where it remains difficult to reproduce the SED and spectral features simultaneously. 
Below 4000\,K, only a handful of empirical spectra find roughly acceptable matches in the 
synthetic collection, and we have therefore refrained from discussing trends there. 
We refer to \citet{LanconIAU18} for a preliminary discussion 
of the coolest, frequently variable giants.

The XSL-GSL discrepancies between 4000 and 5000\,K
manifest as differences between the parameters obtained from the three X-shooter arms 
taken separately and as features in the low or intermediate resolution residuals. 
Models selected by fitting only one arm are not usually a good representation 
of the observed spectrum as a whole and this is not explained by flux 
calibration errors alone. There is a real inconsistency between the theoretical
spectral features and SEDs, for which potential causes are multiple (inadequate 
element abundances, possible departures from LTE, simple assumptions in the spherical 
models for properties such as microturbulence, atmospheric extension or stellar mass as a function of 
position in the HR diagram, shortcomings in fundamental data such as  line lists, etc.).

We summarize a few of the differences between arms as follows. 
Between 4000 and 5000\,K, 
$\Delta A_V \equiv A_V\mathrm{(UVB)}-A_V\mathrm{(VIS)}$, 
derived from the SED-fits with the standard extinction law of \citet{Cardelli89}, 
is in general positive. This difference is reduced, but not erased, 
when using $R_V=2$ instead of $R_V=3.1$. But we do not expect
such an extreme extinction law to apply to many XSL stars
\citep{Schlafly_etal16} and more generally we do not expect the
modeling of galactic extinction to resolve the inconsistencies. $\Delta A_V$ correlates with 
\teff(UVB)$-$\teff(VIS), but with some dependence on position in the 
HR diagram; \teff(UVB)$-$\teff(VIS) is largest for the metal-poor
giants in the sample. Along much of the main sequence, 
log($g$)(NIR) is found larger than log($g$)(VIS), while the 
opposite trend is true on the low-luminosity giant branch.

The systematic features in the spectral residuals (difference XSL$-$GSL at
R=500 or R=3000) are more easily isolated at low metallicity where
features are not so numerous. Between 4000 and 5000\,K, the
models selected based on matches at all X-shooter wavelengths tend to lack
opacity in between the strongest absorption features at the blue end 
of the UVB range ($<$400\,nm), while their strongest individual 
metal lines at the red end of the UVB range  ($>490$\,nm) are 
generally too deep. Also, differences in the depth of molecular bands
and Ca or Mg lines suggest that the abundances of light elements
and $\alpha$-elements may need tuning.

\smallskip

(iii) There are systematic offsets between the parameters
of the best-fitting SEDs and those based on the comparison 
with ELODIE/MILES in \citet{Arentsen_PP_19}, even when one accounts for
the uncertainties associated with the differences between arms mentioned above.
These systematic offsets however depend in a complex way on position
in the HR-diagram and they cannot be summarized with a small set of constants.
For example, at non-extreme temperatures along the main sequence
the best-fitting reddened GSL-SEDs have higher temperatures and gravities
than the DR2-values; on the low-luminosity half of the giant branch they have
higher temperatures, gravities and metallicities; on the more luminous
half of the giant branch they have higher temperatures but
lower gravities and metallicities.

Because similar trends with position in the HR diagram appear 
independently of the spectral arm used (UVB, VIS, NIR, or ALL), 
it is unlikely that they result from a data artifact, or from modeling 
issues related to a specific spectral 
feature, or even a difference in individual abundances. 
Difference in the parameter scales of the two
reference collections seem to play a dominant role.

It is interesting that the trends found here
resemble those of other work with empirical
and theoretical reference libraries rather well.
\citet{Maraston_etal20} provide a recent example, 
where the MILES library was used as an 
empirical reference, and ATLAS and MARCS models for 
the theoretical one. 
Jumping to the conclusion that the parameters of ELODIE/MILES 
should be re-assessed would be premature. Indeed, 
the parameters from the ``theoretical approach" also suffer from a number
of caveats that lead to large internal errors, as recalled in (ii) 
above. We refrain from publishing individual parameters
here for this very reason. What our study highlights again, 
is the need for improvement in the synthetic spectral libraries, 
which in fact underly both 
parameter estimate methods in different ways. On one hand, 
the so-called "empirical" libraries have associated parameters 
that come from the analysis of modeled and excitation equilibria 
using isolated lines of metals in high-resolution spectroscopic data, 
or for cool stars on stellar radii
or the infrared flux method. Hence they use only a small part of the full 
information present in modern empirical spectra. 
More generic ``full spectrum fitting methods", on the other hand, 
exploit a broader range of data, 
but that makes them more sensitive to the physical completeness and 
the internal consistency of the spectral calculations.

(iv) We have explored only models at [$\alpha$/Fe]=0 and [$\alpha$/Fe]=0.4. 
As expected, low metallicity giant star
spectra, even at low resolution, on average prefer $\alpha$-enhanced models. 
Because of degeneracies between
the effects of [$\alpha$/Fe], [Fe/H], log($g$) and \teff, 
the differences in the quality-of-fit measurements are usually
small between the model-fits at the two values of [$\alpha$/Fe]. 
Differences between the parameters derived 
from various arms are in general larger than the differences due 
to the change of 0.4 in [$\alpha$/Fe]. Hence our
method by itself does not allow us to provide automatic [$\alpha$/Fe] 
estimates for individual stars. This would
require a one-by-one adjustment of the chemical abundances, 
including those of C,N, and O which shape the 
molecules that are responsible for the largest residuals in 
the data-model comparisons.

(v) Consequences of the highlighted 
trends are discussed via bolometric corrections. As illustrated
recently by \citet{Coelho_BC_2020} or \citet{Maraston_etal20}, 
local dispersion in estimated parameters
(random errors near a given position of the HR diagram) 
have little impact on predicted properties of stellar populations,
but systematic effects do \citep{Percival_Salaris_2009}. The systematic offsets between parameters
obtained with empirical or theoretical reference libraries translate
into differences in bolometric corrections, that directly affect the
relative contributions of different stars at different wavelengths. 
When the differences in bolometric corrections depend on position in
the HR-diagram in a systematic way, they will affect the 
interpretation of galaxy spectra. From the trends observed, we 
expect uncertainties on BC$_U$ and BC$_K$ (due to  method-dependencies of 
stellar parameters) to affect mainly the estimated proportions of young
and old stars, and those on BC$_V$ and BC$_I$ the proportions of stars of
different metallicities. 
\smallskip

Discrepancies between synthetic and empirical spectra are a limiting factor
in our interpretation of the light received from objects at all distances
in the universe and much investment in stellar astrophysics is still
needed. Collections such as XSL should be expanded and used
to assess new model grids. But in parallel we believe it will be necessary to focus on 
small samples of reference stars in key locations of the HR diagram,
and attempt to optimize models for these objects. For instance, an
extension of the Gaia FGK Benchmark Star project \citep{BlancoC_etal14} 
to the near-infrared would be most welcome.

\begin{acknowledgements}
This work was supported in part by grant ANR-19-CE31-0022 (POPSYCLE) of the French
Agence Nationale de la Recherche, as well as by the Programme National Cosmology et Galaxies (PNCG) of CNRS/INSU with INP and IN2P3, co-funded by CEA and CNES,
and by the Programme National de Physique Stellaire (PNPS) of CNRS/INSU.
A.V and J.~F-B  acknowledge support through the RAVET project by the grant AYA2016-77237-C3-1-P from the Spanish Ministry of Science, Innovation and Universities (MCIU) and through the IAC project TRACES which is partially supported through the state budget and the regional budget of the Consejer\'ia de Econom\'ia, Industria, Comercio y Conocimiento of the Canary Islands Autonomous Community.
RFP acknowledges financial support from the European Union's Horizon 2020 research and innovation program under the Marie Sk\l{}odowska-Curie grant agreement No. 721463 to the SUNDIAL ITN network.
We made extensive usage of the Topcat and Stilts tools \citep{TopcatRef} and we thank
M. Taylor for his attention to questions from users. We also made use of the VizieR catalogue access tool, CDS, Strasbourg, France (DOI : 10.26093/cds/vizier); 
the original description of the VizieR service was published by 
\citet{VizierRef}. We thank C.\,No\^us and the Cogitamus Institute, France (http://www.cogitamus.fr/), and
through them the research community as a whole, for their contributions to this work.
\end{acknowledgements}

\bibliographystyle{aa}   
\bibliography{AL_XSLhusser}.  

\begin{appendix} 

\section{Example of the diagnostic plots produced for model-data comparisons}
\label{app:fits_with_bestfit}

\begin{figure*}
\includegraphics[clip=,width=0.95\textwidth]{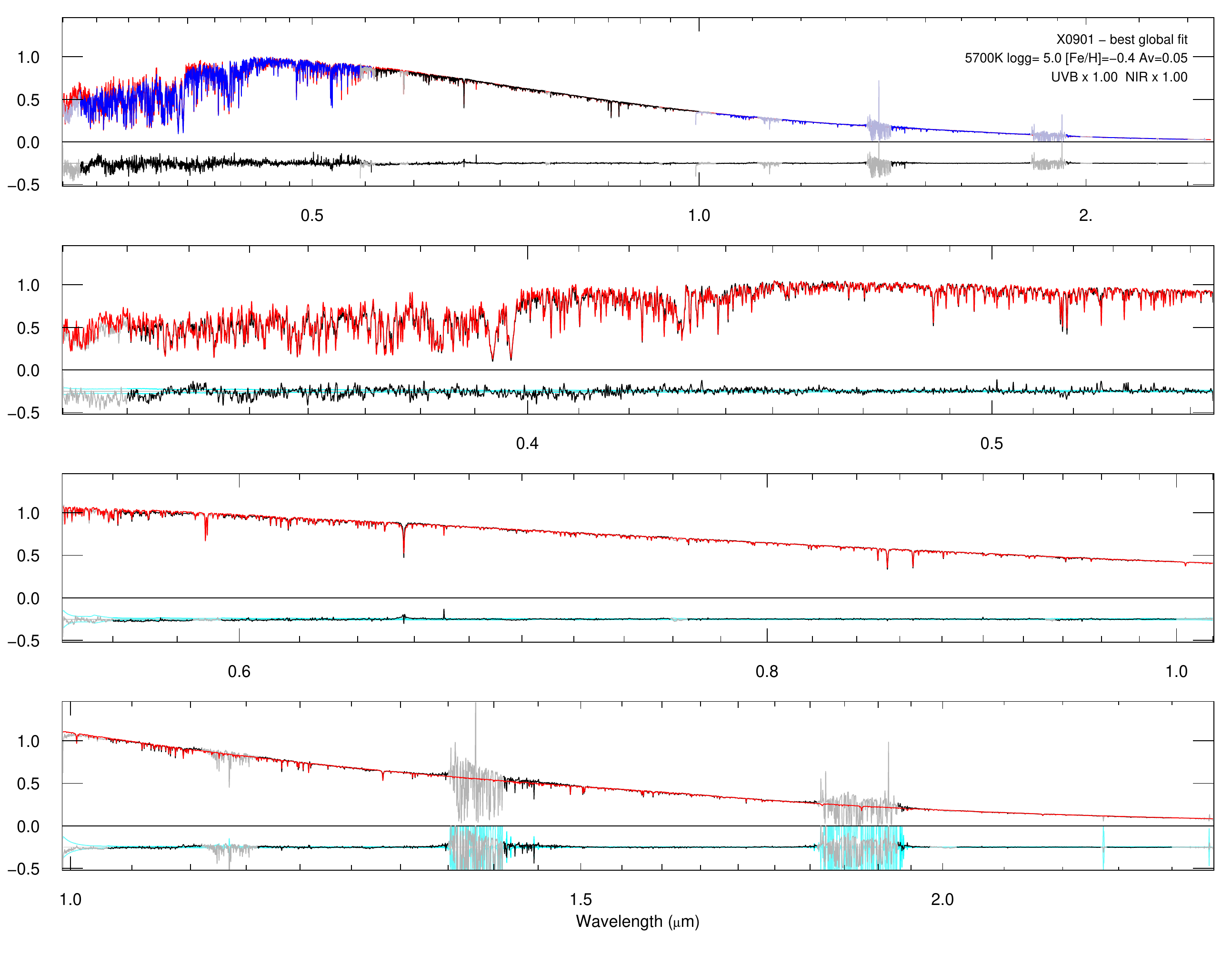} 
\caption[]{Best fit to the XSL spectrum of HD\,111515 (observation X0901), 
at R=3000, using all (but masked) wavelengths. Each XSL-arm is 
rescaled independently to optimize the global quality of the fit. The model is 
shown in red. In the top panel, the blue-black-blue color-scheme identifies the
three arms of X-shooter spectra, which are then shown separately (in black) in
the three lower panels. Masked wavelength regions in the XSL data are plotted in gray.
The noise spectra from the XSL pipeline are in cyan.}
\label{fig:X0901_bestglob}
\end{figure*}

\begin{figure*}
\includegraphics[clip=,width=0.95\textwidth]{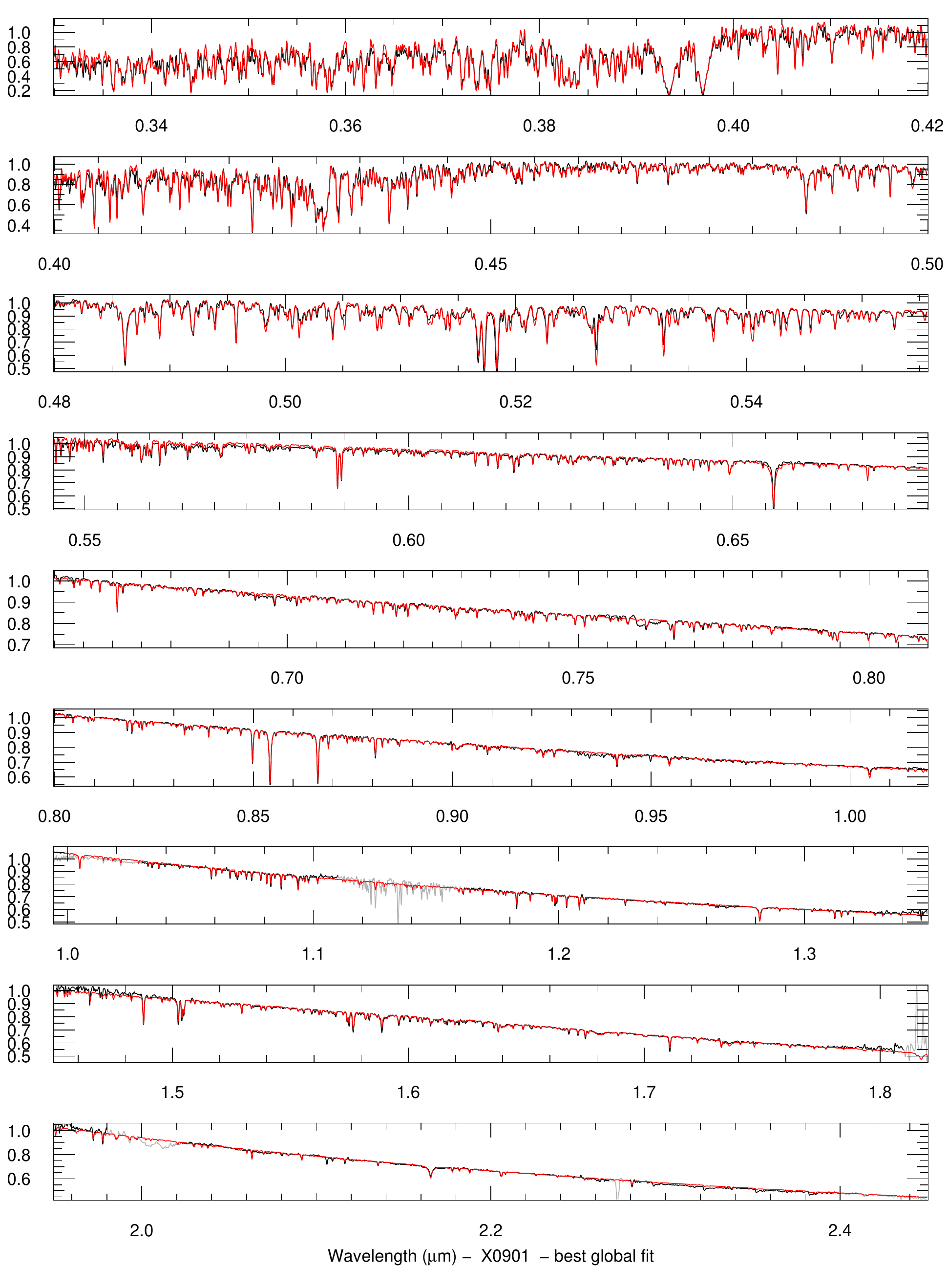} 
\caption[]{Zooms into various wavelength regions, for the match shown in Fig.\,\ref{fig:X0901_bestglob} 
(HD\,111515, observation X0901), at R=3000. The model is 
shown in red. Masked wavelength regions in the XSL data are plotted in gray.}
\label{fig:X0901_zooms}
\end{figure*}

\begin{figure*}
\includegraphics[clip=,width=0.95\textwidth]{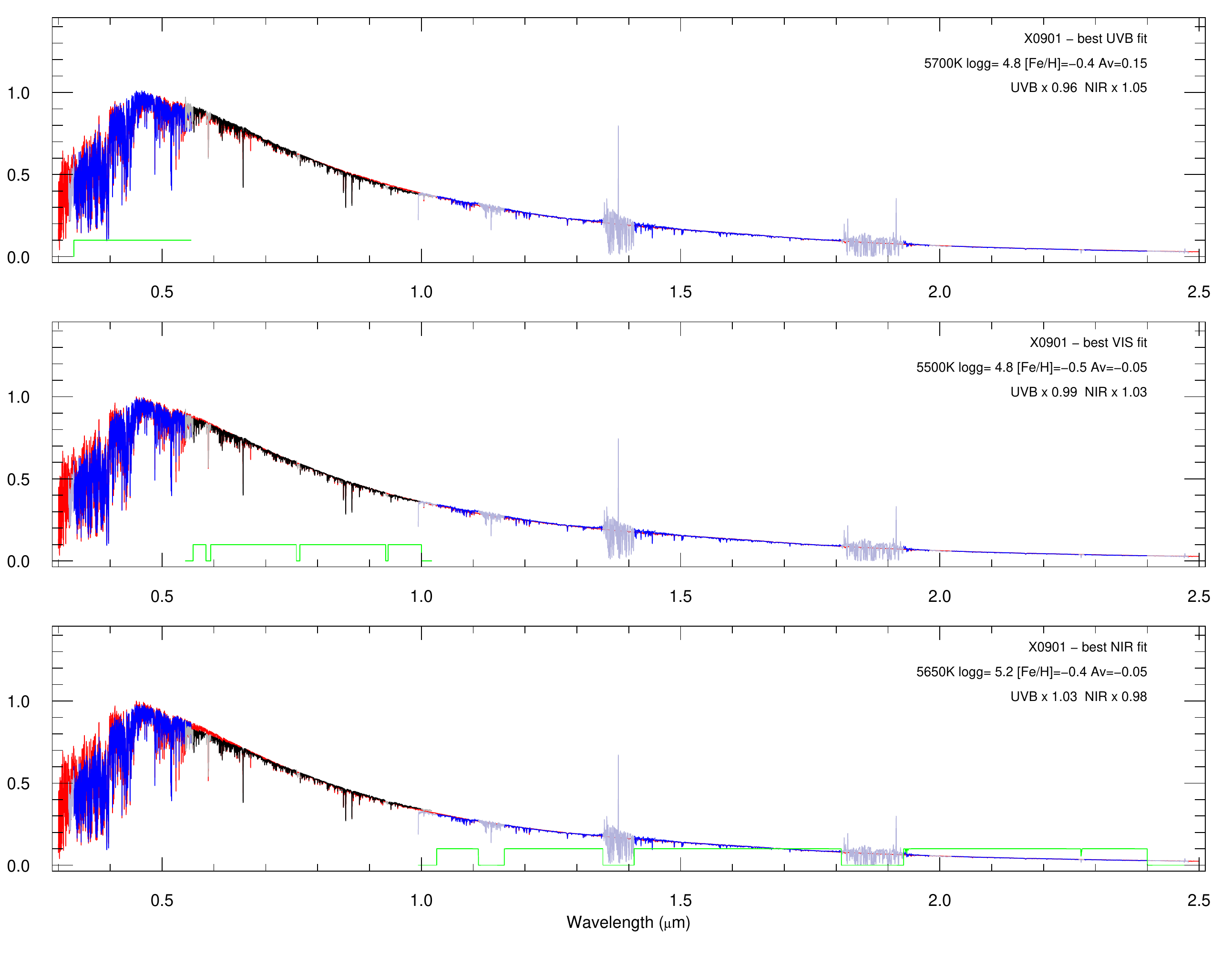} 
\caption[]{Best fits to the XSL spectrum of HD\,111515 (observation X0901), 
but this time using only {\em one}\ of
the three XSL arms in the optimization process :
the UVB arm in the top panel, the VIS arm in the middle, and
the NIR arm in the lower panel. The best-fit synthetic spectrum is shown in red,
the empirical spectra in blue and black, or in gray for the masked regions 
(the mask, rescaled for display purposes, is displayed in green).
In each panel, the arm-spectra {\em not} used in the fit are simply overlaid,
after independent rescaling to the level of the synthetic spectrum. 
The scaling factors in the legends are those applied
to the UVB and NIR observations
relative to the VIS observations. The parameters vary between the panels. }
\label{fig:X0901_best1armfits}
\end{figure*}

\begin{figure*}
\includegraphics[clip=,width=0.95\textwidth]{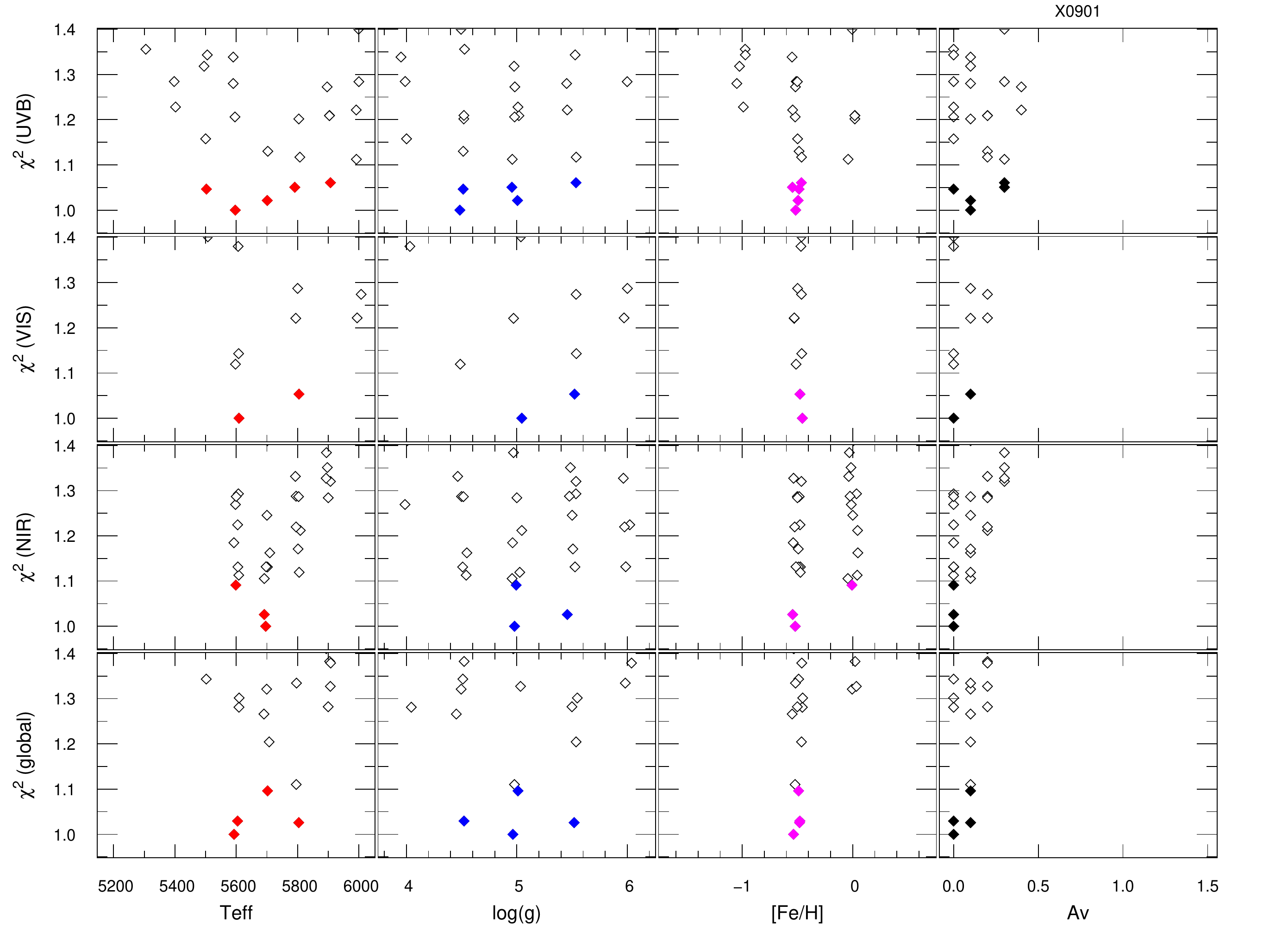}
\caption[]{$\chi^2$ distributions around the minimum, as a function of
the wavelength range used to constrain the fits and as a function of the
four model parameters (here, [$\alpha$/Fe]$\,=\,0$ is assumed).
The $\chi^2$ values are rescaled to a minimum of 1 for display; 
only values smaller than 1.4 are shown. \\
For this star, the parameter estimates based on each arm separately or on all
arms jointly are consistent with each other. The bottom left panel illustrates
effects of the discrete sampling of the model grid\,: at 5700\,K, the available
values of the other parameters lead to a higher global $\chi^2$ than at 
5600\,K or 5800\,K. To produce figures 
\ref{fig:X0901_bestglob}, \ref{fig:X0901_zooms}, 
\ref{fig:X0901_best1armfits} we have interpolated in the original
model grid around the minima of the present figure -- hence 
the slightly different parameters in those figures.}
\label{fig:X0901_chi2map}
\end{figure*}

Figures \ref{fig:X0901_bestglob}, \ref{fig:X0901_zooms}, 
\ref{fig:X0901_best1armfits}, and \ref{fig:X0901_chi2map} 
show the match between the models and the data,
in the case of HD\,111515 (XSL spectrum X0901), plotted at R=3000. 
This is a main sequence star with parameters relatively close 
to those of the sun. The fits to individual arms
and the global fit provide consistent results considering
the errors apparent as dispersion among the parameters
with low-$\chi^2$ values in Fig.\,\ref{fig:X0901_chi2map};
these parameters are slightly offset from to those of 
\citet{Arentsen_PP_19} (5383\,K, log($g$)=4.4, [Fe/H]$=-0.6$).
The value of $D$ (Eq.\,\ref{eq:D}) with the latter parameters,
for ALL wavelengths and at R=500, was 0.073,
and it is 0.059 for the best fit. In 
Sect.\,\ref{sec:results_forced} we describe such fits as consistent 
with flux-calibration errors at the 2-$\sigma$ level.

\clearpage

\section{Differences between neighboring models}
\label{app:TOHvsTOH}

\begin{figure*}
\begin{center}
\includegraphics[clip=,width=0.40\textwidth]{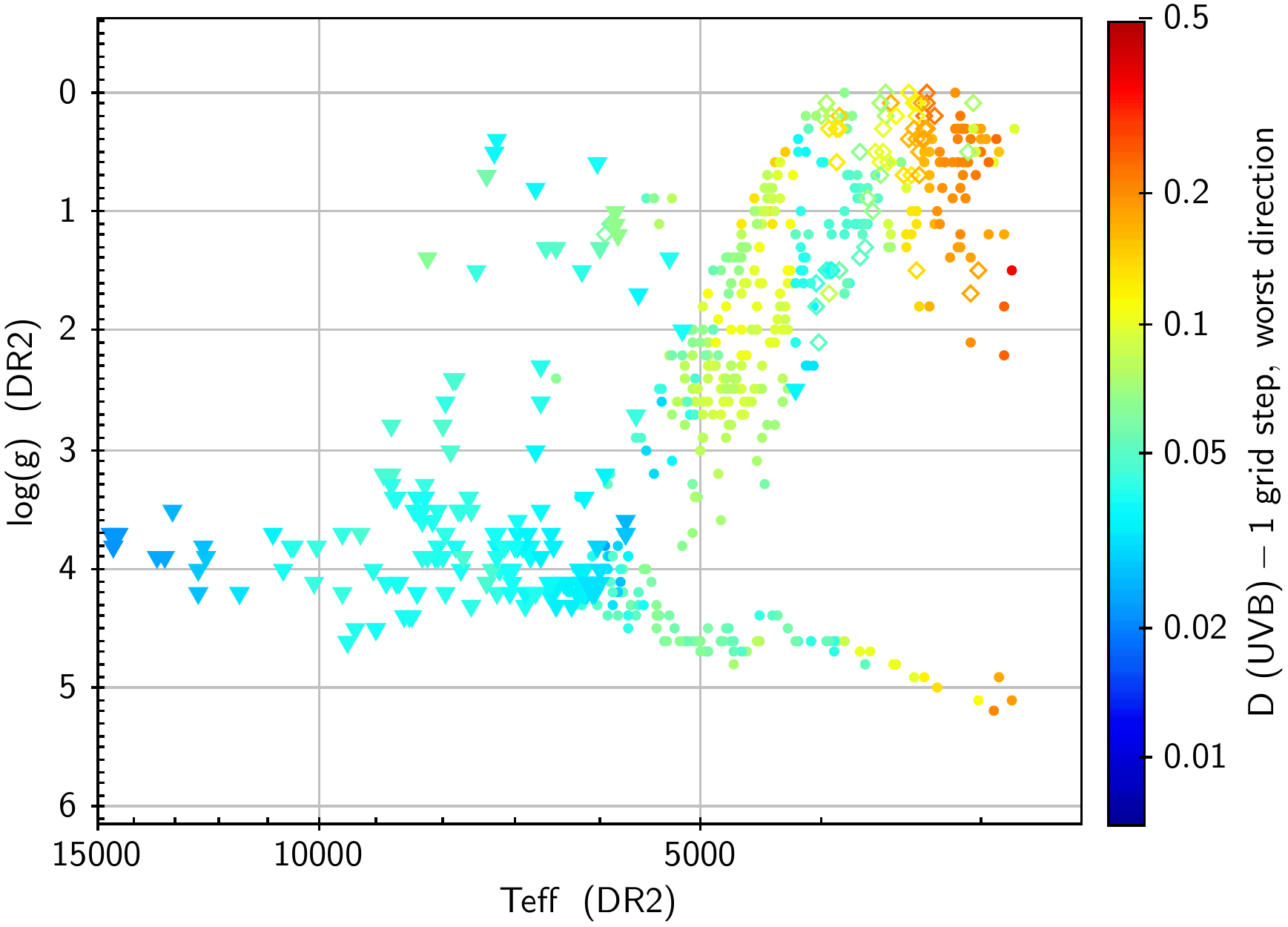}\hspace{0.5cm}
\includegraphics[clip=,width=0.40\textwidth]{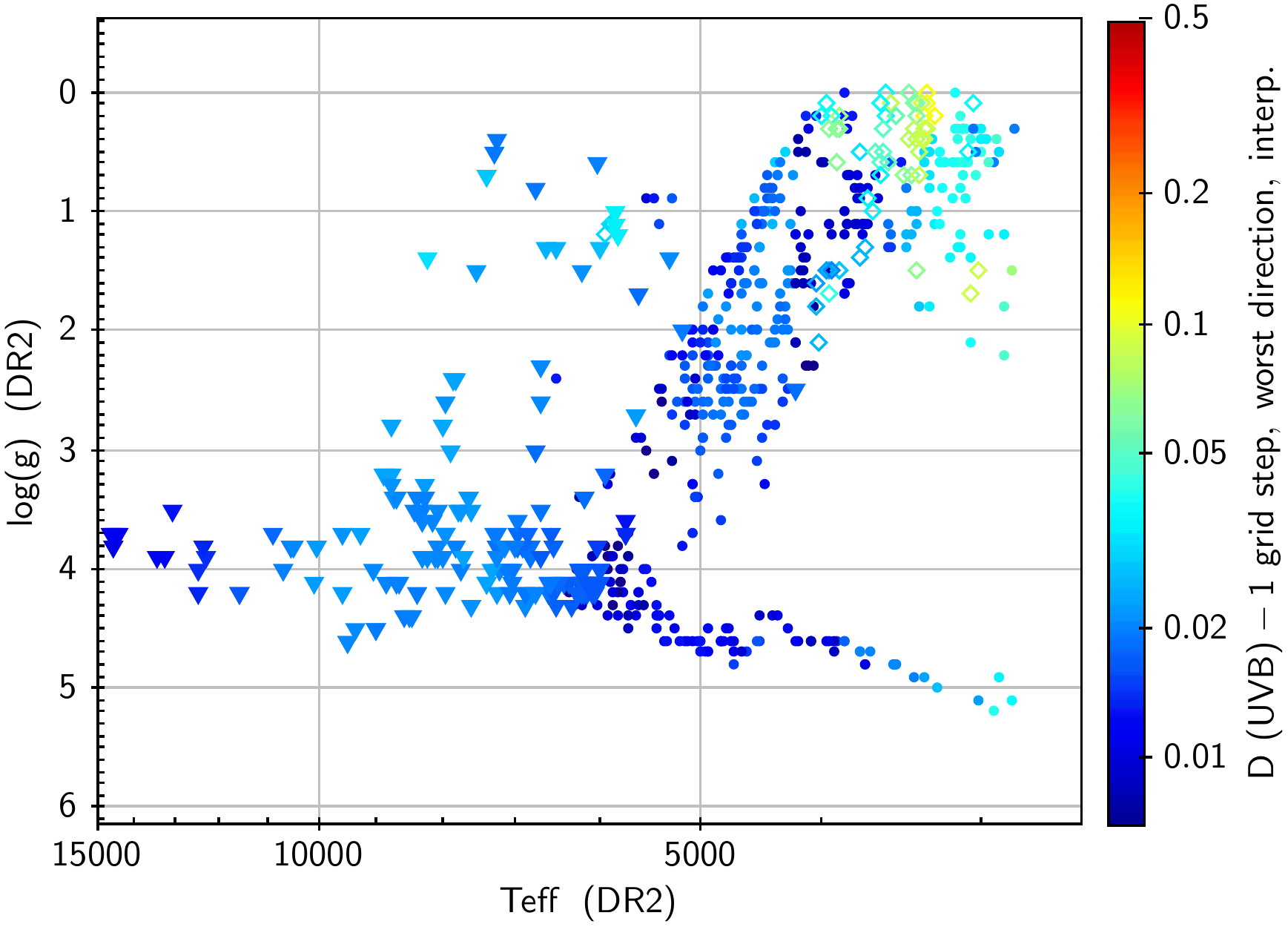}\\
\includegraphics[clip=,width=0.40\textwidth]{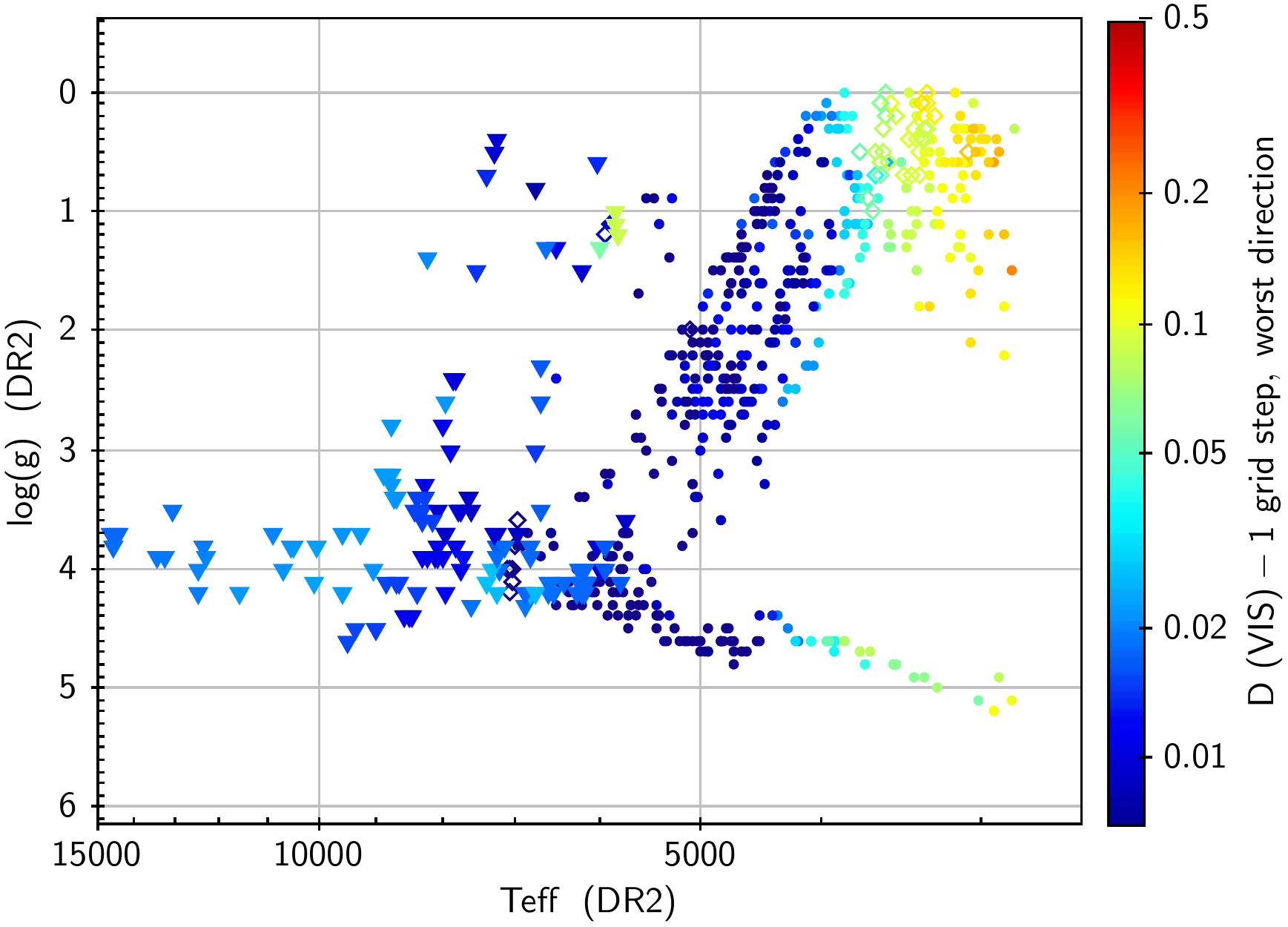}\hspace{0.5cm}
\includegraphics[clip=,width=0.40\textwidth]{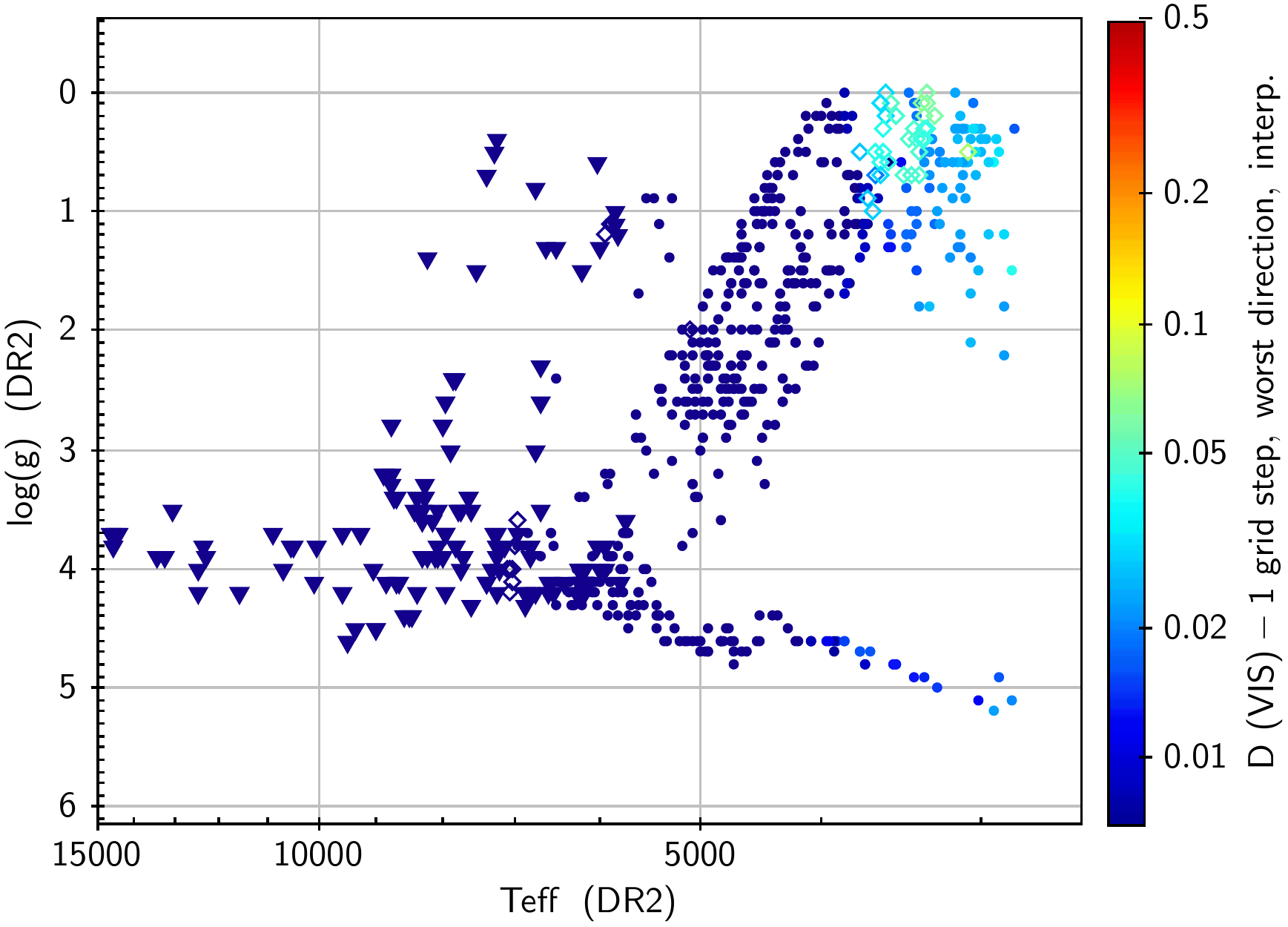}\\
\includegraphics[clip=,width=0.40\textwidth]{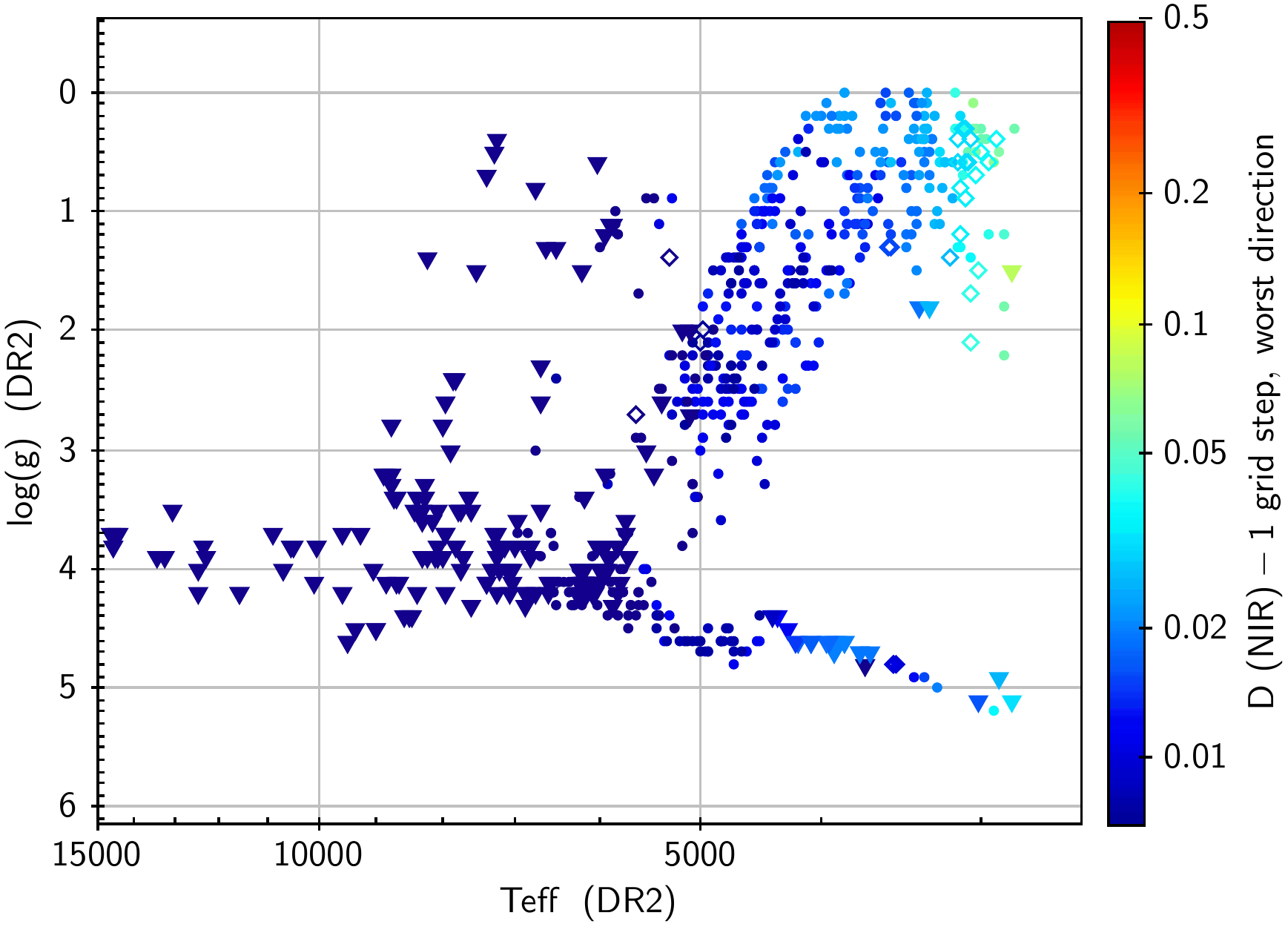}\hspace{0.5cm}
\includegraphics[clip=,width=0.40\textwidth]{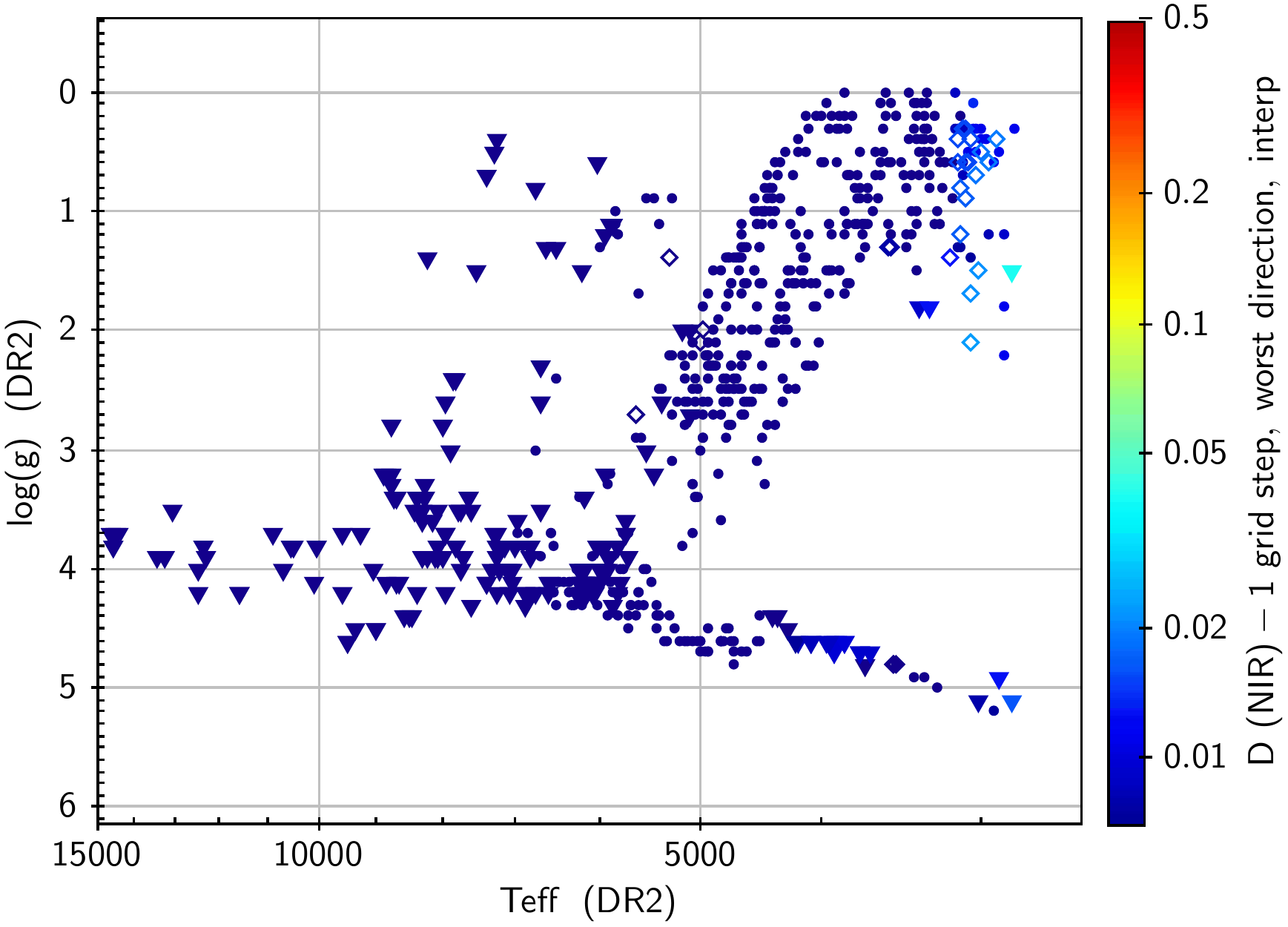} \\
\includegraphics[clip=,width=0.40\textwidth]{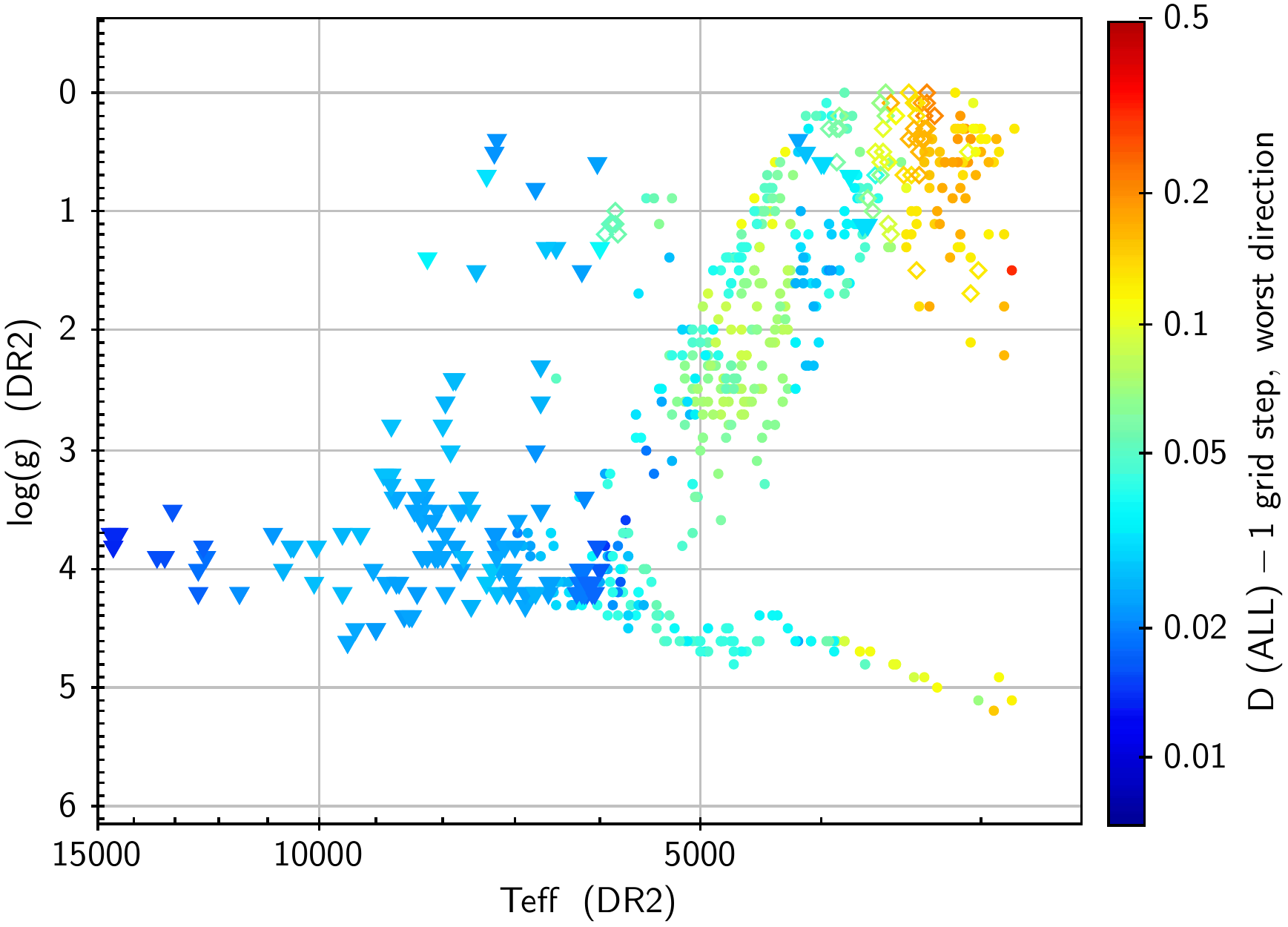}\hspace{0.5cm}
\includegraphics[clip=,width=0.40\textwidth]{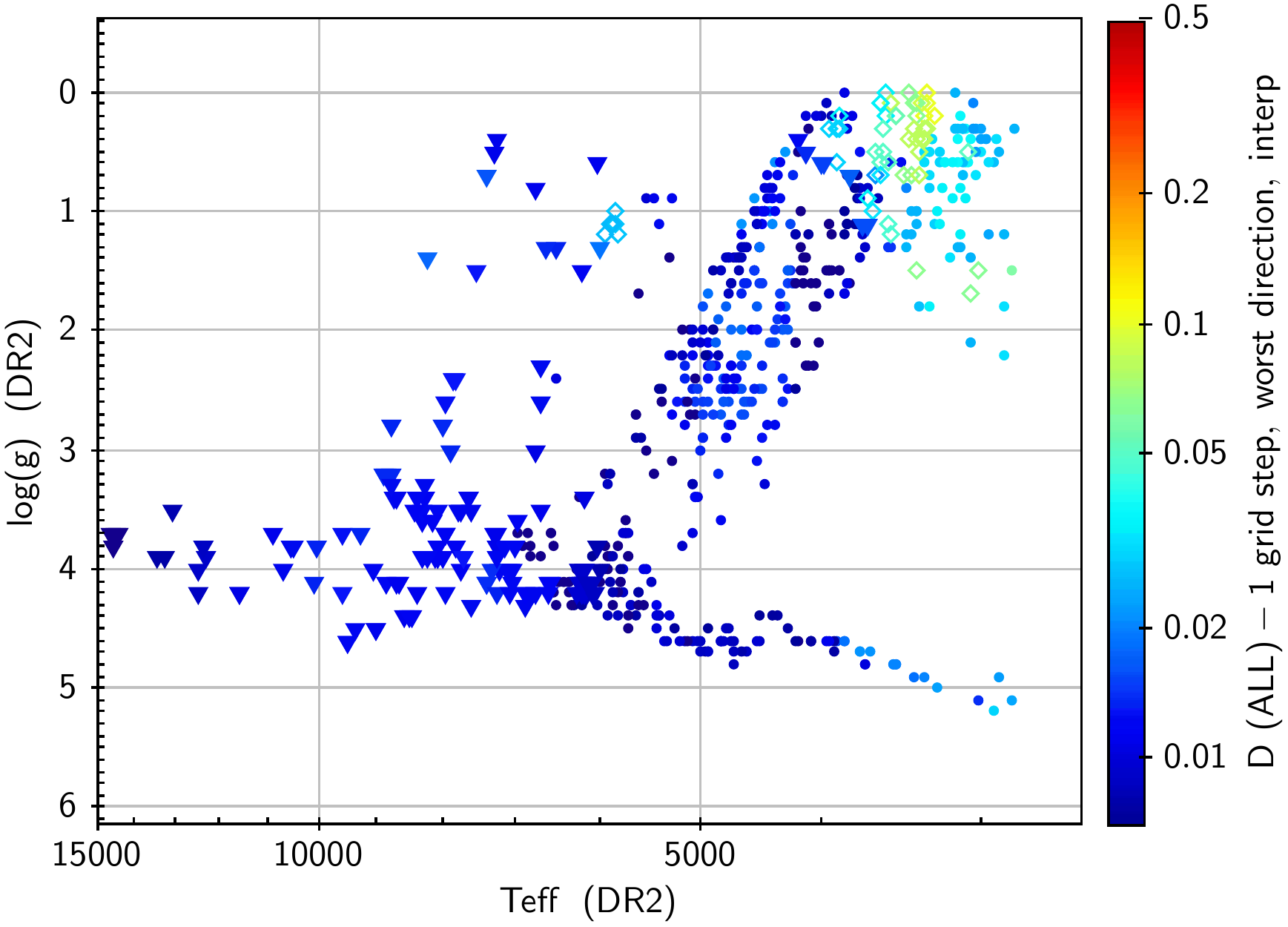}
\end{center}
\caption[]{Sensitivity of $D$ (Eq.\,\ref{eq:D}), measured at $R=500$, 
to one step in \teff, log($g$) or [Fe/H] in the {\em worst} direction in
the grid of theoretical spectra.
One symbol is plotted for each XSL spectrum. It is positioned according to the parameters
of Arentsen et al. 2019 (as elsewhere in this article), and colored according to the {\em largest} 1-step 
difference $D$ between the model closest to the XSL-spectrum's parameters on
one hand, and the 6 models surrounding that one on the other. 
In the {\em left} panels, the original grid of GSL
models is used, while in the {\em right} panels interpolation was used to reduce 
the grid steps as detailed in 
Sect.\,\ref{sec:method_forced}. Triangles, small disks, or 
open diamonds indicate,
respectively, that the largest $D$ between neighboring models in the
grid corresponds to a change in log($g$), in [Fe/H] or in \teff. 
From top to bottom, the wavelength ranges used are the UVB, the VIS, and the NIR arms, and 
finally all X-shooter wavelengths (ALL). 
}
\label{fig:TOHvsTOH_nearestParams}
\end{figure*}

As a reference for the interpretation of Fig.\,\ref{fig:forcedfits},
we show here how the finite sampling of the
original grid of theoretical spectra may affect $D$. To this purpose,
we have compared models chosen to lie close to XSL
observations, with models one grid-step away in \teff,
log($g$), or [Fe/H], and we have produced Fig.\,\ref{fig:TOHvsTOH_nearestParams}
with the largest $D$ among these. In the left panels, the 
model grid is used as it is originally, and in the right panels
the grid is interpolated as described in Sect.\,\ref{sec:method_forced}.

Among \teff, log($g$), and [Fe/H], the single parameter producing the 
largest difference $D$ over one grid step of the original grid is
log($g$) for hot stars (via hydrogen absorption features, in particular
the shape of the Balmer lines and of the Balmer jump; triangular symbols), 
and  [Fe/H] for most of the red giant branch (small disks).
Interpolation between models increases the significance of the trends discussed
in Sect.\,\ref{sec:results_forced} and \ref{sec:results_best}, without
changing the nature of the trends.

\clearpage

\section{Direct XSL--GSL comparisons with the parameters from DR2}
\label{app:sec:fits_with_DR2params}

In this section, 
only $A_V$ is a fitted parameter; other parameters are taken from
\citet{Arentsen_PP_19}. We present examples in order 
of decreasing effective temperature.
To be readable in small format, plots are all at $R=500$. 

\begin{figure}
\includegraphics[clip=,width=0.48\textwidth]{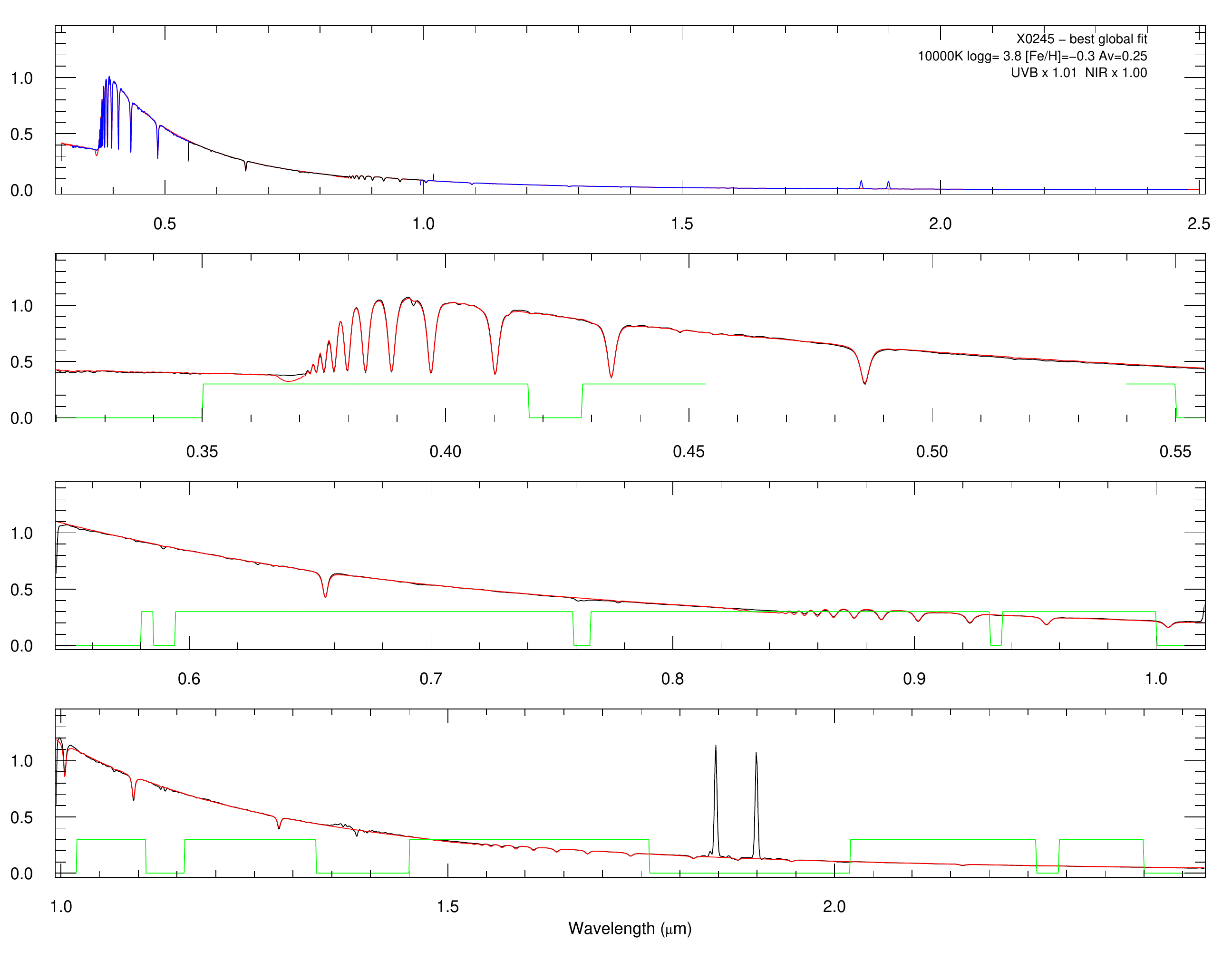} 
\caption[]{Comparison of the SED of HD\,147550 (observation X0245) with
the (interpolated) GSL model nearest the parameters of DR2
\citep{Arentsen_PP_19}, at R=500. 
In the uppermost panel, XSL data are shown in blue, black, and blue for the 
UVB, VIS, and NIR arms, over the model in red. In the lower panels, the 
model in red is overlaid on data in black.
At this resolution the main disagreements are around the
ends of the Hydrogen absorption line series, in particular around 370\,nm.
With $A_V=0.25$, the extinction law of \citet{Cardelli89} with
$R_V=3.1$ behaves adequately.
}
\label{fig:X0245_forced_R500}
\end{figure}

\begin{figure}
\includegraphics[clip=,width=0.48\textwidth]{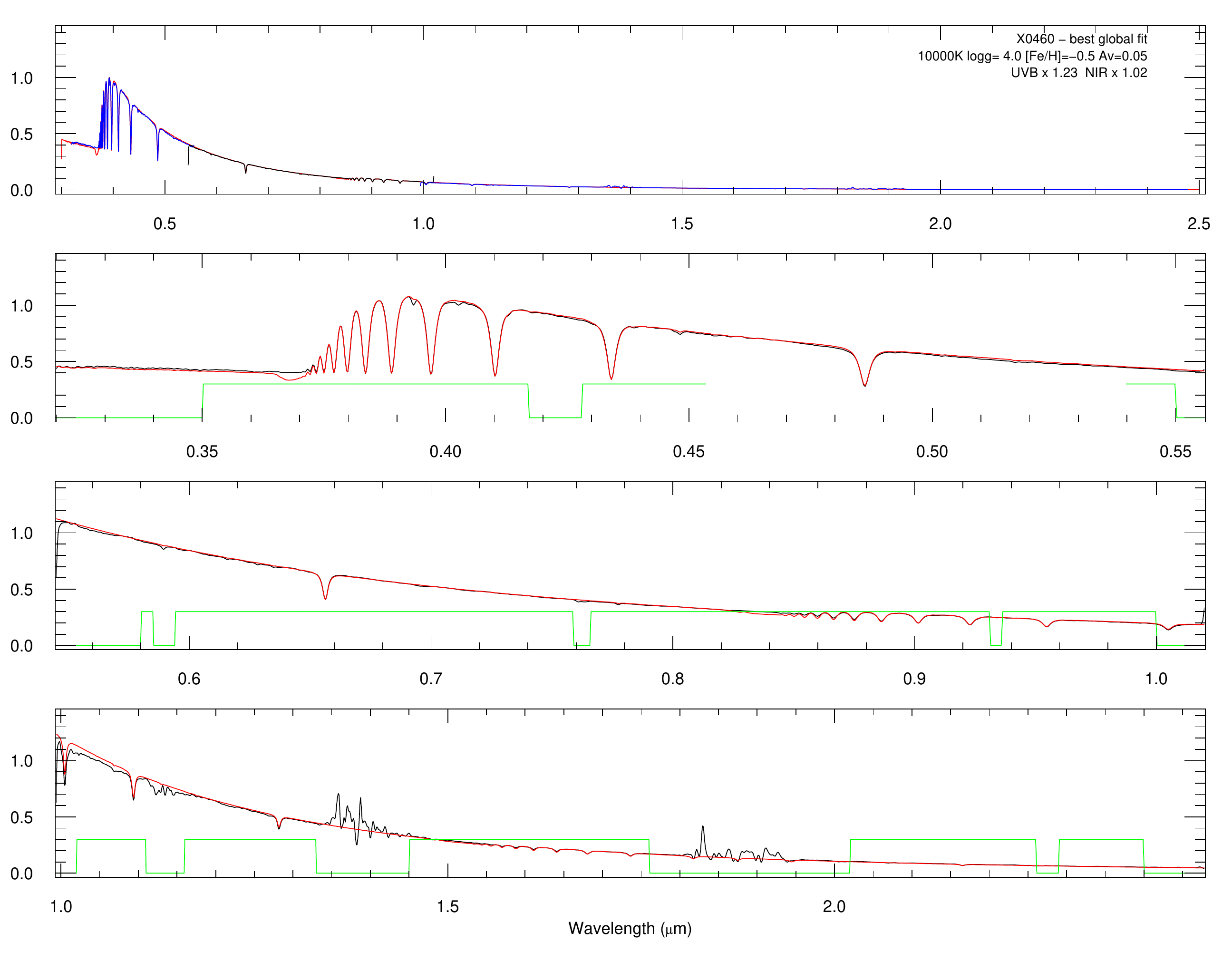} 
\caption[]{Same as Fig.\,\ref{fig:X0245_forced_R500}, but for HD\,167946 (observation X0460). 
Wavelengths below 1.03\,$\mu$m are affected by large flux 
calibration errors. Between 1.03 and 1.1\,$\mu$m the situation is
unclear: while flux calibration errors may be responsible for some of the
discrepancy seen for this spectrum, we note that
a similar discrepancy is seen in most hot stars (out of 21 XSL spectra between
9000\,K and 13000\,K, only two have fluxes above those of the model in that
region), while the discrepancy tends to be the other way round in cool
stars (empirical fluxes higher than the model fluxes).}
\label{fig:X0460_forced_R500}
\end{figure}

As discussed in Sect.\,\ref{sec:results_forced}, models and observations agree
well above 5000\,K, with only few exceptions. 
Figure\,\ref{fig:X0245_forced_R500} illustrates the excellent match obtained for the hot
main sequence star HD\,147550 with the parameters of
Arentsen et al. (X0245; \teff $= 10\,000$\,K). The dip in the model
shortwards of the Balmer series is a known peculiarity of the GSL
library. When zooming into the figure (or at higher spectral resolution),
one can see that the metal lines in the XSL spectra are on average 
stronger than in the synthetic spectrum, and also that the relative
central depths of low- and high-order hydrogen lines in the Paschen 
and Brackett series are not precisely reproduced, even though
the Balmer lines and the Balmer jump are modeled well.
An example of a comparatively poor match to a warm star
is given in fig.\,\ref{fig:X0460_forced_R500}: the slope at 
the blue end of the NIR arm is slightly off. Taken individually,
this offset would be consistent with flux calibration errors; but
an offset in the other direction is seen for only 2 out of 21 warm
stars in XSL. Nevertheless, the average offset is small compared
to the dispersion between objects and we do not discuss it
further.

\begin{figure}
\includegraphics[clip=,width=0.48\textwidth]{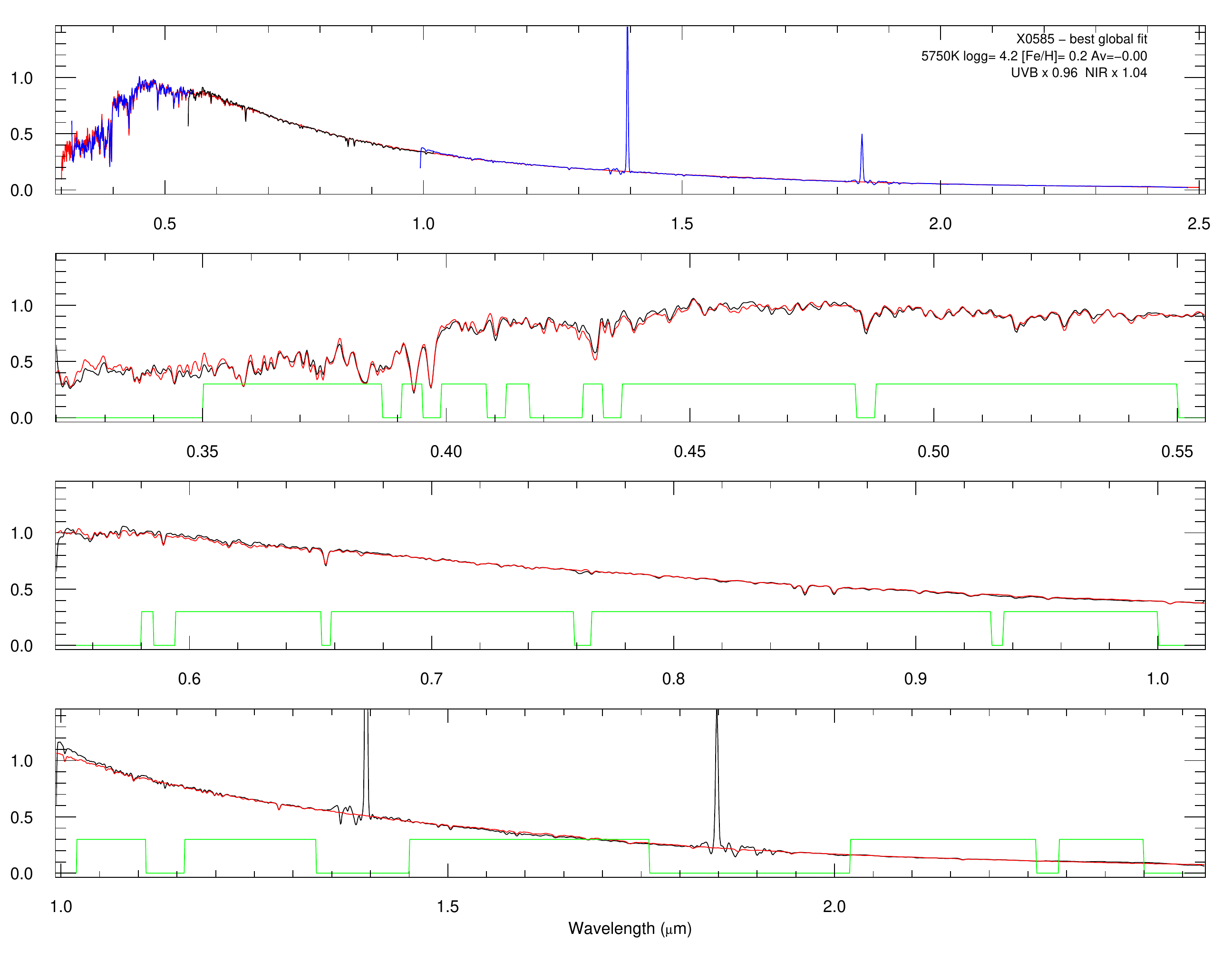} 
\caption[]{Comparison of the SED of the solar analog 
HD\,55693 (observation X0585) with
the (interpolated) GSL model nearest the parameters of 
\citet{Arentsen_PP_19}, at $R=500$.
We note that a slightly higher \teff\ would provide an even better match. 
}
\label{fig:X0585_forced_R500}
\end{figure}

\begin{figure}
\includegraphics[clip=, width=0.48\textwidth]{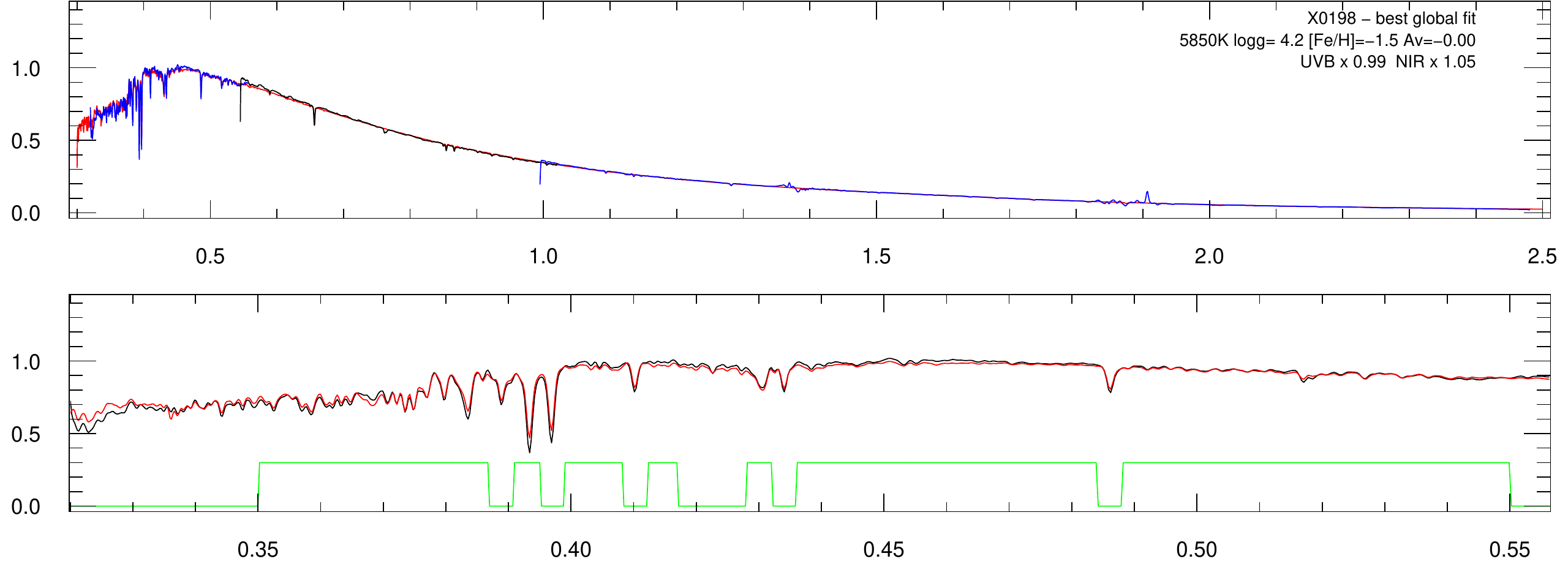} 
\includegraphics[clip=, width=0.48\textwidth]{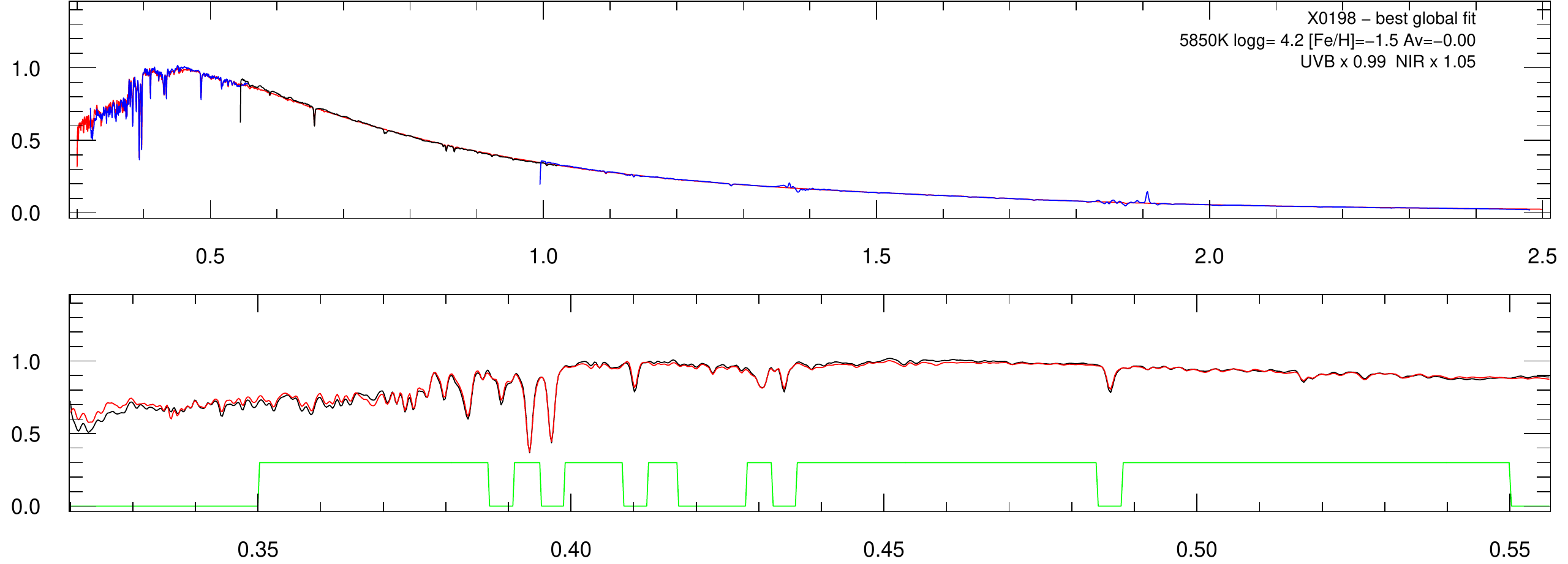} 
\vspace{-3pt}

\includegraphics[clip=, trim=0 0 0 1,width=0.48\textwidth]{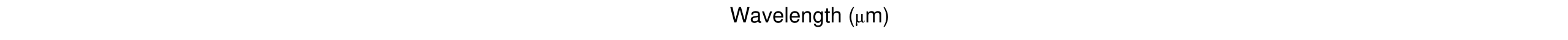}
\caption[]{Illustration of the small effects of a change of 0.4 in [$\alpha$/Fe] at $R=500$, 
using the low metallicity near-solar 
star G187-40 (observation X0198) with
the parameters of \citet{Arentsen_PP_19}.
The top two panels show that nearest model 
at  [$\alpha$/Fe] of $0$, the bottom two panels at [$\alpha$/Fe]=$+0.4$.}
\label{fig:X0198_forced_R500}
\end{figure}

For solar analogs, the fits are also satisfactory, with some discrepancies
appearing (at $R=500$) mostly in UVB features sensitive to abundances of
light elements or $\alpha$-elements. An example is given in
Fig.\,\ref{fig:X0585_forced_R500} for HD\,55693 (X0585; \teff $=5750$\,K, 
log($g$)=4.2, [Fe/H]=0.2). We note that the best-fit model in GSL 
(not shown)
has as somewhat higher \teff, of  6000\,K.
An example at subsolar metallicity ([Fe/H]$=-1.5$), for which [$\alpha$/Fe]$=+0.4$ improves
the match from good to excellent, is provided in
Fig.\,\ref{fig:X0198_forced_R500} with star G187-40 (X0198).

\begin{figure}
\includegraphics[clip=, width=0.48\textwidth]{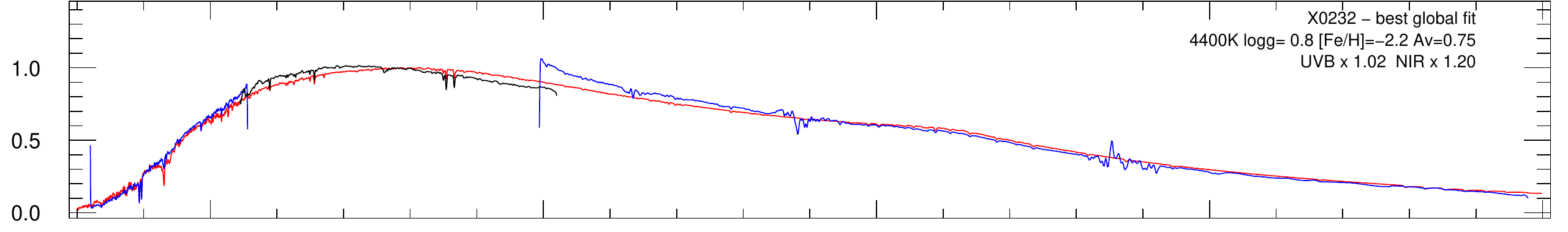} 
\includegraphics[clip=, width=0.48\textwidth]{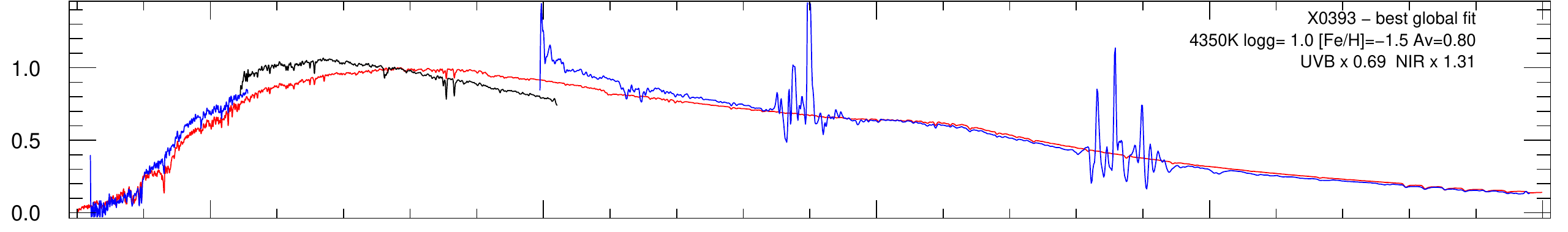} 
\includegraphics[clip=, width=0.48\textwidth]{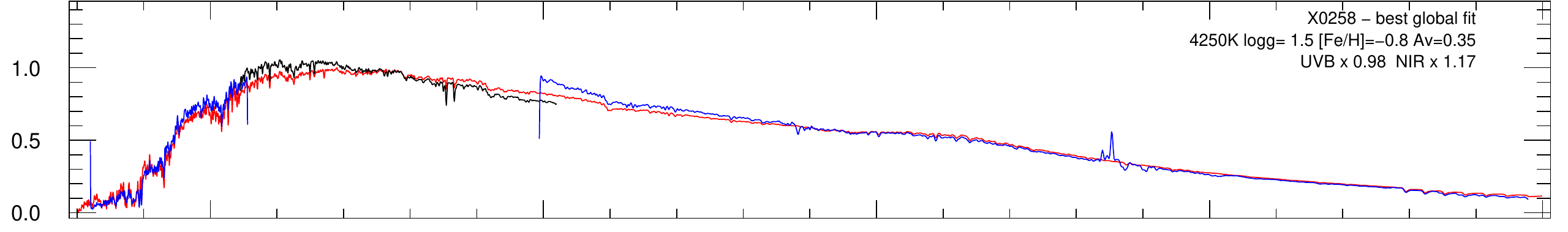} 
\includegraphics[clip=, width=0.48\textwidth]{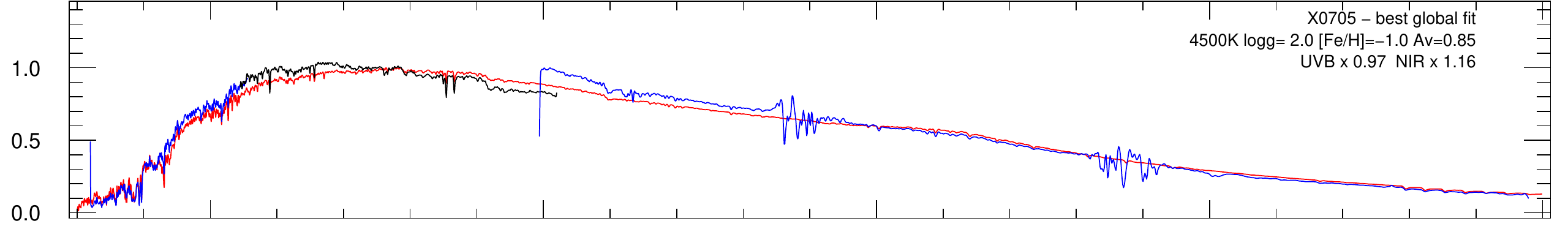} 
\includegraphics[clip=, width=0.48\textwidth]{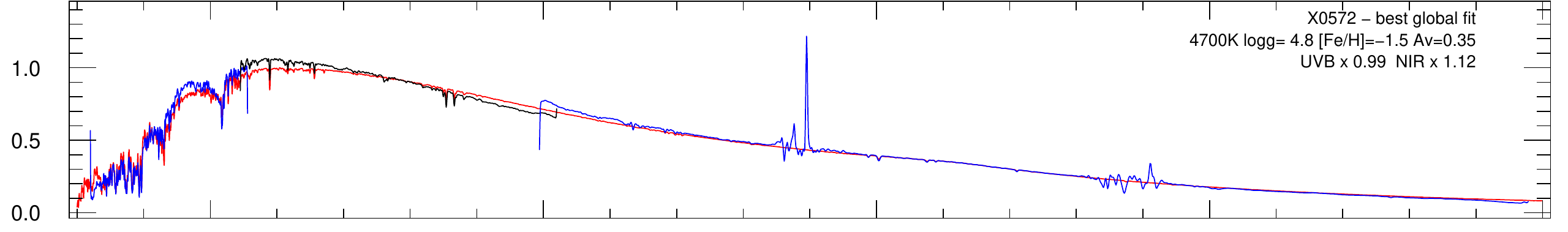} 
\includegraphics[clip=, width=0.48\textwidth]{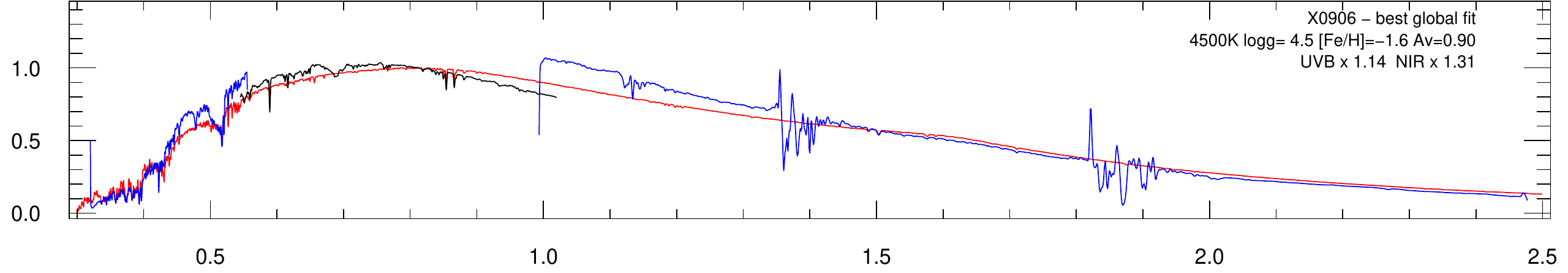} 
\includegraphics[clip=, trim=0 0 0 555,width=0.48\textwidth]{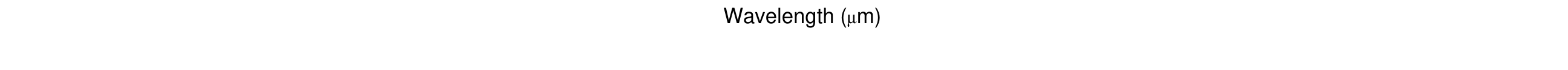}
\caption[]{Typical comparisons between XSL and GSL energy 
distributions for low metallicity stars with estimated \teff\ between 4000\,K and 5000\,K, at $R=500$,
when adopting the parameters of \citet{Arentsen_PP_19} unchanged. 
Gravity increases from top to bottom: HD\,165195 (X0232), HD\,1638 (X0258), 
NGC\,6838\,1037 (X0705), LHS\,1841 (X0572), LHS\,343 (X0906). The
discontinuities between the UVB, VIS and NIR segments of the XSL data
appear when the general energy distribution is not well matched by the
models, because our method scales each arm independently: no other
$A_V$+scaling combination would provide a smaller discrepancy $D$ (Eq.\,\ref{eq:D})
across all wavelengths. In other words, $D$ would be worse if the 
UVB, VIS and NIR segments were first merged and $A_V$ was estimated
subsequently.
}
\label{fig:X0705etal_forced_R500}
\end{figure}

\begin{figure}
\includegraphics[clip=, trim=0 420 0 10,width=0.48\textwidth]{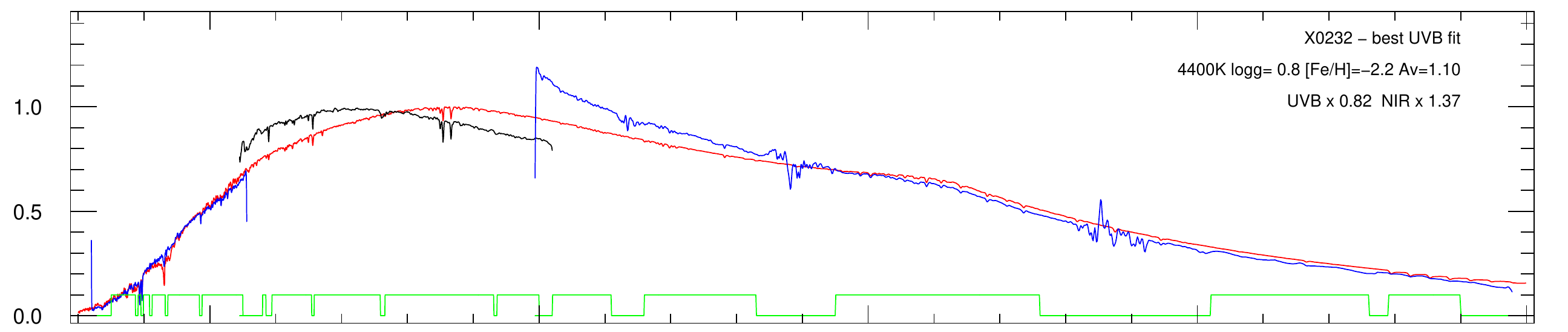} 
\includegraphics[clip=, trim=0 420 0 10,width=0.48\textwidth]{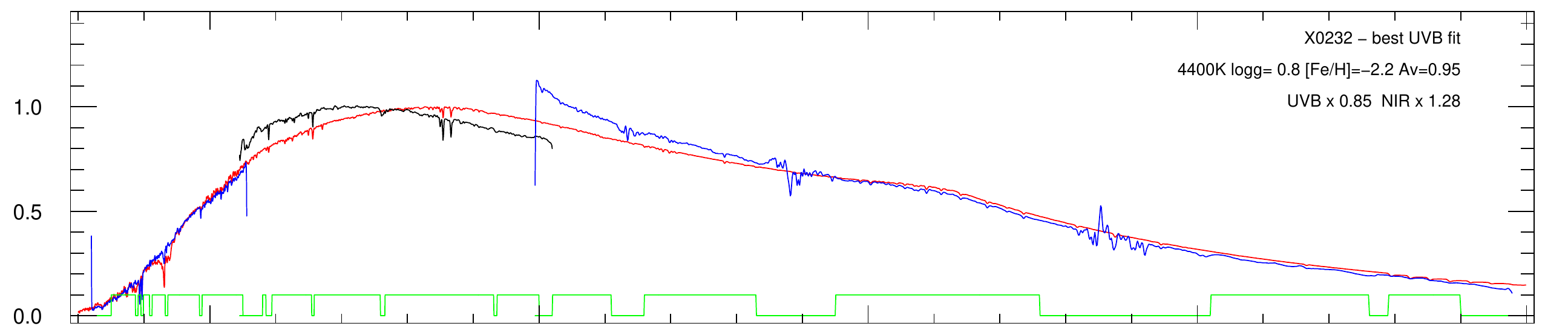} 
\includegraphics[clip=, trim=0 420 0 10,width=0.48\textwidth]{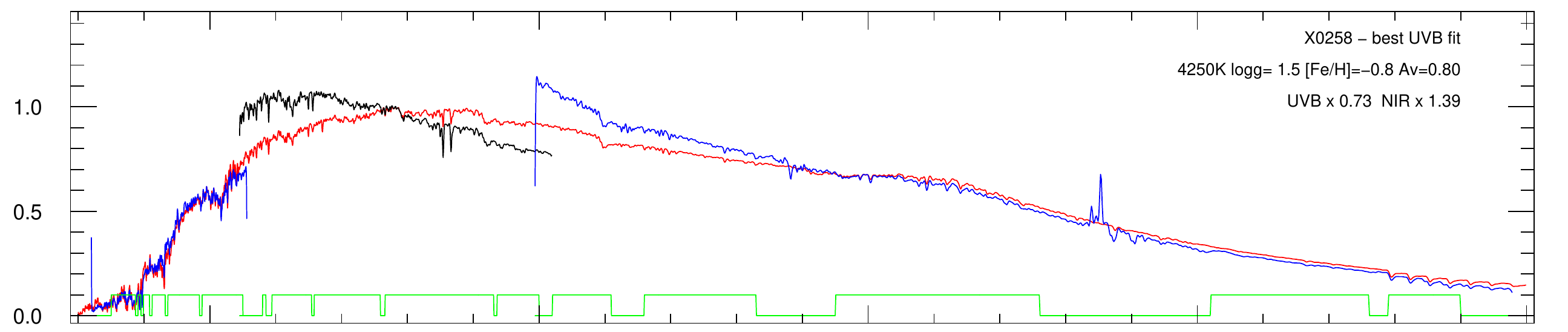} 
\includegraphics[clip=, trim=0 420 0 10,width=0.48\textwidth]{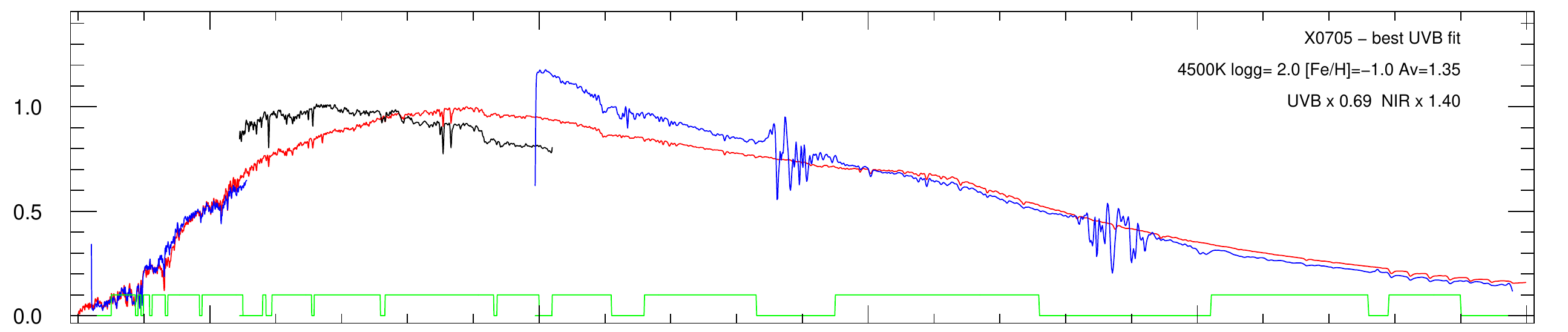} 
\includegraphics[clip=, trim=0 420 0 10,width=0.48\textwidth]{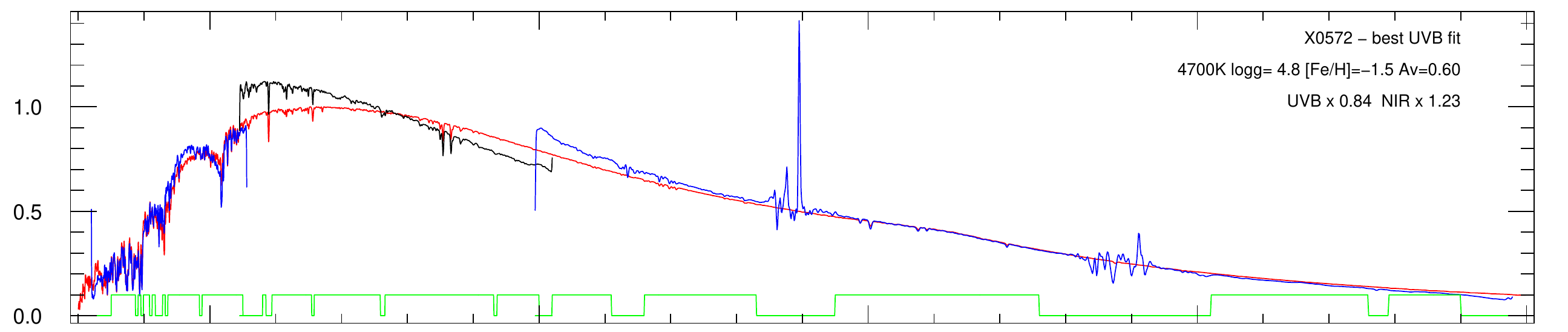} 
\includegraphics[clip=, trim=0 400 0 10,width=0.48\textwidth]{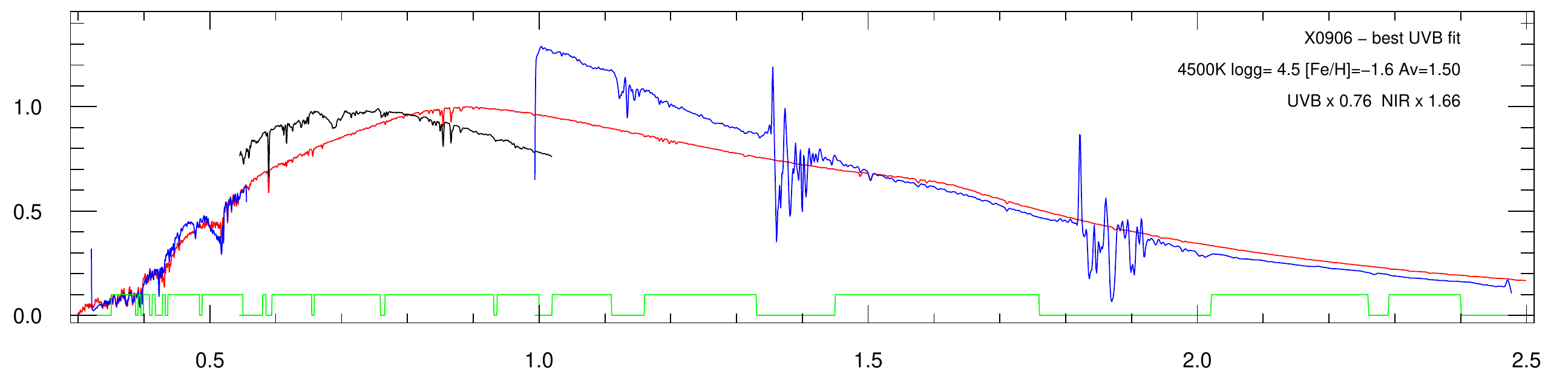} 
\includegraphics[clip=, trim=0 0 0 555,width=0.48\textwidth]{figs_spectra_forced/plt_bestglobalfit_X0258_aFe0p0.pdf}
\caption[]{Same as Fig.\,\ref{fig:X0705etal_forced_R500}, except that
$A_V$ is estimated using only UVB data.}
\label{fig:X0705etal_forcedUVB_R500}
\end{figure}

Below 5000\,K, the agreement between models and observations is progressively
lost in a systematic way. At low metallicity, discrepancies are obvious to the
eye even when all wavelengths are used to optimize $A_V$, as in Fig.\,\ref{fig:X0705etal_forced_R500}:
the observed energy distributions are more peaked around 550\,nm than the 
models, and the choice to rescale each XSL arm-spectrum independently to
the local level of the model leads to systematic discontinuities at 1\,$\mu$m.
The discrepancies are even more striking when the best $A_V$ is determined
from one of the spectral arms only (Figs.\,\ref{fig:X0705etal_forcedUVB_R500},
 \ref{fig:X0705etal_forcedVIS_R500} and \ref{fig:X0705etal_forcedNIR_R500}
 for estimates of $A_V$ based on the UVB, the VIS, and the NIR ranges only).
 It is noteworthy that the attempt to match UVB observations and models 
 \citep[with the parameters of][]{Arentsen_PP_19} leads to positive $A_V$ in all cases,
 and consequently also when all wavelengths are exploited. In general,
 the value of the preferred $A_V$ decreases with wavelength.

\begin{figure}
\includegraphics[clip=, trim=0 236 0 194,width=0.48\textwidth]{figs_spectra_forced/plt_best1armfits_X0232_aFe0p4.pdf} 
\includegraphics[clip=, trim=0 236 0 194,width=0.48\textwidth]{figs_spectra_forced/plt_best1armfits_X0393_aFe0p0.pdf} 
\includegraphics[clip=, trim=0 236 0 194,width=0.48\textwidth]{figs_spectra_forced/plt_best1armfits_X0258_aFe0p0.pdf} 
\includegraphics[clip=, trim=0 236 0 194,width=0.48\textwidth]{figs_spectra_forced/plt_best1armfits_X0705_aFe0p0.pdf} 
\includegraphics[clip=, trim=0 236 0 194,width=0.48\textwidth]{figs_spectra_forced/plt_best1armfits_X0572_aFe0p0.pdf} 
\includegraphics[clip=, trim=0 216 0 194,width=0.48\textwidth]{figs_spectra_forced/plt_best1armfits_X0906_aFe0p0.pdf} 
\includegraphics[clip=, trim=0 0 0 555,width=0.48\textwidth]{figs_spectra_forced/plt_bestglobalfit_X0258_aFe0p0.pdf}
\caption[]{Same as Fig.\,\ref{fig:X0705etal_forced_R500}, except that
$A_V$ is estimated using only VIS data.}
\label{fig:X0705etal_forcedVIS_R500}
\end{figure}

\begin{figure}
\includegraphics[clip=, trim=0 55 0 372,width=0.48\textwidth]{figs_spectra_forced/plt_best1armfits_X0232_aFe0p4.pdf} 
\includegraphics[clip=, trim=0 55 0 372,width=0.48\textwidth]{figs_spectra_forced/plt_best1armfits_X0393_aFe0p0.pdf} 
\includegraphics[clip=, trim=0 55 0 372,width=0.48\textwidth]{figs_spectra_forced/plt_best1armfits_X0258_aFe0p0.pdf} 
\includegraphics[clip=, trim=0 55 0 372,width=0.48\textwidth]{figs_spectra_forced/plt_best1armfits_X0705_aFe0p0.pdf} 
\includegraphics[clip=, trim=0 55 0 372,width=0.48\textwidth]{figs_spectra_forced/plt_best1armfits_X0572_aFe0p0.pdf} 
\includegraphics[clip=, trim=0 35 0 372,width=0.48\textwidth]{figs_spectra_forced/plt_best1armfits_X0906_aFe0p0.pdf} 
\includegraphics[clip=, trim=0 0 0 555,width=0.48\textwidth]{figs_spectra_forced/plt_bestglobalfit_X0258_aFe0p0.pdf}
\caption[]{Same as Fig.\,\ref{fig:X0705etal_forced_R500}, except that
$A_V$ is estimated using only NIR data.}
\label{fig:X0705etal_forcedNIR_R500}
\end{figure}

\begin{figure}
\includegraphics[clip=,width=0.48\textwidth]{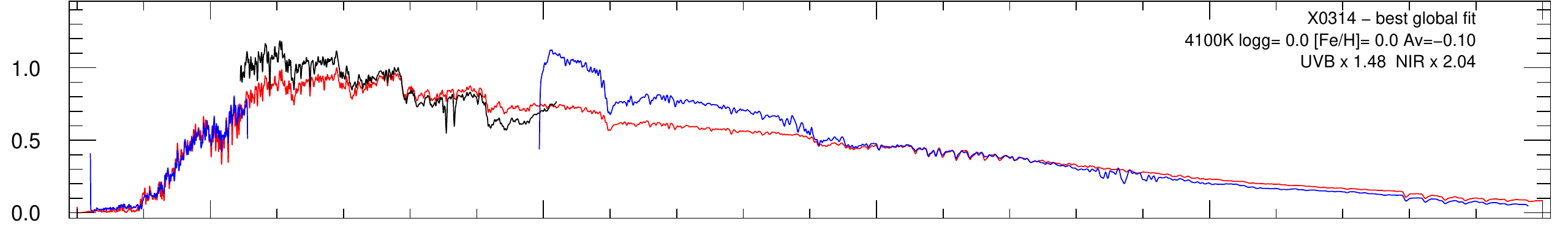}
\includegraphics[clip=, width=0.48\textwidth]{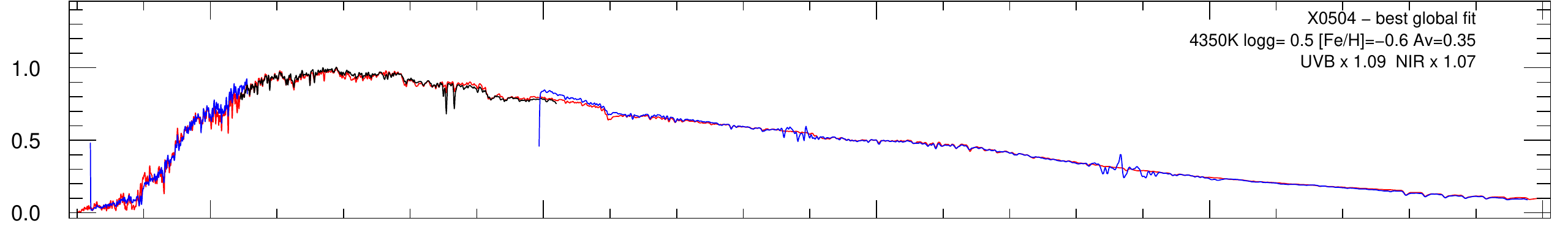} 
\includegraphics[clip=, width=0.48\textwidth]{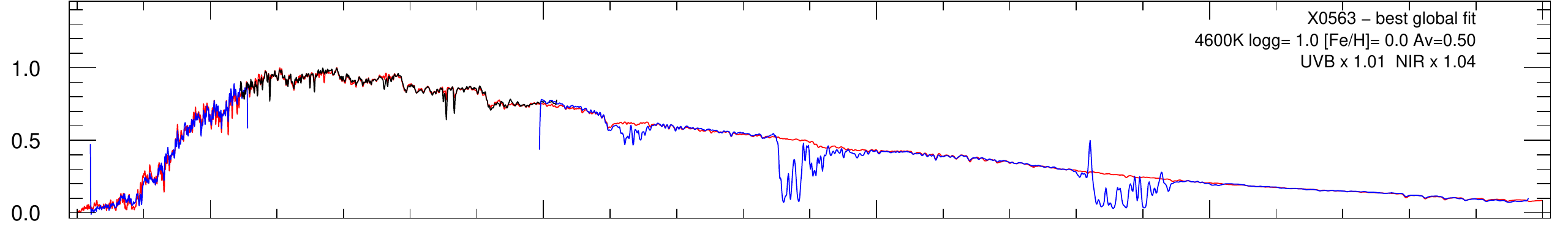} 
\includegraphics[clip=, width=0.48\textwidth]{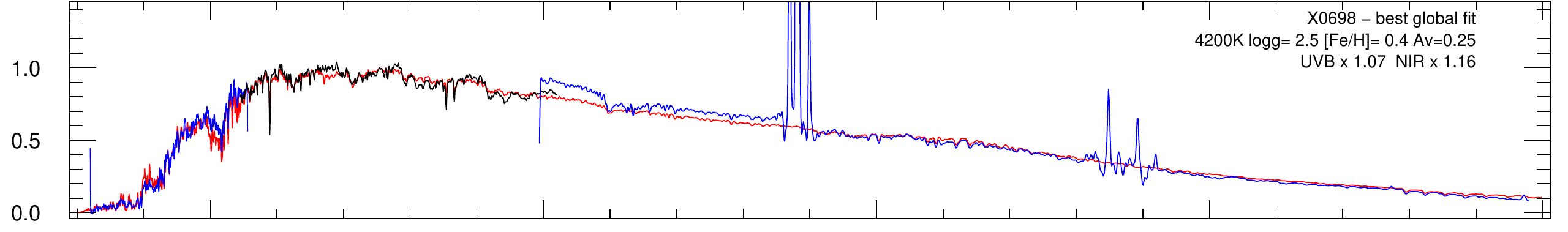} 
\includegraphics[clip=, width=0.48\textwidth]{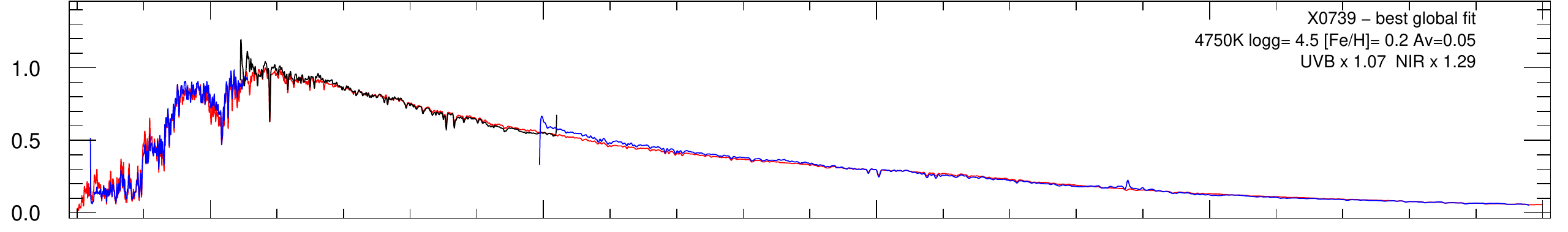} 
\includegraphics[clip=, width=0.48\textwidth]{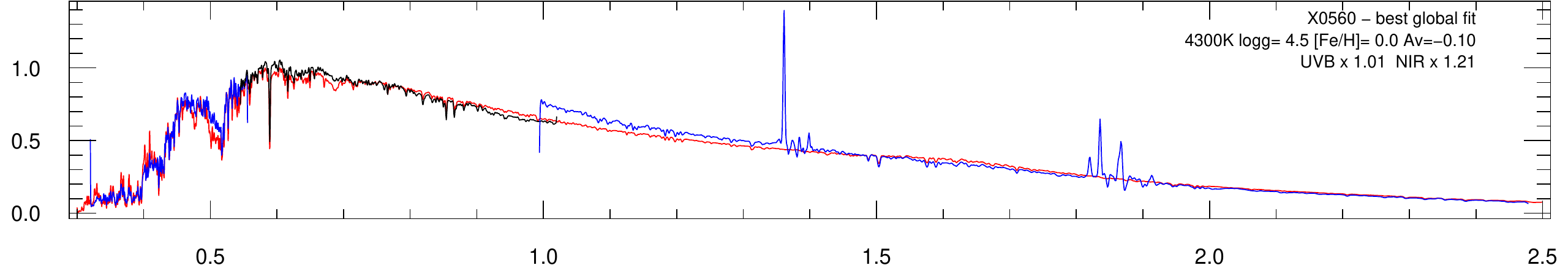} 
\vspace{-3pt}

\includegraphics[clip=, trim=0 0 0 1,width=0.48\textwidth]{figs_spectra_forced/waveBanner.pdf}
\caption[]{Typical comparisons between XSL and GSL energy 
distributions for stars with relatively high metallicities and 
with estimated \teff\ between 4000\,K and 5000\,K, at $R=500$,
when adopting the parameters of \citet{Arentsen_PP_19} unchanged. 
Gravity increases from top to bottom: HD\,50877 (X0314), 
BBB\,SMC\,104 (X0504), HD\,44391 (X0563), 2MASS\,J18351420-3438060 (X0698),
HD\,218566 (X0739), HD\,21197 (X0560).  As in Fig.\,\ref{fig:X0705etal_forced_R500},
discontinuities between the UVB, VIS, and NIR segments of the XSL data
appear when the general energy distribution is not well matched by the
models.}
\label{fig:X0314etal_forced_R500}
\end{figure}

\begin{figure}
\includegraphics[clip=, trim=0 420 0 10,width=0.48\textwidth]{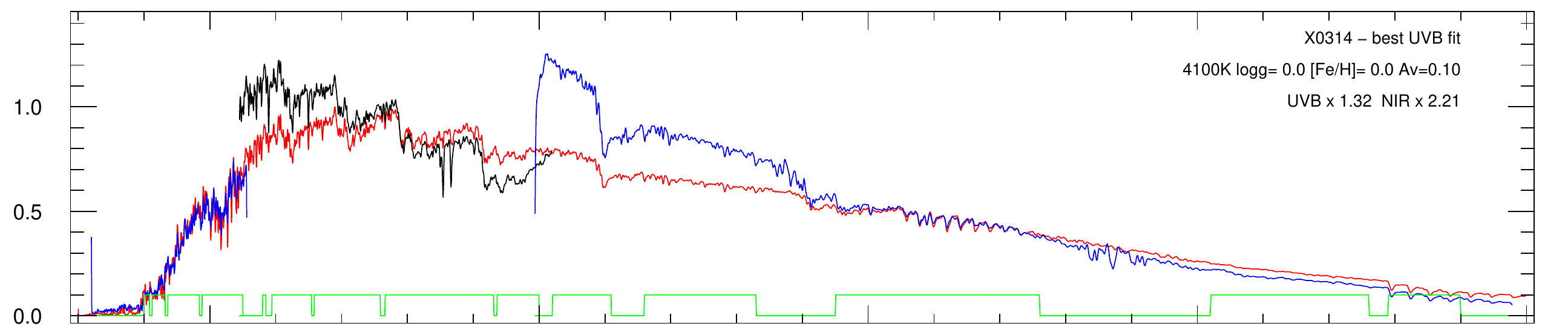} 
\includegraphics[clip=, trim=0 420 0 10,width=0.48\textwidth]{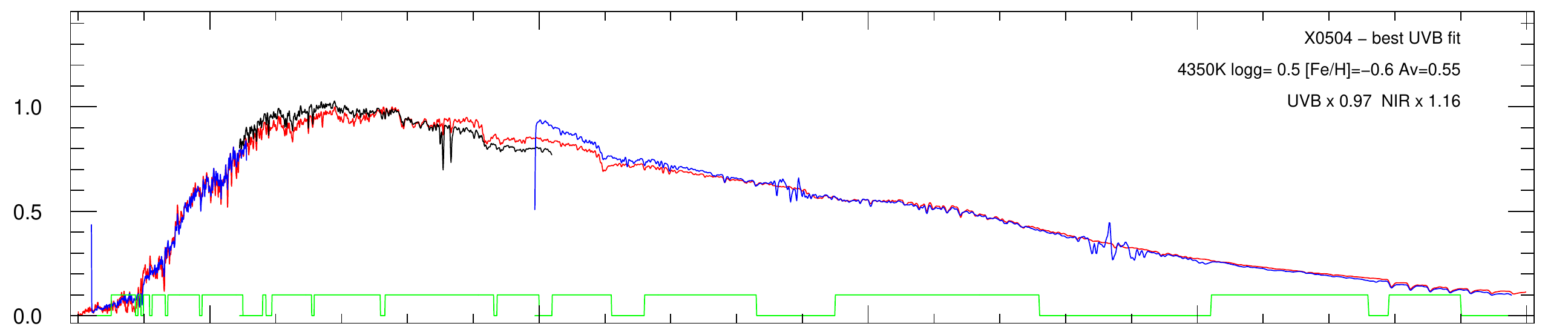} 
\includegraphics[clip=, trim=0 420 0 10,width=0.48\textwidth]{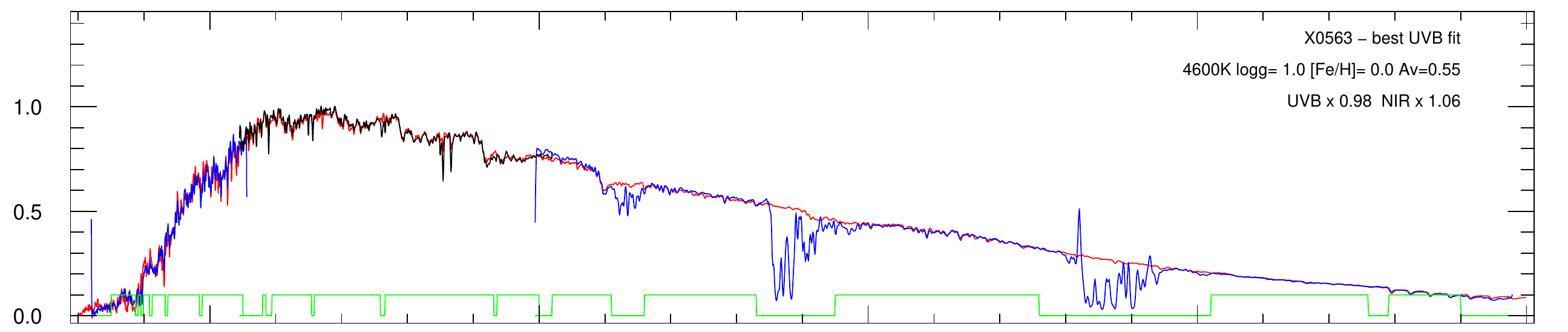} 
\includegraphics[clip=, trim=0 420 0 10,width=0.48\textwidth]{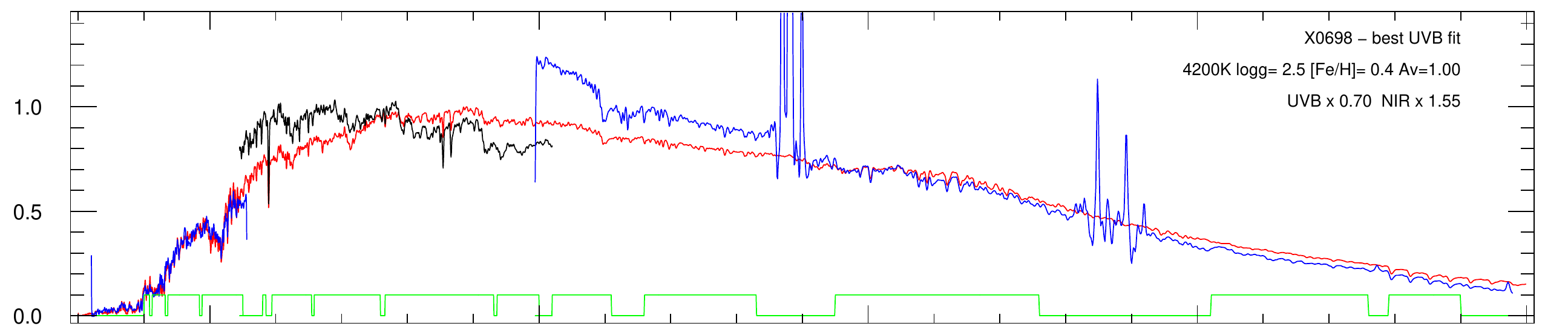} 
\includegraphics[clip=, trim=0 420 0 10,width=0.48\textwidth]{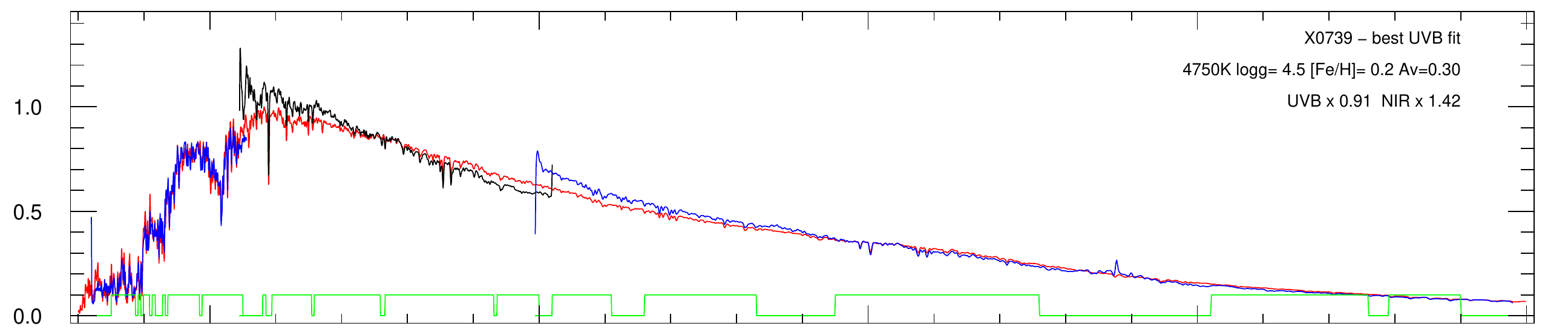} 
\includegraphics[clip=, trim=0 400 0 10,width=0.48\textwidth]{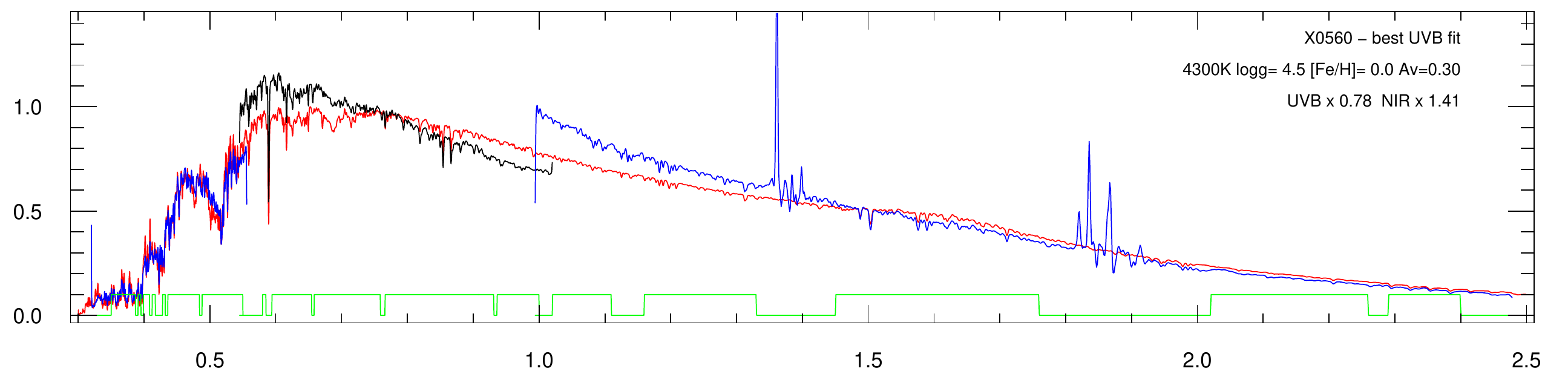} 
\includegraphics[clip=, trim=0 0 0 555,width=0.48\textwidth]{figs_spectra_forced/plt_bestglobalfit_X0258_aFe0p0.pdf}
\caption[]{Same as Fig.\,\ref{fig:X0314etal_forced_R500}, except that
$A_V$ is estimated using only UVB data.
}
\label{fig:X0314etal_forcedUVB_R500}
\end{figure}

\begin{figure}
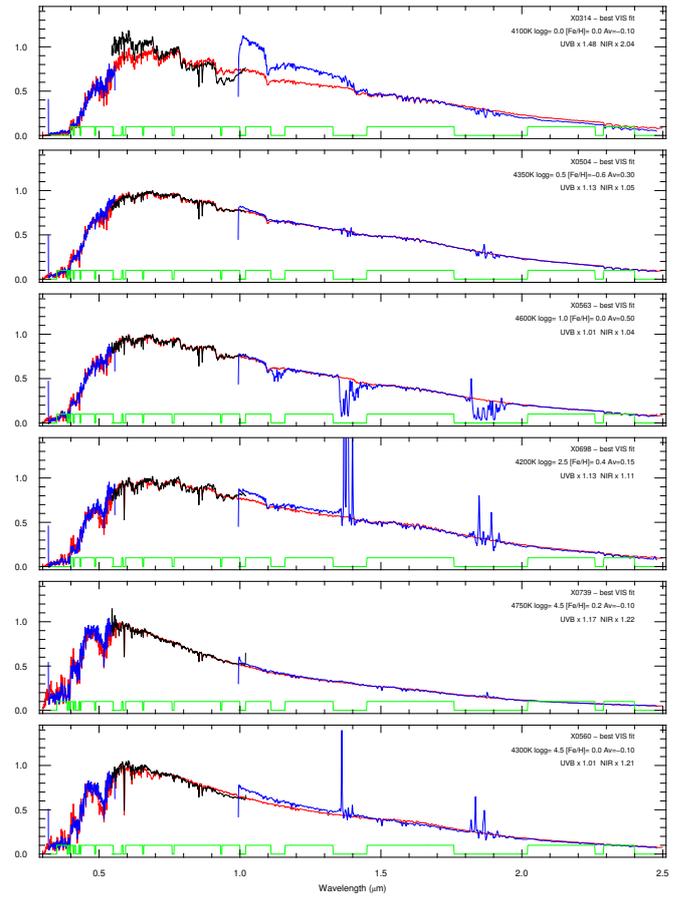

\includegraphics[clip=, trim=0 236 0 194,width=0.48\textwidth]{figs_spectra_forced/plt_best1armfits_X0314_aFe0p0.pdf} 
\includegraphics[clip=, trim=0 236 0 194,width=0.48\textwidth]{figs_spectra_forced/plt_best1armfits_X0504_aFe0p0.pdf} 
\includegraphics[clip=, trim=0 236 0 194,width=0.48\textwidth]{figs_spectra_forced/plt_best1armfits_X0563_aFe0p0.pdf} 
\includegraphics[clip=, trim=0 236 0 194,width=0.48\textwidth]{figs_spectra_forced/plt_best1armfits_X0698_aFe0p0.pdf} 
\includegraphics[clip=, trim=0 236 0 194,width=0.48\textwidth]{figs_spectra_forced/plt_best1armfits_X0739_aFe0p0.pdf} 
\includegraphics[clip=, trim=0 216 0 194,width=0.48\textwidth]{figs_spectra_forced/plt_best1armfits_X0560_aFe0p0.pdf} 
\includegraphics[clip=, trim=0 0 0 555,width=0.48\textwidth]{figs_spectra_forced/plt_bestglobalfit_X0258_aFe0p0.pdf}
\caption[]{Same as Fig.\,\ref{fig:X0314etal_forced_R500}, except that
$A_V$ is estimated using only VIS data.
}
\label{fig:X0314etal_forcedVIS_R500}
\end{figure}

Still in the range between 4000\,K and 5000\,K, the comparisons of the empirical
and theoretical energy distributions for stars with solar-like metallicities 
(Fig,\,\ref{fig:X0314etal_forced_R500}) is, in fact, 
generally slightly better than at subsolar metallicities.
The outlier in the top panel of that figure is a red supergiant for which 
the DR2-parameters seem far off, possibly as a result of variability; other red 
supergiants are not this drastically discrepant. 
The other panels are representative.
Although figures with $A_V$ determined from single arms
display trends similar in nature to those shown previously at lower metallicity, the 
discrepancies have a somewhat smaller amplitude, on average (only two
of these figures is shown, in Fig.\,\ref{fig:X0314etal_forcedUVB_R500} and \ref{fig:X0314etal_forcedVIS_R500}).

\clearpage

\section{Improvements obtained when replacing DR2 parameters with best fit parameters.}
\begin{figure*}
\begin{center}
\includegraphics[clip=,width=0.45\textwidth]{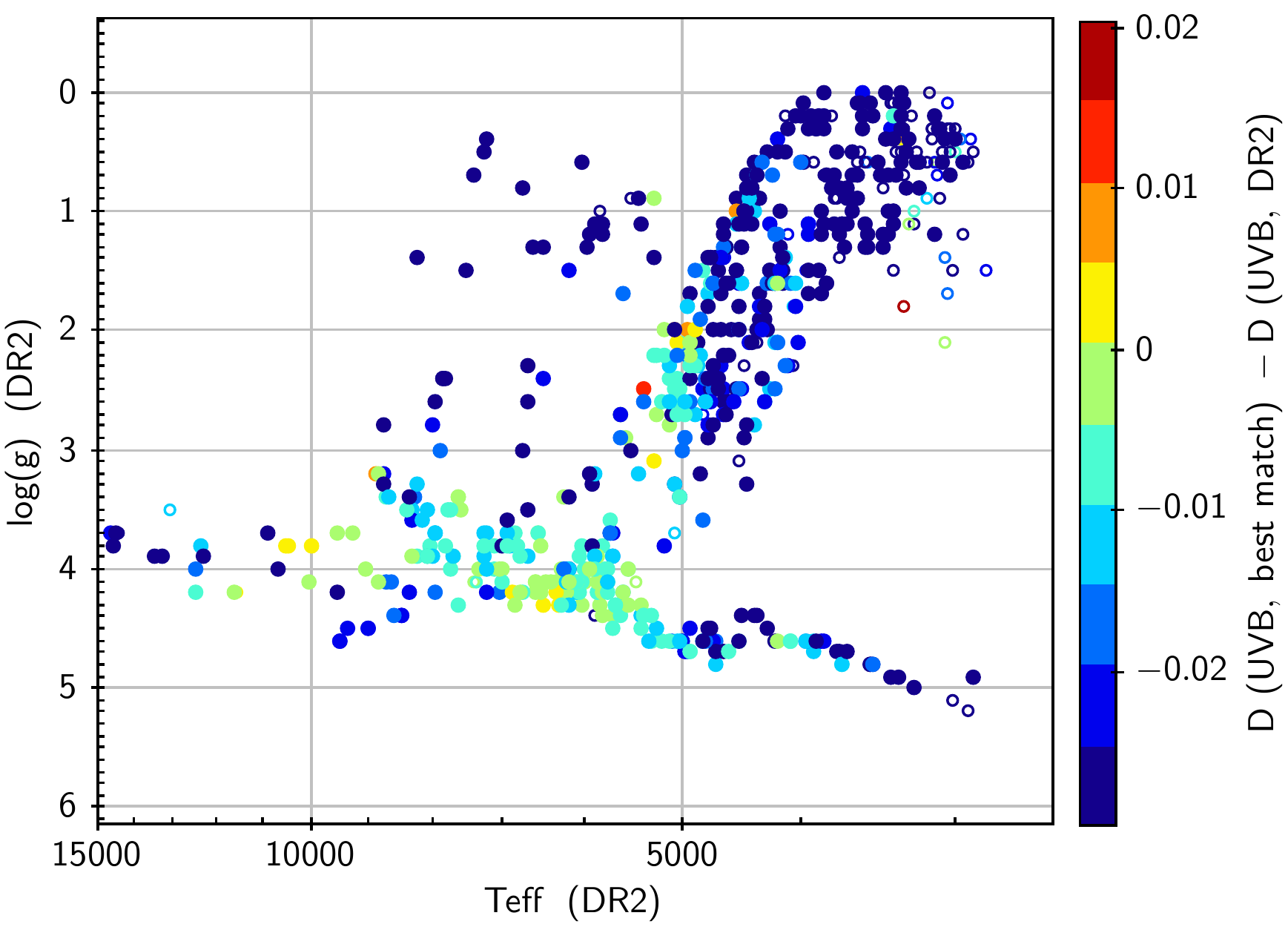}
\includegraphics[clip=,width=0.45\textwidth]{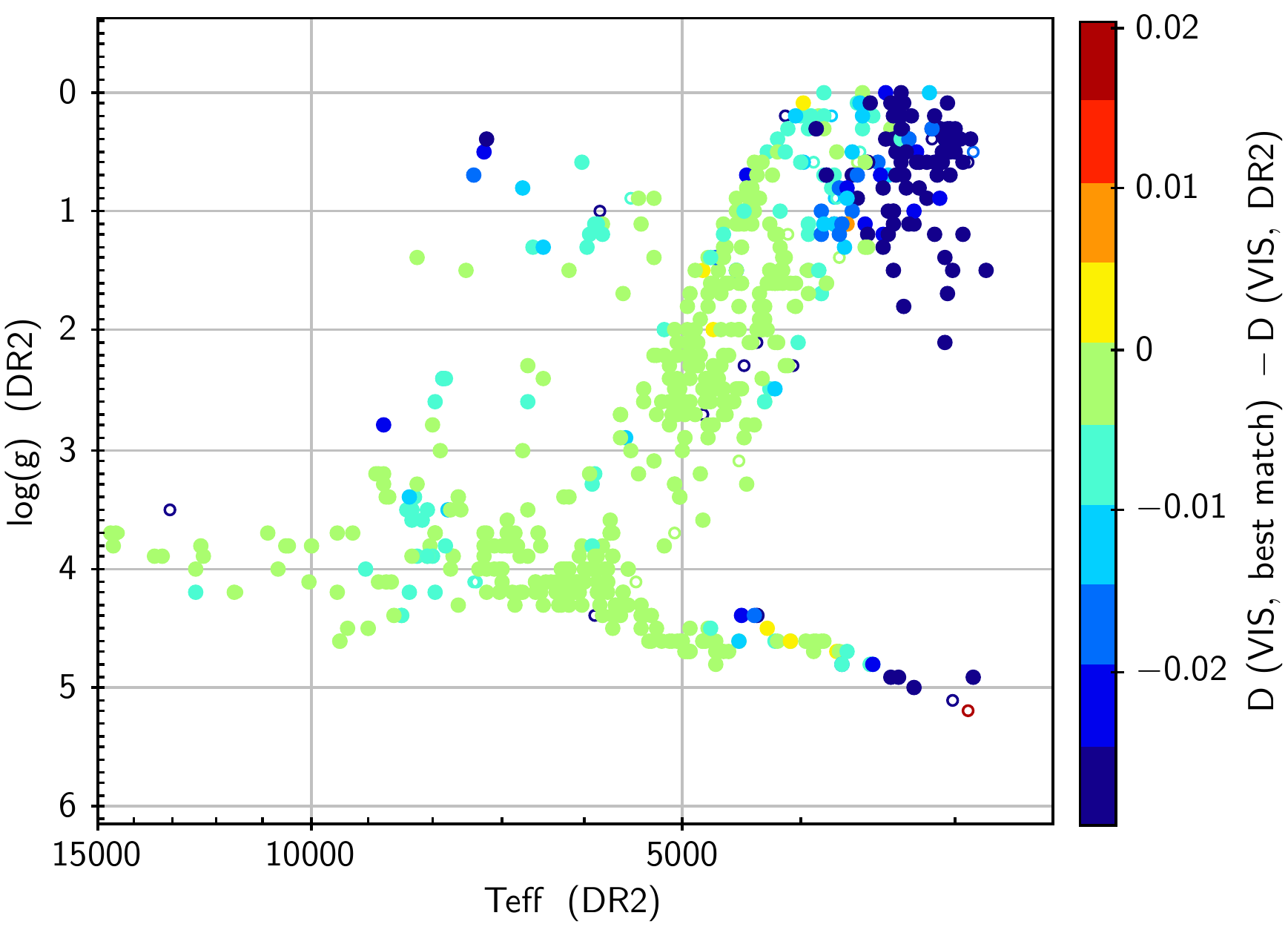} \\
\includegraphics[clip=,width=0.45\textwidth]{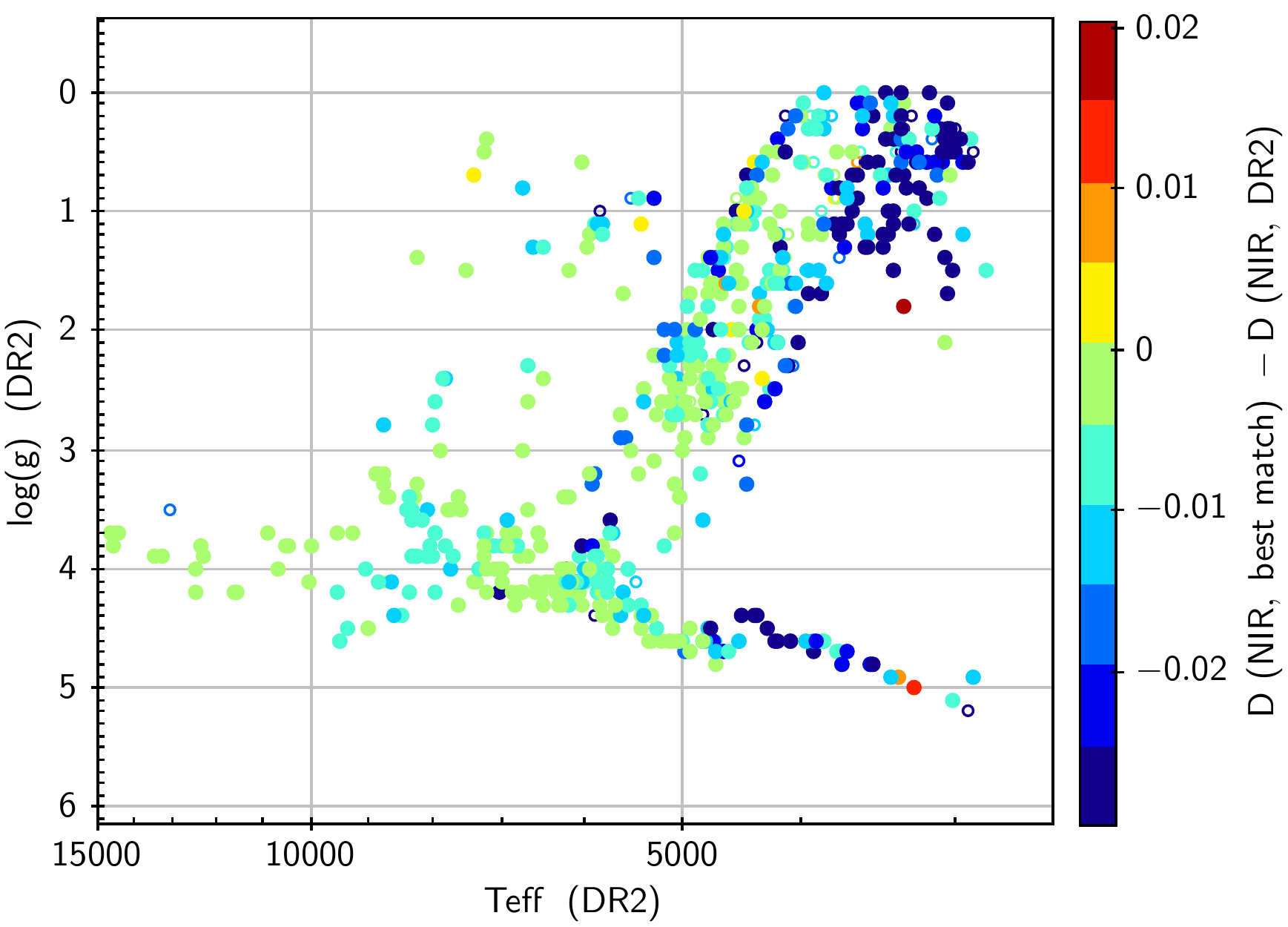}
\includegraphics[clip=,width=0.45\textwidth]{figs_bestfit_R500_nonoise_aFe_26mar20/HRdiag_by_D_ALL_FitMinusForced.pdf}
\end{center}
\caption[]{Improvement in the XSL-GSL match when instead of adopting 
stellar parameters from \citet{Arentsen_PP_19}, the parameters are optimized. Only 
the bottom right panel was shown in the main body of the paper (Fig.\,\ref{fig:HRdiag_D_best}). 
}
\label{fig:HRdiag_Dimproved_allArms}
\end{figure*}

\begin{figure*}
\begin{center}
\includegraphics[clip=,width=0.45\textwidth]{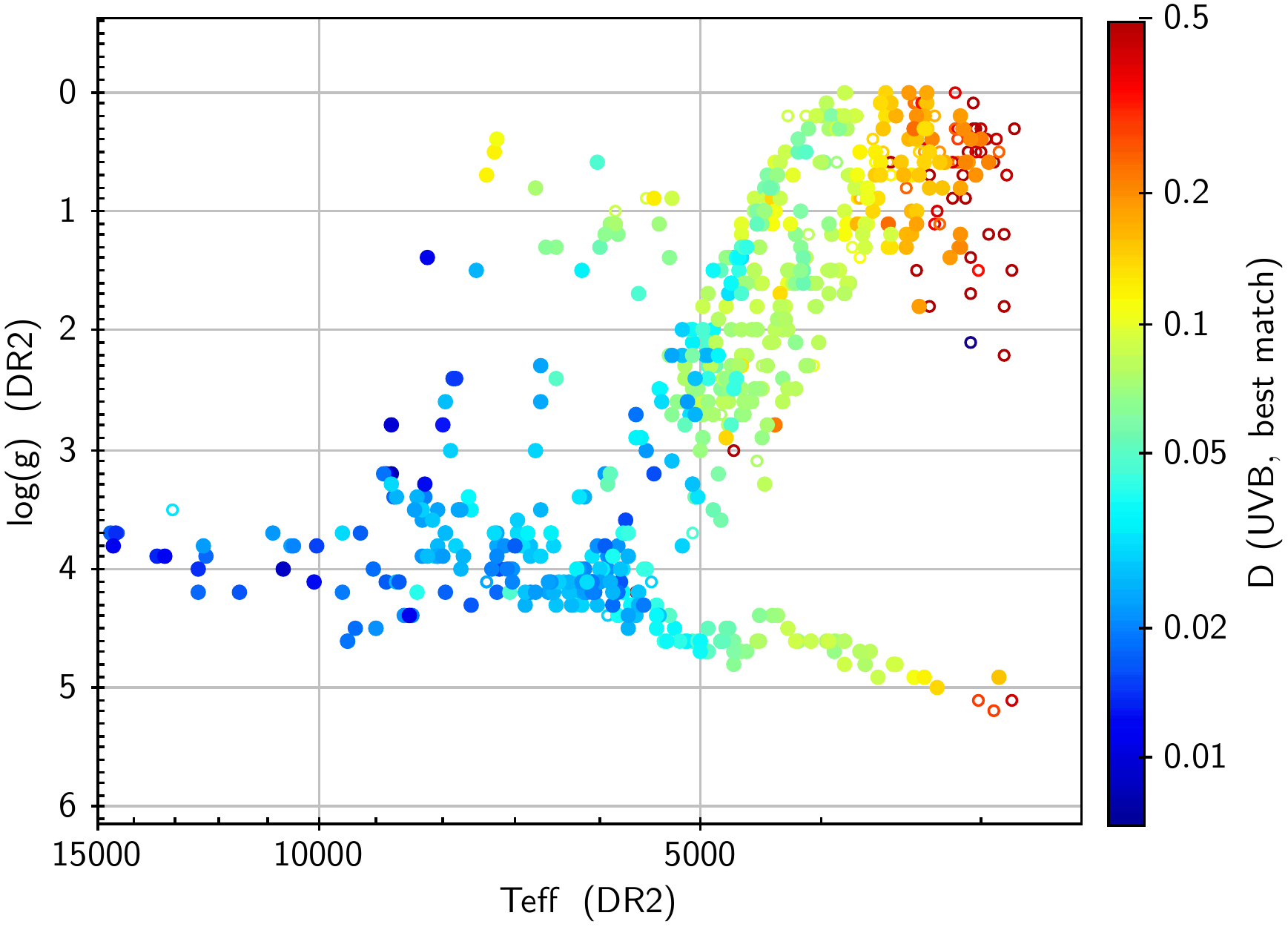}
\includegraphics[clip=,width=0.45\textwidth]{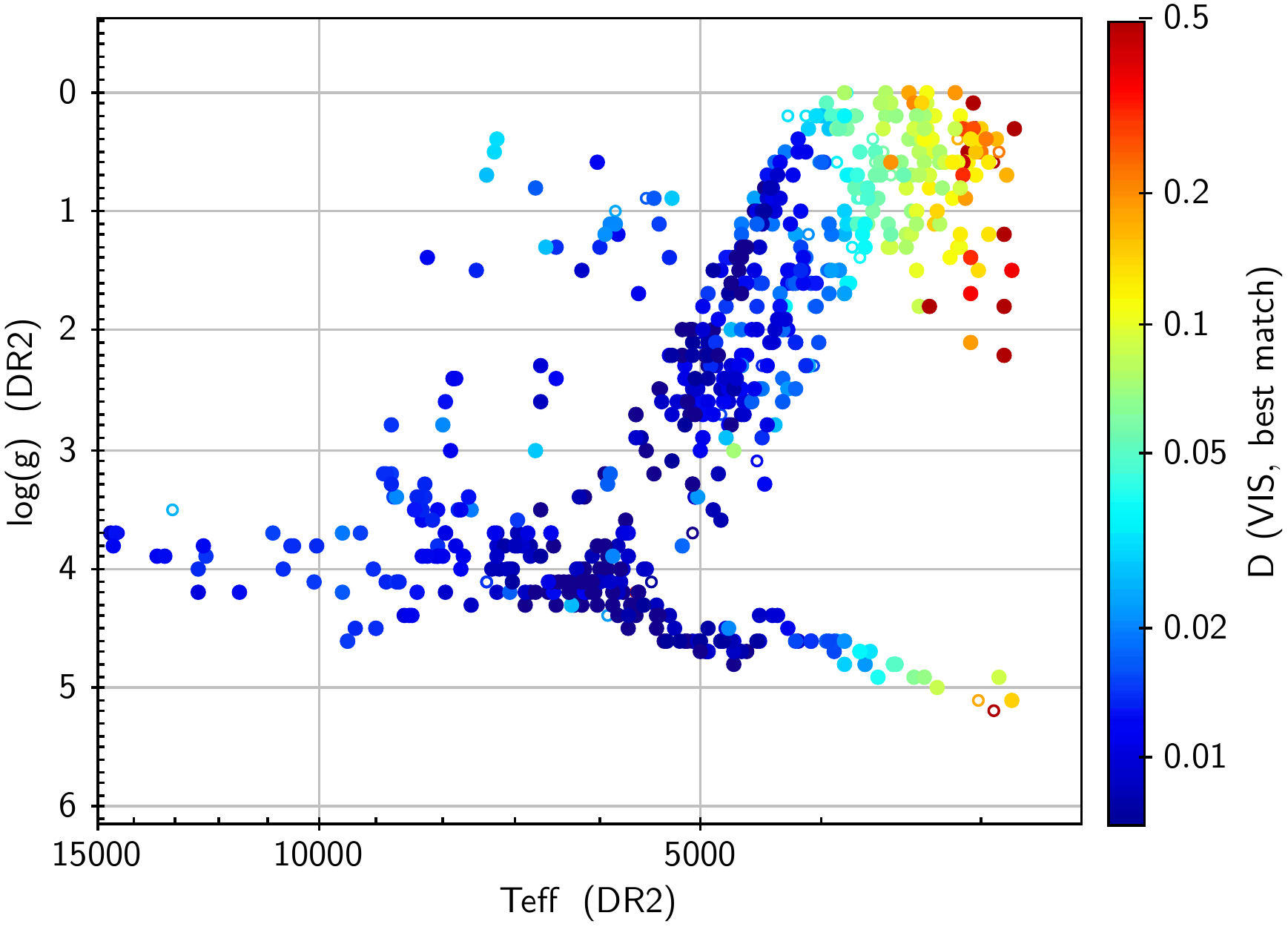}\\
\includegraphics[clip=,width=0.45\textwidth]{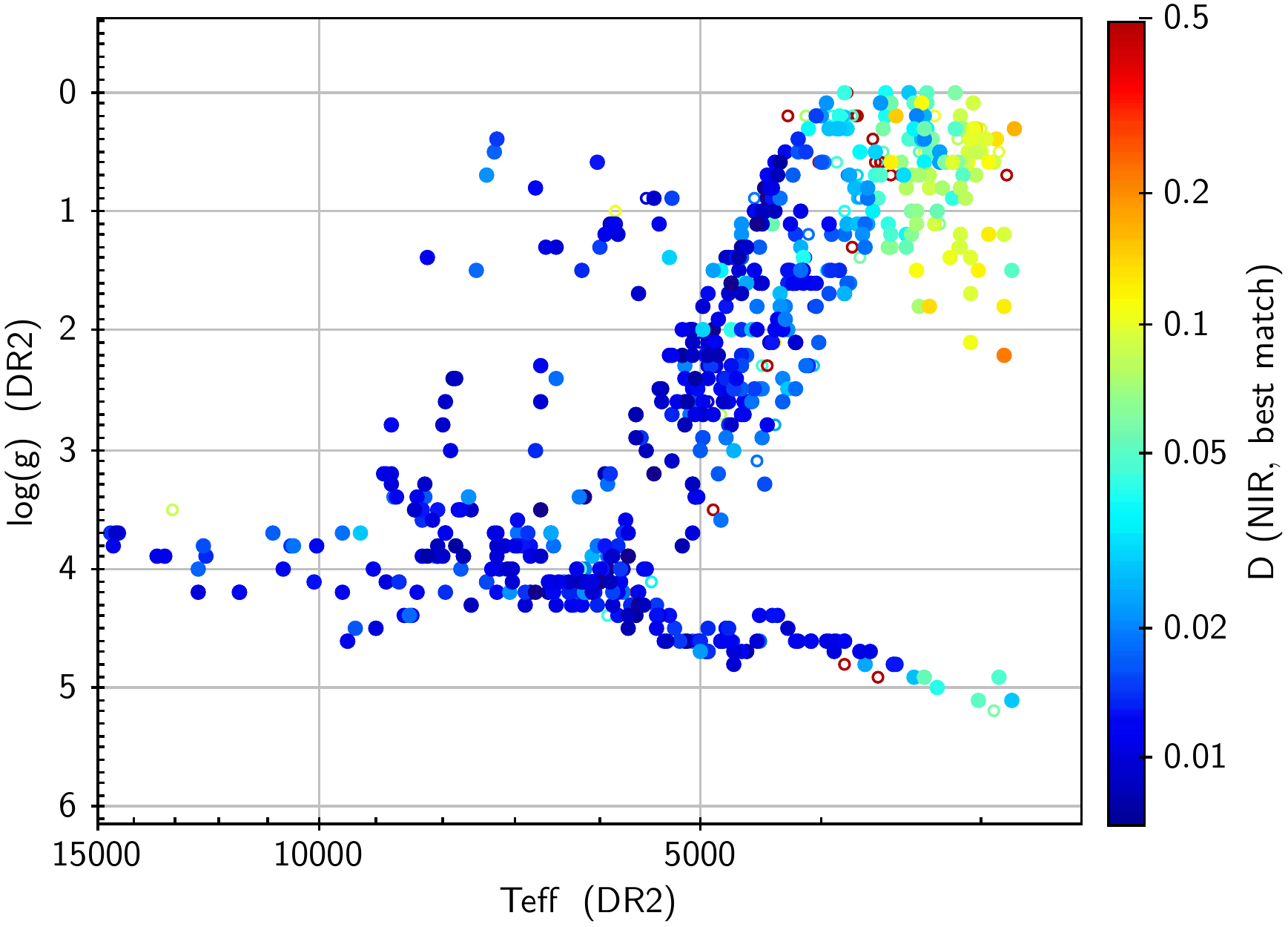}
\includegraphics[clip=,width=0.45\textwidth]{figs_bestfit_R500_nonoise_aFe_26mar20/HRdiag_bestfit_by_ALLchi2_glob.pdf}
\end{center}
\caption[]{Discrepancy measure $D$ between the XSL and GSL spectra, as 
obtained for the best-match synthetic spectra. The
four panels correspond best-fit determinations and calculations of $D$ based respectively
on the UVB, VIS, and NIR arm of X-shooter, and on the three arms combined.
To be compared to  Fig.\,\ref{fig:forcedfits}, where the stellar parameters of XSL-DR2 were assumed.
}
\label{fig:HRdiag_D_best_allArms}
\end{figure*}

In Fig.\,\ref{fig:HRdiag_D_best}, we showed how a free choice 
of best-fit parameters improves the goodness-of-fit measure $D$, in the case of a fit
to all wavelengths. Here, Figs.\,\ref{fig:HRdiag_Dimproved_allArms} and
\ref{fig:HRdiag_D_best_allArms} provide this information for the four variants of 
the fits: UVB only, VIS only, NIR only or ALL wavelengths.

\clearpage

\section{Parameter estimates : differences between arms}
\label{app:diffs_between_arms}

\begin{figure}
\includegraphics[clip=,width=0.48\textwidth]{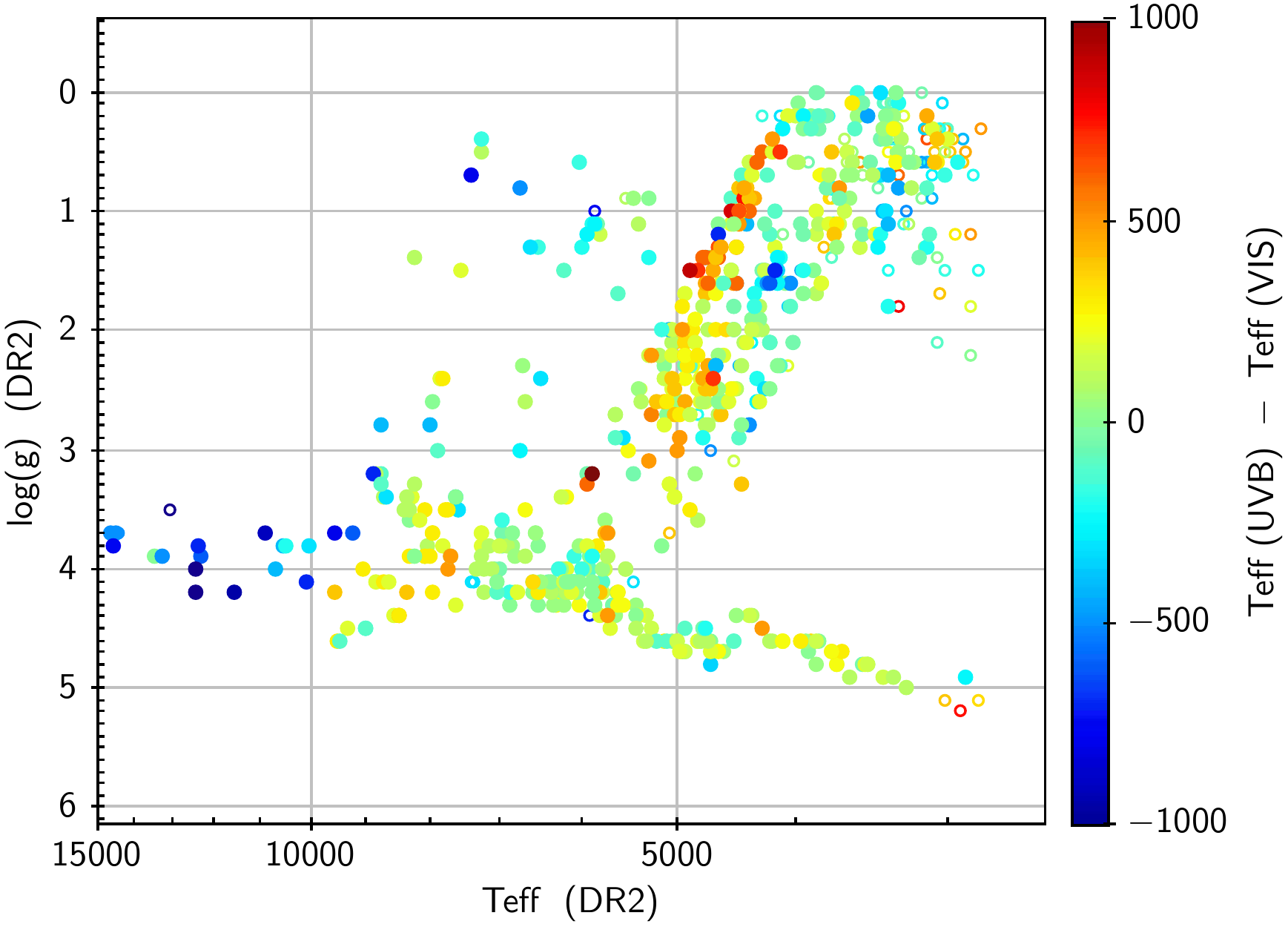}
\includegraphics[clip=,width=0.48\textwidth]{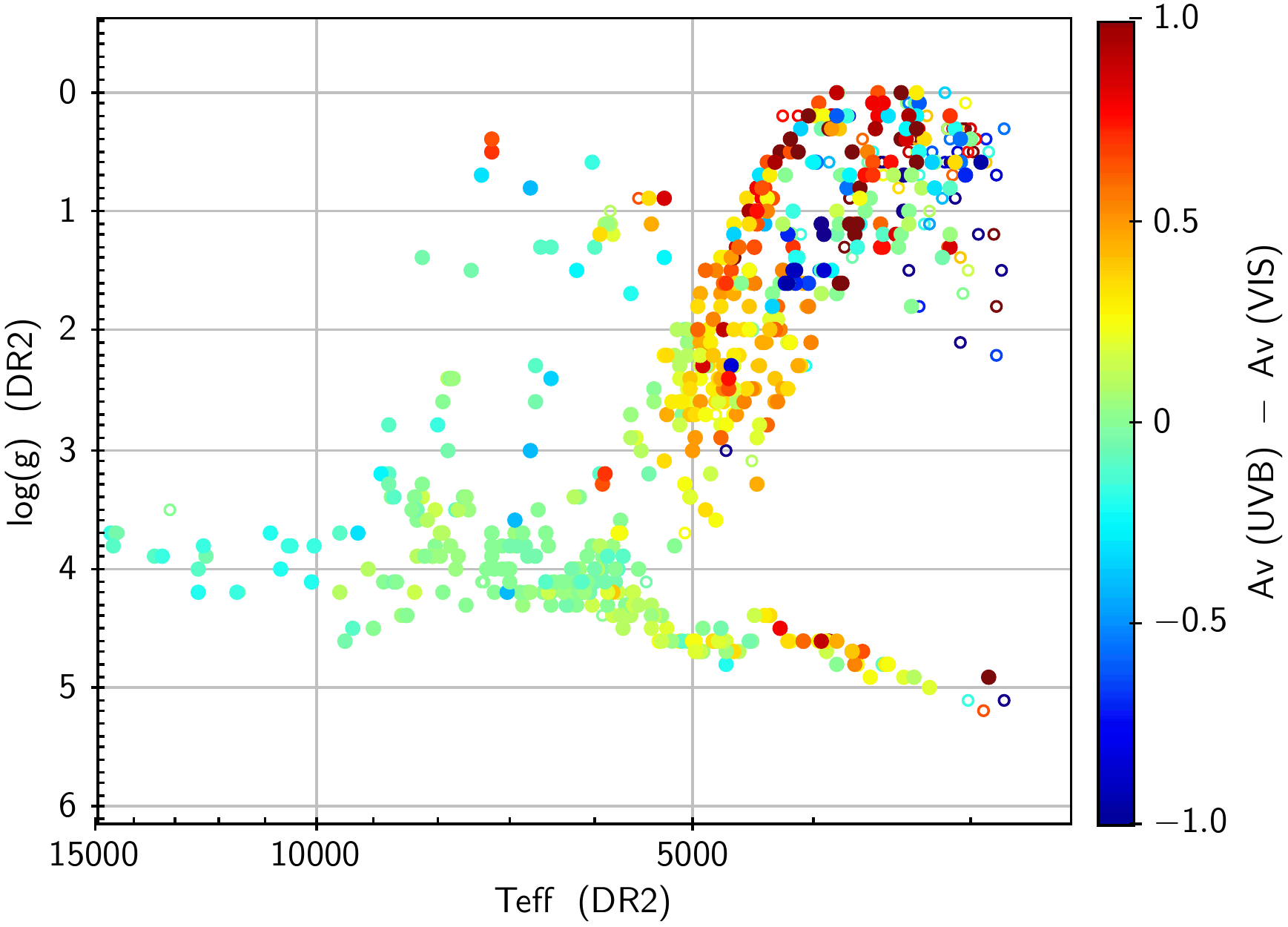}
\caption[]{HR diagrams colored by $\Delta \teff \equiv \teff \mathrm{(UVB)}-\teff \mathrm{(VIS)} $
(top) and by $\Delta A_V \equiv A_V \mathrm{(UVB)}-A_V \mathrm{(VIS)}$\ (bottom),
based on inverse-variance weighted $\chi^2$ comparisons at R=500.}
\label{fig:HRD_by_TeffUVB_m_TeffVIS}
\end{figure}

\begin{figure}
\includegraphics[clip=,width=0.48\textwidth]{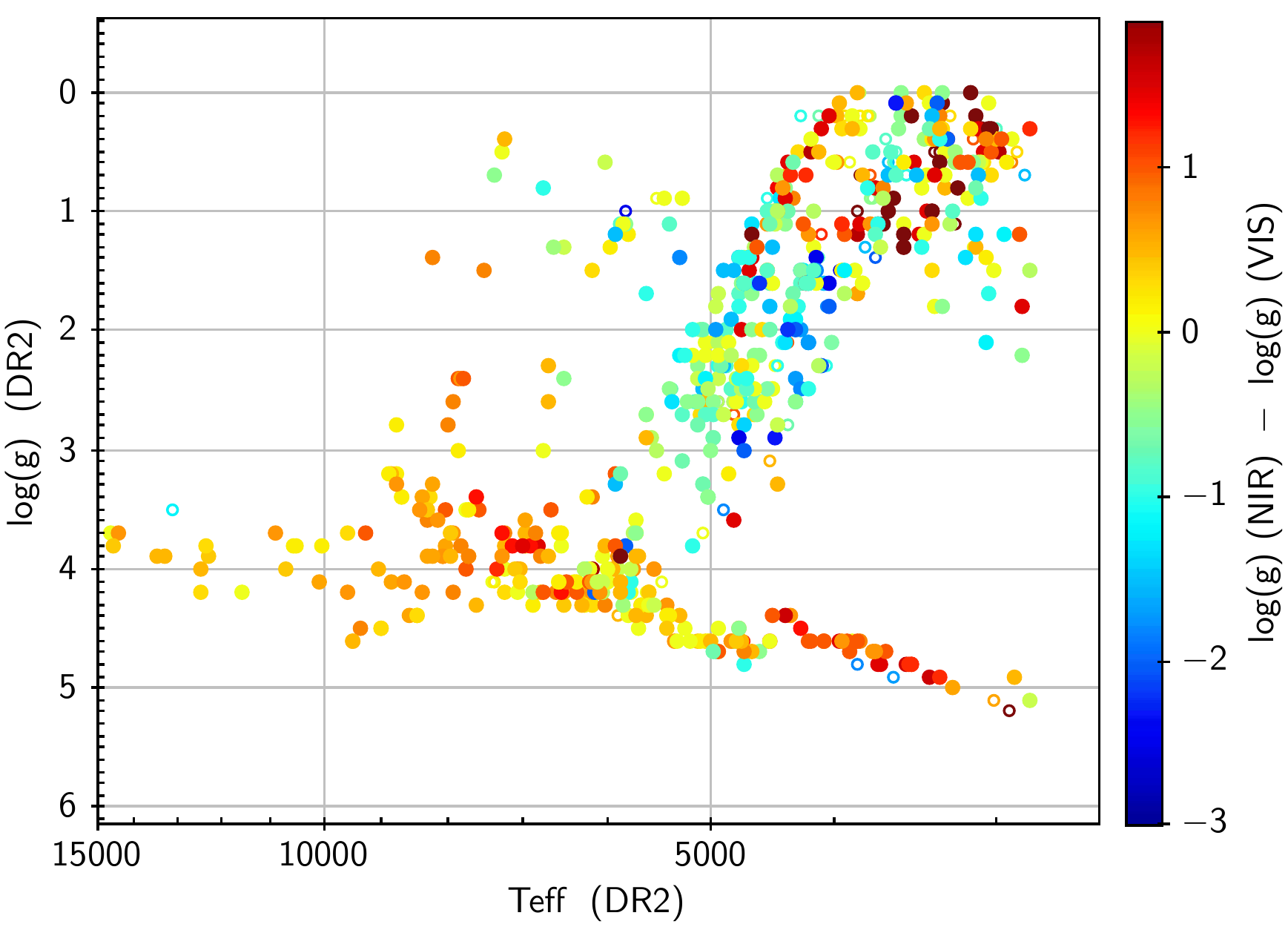}
\caption[]{HR diagram colored by $ \log(g) \mathrm{(NIR)}-\log(g) \mathrm{(VIS)}$,
based on inverse-variance weighted $\chi^2$ comparisons at R=500.}
\label{fig:HRD_by_loggNIR_m_loggVIS}
\end{figure}

As discussed in Section\,\ref{sec:results_best_residualDiscrepancies}, differences
between the parameters derived separately from the three X-shooter arms
are an indication of imperfection in the models, when they are systematic 
(otherwise they may result from flux calibration errors or other anecdoticalgray
artifacts). The main trends have been discussed in the main text.
A few more localized systematics are mentioned here.

A first example
is the difference  $\Delta \teff \equiv \teff \mathrm{(UVB)}-\teff \mathrm{(VIS)}$,
which correlates with $\Delta A_V \equiv A_V \mathrm{(UVB)} - A_V \mathrm{(VIS)}$
on the whole, but shows a more discriminant 
behavior in the HR diagram. For instance, Fig.\,\ref{fig:HRD_by_TeffUVB_m_TeffVIS}
highlights that while the majority of giants between 4000 and 5000\,K lead 
to a positive $\Delta A_V$, a smaller fraction has a positive $\Delta \teff$. Those are
typically the low-metallicity giants:
90\% of the giant branch stars with [Fe/H]$<-1.5$ \citep[according to][]{Arentsen_PP_19} have 
best-fit temperatures with
$\Delta \teff \geqslant 100$\,K, while this fraction is about 45\,\% for 
more metal-rich giants between 3800 and 7000\,K (or also for the whole sample of spectra).
The median $\Delta \teff$ for metal-poor giants is $\sim 400$\,K.
These numbers remain the same whether one assumes [$\alpha$/Fe]=0 or [$\alpha$/Fe]=+0.4.
%
On the main sequence at temperatures between 10\,000 and 15\,000\,K (20 stars), $\Delta \teff$  is smallest : it 
is systematically negative with typical values near $-600$\,K. Extinction is not involved here, since
these stars are in general reddened very little.

In Fig.\,\ref{fig:HRD_by_loggNIR_m_loggVIS}, the HR diagram is colored by the difference between
arms that shows the strongest local systematics when we require that the NIR arm be involved.
This turns out to be the difference between the surface gravities derived from the NIR and from
another arm. In the comparison with the GSL models, the NIR arm pulls towards high gravities
on the main sequence, while it pulls towards low gravities for many intermediate luminosity red 
giants.

\clearpage

\section{Effects of [$\alpha$/Fe] on parameters estimated here}
\label{app:aFe_bestfit}

\begin{figure*}
~\hfill $\chi^2$ \hfill \teff \hfill log($g$) \hfill [Fe/H] \hfill ~\\
\includegraphics[clip=,width=0.24\textwidth]{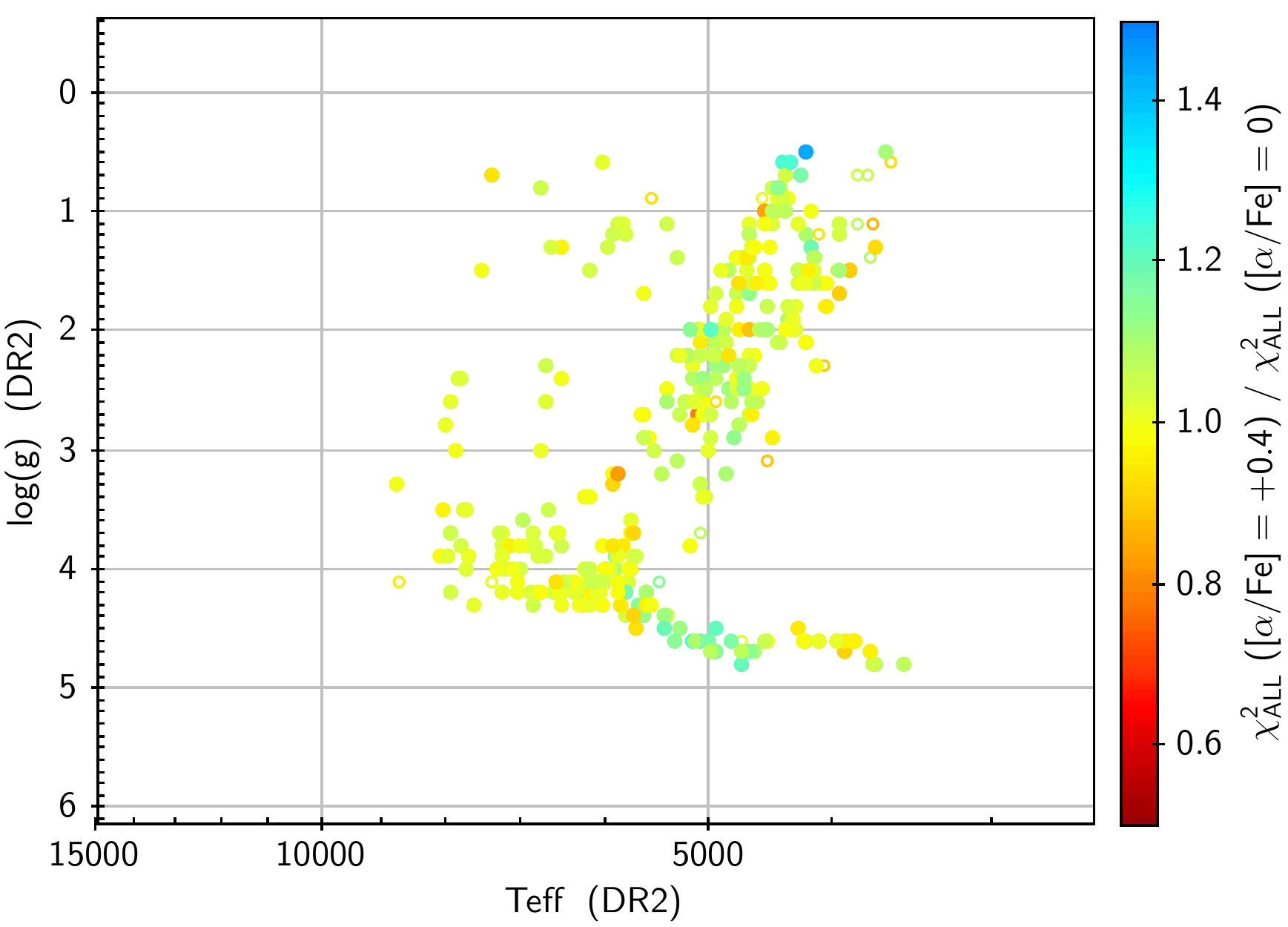}
\includegraphics[clip=,width=0.24\textwidth]{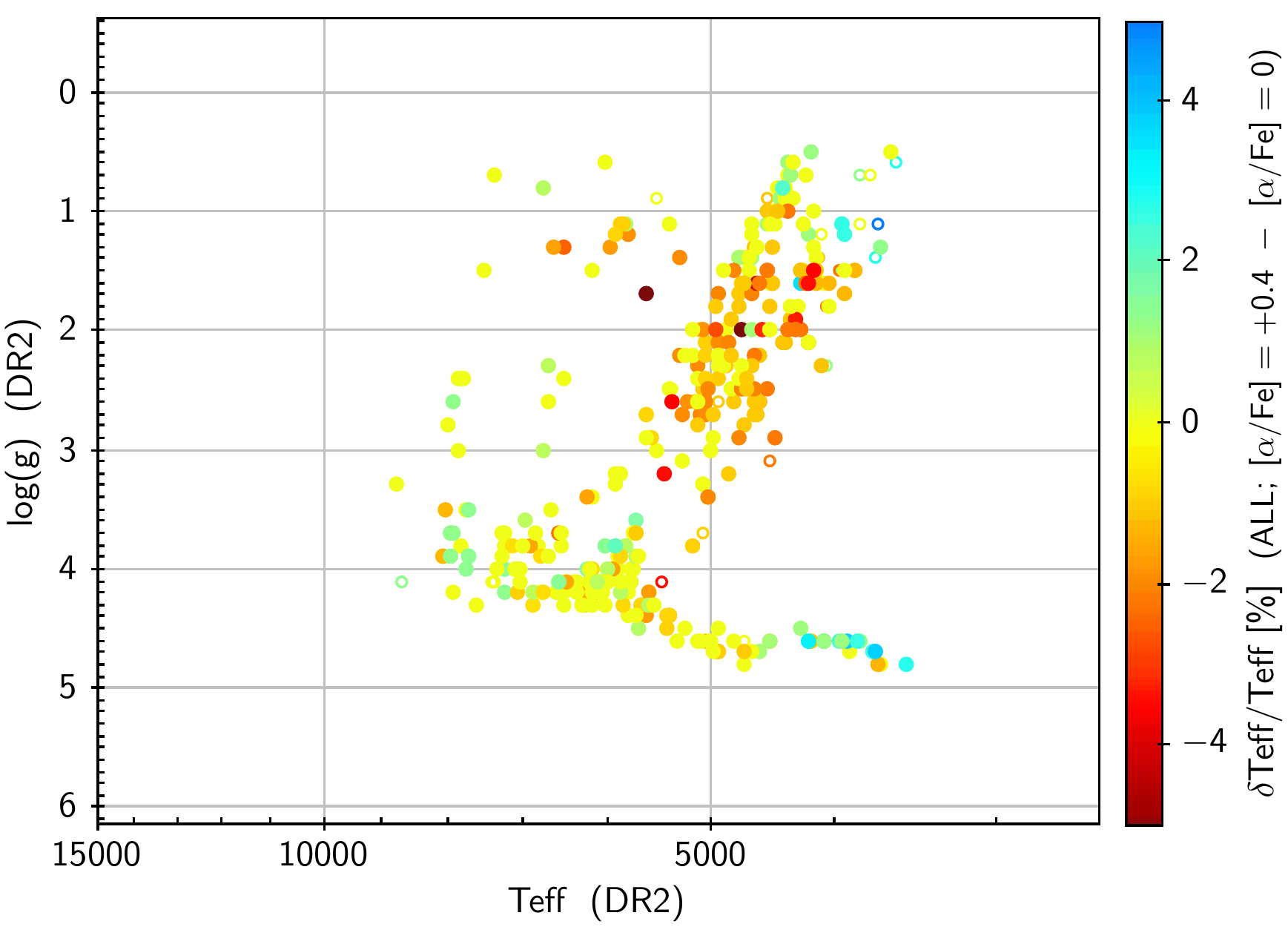}
\includegraphics[clip=,width=0.24\textwidth]{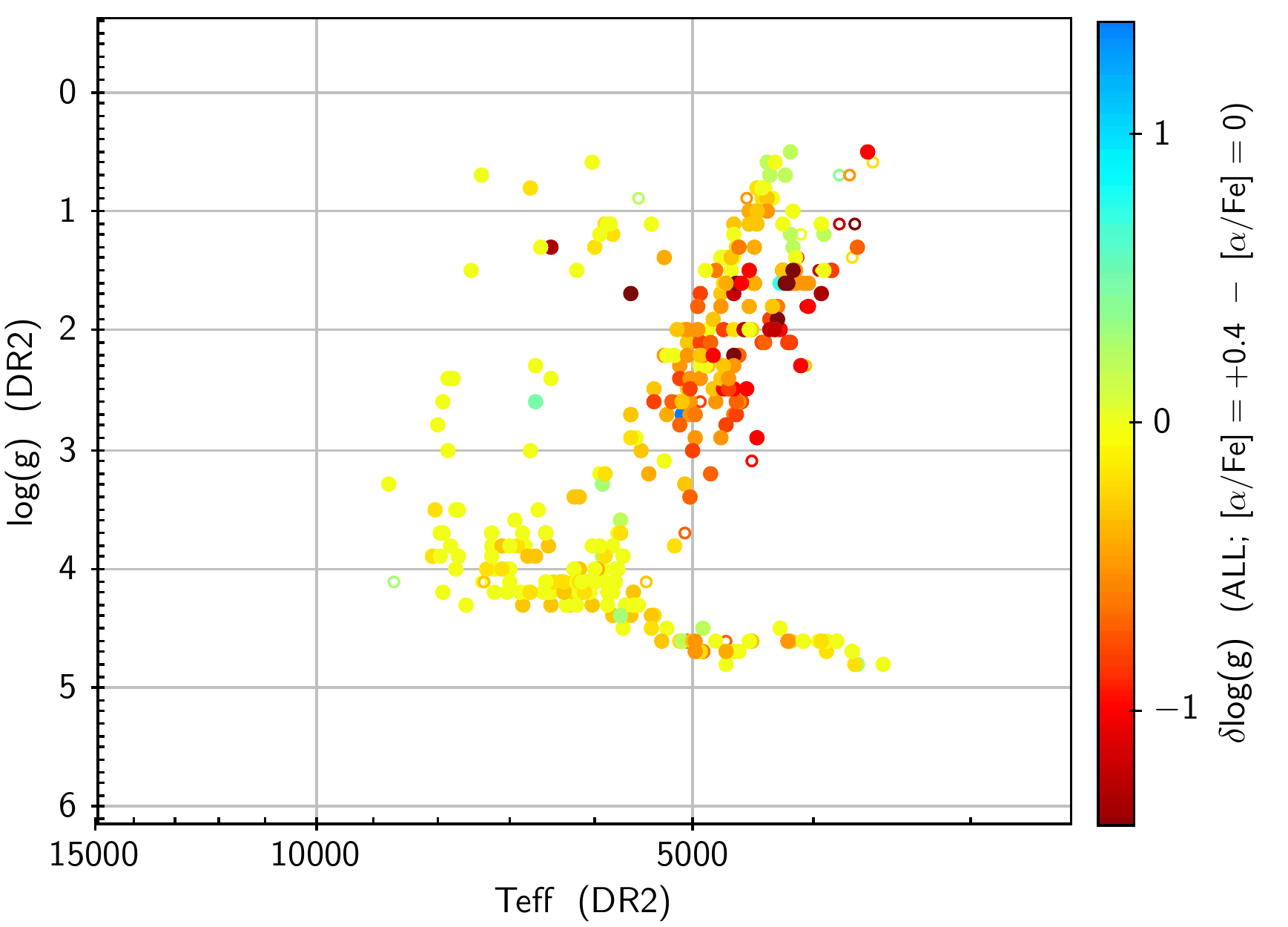}
\includegraphics[clip=,width=0.24\textwidth]{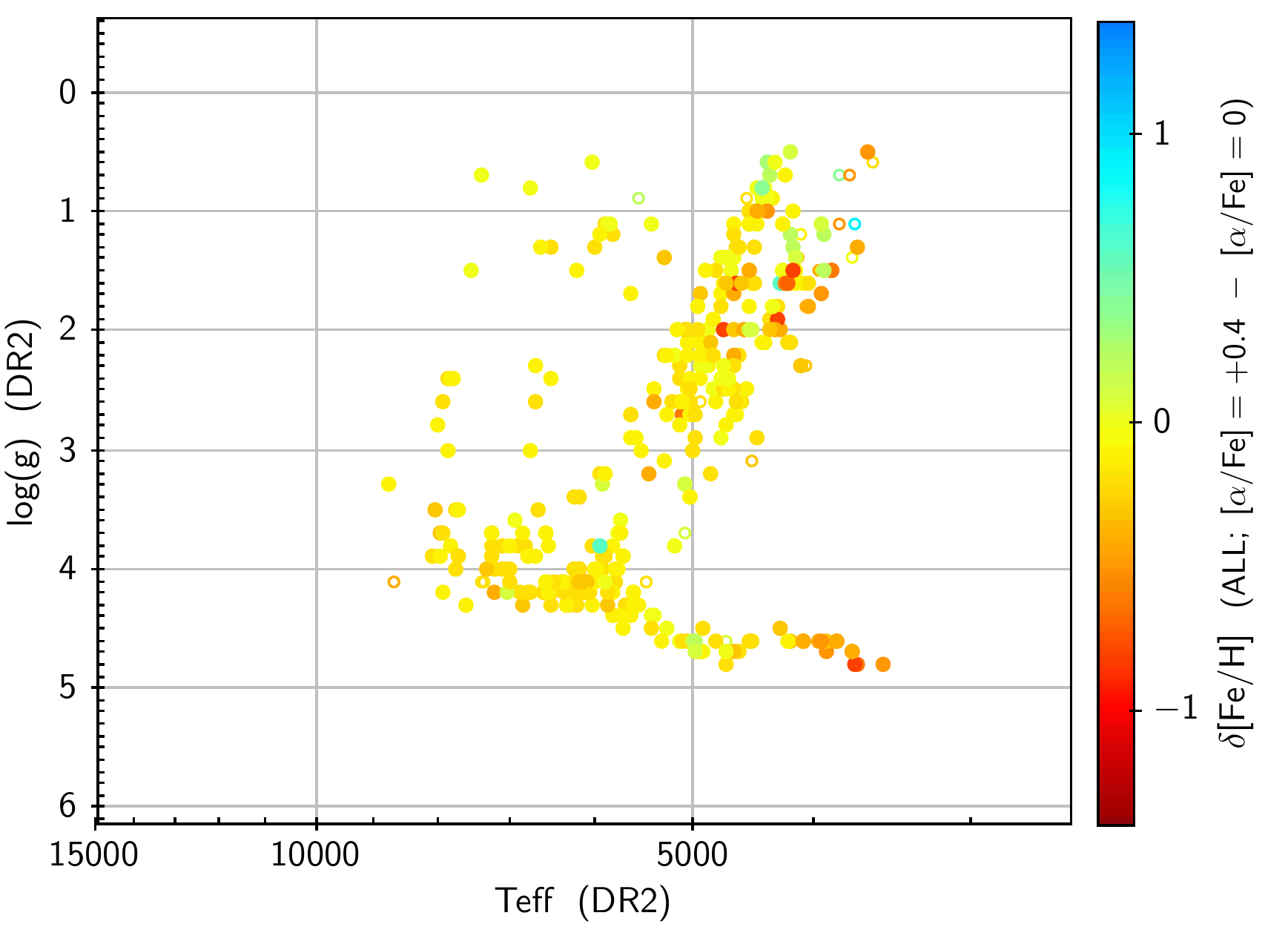}\\
\includegraphics[clip=,width=0.24\textwidth]{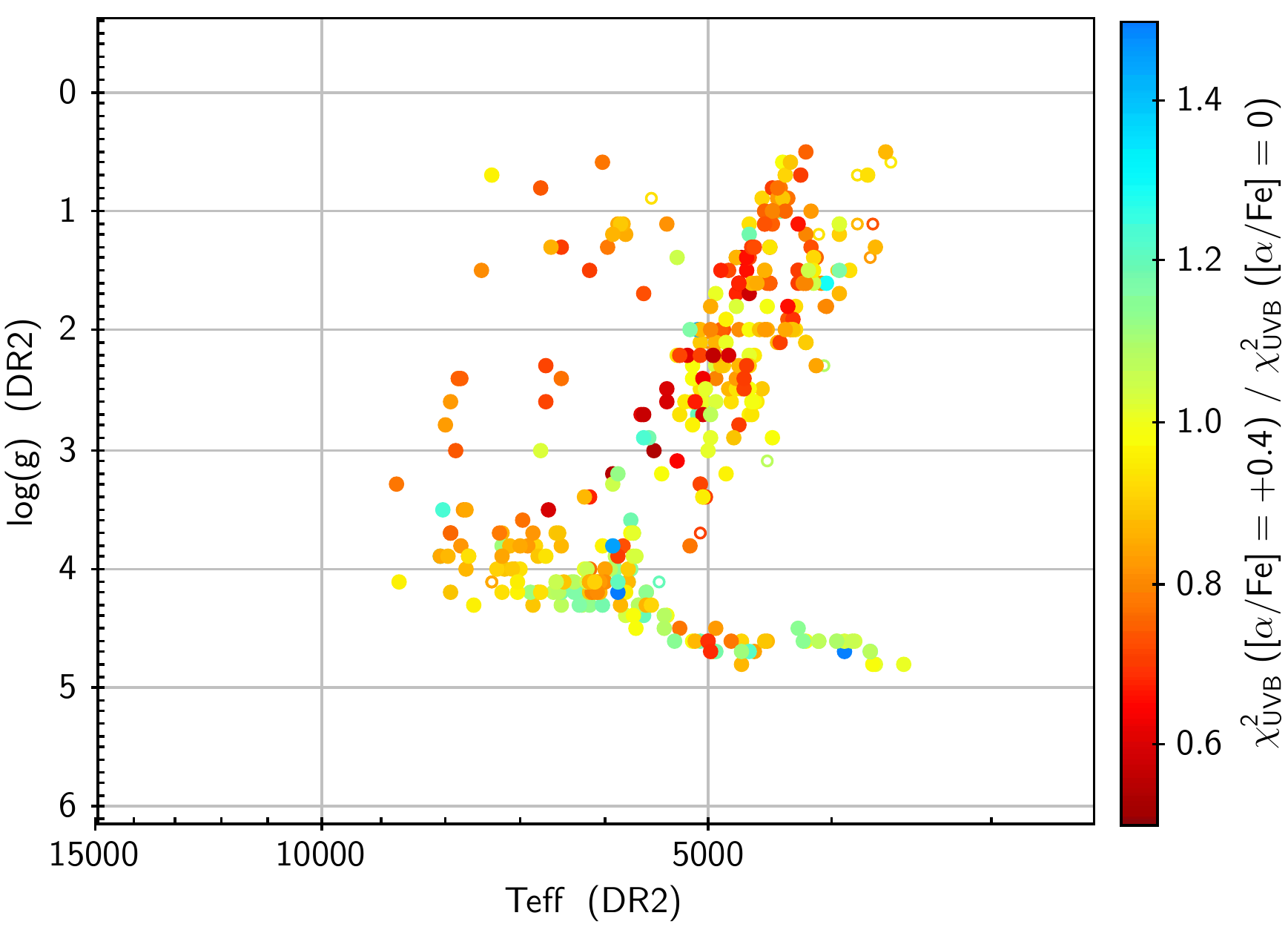}
\includegraphics[clip=,width=0.24\textwidth]{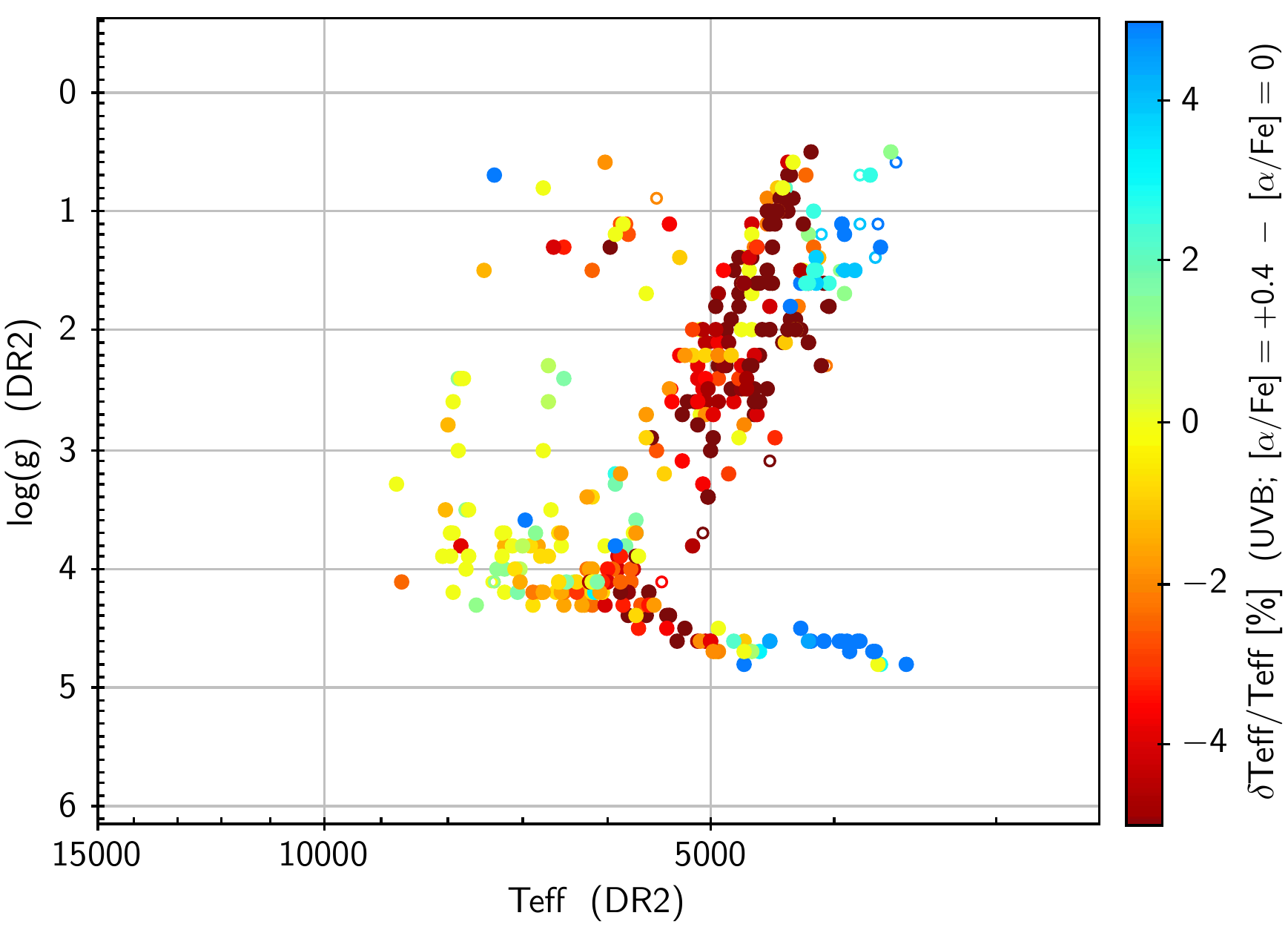}
\includegraphics[clip=,width=0.24\textwidth]{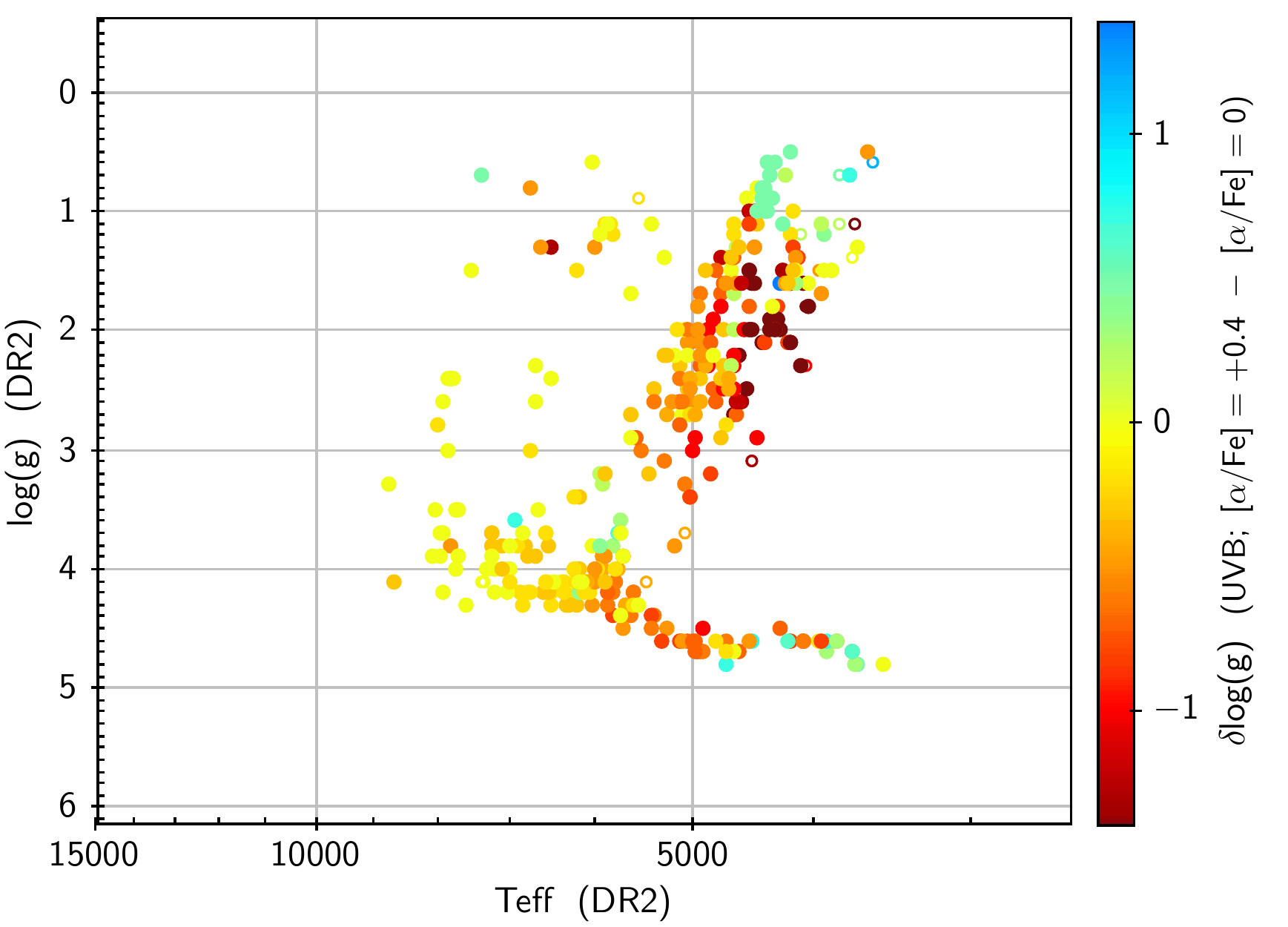}
\includegraphics[clip=,width=0.24\textwidth]{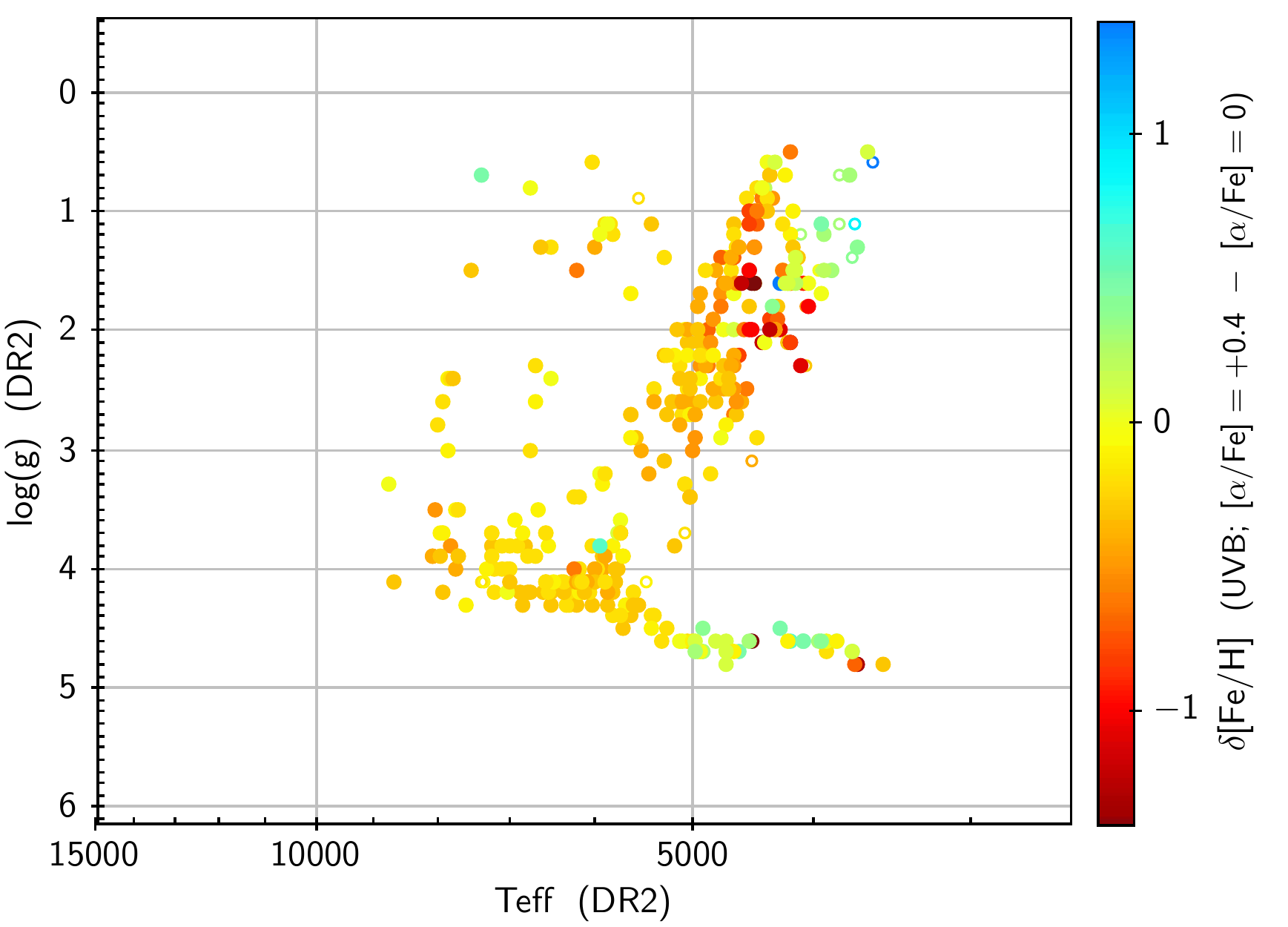}\\
\includegraphics[clip=,width=0.24\textwidth]{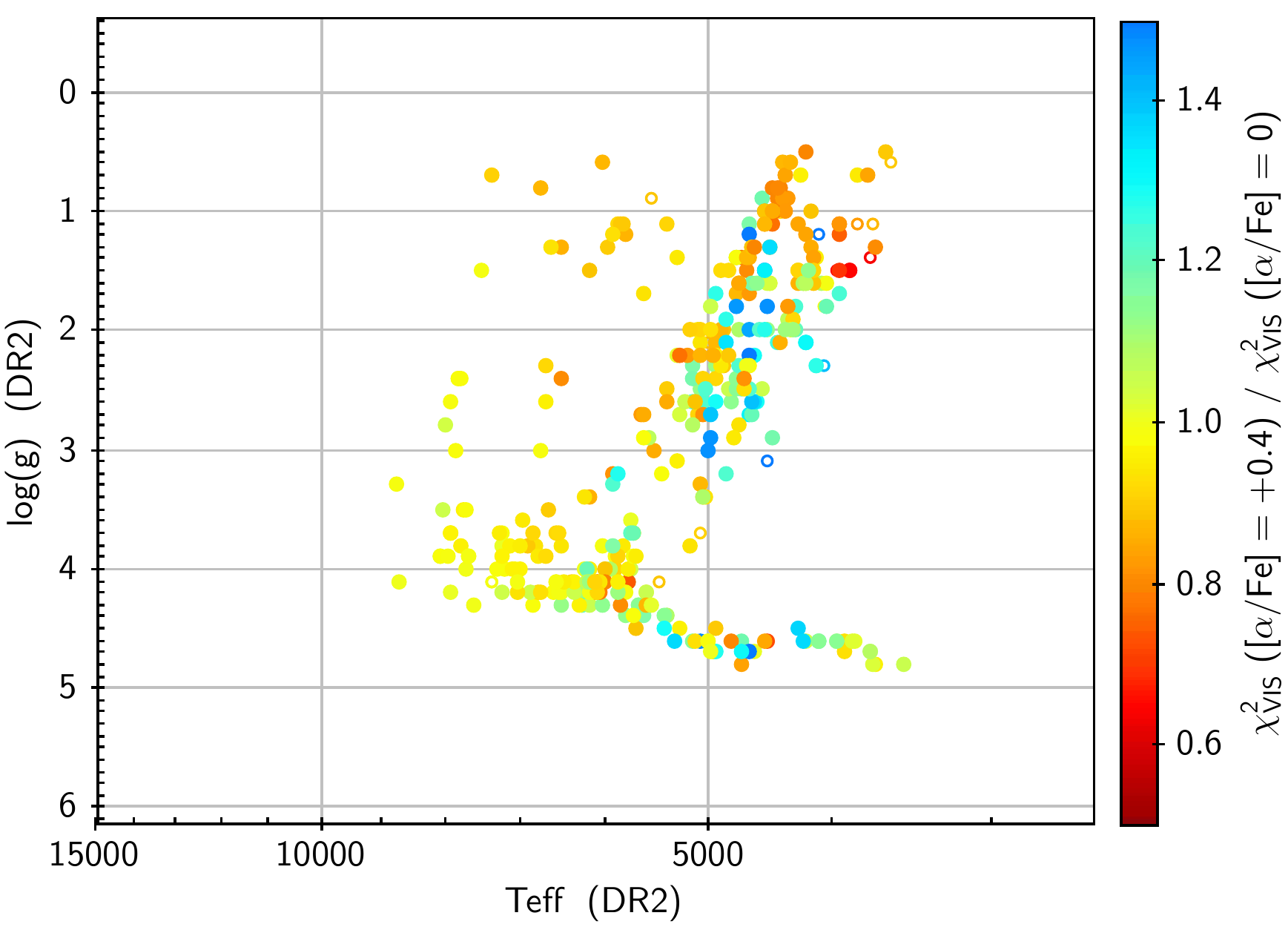}
\includegraphics[clip=,width=0.24\textwidth]{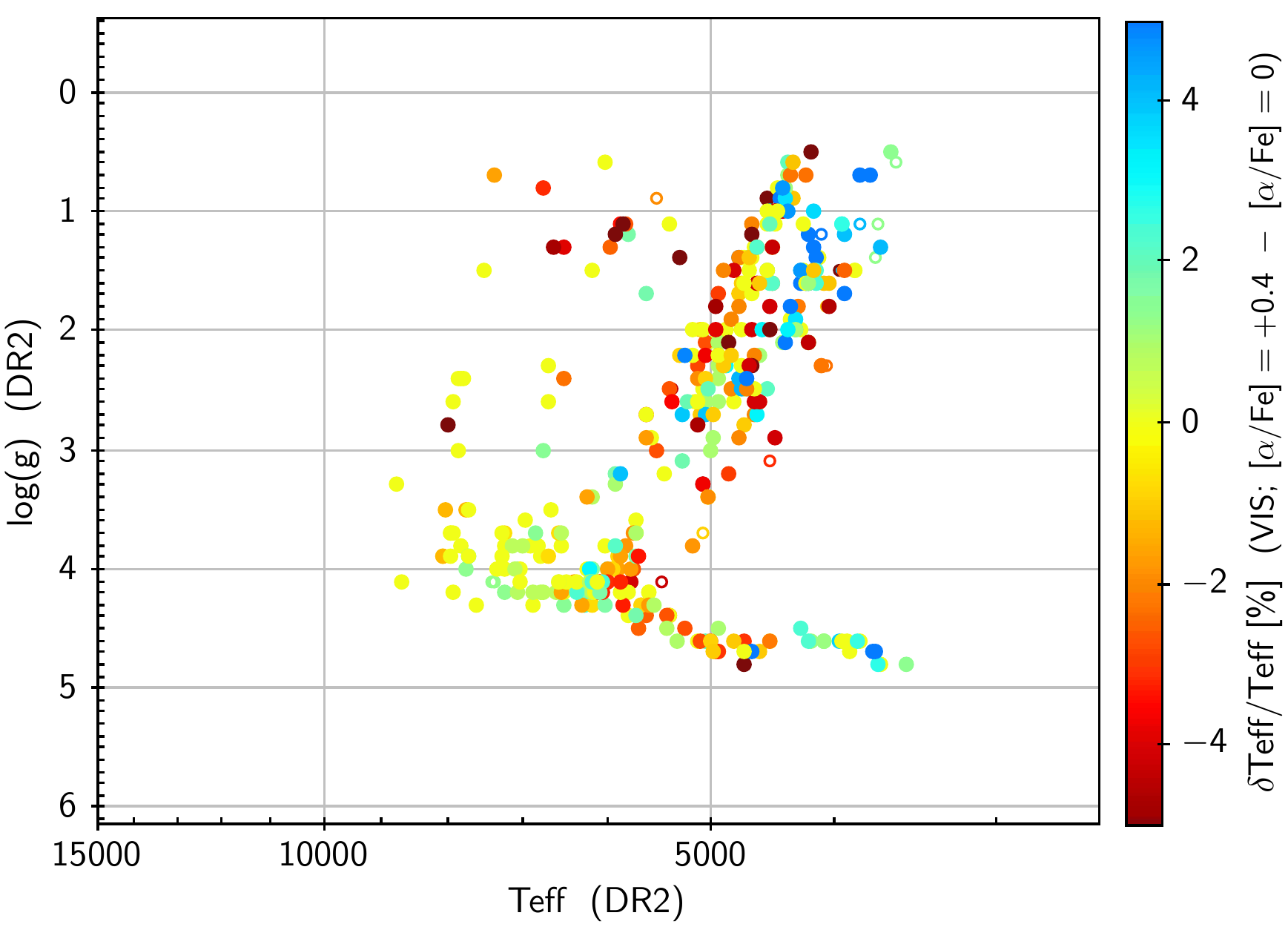}
\includegraphics[clip=,width=0.24\textwidth]{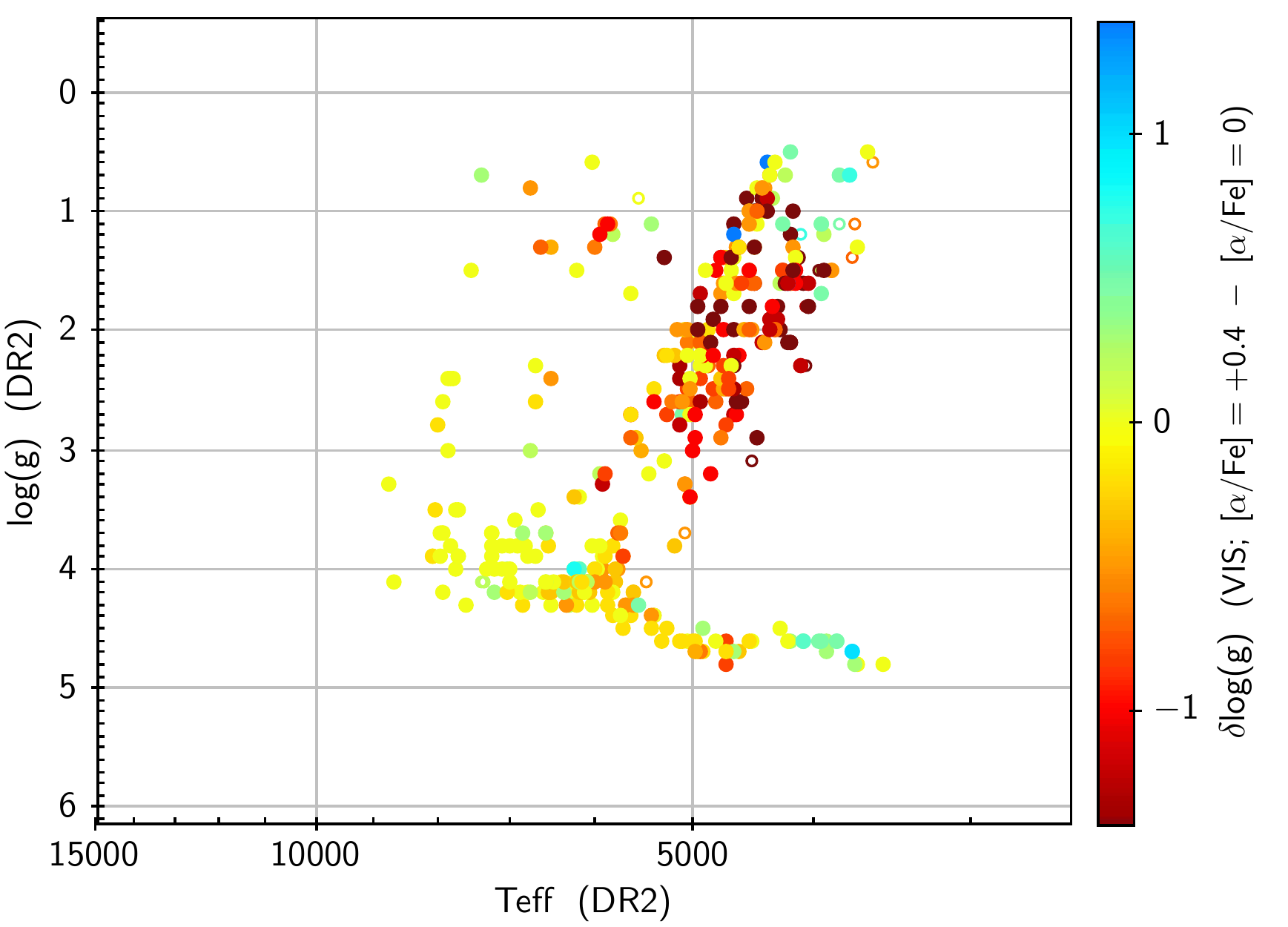}
\includegraphics[clip=,width=0.24\textwidth]{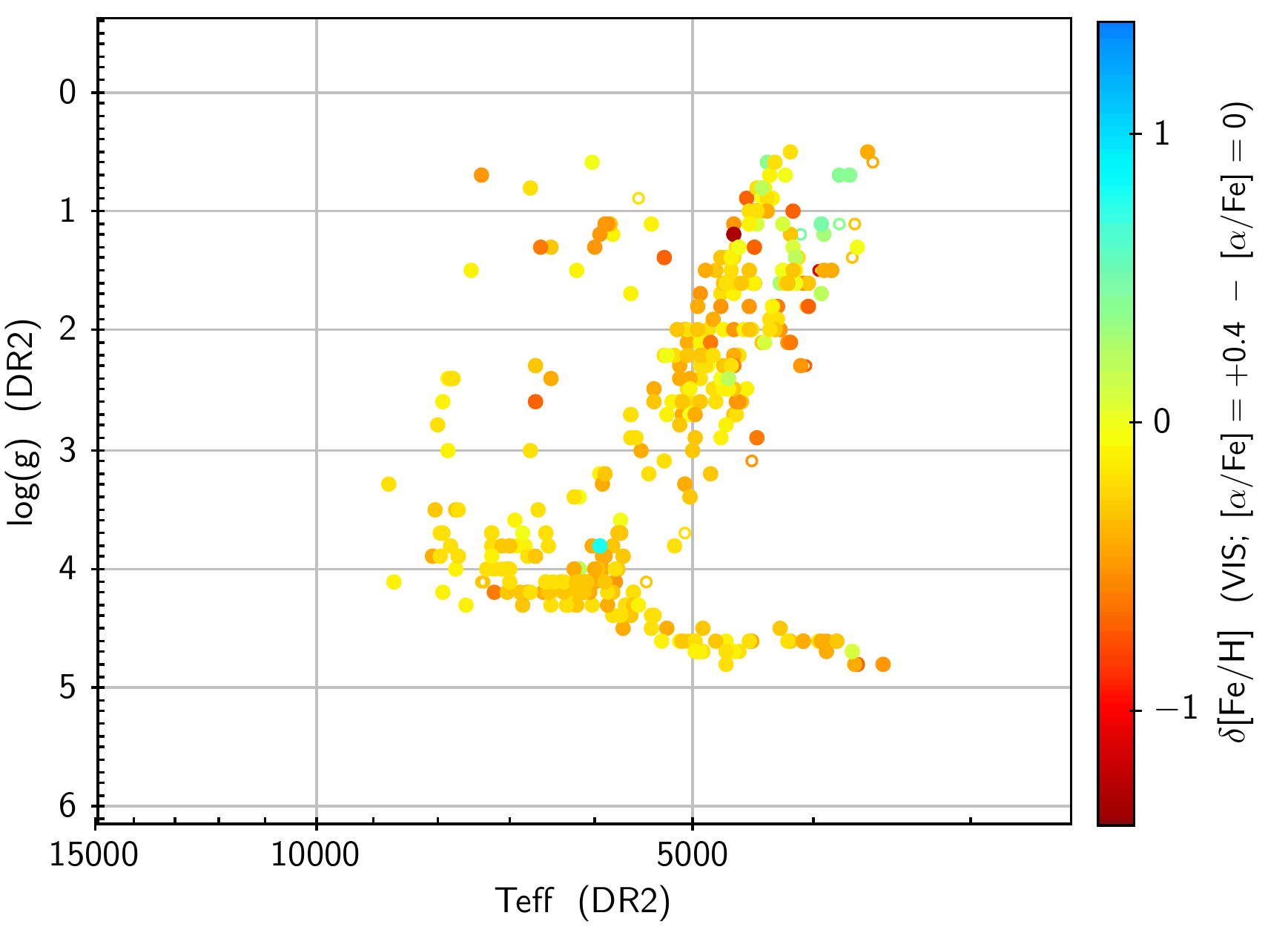}\\
\includegraphics[clip=,width=0.24\textwidth]{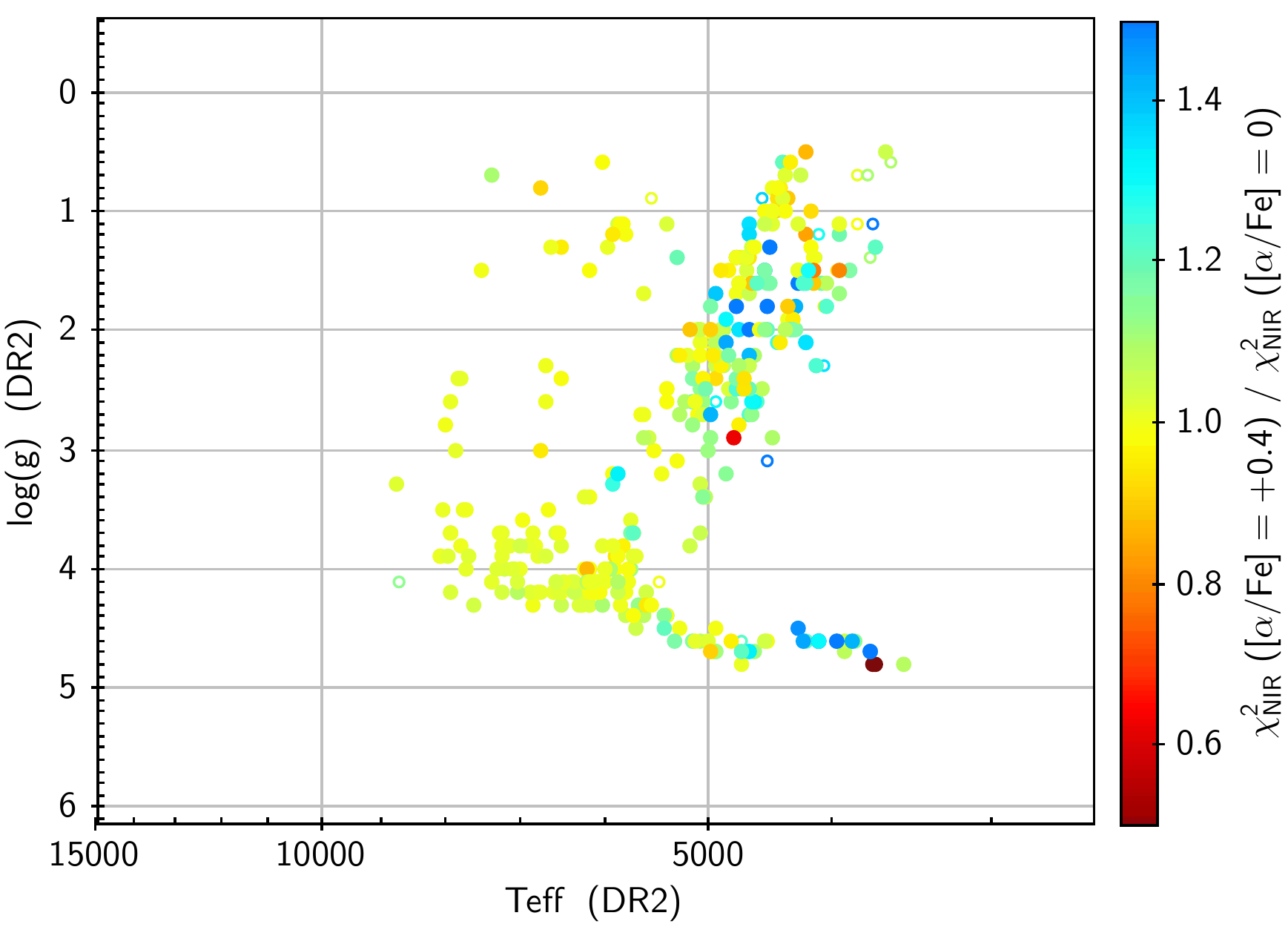}
\includegraphics[clip=,width=0.24\textwidth]{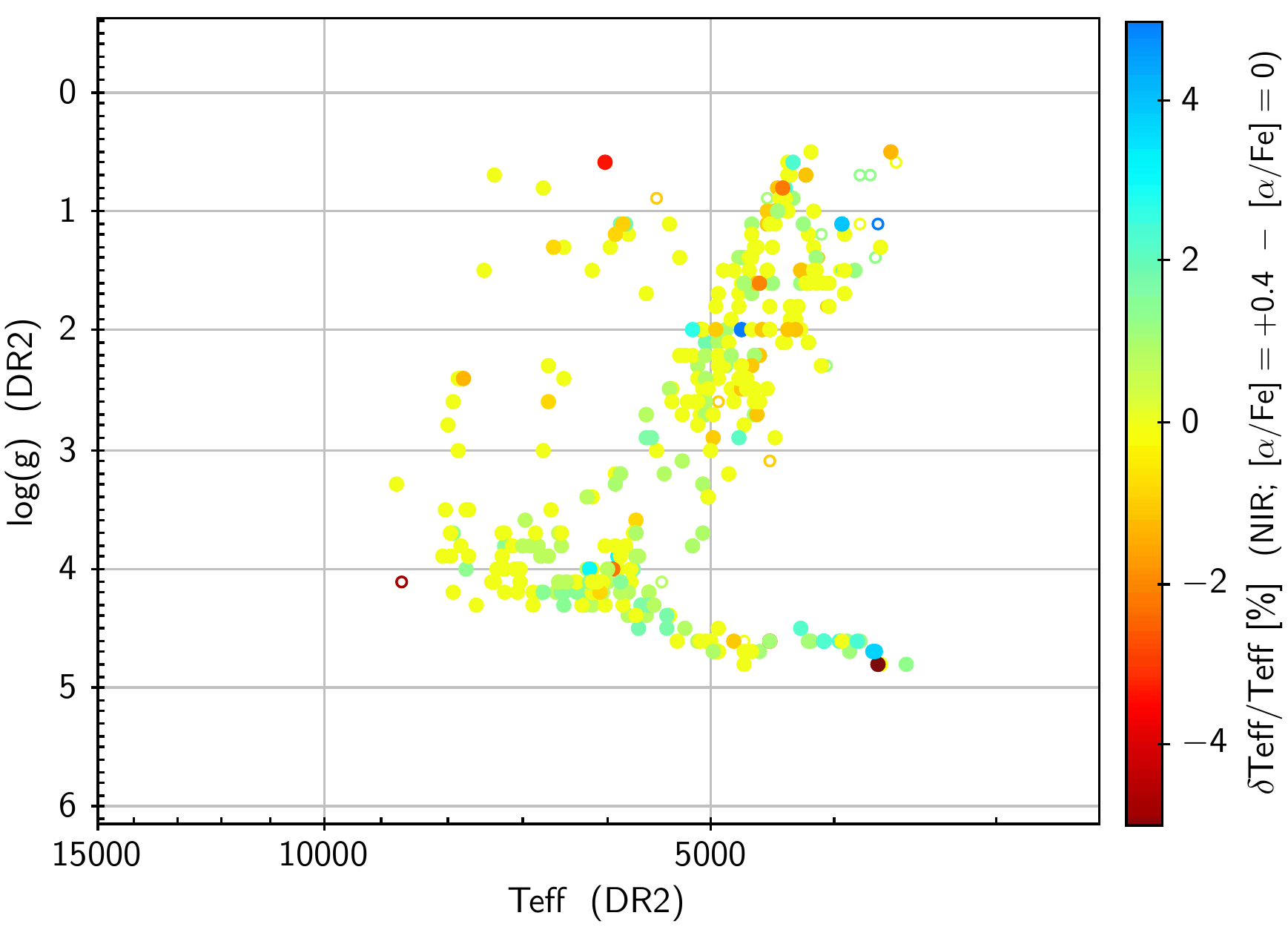}
\includegraphics[clip=,width=0.24\textwidth]{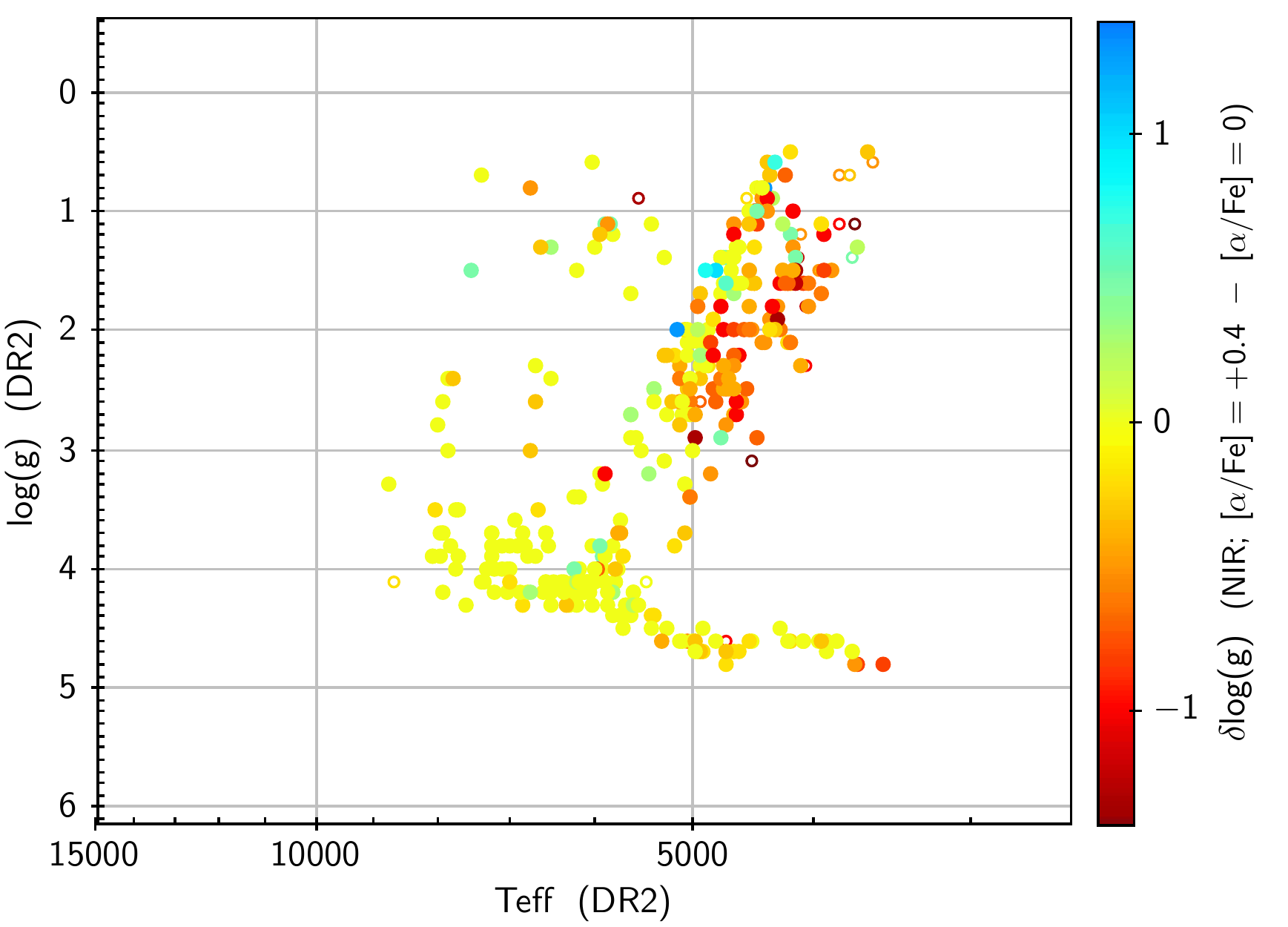}
\includegraphics[clip=,width=0.24\textwidth]{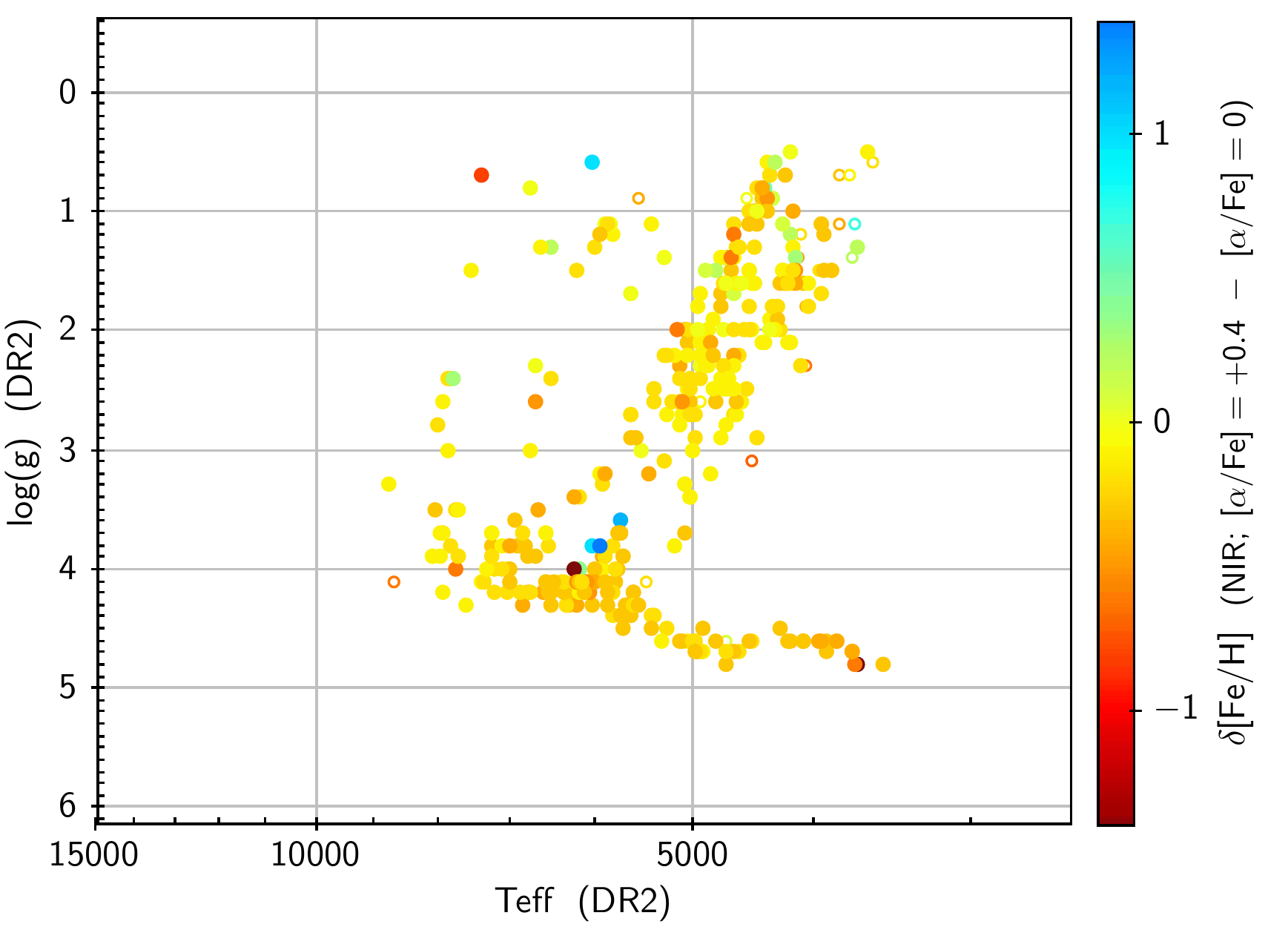}\\
\caption[]{Effects of a switch from models at [$\alpha$/Fe]=0 to models
at [$\alpha$/Fe]=$+0.4$. The model-data comparison is carried out 
at $R=3000$, with interpolation in the model grid, 
using the inverse-variance weighted $\chi^2$.
The restricted sample shown in these figures corresponds to DR2-parameters
in the range for which models are available at [$\alpha$/Fe]=$+0.4$.}
\label{fig:effect_aFe_x16}
\end{figure*}

Figure\,\ref{fig:effect_aFe_x16} summarizes the effect of a change
in the assumed [$\alpha$/Fe] on the best-fit parameters from
the comparison of the XSL spectra, at $R=3000$, with the
PHOENIX models from \citet{Husser_etal2013}, using the
inverse-variance weighted $\chi^2$ over different wavelength
ranges. Two values of [$\alpha$/Fe] are considered: 0 and +0.4. 

\begin{figure}
\includegraphics[clip=,width=0.49\textwidth]{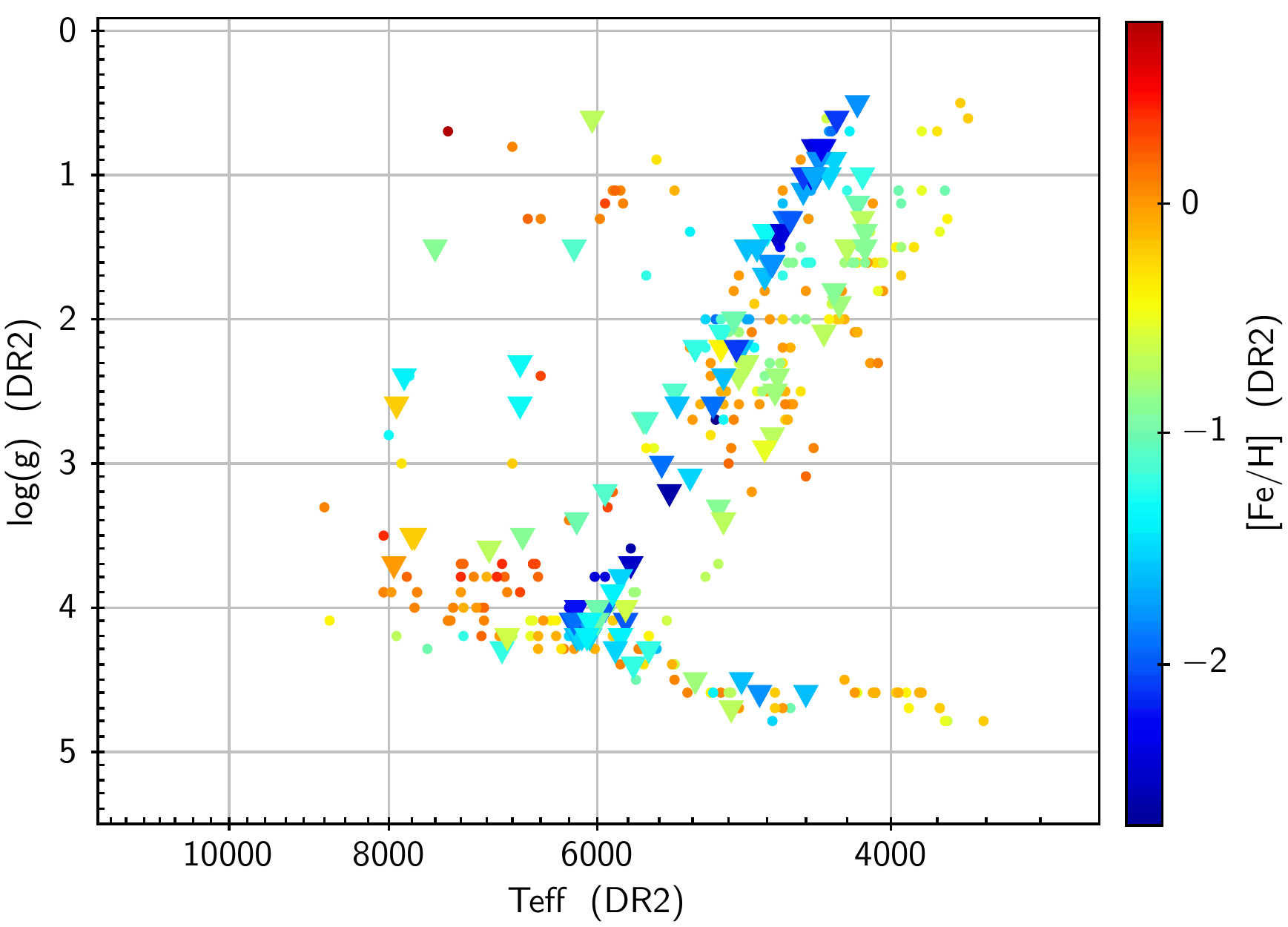}
\caption[]{Large triangles: stars classified in this paper 
as favoring models with [$\alpha$/Fe]=$+0.4$ over solar-scaled models, 
in  model-data comparisons based on the inverse-variance $\chi^2$
in which parameters are free. They have the following XSL identifiers:
\footnotesize 
X0007, X0023, X0024, X0065, X0070, X0072,
X0075, X0077, X0089, X0092, X0122, X0123,
X0137, X0140, X0143, X0151, X0157, X0171,
X0176, X0198, X0200, X0206, X0213, X0225,
X0232, X0238, X0282, X0286, X0290, X0350,
X0352, X0359, X0361, X0366, X0380, X0388,
X0389, X0392, X0394, X0408, X0415, X0417,
X0418, X0419, X0432, X0442, X0445, X0446,
X0449, X0464, X0477, X0480, X0490, X0496,
X0500, X0521, X0522, X0536, X0547, X0548,
X0549, X0566, X0598, X0600, X0601, X0604,
X0611, X0613, X0618, X0619, X0621, X0623,
X0627, X0630, X0634, X0640, X0643, X0670,
X0678, X0681, X0682, X0687, X0700, X0702,
X0728, X0733, X0734, X0757, X0770, X0775,
X0827, X0843, X0848, X0856, X0857, X0863,
X0866, X0867, X0870, X0878, X0882, X0884,
X0886, X0893, X0897, X0899, X0900, X0906
\normalsize 
}
\label{fig:HRD_preferred_aFe}
\end{figure}

The leftmost column shows the ratio $\chi^2 (0)/\chi^2(+0.4)$. The fit
to the whole energy distribution and the fit to the NIR arm are 
not very sensitive to [$\alpha$/Fe], but if anything would mildly favor
the solar ratio (see Table~\ref{tab:pref_aFe_bestfit}). 
The fit to the UVB on the other hand favors 
a super-solar [$\alpha$/Fe] everywhere except on the lower main sequence,
and a comparison with the metallicity distribution in 
Fig.\,\ref{fig:HRdiag0_byFeH0} shows this to be essentially independent
of metallicity. The anticorrelation between [$\alpha$/Fe] and
[Fe/H] characteristic of the Milky Way is seen at a significant
level only in the fit to VIS wavelengths. In the main text of this
paper, the criterion adopted for saying that a spectrum ``prefers
[$\alpha$/Fe]=$+0.4$" is that it reduces the $\chi^2$ in the UVB
without degrading the $\chi^2$ in the VIS or NIR ranges, 
which in effect confers the VIS an important role. The
resulting subset is shown in Fig.\,\ref{fig:HRD_preferred_aFe}.

\begin{table}
\caption[]{Fractions of spectra that prefer [$\alpha$/Fe]=+0.4,
for metal-poor and metal-rich stars. Fits are done at $R=3000$,
using the inverse-variance weighted $\chi^2$.}
\label{tab:pref_aFe_bestfit}
\begin{tabular}{ccc} \hline \hline 
Wavelength &  &  \\
range & [Fe/H]$<-0.5$ & [Fe/H]$\geqslant -0.5$ \\ \hline
UVB & 90\% & 70\% \\ 
VIS & 90\% & 46\% \\
NIR & 50\% & 35\% \\
ALL & 35\% & 42\% \\ \hline
\end{tabular}
\end{table}

The change from solar to super-solar [$\alpha$/Fe] leads, in most areas of 
the HR diagram, to a decrease in \teff, a decrease in log($g$)
and a decrease in metallicity for a given XSL spectrum. 
The decrease in [Fe/H] logically compensates
the increase in total metallicity [M/H] in the $\alpha$-enhanced
models (the $\alpha$-elements being added at a given [Fe/H]). 
The systematic changes in \teff and log($g$) occur in addition
to that partial compensation. 
The trends are, in general, seen whether one fits the UVB, the 
VIS or the NIR wavelength ranges, with only few exceptions.
The trend towards lower \teff\ at higher [$\alpha$/Fe] is
not seen in the NIR range, although that wavelength range by 
itself recovers the changes in log($g$) or [Fe/H] with [$\alpha$/Fe], 
found in the other spectral ranges.

\end{appendix}

\end{document}